\documentclass{article}
\usepackage[a4paper, 
        left=1in,
        right=1in,
        top=1in,
        bottom=1in,
]{geometry}
\usepackage{graphics}
\usepackage{amsmath}
\usepackage{stmaryrd}
\usepackage{bm}
\allowdisplaybreaks
\usepackage{fancyhdr}
\usepackage{amssymb}
\usepackage{latexsym}
\usepackage{color}
\usepackage{graphicx}
\usepackage{hyperref}
\hypersetup{
colorlinks = true,
citecolor  = blue,
linkcolor = blue
}
\usepackage{epsfig}
\usepackage{setspace}
\usepackage{graphicx}
\usepackage{epstopdf}
\usepackage{subfigure}
\usepackage{float}
\usepackage[titletoc,toc,title]{appendix}
\usepackage[table]{xcolor}
\usepackage{changes}
\definechangesauthor{red} 
\colorlet{Changes@Color}{red}
\usepackage{authblk}
\usepackage{citesort}
\usepackage{cite}

\begin{document}
\title{A tutorial on the stability and bifurcation analysis of the electromechanical behaviour of soft materials}
\author{Shengyou Yang$^1$ and Pradeep Sharma$^{2,}$$^3$$^*$}
\date{%
{\small \it   $^1$Department of Engineering Mechanics, School of Civil Engineering, Shandong University, Jinan, 250061, China\\%
$^2$Department of Mechanical Engineering, University of Houston, Houston, TX 77204, USA\\%
    $^3$Department of Physics, University of Houston, Houston, TX 77204, USA\\
    $^*$E-mail: psharma@uh.edu; Fax: +1-713-743-4503; Tel: +1-713-743-4502
}}

    \maketitle

\tableofcontents
    
\begin{abstract}    
Soft materials, such as liquids, polymers, foams, gels, colloids, granular materials, and most soft biological materials, play an important role in our daily lives. From a mechanical viewpoint, soft materials can easily achieve large deformations due to their low elastic moduli; meanwhile, surface instabilities, including wrinkles, creases, folds, and ridges, among others, are often observed. In particular, soft dielectrics subject to electrical stimuli can achieve significantly large deformations that are often accompanied by instabilities. While instabilities are often thought to cause failures in the engineering context and carry a negative connotation, they can also be harnessed for various applications such as surface patterning, giant actuation strain, and energy harvesting. In the biological world, instability and bifurcation phenomena often precede important events such as endocytosis, cell fusion, among others. Stability and bifurcation analysis (especially for soft materials) is challenging and often presents a formidable barrier to entry in this important field. A multidisciplinary audience may lack the background in one or more areas that are needed to carry out the requisite modeling or even understand papers in the literature. Furthermore, combining electrostatics together with large deformations brings its own challenges. In this article, we provide a tutorial on the basics of stability and bifurcation analysis in the context of soft electromechanical materials. The aim of the article is to use simple examples and ``gently" lead a reader, unfamiliar with either stability analysis or electrostatics of deformable media, to develop the ability to understand the pertinent literature that already exists and position them to embark on state-of-the-art research on this topic. 
\end{abstract}

\section{Introduction}%
\begin{figure}[t] %
    \centering
\includegraphics[width=0.6\textwidth]{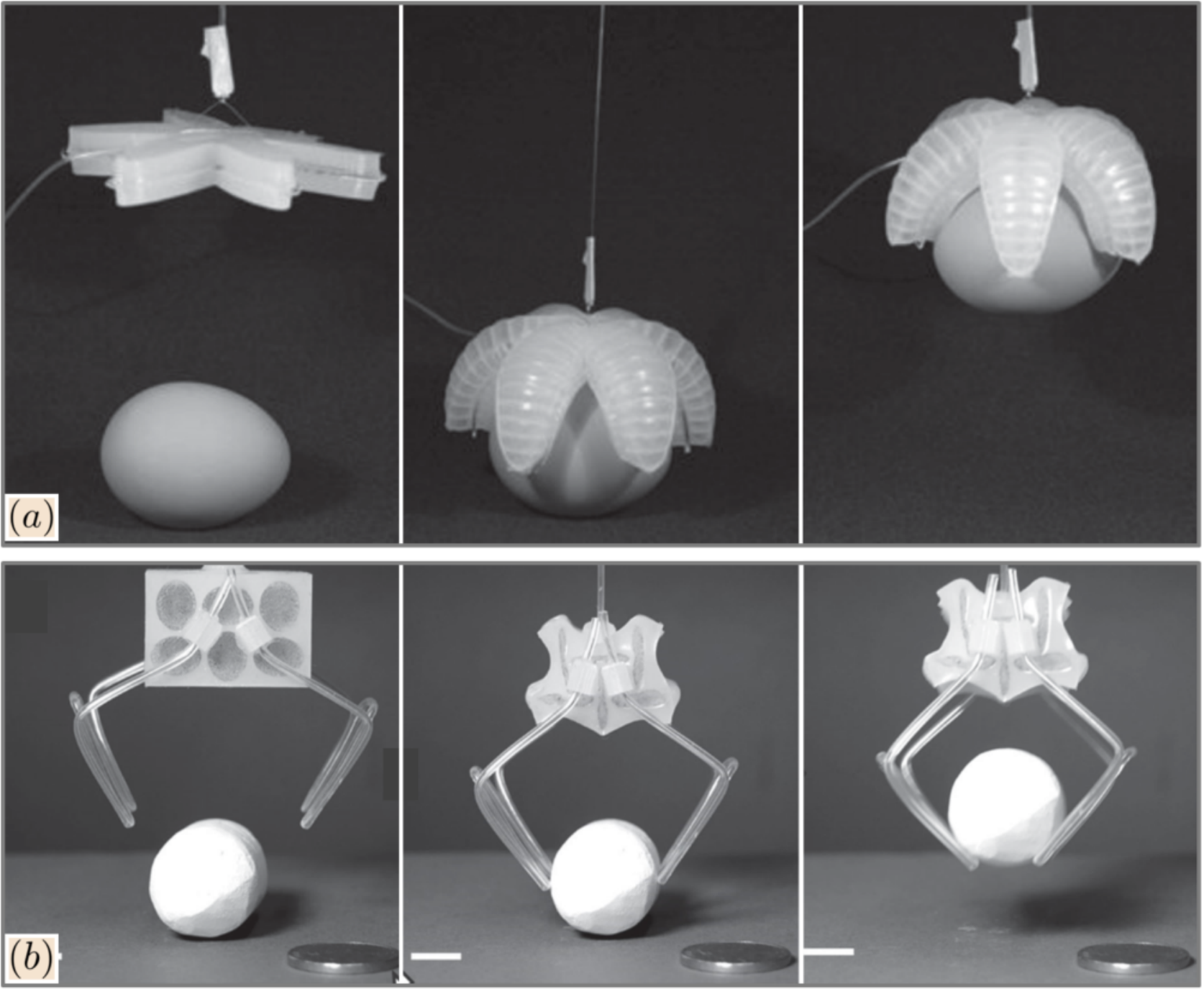}
    \caption {The utility of nonlinear behavior of soft materials. (a) A 9 cm tip-to-tip PneuNet actuator gripping a raw chicken egg.The actuator legs curled downward when the pressure is applied to the PneuNets in the top active layer, and the soft gripper did not damage the egg (Reproduced with permission from F. Ilievski et al.\cite{ilievski2011soft}. Copyright 2011 by John Wiley and Sons). (b) A soft gripper made of a buckling actuator that can pick up a piece of chalk (Reproduced with permission from D. Yang et al.\cite{yang2015buckling}. Copyright 2015 by John  Wiley and Sons).}
    \label{intro-1}
\end{figure}

\begin{figure}[h] %
    \centering
    \includegraphics[width=0.6\textwidth]{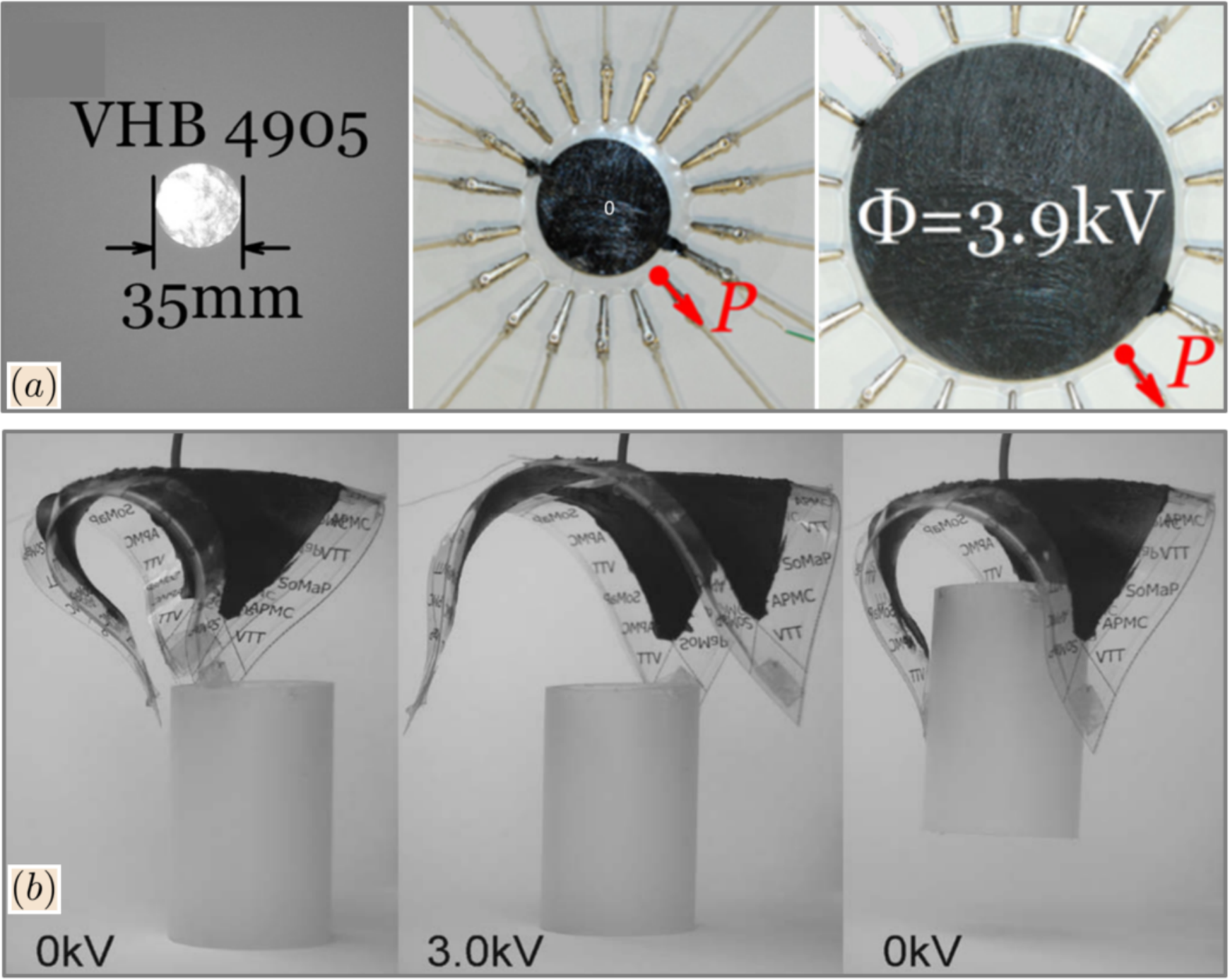}
    \caption{(a) Large deformation of a circular disc of VHB induced by an applied voltage on the upper and bottom surfaces coated with electrodes of carbon grease (Reproduced with permission from J. Huang et al.\cite{huang2012giant}. Copyright 2011 by AIP Publishing). (b) An elastomer gripper opens at a voltage of 3.0 kV, which can grasp objects when the voltage is removed (Reproduced with permission from G. Kofod et al.\cite{kofod2007energy}. Copyright 2007 by AIP Publishing).}
    \label{intro-2}
\end{figure}

\begin{figure}[h] %
    \centering
    \includegraphics[width=0.6\textwidth]{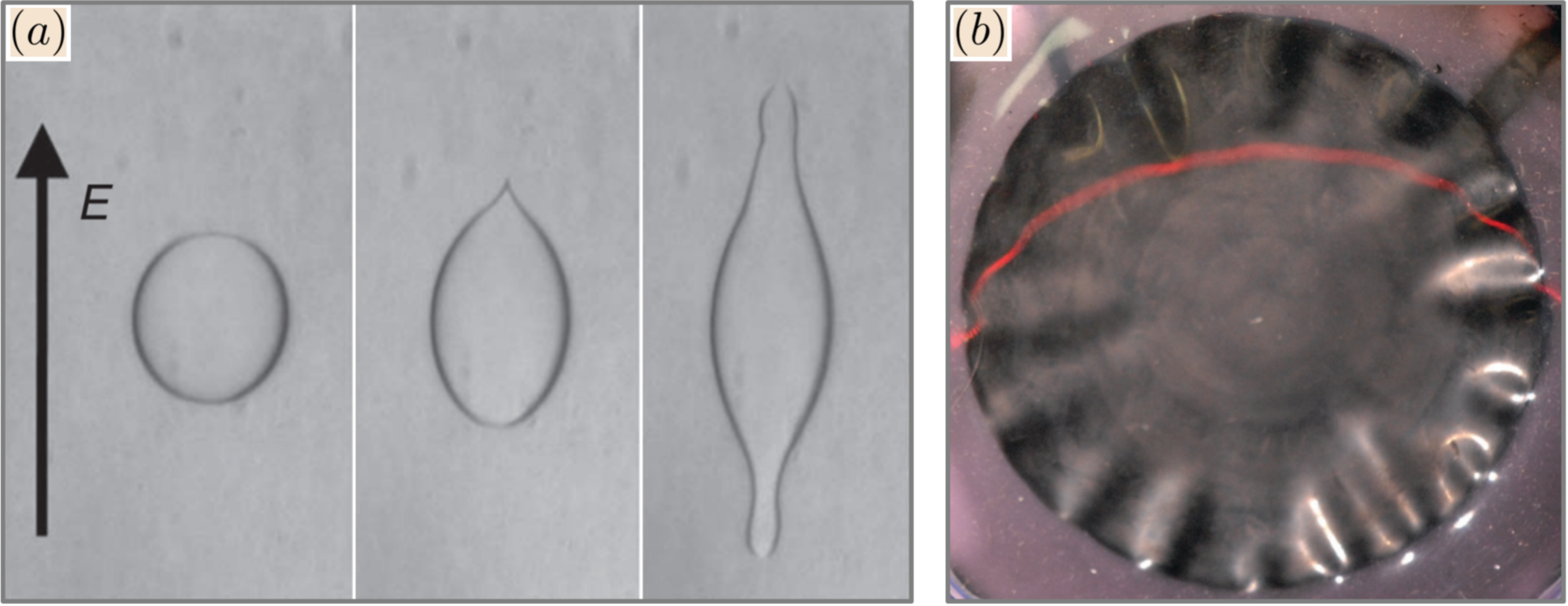}
    \caption{(a) Spherical drops of conductive liquid in solid dielectrics become unstable and burst under sufficiently high electric fields (Reproduced with permission from Q. Wang et al. \cite{wang2012bursting}. Copyright 2012 by Springer Nature). (b) Azimuthal wrinkles in dielectric elastomer sheets clamped on a circular rigid frame (Reproduced with permission from H. Bense et al.\cite{bense2017buckling}. Copyright 2017 by Royal Society of Chemistry).}
    \label{intro-3}
\end{figure}

The simple act of picking an object by a human-like robot would not be possible without the ability of the underlying material to sustain large deformation. This example underscores the motivation for studying soft materials. Beyond robotics \cite{ilievski2011soft, yang2015buckling, Rus2015Design, bauer201425th} (Fig.~\ref{intro-1}), in contexts ranging from biotechnology to electronics, soft materials are important wherever we need large deformations. Often, just the capacity to sustain large deformation in response to mechanical forces is not enough. Ideally, we hope to influence their behavior by applying an external electric field that permits numerous applications such as the already mentioned soft robotics, energy harvesting \cite{bauer201425th, pelrine2001dielectric, koh2009maximal, koh2011dielectric, invernizzi2016energy, DENG20143218}, sensors and actuators \cite{kofod2007energy, Shankar2007Dielectric, carpi2010stretching, keplinger2010rontgen, Brochu2010Advances, shao2018bioinspired, Banet2021Evaluation, Pu2022unimorph, rahmati2019nonlinear}, among others \cite{Ende2013voltage, Shian2015gripper, RevModPhys.94.025003}. Soft dielectrics, when subjected to electrical stimuli, can achieve significantly large deformations (Fig.~\ref{intro-2}) due to the highly nonlinear electromechanical coupling between mechanical and electric fields \cite{pelrine2000high, huang2012giant, zhao2010theory, dorfmann2017nonlinear, LU2020100752}. Large deformations invariably lead to the possibility of instabilities which include the phenomena of surface wrinkles \cite{biot1963surface, bowden1998spontaneous, yang2010harnessing, wang2013creasing, liu2016voltage, godaba2017dynamic, dorfmann2010nonlinear}, creases \cite{gent1999surface, trujillo2008creasing, hong2009formation, hohlfeld2011unfolding, wang2011creasing, wang2012dynamic, zurlo2017catastrophic}, folds \cite{pocivavsek2008stress, kim2011hierarchical}, electro-buckling \cite{tavakol2014buckling, tavakol2016voltage, bense2017buckling, yang2017revisiting}, pull-in instability \cite{stark1955electric, plante2006large, zhao2007method, zhao2009electromechanical, huang2012giant, yang2017avoiding, ChenLL2020nonlinear}, symmetry-breaking instability \cite{ChenLL2021interplay, Chen2022Symmetry}, bursting drops in solid dielectrics \cite{wang2012bursting}, among others \cite{dorfmann2019instabilities, CHEN2022109995}. Two types of electromechanical instabilities observed in experiments are shown in Fig.~\ref{intro-3}: morphological instability of a single drop in a solid dielectric polymer under electric fields and regular azimuthal wrinkles in a dielectric membrane.\\

While instabilities are generally often avoided in engineering applications as they are considered to be a sign of, or precursor to, ``failure", as well reviewed recently \cite{zhao2014harnessing, hu2015buckling, Pal2021Exploiting}, they may be exploited in a positive manner. For example, Fig.~\ref{intro-1}(b) demonstrates the utility of the buckling of elastomer beams under pressure to achieve a remarkably large actuation in soft machines \cite{yang2015buckling}.  In the neighborhood of threshold of instability, a small perturbation (in this case, pressure) can induce a large output---in this instance, the buckling of pillars and the result that the air chambers collapse in Fig.~\ref{intro-1}(b).  Other examples of the utility of elastic instabilities include inducing negative Poisson's ratio behavior in cellular solids \cite{bertoldi2010negative}, a ``smart'' polymer adhesive based on surface wrinkling \cite{chan2008surface}, the rapid closure of the Venus flytrap through the snap-buckling instability \cite{forterre2005venus}, the advanced functionality within shape changing materials \cite{HOLMES2019118}, the rapid actuation and giant deformation of hard-magnetic soft materials \cite{CHEN2022111607, TAN2022107523}. More examples of  buckling-induced smart applications can be found in the review \cite{hu2015buckling}.\\

A multidisciplinary audience may lack the background in one or more areas that are needed to carry out the requisite modeling of instability and bifurcations in soft electromechanical solids---or even to understand the literature in this field. For example, most books that explain large nonlinear deformation mechanics are difficult for non-mechanicians to follow. Combining electrostatics, together with large deformations, brings on its own challenges. Finally, while several good expositions \cite{strogatz2014nonlinear, seydel2009practical, guckenheimer2013nonlinear, perko2013differential, kuznetsov2013elements, sastry2013nonlinear, timoshenko1961theory, cedolin2010stability, antman2013nonlinear, ogden1997non, dorfmann2014nonlinear, dorfmann2019instabilities, golubitsky2012singularities, van2009wt} do exist on stability and bifurcation analysis, some don't necessarily take a tutorial approach to educate a multidisciplinary audience while others may not cover all the pertinent topics that are central to the theme of this paper. This tutorial article is intended to fill this gap and will gently guide researchers, using simple examples, to the basics of stability analysis and electrostatics of deformable media. Clearly, a mere article (especially intended to be a tutorial) cannot hope to be either comprehensive or have the rigor that leading experts in this field expect. From that viewpoint, our aspirations are quite modest---we expect the tutorial to simply be a convenient starting point to facilitate the reader's ability to undertake a study of the more specialized literature. \\

This paper is organized as follows. In Sec.~\ref{new-addition-section-p1}, we discuss the basics concepts of stability and bifurcation, along with some simple illustrative examples and their key pertinent features. In Sec.~\ref{section-1D-continuum}, we present one-dimensional (1D) nonlinear electroelasticity in continuum media. This section will be a useful starting point for those who wish to decouple the conceptual understanding of electrostatics of deformable media from the tensorial machinery necessary for three-dimensional (3D) continuum mechanics. In addition, the boundary-value problem and its incremental formulation are presented for the bifurcation analysis of the electromechanical behaviour of soft materials. In Sec.~\ref{section-3D-continuum}, we present the 3D theory of electroelasticity and formulate both the boundary-value problem and its incremental formulation. We take pull-in stability and electro-wrinkling as examples to illustrate the incremental method in electroelasticity. In Sec.~\ref{section-e-stability-zhao}, we present the stability and bifurcation analysis of pull-in instability of soft dielectrics by using the energy formulation presented by Zhao and Suo \cite{zhao2007method}. We use these examples to highlight the difference between stability and bifurcation. In Sec.~\ref{section-e-stability-yang}, we use an alternative energy formulation to study pull-in instability of a dielectric film and a disc. The literature is rife with ``different" formulations of electroelasticity. While equivalent in some sense, these theories can often appear to be quite different depending on the independent variable used or the variational principle invoked. Liu \cite{liu2013energy, liu2014energy} recently provided a detailed overview of the various approaches to formulate electro-magneto-elasticity and taking a cue from him, we have highlighted this aspect as well by presenting illustrative examples that use different formulations. In Sec.~\ref{section-cavitation}, we follow the energy formulation by Liu and present a study of electromechanical cavitation and the snap-through instability of a hollow sphere in a dielectric elastomer. In Sec.~\ref{LSK-section}, we briefly introduce the ``Lyapunov-Schmidt-Koiter" (LSK) approach for the initial post-bifurcation analysis. We finally conclude in Sec.~\ref{section-conclusion} by highlighting topics not covered, or inadequately covered, in the tutorial, which the reader may consider for further study.

\section{Stability and bifurcation: the big picture and basic concepts} \label{new-addition-section-p1}

In this section, we introduce the key elements pertaining to stability and bifurcation analysis. This subject is both broad and deep, and its nuances have been extensively covered from both a purely mathematical viewpoint (its natural home) as well as in terms of specific discipline-specific applications that range from biology, economics, physics, engineering, and others. We will, rather quickly, specialize all our discussion primarily to nonlinear physical systems that are \emph{conservative} (as opposed to \emph{dissipative}). This permits the use of powerful and well-established approaches like the calculus of variations to study the stability. In the concluding section, where we discuss topics not covered in this article, we point to the literature for stability and bifurcation analysis for the non-conservative case. This section is strongly influenced by the following books: \cite{strogatz2014nonlinear, guckenheimer2013nonlinear, seydel2009practical, perko2013differential, kuznetsov2013elements, sastry2013nonlinear}.

\subsection{The big picture} \label{new-addition-section-p1-sub1}

The vast majority of the differential equations that purport to describe nature (and often human made problems) are nonlinear. In contrast, 99 $\%$ of the engineering curriculum (and most of the graduate curriculum) is dedicated to studying linear systems. We would not be able to hear music without nonlinear effects in our ears; there will be no turbulence in fluids and plasticity in solids without nonlinearities; self-assembly of nano structures, phase transformations, superconductivity, laser device operation and neuron interaction are but just a few examples of the manifestation of nonlinear phenomena. There are good reasons why our curriculum is focused on linear systems. They are important to develop intuition (---which can, however, got out of the window once we discuss some nonlinear systems). Moreover, we can usually solve linear systems---even if not in closed-form, fairly easily using numerical methods. However, what if we could understand the \emph{nature} of the nonlinear solutions without actually having to solve those nasty equations? What information can we extract from them? This is the subject of stability analysis and is a fascinating field of research in engineering and physics. For example, some nonlinear systems will lead to bifurcations, i.e., the presence of multiple solutions. If we are aware of the nature of the bifurcations, we can often get interesting insights prior to the hard work of numerically solving the equations \emph{and} prevent ourselves from making mistakes in the computational solution of those problems. Before we go further into the technical content, it is worthwhile to examine some of the statements we have just made in more depth. \\

Consider a simple example of a nonlinear system---motion of a pendulum. It's governing equation is $\ddot{y}+\frac{g}{L} \sin y=0$. This equation cannot be solved analytically, at least not in terms of elementary functions. In undergraduate courses, the assumption of $\sin y \sim y$ for $|y(t)| \ll 1$ is usually made, thus linearizing the equation. We can certainly be more sophisticated and use one of the perturbation methods (to obtain an approximate analytical solution) and, therefore, perhaps even extend the range of the solution to $y (t)$ that is not too small. Is there an easier route, however, to extract the basic behavior without actually solving the equations? This system is so simple that even our physical intuition appears to be able to do better than the linear solution. For example, at low energy, the pendulum will simply swing around some location. If high enough energy is imparted, the pendulum will simply whirl over the top. It should be obvious that the linearized or perturbation solution cannot handle the ``whirling over the top" part of the problem. \\

\begin{figure}[h] %
    \centering
    \includegraphics[width=4in]{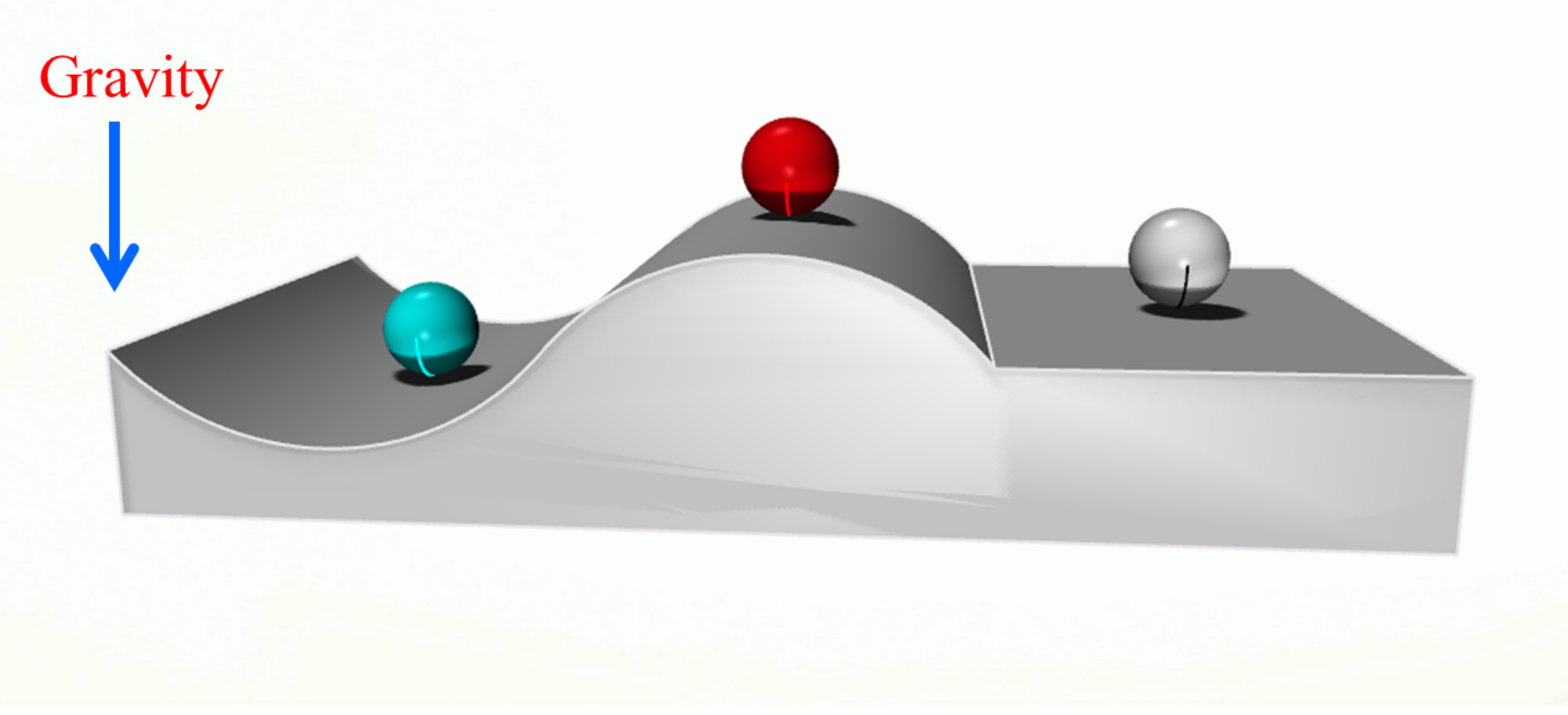}
    \caption{Schematic of the nature of stability of three rigid balls under the influence of gravity: stable, unstable, and neutral (from left to right).}
    \label{stability-1}
\end{figure}

Nonlinear equations (and hence by extension the physical phenomena they represent) often have multiple solutions and so-called bifurcations. One question we may ask is when should we expect multiple solutions and when are they ``stable"? We have not yet defined stability, but here is a simple example that gives a nice intuitive idea---precise definition will be given later on. Consider rigid balls shown in Fig.~\ref{stability-1} under the influence of gravity. The first ball (starting from left) is somewhat ``stable". By the latter, we intuitively mean that under small vibration or perturbations, the ball is likely to get a bit disturbed, but it will stay put in its position. This conforms to our intuition that the ball on the left is in a stable state. In contrast, the middle ball may be considered to be unstable because any small perturbation is likely to move it from its original position.\\

\begin{figure}[h] 
\centering
\includegraphics[width=4in]{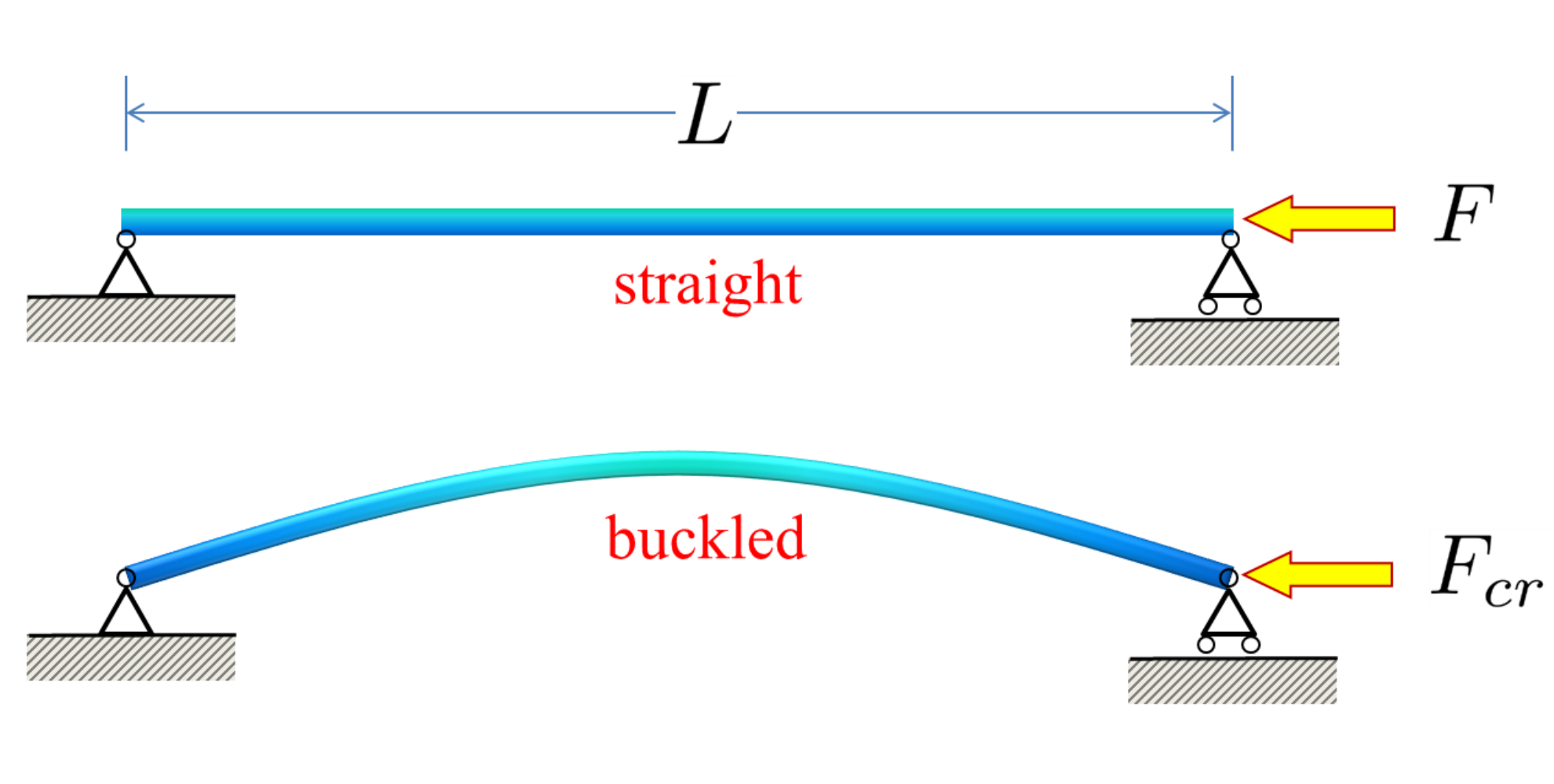}
\caption{Schematic of the buckling of Euler's column with pinned-pinned ends.}
\label{Addition-Stability-1}
\end{figure}

Let us elaborate further on multiple solutions and bifurcations. The latter refers to a qualitative change in the behavior of a system as some parameter changes. For example, consider buckling of a column in Fig.~\ref{Addition-Stability-1} under compression. As long as the ``parameter" (in this case, axial force $F$) is below a certain value $F_{cr}$, the column deforms axially, i.e., its length shortens. In order to find the shortening of this length, we have to solve the partial differential equations of nonlinear elasticity. Above a certain value $F_{cr}$ of the load, the equations of elasticity no longer admit a unique solution and allow for multiple possibilities. As we know from our everyday experience, beyond $F_{cr}$, the column will ``buckle" and adopt either of the two equilibrium states---up or down. The critical load at which the beam buckles is the bifurcation point. The key point is that multiple solutions are quite common in nonlinear systems; however, stability and bifurcation analysis gives us information about how stable those solutions are and under what conditions. Furthermore, bifurcation is related to how both the number and nature of the stable solutions may change as some parameter is varied. \\

\begin{figure}[h] 
\centering
\includegraphics[width=3in]{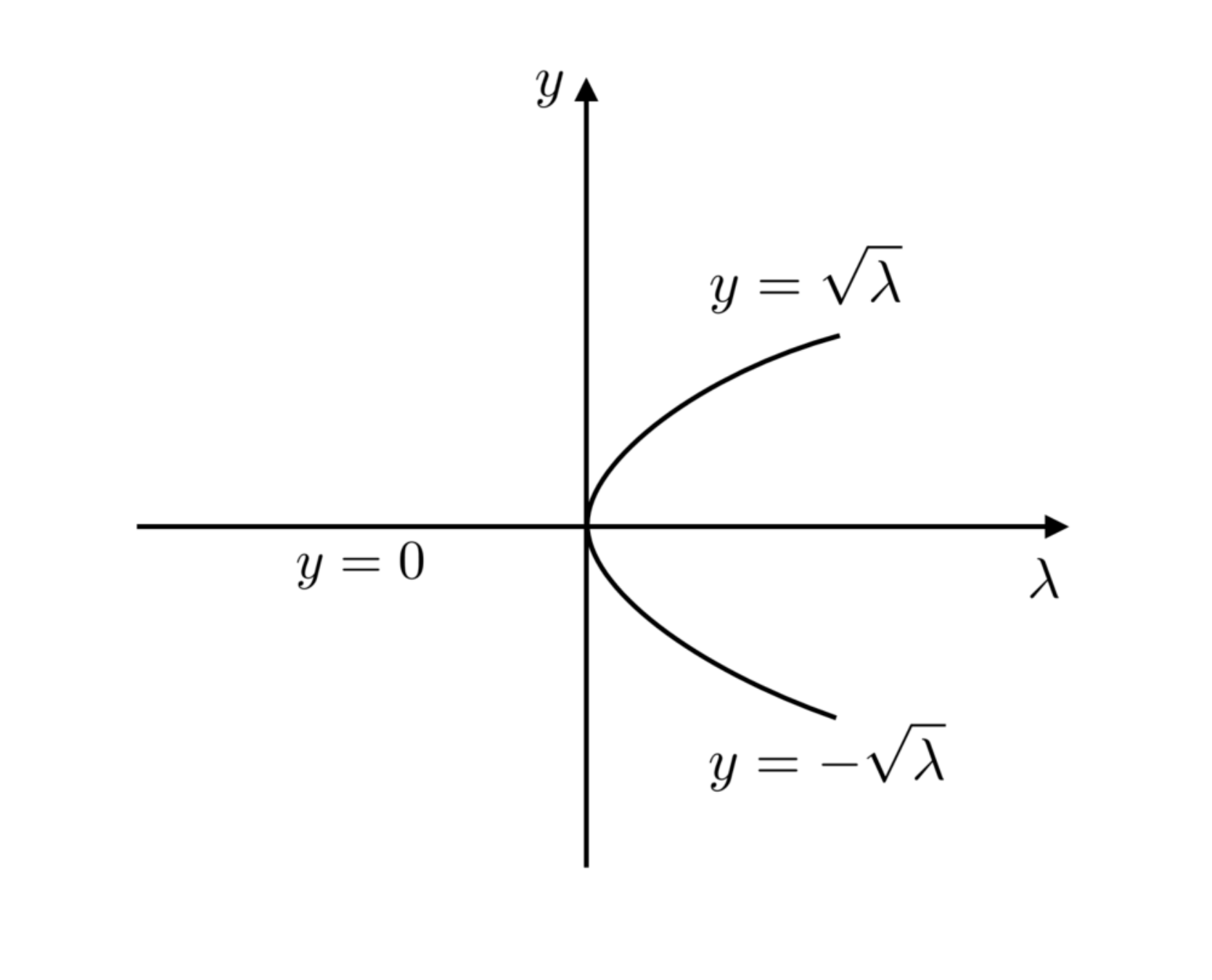}
\caption{Bifurcation diagram for steady-state solutions of Eq.~\eqref{eq1-day-10-10}.}
\label{fig-3-day-10-10}
\end{figure}

Here is an example from the textbook \cite{guckenheimer2013nonlinear} about bifurcation and also underscores the kind of questions we often ask about nonlinear systems
\begin{equation} \label{eq1-day-10-10}
\ddot{y}+2\beta\dot{y}-\lambda y+y^3=0, \quad t > 0,
\end{equation}
where the superimposed $\dot{(\ )} = d()/d t$ denotes the differentiation with respect to time $t$, $\beta$ and $\lambda$ are parameters. Equation \eqref{eq1-day-10-10} represents a Duffing type oscillator which has applications in biology and electrical engineering. The steady-state or equilibrium\footnote{Steady-state conditions are sometimes referred to in the literature as equilibria. This is a somewhat ill-advised use of the word but we think it worthwhile to bring this subtlety to your attention.} equations are straightforward to obtain and are the solutions of $\lambda y-y^3=0$. The value $y=0$ is a solution regardless of the value of the parameter $\lambda$ and $y=\pm \sqrt{\lambda}$ for $\lambda\geq 0$. We can graphically represent the so-called bifurcation diagram (see Fig.~\ref{fig-3-day-10-10}). In this diagram, $(\lambda, y)=(0,0)$ is the bifurcation point. In short, three distinct steady-state solutions are possible. Which specific steady-solution will be realized in practice? Some insights can be obtained from a direct numerical solution (see \cite{guckenheimer2013nonlinear} for details). \\

An interesting example of bifurcation is given by Strogatz \cite{strogatz2014nonlinear} regarding social dynamics. Assume that a small segment of the population has a resolute belief in something, e.g., right of women to vote.\footnote{which, by the way, became national law in the USA in 1920 after much struggle and debate. Individual states began to slowly ratify the 19th amendment over the years to follow. The last state to ratify this law was Mississippi---in 1984.} Imagine four overlapping segments of the populations: (i) Group A: A group that holds resolutely to the belief and nothing can change their mind, (ii) B is the group with the opposing views, (ii) this group agrees with A but are somewhat uncommitted to A. This means that if someone argues against A then B becomes AB (a person who is ambivalent or sees both sides of the arguments, (iv) this fourth group agrees with B but faced with sufficient arguments from A, will also join group AB. Interestingly enough, a simple model of differential equations can be used to represent this dynamics. It can be shown that at a critical population of ``true believers" there is a bifurcation point and everyone will end up agreeing with A. Similar models also exist for population dynamics, i.e., depending on the critical parameters (resources, number of mating pairs, etc.), a population may shrink below some parameters or grow. \\

There are other examples that illustrate the importance of understanding nonlinearities and stability. An aircraft is subjected to a variety of forces as it is flying. Beyond a critical speed and depending on how the aircraft has been designed, the vehicle can end up in a sustained stable oscillations state. This can be disastrous, as you can imagine, for a plane. Indeed actual cases have been documented in the fifties and sixties where planes disintegrated in mid-air due to this phenomenon of ``flutter". In materials science, defect nucleations, the formation of microstructure, endocytosis, frequency entrainment in biology and numerous other physical phenomena are consequences of nonlinear behavior and closely tied to the concept of stability. \\

We now turn to a very important part of the ``big picture". Most books by mathematicians on nonlinear systems tend to focus on the stability analysis of finite-dimensional systems. In other words, discrete or finite degrees of freedom systems like the pendulum or an oscillator. These systems are governed exclusively by a system of ordinary differential equations. On the other hand, engineers have usually focused on infinite-dimensional or continuous systems, e.g., theory of elasticity, Navier-Stokes equation of fluid mechanics, and so forth. The latter involves the study of partial differential equations. Why this disconnect? There is a good reason for this. Most of the rigorous results in the stability analysis are only available for finite degrees of freedom system, and accordingly, mathematicians have tended to focus textbooks only in this setting. Research monographs and papers, of course, are a different story. The truth is that one must understand the stability notions dealt in the host of mathematics books (those dealing with discrete systems) to develop some intuition and an understanding of results that can be proven rigorously. Many of those methods and theorems, if not exactly, at least in spirit, carry over to the continuous case also. Furthermore, a continuous system can often be reduced to a discrete system. However, we want to caution the reader that the transfer of the results proven in most mathematics books on finite degrees of freedom systems to continuous systems is not to be taken for granted. We introduce a simple finite-dimensional system below along with some definitions. \\

Consider the spring-mass-dashpot system. This is essentially a damped oscillator (see Sec.~1.2 in the textbook by Strogatz \cite{strogatz2014nonlinear} for more details). The governing equation is
\begin{equation}  \label{eq2-day-10-10}
m\ddot{y}+b\dot{y}+k y=0.
\end{equation}

This equation is linear and is said to be finite-dimensional with dimension 2. How do we come up with this nomenclature? Note that we need two pieces of information to completely describe the system, $y$ and $\dot{y}$. This can be much easily seen if we express this equation in the standard format of nonlinear systems, which involves expressing equations as coupled first-order equations. To do this for Eq.~\eqref{eq2-day-10-10}, we define two variables, $y_1=y$ and $y_2=\dot{y}$. Then, we can convert Eq.~\eqref{eq2-day-10-10} into
\begin{equation}  \label{eq3-day-10-10}
\dot{y}_1 = y_2, \qquad 
\dot{y}_2 = -\frac{b}{m} y_2-\frac{k}{m} y_1.
\end{equation}

In general, the system can be n-dimensional and, of course, nonlinear:
\begin{equation}   \label{eq4-day-10-10}
\left.
\begin{aligned}
\dot{y}_1 & = f_1(y_1, y_2, \cdots, y_n), \\
\dot{y}_2 & = f_2(y_1, y_2, \cdots, y_n), \\
&\ \, \vdots \\
\dot{y}_n & = f_n(y_1, y_2, \cdots, y_n).
\end{aligned}
\right\}
\end{equation}

The key point is that in the finite-dimensional systems, we are dealing with a countable finite number of variables. We may express the above equations more concisely using vector and matrix notation:

\begin{equation}
\dot{\mathbf y}={\mathbf f} (\mathbf y).
\end{equation}

Here, ${\bf y}(t) \in \mathbb R^n$ is an n-dimensional vector with components $(y_1, y_2, \cdots, y_n)^T$ and ${\bf f}$ is a nonlinear function of ${\bf y}$. In the case of linear systems, we can reduce the system of equations to a matrix form:

\begin{equation}
\dot{\mathbf y}=M {\mathbf y}.
\end{equation}

Evidently, for the simple damped oscillator in Eq.~\eqref{eq3-day-10-10}, ${\mathbf y} (t)=(y_1, y_2)^T$ and $M$ is $\left( \begin{array}{cc}0 & 1 \\ -k/m & -b/m \end{array} \right)$. \\

In an infinite-dimensional system, ${\mathbf y}$ is a {\it field} and depends continuously in some n-dimensional regions. Most infinite-dimensional engineering problems are defined in 3D regions (at most), so in that case, we may say that ${\mathbf y}$ depends on both time and has spatial variation. In that case, ${\mathbf f}$ is an {\it operator}. Simply put, in the case of continuous or infinite-dimensional systems, we must contend with partial differential equations such as the Navier-Stokes equation of fluid mechanics or elasticity. \\

Strogatz's book has a nice table (see Fig.~1.3.1 in \cite{strogatz2014nonlinear}), which discusses various types of physical systems in terms of dimensions and whether the systems are linear or not. The table also covers nonlinear continuum theories of elasticity, plasticity, biological membranes, among others. We note that the ``coupled nonlinear oscillators" in the category of $n\gg 1$ includes molecular dynamics.

\subsection{One-dimensional systems and geometric approach} \label{new-addition-section-p1-sub2}
The 1D system is the simplest dynamical system and permits a rather simple introduction of several essential concepts. We basically contend with systems like $\dot{y}=f(y)$, where $\dot{y} = d y / d t$, $y: I \to \mathbb{R}$, $I$ is an open interval of $\mathbb{R}$, and $f:\mathbb{R}\to \mathbb{R}$. While in an actual physical problem, $y(t)$ may not be position and $t$ may not be time, it is often useful to think of them in such terms as an aid to build intuition. In that case, $\dot{y} = d y / d t$ is the particle velocity whose motion is confined to $\mathbb{R}$. If we plot $\dot{y}$ {\it vs.} $y$, then $\dot{y}$ has the character of a \emph{vector}. At each point $y$, $\dot{y}$ represents a vector with a direction. \\

Consider the following example from Strogatz \cite{strogatz2014nonlinear}: $\dot{y} = \sin y$. What are its equilibrium points or fixed points or steady-state solutions? It is simple to find. We merely set $\dot{y}=0$. This tells us that the fixed points of this system are $y=n\pi$, where $n \in \mathbf Z$ are integers. \\

The plot of $\dot{y}$ {\it vs.} $y$ is called the {\it phase portrait} and represents how the trajectories {\it flow}. Although such plots may appear non-intuitive at first, they provide a wealth of information {\it without solving any equations} in both 1D and 2D; eventually establishing some level of intuition for higher dimensional systems.\\

\begin{figure}[h] 
\centering
\includegraphics[width=5in]{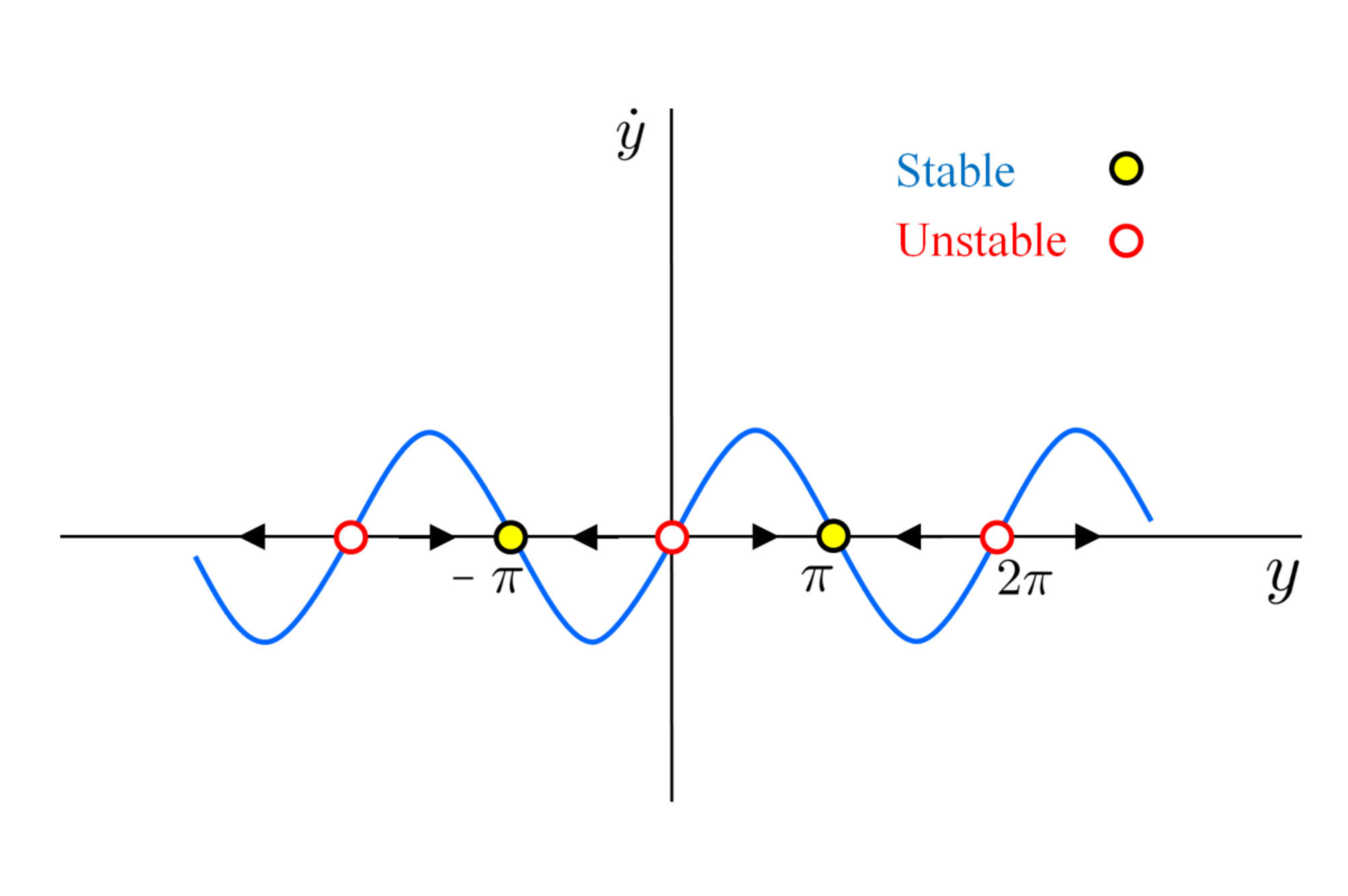}
\caption{Phase portrait of the nonlinear differential equation $\dot y = \sin y$. The arrows point to the right when the velocity is positive $\dot{y} (t) >0$ and to the left when $\dot{y} (t) <0$.}
\label{fig6-day-10-10}
\end{figure}

Keeping in mind the spirit expressed in the preceding paragraph, we attempt to plot the phase diagram without solving the nonlinear differential equation. This case, $\dot y = \sin y$, is rather simple, of course, and the result is shown in Fig.~\ref{fig6-day-10-10}. The fixed or equilibrium or steady-state solutions, $y = n \pi$, are where the curve intersects the $y$-axis. \\

Now we draw arrows to indicate the \emph{direction} of the velocity vector. We point the arrow towards the right when velocity is positive and vice-versa. We choose to draw the arrows on the $y$-axis in Fig.~\ref{fig6-day-10-10}, although it can be done anywhere. Notice that there are two types of fixed points: those that have arrows diverging from them and those which have arrows pointing towards them. Although we will define stability more formally in due course, the figure graphically shows that stable points (marked by a solid circle) are those towards which trajectories gravitate towards, and the reverse is true for unstable fixed points (marked by a hollow circle) in Fig.~\ref{fig6-day-10-10}. \\

The phase portrait can be used to construct the actual solution qualitatively. To see this, consider Fig.~\ref{fig6-day-10-10}. Assume that our initial condition is $y=0$. The phase portrait tells us that at that point, the velocity is zero. In other words, the particle will not move. Let the initial condition be $y = \pi/4$. In the beginning, the velocity $y$ will increase due to $\dot y > 0$, then after reaching a maximum value, there will be a period of deceleration ($\dot y >0$ but $\dot y$ gradually decreases) until it approaches the fixed point $y=\pi$. Other trajectories, starting from different initial points, can be similarly drawn. For more examples of the phase portrait, for example, n-dimensional systems, the reader is referred to the textbooks \cite{strogatz2014nonlinear, guckenheimer2013nonlinear, seydel2009practical, perko2013differential, kuznetsov2013elements, sastry2013nonlinear}.

\subsection{Definition of stability and the importance of norm}  \label{new-addition-section-p1-sub3}
It is now time to define stability more formally. Although we have so for explicitly only given examples in the context of 1D systems, the definitions articulated here is valid for a general finite-dimensional system. In general, the concept of stability is defined by the motion of the physical system. Thus it is a dynamic concept, and we have the following definition. \\

{\bf Definition of Stability}: An equilibrium solution ${\mathbf y}_0$ at $t=t_0$ of the system $\dot{\mathbf y}={\mathbf f}({\mathbf y})$ is Lyapunov stable if given a small number $\epsilon>0$, there exists $\delta(\epsilon)>0$ for any small perturbation, such that the subsequent state ${\mathbf y}(t)$, $t_0 < t < \infty$, satisfies $\Vert {\mathbf y}(t) - {\mathbf y}_0 \Vert < \epsilon$.\\

This implies that for any equilibrium point, any solution that is close to it initially (within some regions $\delta$ of ${\mathbf y}_0$), remains so and is confined within some regions $\epsilon$ of ${\mathbf y}_0$ for all positive time.\\

The concept of asymptotic stability requires that, eventually, any solution that starts close to the equilibrium point must approach ${\mathbf y}_0$ as $t \to \infty$ or $\Vert {\mathbf y}(t) - {\mathbf y}_0 \Vert \to 0$. Detailed mathematical discussion of the definition of stability may be found in \cite{lyapunov1992general, triantafyllidis2011stability, thompson1973general, budiansky1974theory, knops1973theory}. \\

{\bf The issue of the norm and finite-dimensional systems {\it vs.} infinite/continuous systems}: The symbol $\Vert \cdot \Vert$ is a suitable norm. This measures how close or far the two solutions are. The Euclidean norm\footnote{The Euclidean norm of a vector ${\bf v}= (v_1, v_2, ..., v_n) \in \mathbb R^n$, for example, is defined as $\Vert {\bf v} \Vert := \sqrt{v_1^2 + v_2^2 + ... + v_n^2}$.} is frequently used infinite-dimensional systems; however, a very key result is that once a theorem has been proven in one norm, it also holds true in {\it any} other norm \cite{como1995theory}. The proof of this very powerful statement is beyond the scope of the present article. Stated slightly differently, we can say that in the context of a finite-dimensional system, if a solution is stable/unstable in one particular norm, then it will be so in any other norm also. This result is {\it not true} for infinite-dimensional or continuous systems. This, among other issues, is one of the key issues that make the stability analysis of continuous systems more complicated. We refer the reader to Como and Grimaldi \cite{como1995theory} for a very nice example of how the change of norm can lead to differing conclusions in a structural mechanics problem.

\subsubsection{Stability analysis by linearization: one-dimensional examples}
Let us consider the case of a single differential equation, i.e., $\dot{y} (t) = f (y, \lambda)$, where $\lambda$ is the load parameter. The Taylor expansion of $f(y, \lambda)$ about a stationary state $y=y_0$ at a fixed load parameter $\lambda$ reads
\begin{equation} \label{eq-2-fff}
\dot{y} (t) = f (y_0, \lambda) +\frac{\partial f}{\partial y}\Big|_s (y-y_0) + o(|y-y_0|).
\end{equation}

Here $\frac{\partial f}{\partial y}\Big|_s$ denotes the partial derivative of $f (y, \lambda)$ with respect to $y$ at $(y_0, \lambda)$, i.e., $\frac{\partial f}{\partial y}\Big|_s = \frac{\partial f}{\partial y} (y_0, \lambda)$. The notation $o(|y-y_0|)$ denotes a function with the limiting behavior, i.e., the ratio $\frac{|o|}{|y-y_0|}$ approaches zero as $|y-y_0| \to 0$. By dropping the zero term, $f (y_0, \lambda)=0$ for a stationary state $y=y_0$, and introducing a new variable $h(t) = y(t) - y_0$, the linearized form of Eq.~\eqref{eq-2-fff} is
\begin{equation} \label{eq-3-fff}
\dot{h} (t) = \frac{\partial f}{\partial y}\Big|_s h(t).
\end{equation}

By inserting the ansatz $h(t) = h_0 e^{\mu t}$ into Eq.~\eqref{eq-3-fff}, we can easily get $\mu = \frac{\partial f}{\partial y}\Big|_s$. For a negative $\mu = \frac{\partial f}{\partial y}\Big|_s < 0$, the perturbed state $h(t) = h_0 e^{\mu t}$ decreases as the time $t$ increases, thus the stationary state $y=y_0$ is stable. Otherwise, the stationary state $y=y_0$ is unstable for $\mu >0$ and a saddle point for $\mu=0$. \\

Consider the nonlinear differential equation $\dot y = \sin y$ whose stability has already been investigated by using the geometric approach in Sec.~\ref{new-addition-section-p1-sub2}. The fixed point is denoted by $y_0 = \pm n \pi$, $n \in \mathbf Z$, at which $\sin y_0 =0$. By Eqs.~\eqref{eq-2-fff} and \eqref{eq-3-fff}, the linearized form of $\dot y = \sin y$ at $y = y_0$ is denoted by $\dot h (t) = (\cos y_0) \, h(t)$, where $h(t) = y(t) - y_0$. For example, at $y_0 = 0, \pm 2\pi$, $\cos y_0 = 1 >0$, and the fixed points $y_0 = 0, \pm 2\pi$ are unstable; in contrast, $\cos y_0 = -1 <0$ at $y_0 = \pm \pi$ and the fixed points $y_0 = \pm \pi$ are stable. Their properties of stability are also shown in Fig.~\ref{fig6-day-10-10} by using the geometric approach. \\

\begin{figure}[h] %
    \centering
    \includegraphics[width=6in]{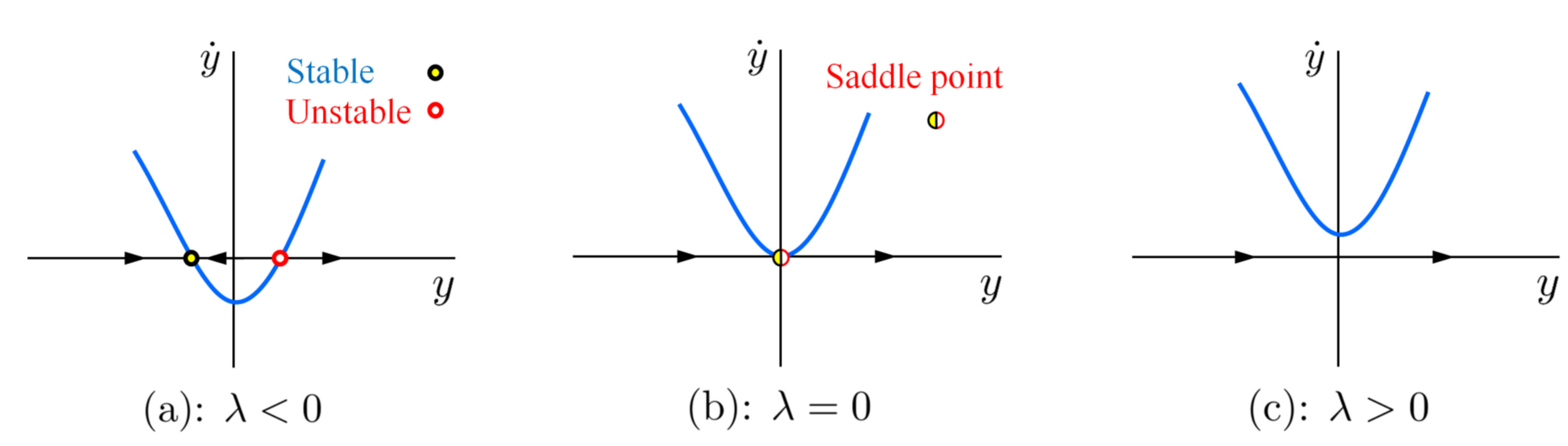}
    \caption{Phase portraits of the differential equation $\dot{y} (t) = y^2 + \lambda$ for different load parameter $\lambda$. The arrows in each figure point to the right when $\dot{y} (t) >0$ and to the left when $\dot{y} (t) <0$.}
    \label{fig2-bifurcation}
\end{figure}
 
Consider another example of the differential equation $\dot{y} (t) = f (y, \lambda) = y^2 + \lambda$. The stationary state $y=y_0$ satisfies $f (y_0, \lambda) =0$, which implies three cases (see Fig.~\ref{fig2-bifurcation}): (a) two stationary points $y_0 = \pm \sqrt{-\lambda}$ for $\lambda <0$; (b) one repeated stationary point $y_0 = 0$ for $\lambda =0$; (c) none stationary point for $\lambda >0$. By $\frac{\partial f}{\partial y}\Big|_s$ in Eq.~\eqref{eq-3-fff}, i.e., $\frac{\partial f}{\partial y}\Big|_s = 2y_0$, we can conclude that: (a) the stationary point $y_0 = -\sqrt{-\lambda}$ is stable while $y_0 = \sqrt{-\lambda}$ is unstable for $\lambda<0$; (b) the stationary point $y_0 = 0$ is a saddle point for $\lambda=0$. The stationary points and their stabilities can be found in Fig.~\ref{fig2-bifurcation}. For further examples, see, the books \cite{strogatz2014nonlinear, seydel2009practical}.
 
\subsubsection{Principle of minimum energy: a conservative system}
The preceding sections highlight stability as a dynamical concept. A stability analysis is related to the motion of the system under small perturbations, and the type of analysis illustrated so far is valid for a general system, including both nonconservative and conservative ones. \\

For a conservative system---which will become the primary focus of this tutorial, the stability analysis of the motion of differential equations is identical to the discussion of the minimization of potential energy. The proof can be found in many textbooks, for example, in the exposition by Triantafyllidis \cite{triantafyllidis2011stability}. A state, denoted by the general state variable ${\bf x}$, is stable if its potential energy $I ({\bf x})$ is lower, interpreted as not greater, energy than that of all the neighborhood states, ${\bf x}+ \delta {\bf x}$, such that 
\begin{equation} \label{2-stability-1}
I ({\bf x}) \le I ({\bf x}+ \delta {\bf x}),
\end{equation}
where $\Vert \delta {\bf x} \Vert$ is sufficiently small and $\Vert \cdot \Vert$ is a properly defined norm for the space of deformation functions.
\subsubsection{Other stability criteria}

In contrast to the linear stability analysis and the principle of minimum energy, other stability criteria are also often used \cite{timoshenko1961theory, cedolin2010stability}. The adjacent equilibrium stability criterion, for example, asserts that a primary equilibrium state becomes unstable when there are other equilibrium states nearby. A detailed discussion of the difference between the energy method and the adjacent equilibrium stability criterion can be found in a recent paper \cite{chen2018surface}. For nonconservative systems, the Lyapunov stability criterion \cite{kuznetsov2013elements, sastry2013nonlinear, lyapunov1992general} can be used to discuss the stability of the motion of a dissipative system \cite{cedolin2010stability}. Furthermore, for the initial post-buckling of continuum media, the ``Lyapunov-Schmidt-Koiter'' (LSK) approach is widely used \cite{van2009wt, budiansky1974theory, triantafyllidis2011stability, cedolin2010stability}. The LSK approach will be discussed later in this tutorial, and a simple example of the initial post-buckling of a dielectric column is introduced.

\subsection{The concept of bifurcation}\label{subsection-bifurcation}
Let us assume that the state of a physical system can be represented by a generalized state variable $u$.\footnote{This generalized state variable $u$ could be either a scalar/tensor variable or a scalar/tensor-valued function, which depends on the actual physical system. For the latter case, we have to use the functional derivative in the calculus of variations \cite{courant1953methods, weinstock1974calculus, antman2013nonlinear, ball1998calculus}.} The generalized load on the system is simply denoted by $\lambda$. The equation
\begin{equation} \label{equation-5}
f (u, \lambda) = 0
\end{equation}
determines the stationary state of a physical system with the state variable $u$ and the bifurcation parameter $\lambda$. 

\subsubsection{Bifurcation points} \label{section-bifurcation-points}

\begin{figure}[h] %
    \centering
    \subfigure[]{%
\includegraphics[width=2.9in]{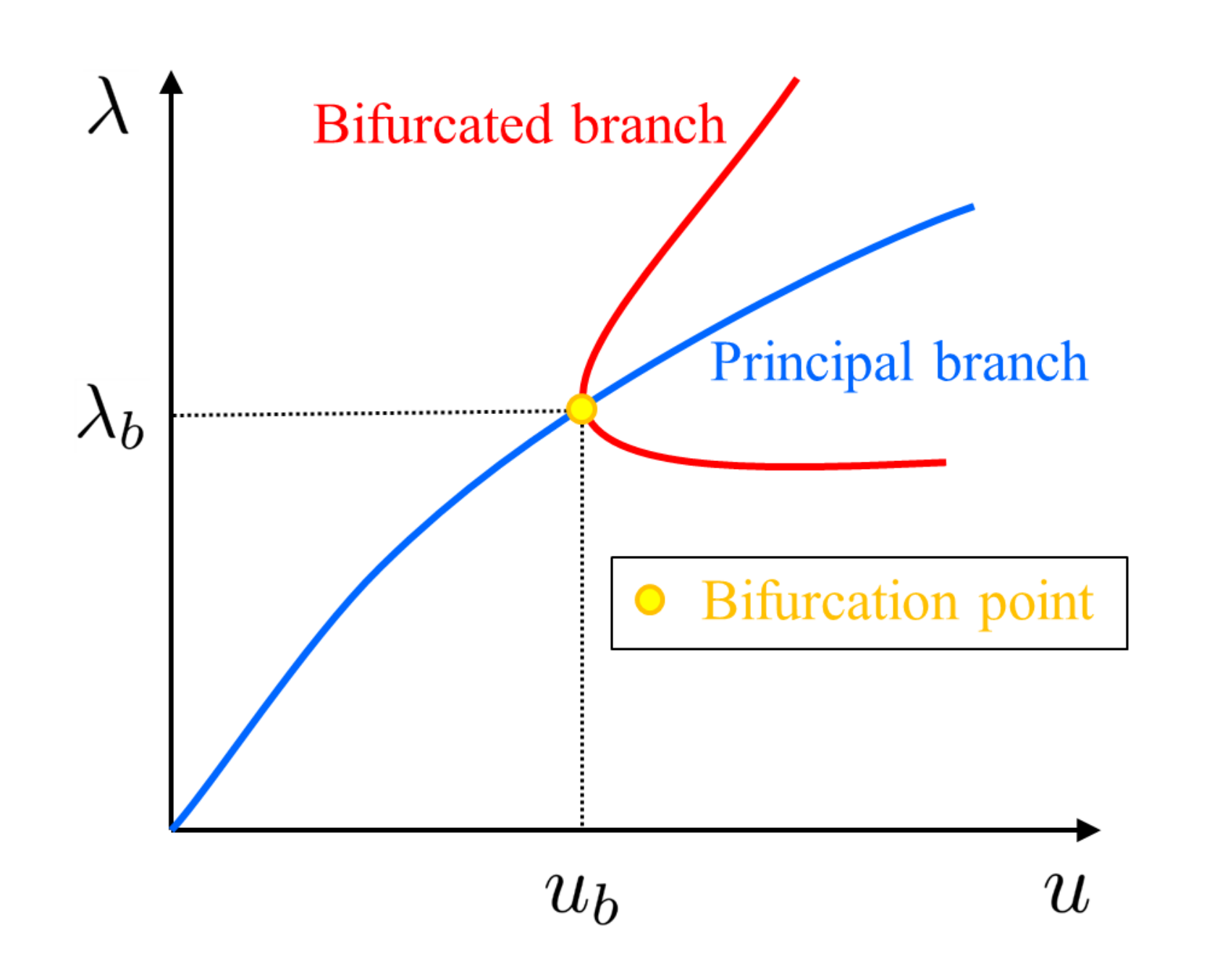}
\label{bifur-1a}}
\subfigure[]{%
\includegraphics[width=2.9in]{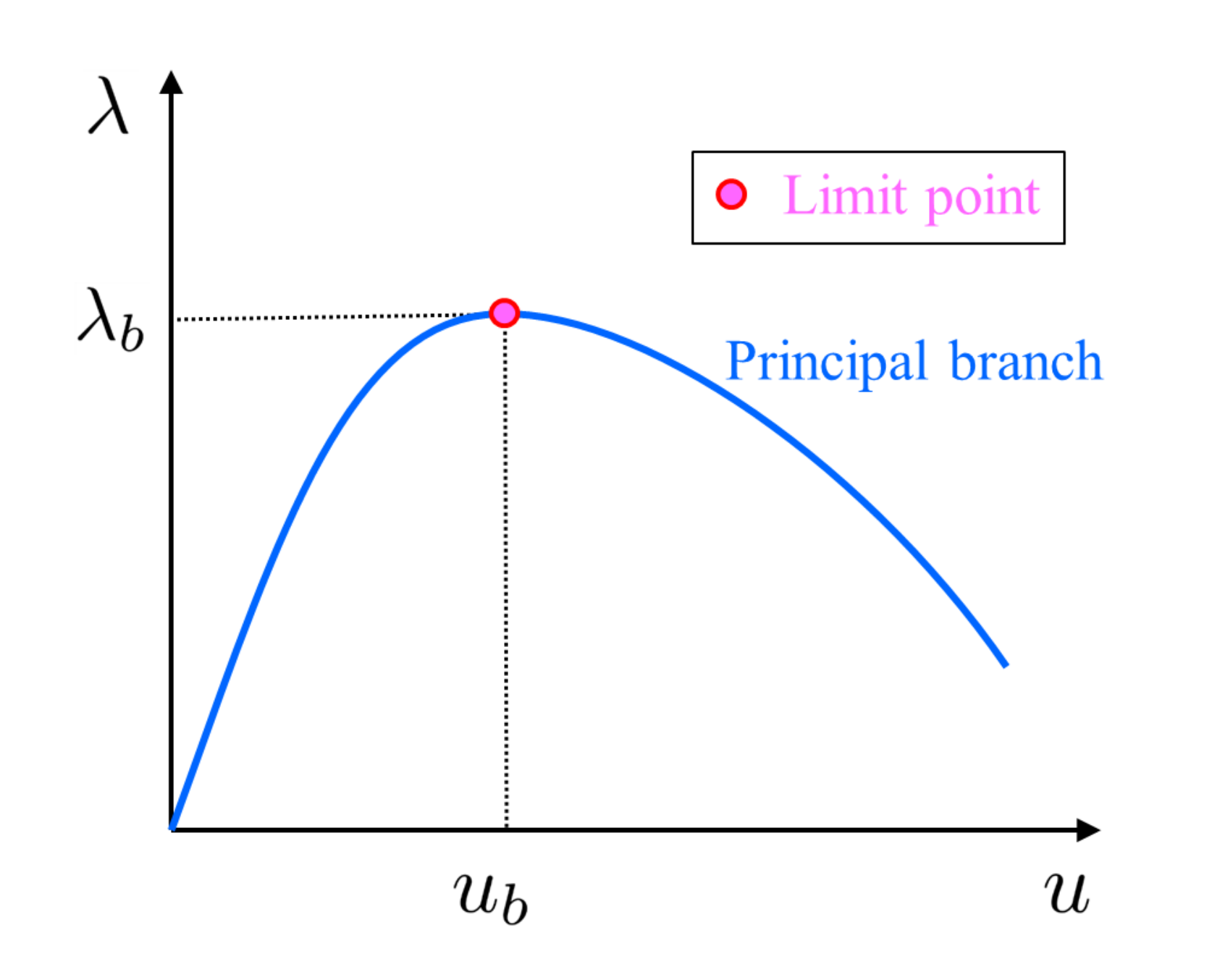}
\label{bifur-1b}}
    \caption{Schematic of the bifurcation states of a physical system $f (u, \lambda) = 0$. $u$ is the generalized state variable of a physical system, and $\lambda$ is the bifurcation (load) parameter. (a) A bifurcation diagram. (b) A limit point.}
    \label{bifur-1}
\end{figure}

We first assume that there is only one solution (branch) $u_0$, the so-called principal branch, as the load parameter $\lambda$ increases at the beginning, such that $f(u_0, \lambda) = 0$ in Eq.~\eqref{equation-5}. In Fig.~\ref{bifur-1}, the principal branch $u_0$ bifurcates to a bifurcated branch $u_0 + v$ when the bifurcation parameter $\lambda$ increases to $\lambda_b$ that corresponds to the state $u_b$. The pair $(u_b, \lambda_b)$ is called a bifurcation point. For example, the unbuckled state of Euler's column is usually called the principal branch, while the buckled state is the bifurcated branch. \\

Now the questions are how to determine the existence of the bifurcated branch and to find the bifurcation point. To answer these questions, we have to introduce the {\it implicit function theorem whose discussion}, especially in the context of interest, may be found in many books \cite{seydel2009practical, kuznetsov2013elements, antman2013nonlinear}.\footnote{In the beginning, we just illustrate the bifurcation theory by using the derivative of a function rather than the functional derivative. This arrangement highly simplifies the discussion and makes the spirit of the bifurcation theory clear.} \\

{\bf Theorem} ({\it Implicit function theorem}). Let $\mathcal S \subseteq \mathbb R^{1+1}$ and write points in the set $\mathcal S$ as $(u, \lambda)$ with $u \in \mathbb R$, $\lambda \in \mathbb R$. Let $f: \mathbb R^{1+1} \to \mathbb R$ be $\mathcal C^1$ on $\mathcal S$. Let $(u_0, \lambda_0) \in \mathcal S$ with $f (u_0, \lambda_0) = 0$. If the derivative
\[ \frac{\partial f}{\partial u} (u_0, \lambda_0) \neq 0, \]
then there exists an $r>0$ and a $\mathcal C^1$ function $g: B(\lambda_0; r) \to \mathbb R$ with $g (\lambda_0) = u_0$ such that
\[ f (g (\lambda), \lambda) = 0 \quad
{\rm and} \quad
\frac{d \, g(\lambda)}{d \, \lambda} = - \Big( \frac{\partial f}{\partial u} (u, \lambda) \Big)^{-1}  \Big( \frac{\partial f}{\partial \lambda} (u, \lambda) \Big) \Big|_{u=g(\lambda)}\]
for all $\lambda \in B(\lambda_0; r) = \{ \lambda \in \mathbb R: |\lambda - \lambda_0| < r \}$. $\Box$ \\

Here we only consider one equation $f (u, \lambda) = 0$ with two unknowns $u$ and $\lambda$. The unknown $u$ is called the {\it dependent variable} while the unknown $\lambda$ is called either the {\it free variable} or {\it independent variable}. We express the unknown $u$ in terms of $\lambda$, such that $u=g(\lambda)$. The implicit function theorem gives us the condition under which we can do this locally and gives us a formula for computing $\frac{d \, g(\lambda)}{d \, \lambda}$. \\

In particular, if $f (u_b, \lambda_b) = 0$ and $\frac{\partial f}{\partial u} (u_b, \lambda_b) = 0$, the point $(u_b, \lambda_b)$ is called the bifurcation point, see Fig.~\ref{bifur-1}, at which the uniqueness of the solution branch $u = g(\lambda)$ does not hold.

\subsubsection{Bifurcation in a continuum media}
In this section, we briefly discuss bifurcations in a continuum media. More classic bifurcation problems of elasticity (e.g., buckling of a naturally straight rod, a whirling rod, a beam, a plate, a cylindrical shell, a spherical cap, etc.) can be found in Chapter 5 in the book by Antman \cite{antman2013nonlinear}. \\

Let us consider a 1D continuum media, $\mathcal B_1 = \{ X \in \mathbb R: 0 \le X \le L\}$. The generalized state variable of the 1D physical system is denoted by $u$, which is a $\mathcal C^2$ mapping $u: \mathcal B_1 \to \mathbb R^n$, i.e., $u  \in \mathcal C^2 ([0, L]; \mathbb R^n)$. With a load parameter $\lambda \in \mathbb R$, the stationary state of the physical system is represented by
\begin{equation} \label{bifur-con-1}
f (u(X), \lambda) = 0.
\end{equation}

At a load parameter $\lambda_0$, there exists a state $u_0$, such that $f (u_0, \lambda_0) = 0$. Consider a small variation $\delta u(X)$ with sufficiently small $\Vert\delta u(X)\Vert$ in the neighborhood of $u_0 (X)$. The expansion of the function $f(u, \lambda_0)$ at $(u_0, \lambda_0)$ reads
\begin{equation} \label{bifur-con-2}
f (u_0 + \delta u, \lambda_0) = f (u_0, \lambda_0) + \frac{\partial f}{\partial u} (u_0, \lambda_0) \delta u + o (\delta u),
\end{equation}
where $\frac{\partial f}{\partial u} (u_0, \lambda_0)$ is called the (Fr\'{e}chet) derivative of $f$ at $(u_0, \lambda_0)$, and $\lim_{\delta u \to 0} \frac{\| o (\delta u) \|}{\|\delta u \|} = 0$. A much weaker notation of a derivative is that of a directional derivative, i.e., the (G\^{a}teaux) derivative. In this tutorial, we don't distinguish the difference between these two derivatives. Thus, we take
\begin{equation} \label{bifur-con-2a}
\frac{\partial f}{\partial u} (u_0, \lambda_0) \delta u \equiv \left. \frac{d}{d \tau} f (u_0 + \tau \delta u, \lambda_0)\right|_{\tau=0} = \lim_{\tau \to 0} \frac{f (u_0 + \tau \delta u, \lambda_0) - f (u_0, \lambda_0)}{\tau}.
\end{equation}

By the implicit function theorem, a necessary condition for the bifurcation of $f(u,\lambda)=0$ at $(u_0, \lambda_0)$ is that the (Fr\'{e}chet) derivative of $f$ at $(u_0, \lambda_0)$ is not invertible, namely
\begin{equation} \label{bifur-con-3}
f (u_0, \lambda_0) = 0 \quad {\rm and} \quad \left. \frac{d}{d \tau} f (u_0 + \tau \delta u, \lambda_0)\right|_{\tau=0} =0 \quad {\text {for sufficiently small}} \ \Vert \delta u \Vert.
\end{equation}

\subsection{More examples}
\subsubsection{Stability examples}
The following simple example just shows how to use the principle of minimum energy to study the stability of a physical system. \\

\begin{figure}[h] %
    \centering
    \includegraphics[width=4in]{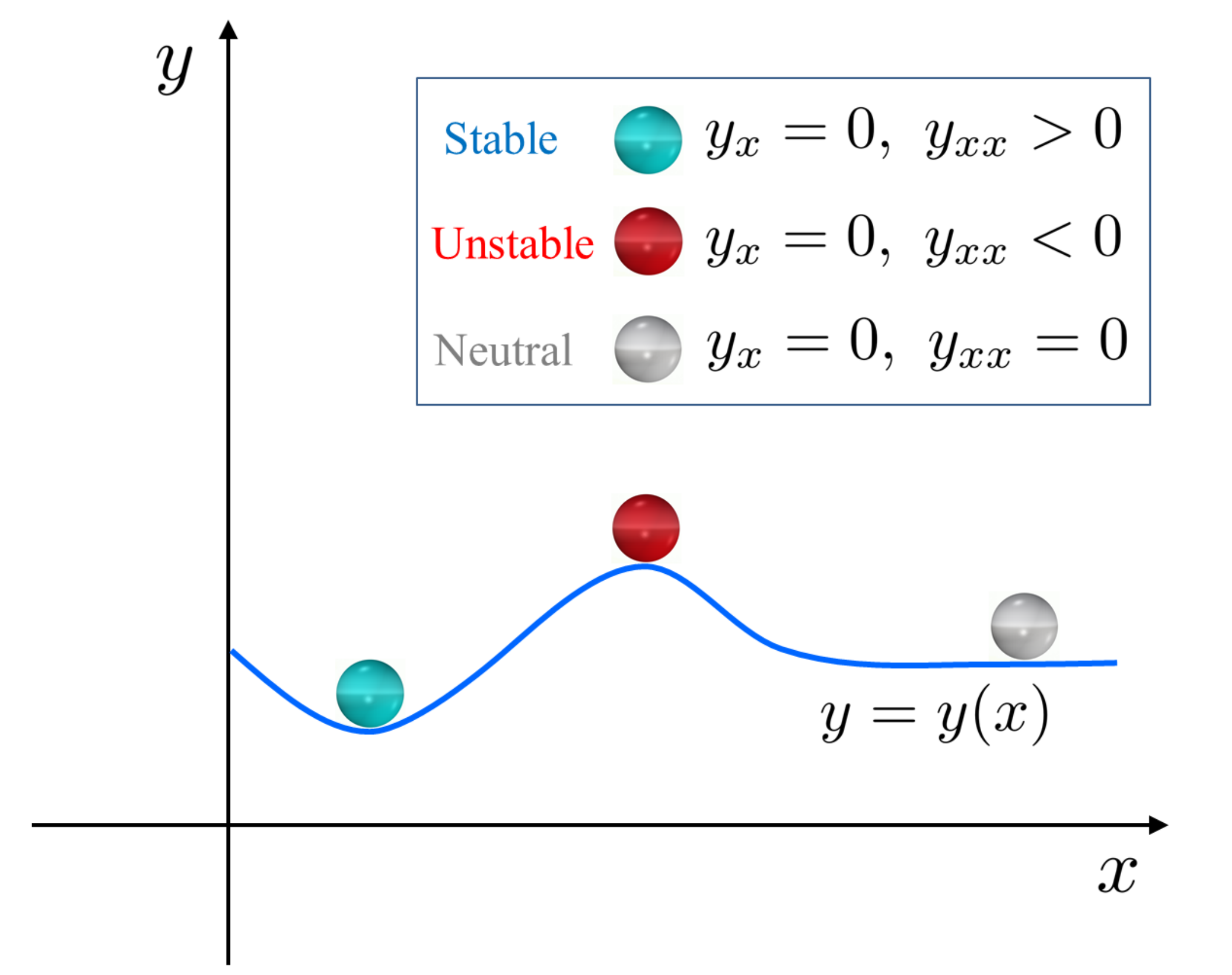}
    \caption{The stability of a rigid ball at different positions of the curve $y=y(x)$ subjected to the gravity. The potential energy of the ball at state $x$ is given by $I(x) = mg y(x)$. The 3D vision of the rigid ball on a curved surface can refer to Fig.~\ref{stability-1}.}
    \label{stability-1a}
\end{figure}

{\it Example}: Consider the stability of rigid balls in Fig.~\ref{stability-1}. We assume that the rigid ball can only move on a smooth curve, $y = y(x), 0 \le x \le a, $ see Fig.~\ref{stability-1a}. Choose the horizontal axis as the reference at which the rigid ball has zero potential energy. Then the potential energy of the ball subjected to the gravity at state $x$ is denoted as
\begin{equation} \label{2-stability-1b}
I (x) = mg y(x),
\end{equation}
where $m$ is the ball mass and $g$ is the gravitational acceleration. By the stability condition \eqref{2-stability-1} and the potential energy \eqref{2-stability-1b}, the stability of a rigid ball at state $x$ indicates
\begin{equation} \label{2-stability-1c}
y(x) \le y (x + \delta x),
\end{equation}
which, for arbitrary $\delta x$ with sufficiently small $|\delta x|$, gives 
\begin{equation} \label{2-stability-1d}
\frac{\partial y}{\partial x} = 0 \quad {\rm and} \quad \frac{\partial^2 y}{\partial x^2} \ge 0.
\end{equation}

Therefore, the left rigid ball is stable, the middle ball is unstable, and the right ball is neutral stable. In Fig.~\ref{stability-1a}, we plot the three states and their stabilities corresponding to the minimization of the potential energy \eqref{2-stability-1b}. This stability example simply involves finding the extremum of a scalar function with one scalar variable. \\

Note that the state variable $x$ in this example is a scalar variable due to its finite-dimensional nature (a single rigid ball). In contrast, continuous physical systems often involve scalar/tensor functions rather than scalar/tensor variables as state variables. These functions, for example, include the displacement in deformable solids, the electric displacement in electrostatics, and the temperature in thermodynamics, etc. Thus the calculus of variations \cite{courant1953methods, weinstock1974calculus, antman2013nonlinear, ball1998calculus}, which is also very useful in nonlinear elasticity, is often used in the stability analysis of deformable solids.

\subsubsection{Bifurcation examples} \label{subsection-bifurcation-exs}
We present a few examples in this sub-section for a better understanding of the concept of bifurcation, especially the knowledge of nonuniqueness of solution and bifurcation points by using the implicit function theorem. For more advanced examples and materials of bifurcation analysis, a few literature are listed here \cite{antman2013nonlinear, strogatz2014nonlinear, chen2001singularity, kuznetsov2013elements, golubitsky2012singularities, ogden1997non}. \\

\begin{figure}[h] 
\centering
\subfigure[]{%
\includegraphics[width=2.9in]{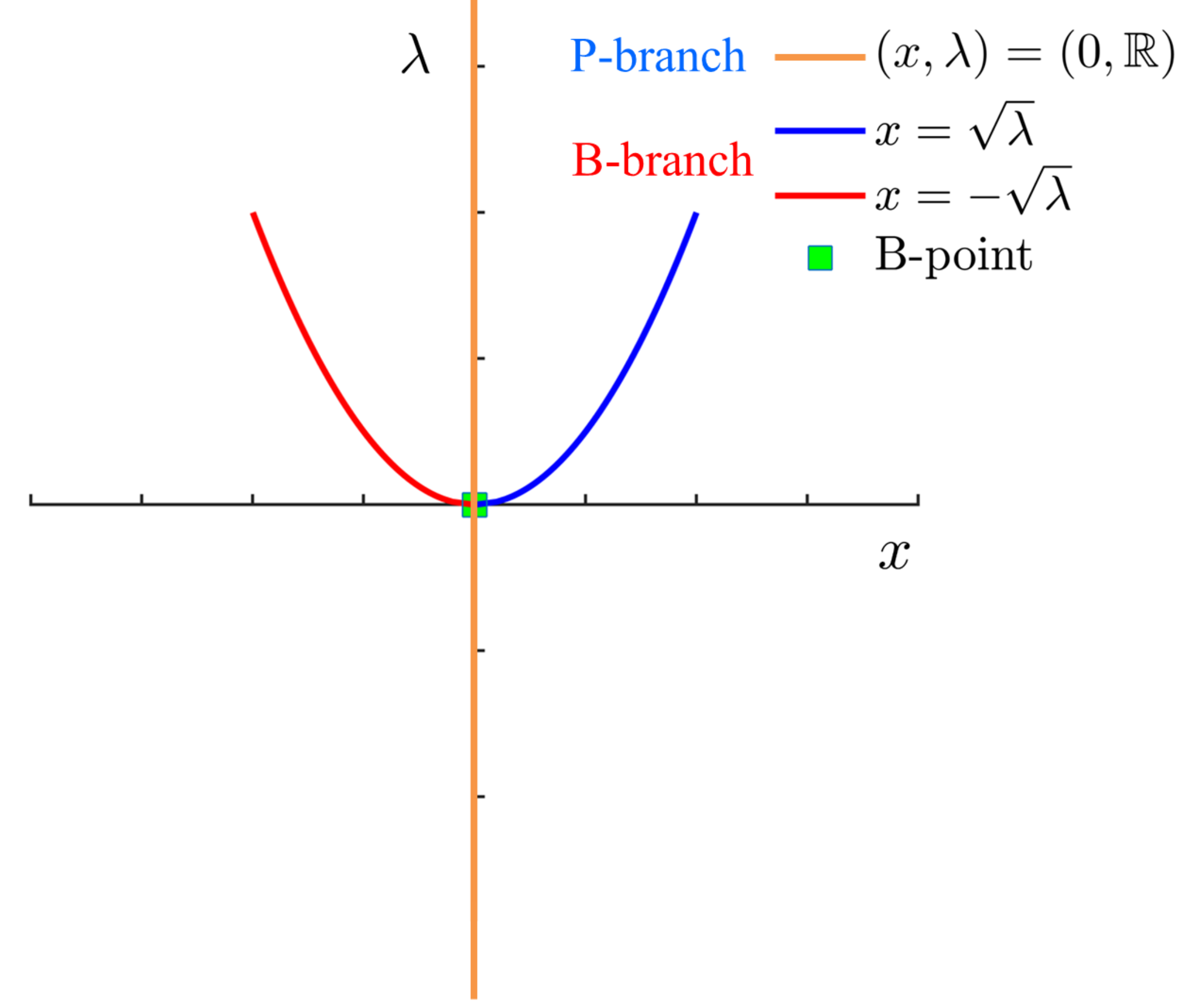}
\label{bifur-2a}}
\subfigure[]{%
\includegraphics[width=2.9in]{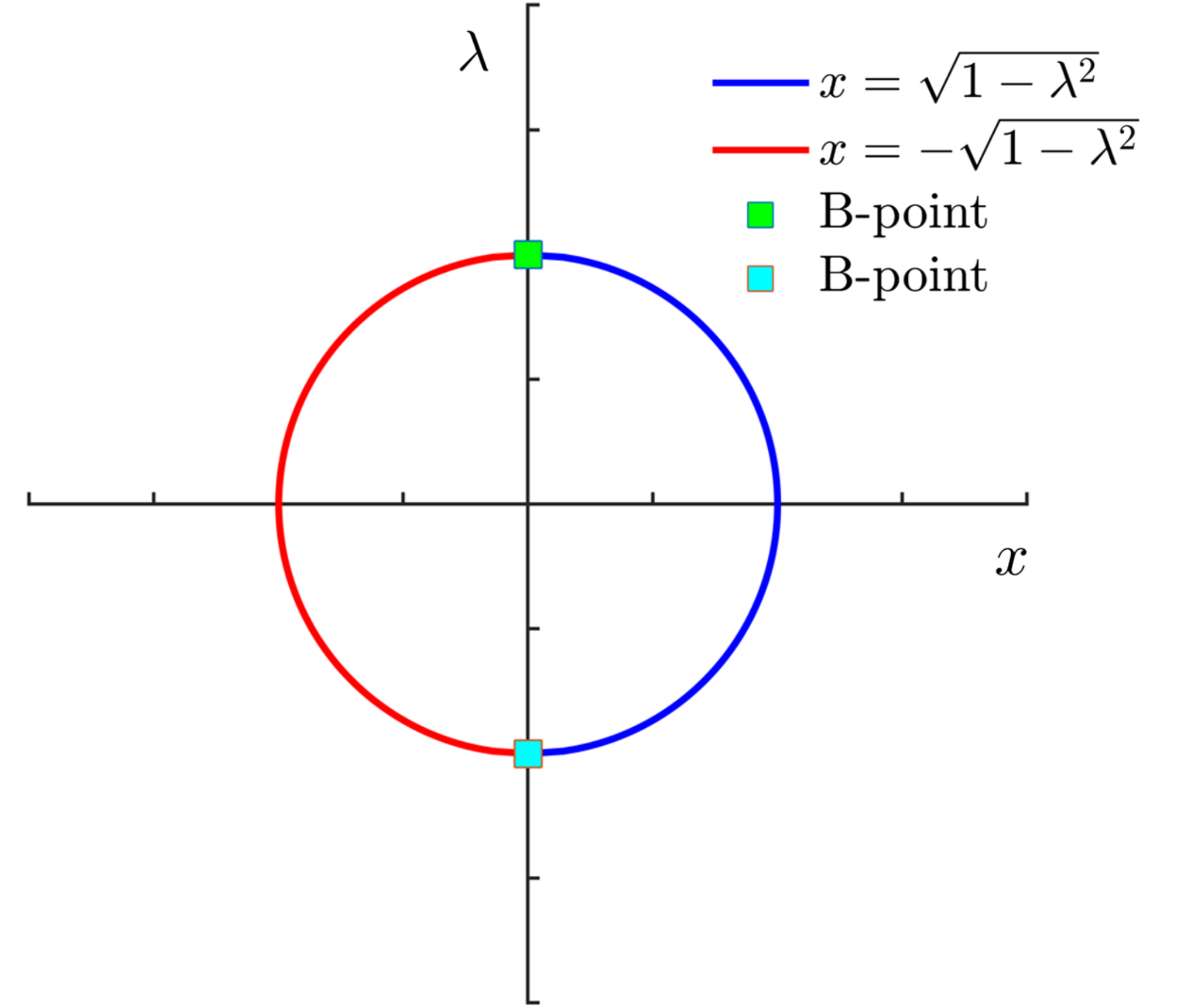}
\label{bifur-2b}}
\caption{Schematic of bifurcation diagrams. (a) Pitchfork type, from the equation $\lambda x - x^3 = 0$ in Eq.~\eqref{bifur-ex-1}. (b) From the equation $\lambda^2 + x^2 - 1 = 0$ in Eq.~\eqref{bifur-ex-2}.}
\label{bifur-2}
\end{figure}

{\it Example}: Consider the equation 
\begin{equation} \label{bifur-ex-1}
f_{1} (x, \lambda) = \lambda x - x^3 = 0, \quad x, \lambda \in \mathbb R. 
\end{equation}

Apparently, $(x, \lambda) = (0, \mathbb R)$ is trivial solution, i.e., the principal branch. At point $(x_0, \lambda_0) = (0,0)$, we have $f_1 (0,0) = 0$ and $\frac{\partial f_1}{\partial x} (0,0) =0$, indicating that $(x_0, \lambda_0) = (0,0)$ is a bifurcation point at which a unique solution branch $(x, \lambda) = (0, \mathbb R)$ does not hold. From the bifurcation diagram in Fig.~\ref{bifur-2a}, we can find that two bifurcated branches initiate at the bifurcation point $(0, 0)$. \\

{\it Example}: Consider the equation 
\begin{equation} \label{bifur-ex-2}
f_2 (x, \lambda) = \lambda^2 + x^2 - 1 = 0,  \quad x, \lambda \in \mathbb R.
\end{equation}

At two points $(x, \lambda) = (0, -1)$ and $(0, 1)$, both $f_2 = 0$ and $\frac{\partial f_2}{\partial x} = 0$ are satisfied. The bifurcation diagram of this example is plotted in Fig.~\ref{bifur-2b}. \\

\begin{figure}[h] 
\centering
\subfigure[]{%
\includegraphics[width=2.9in]{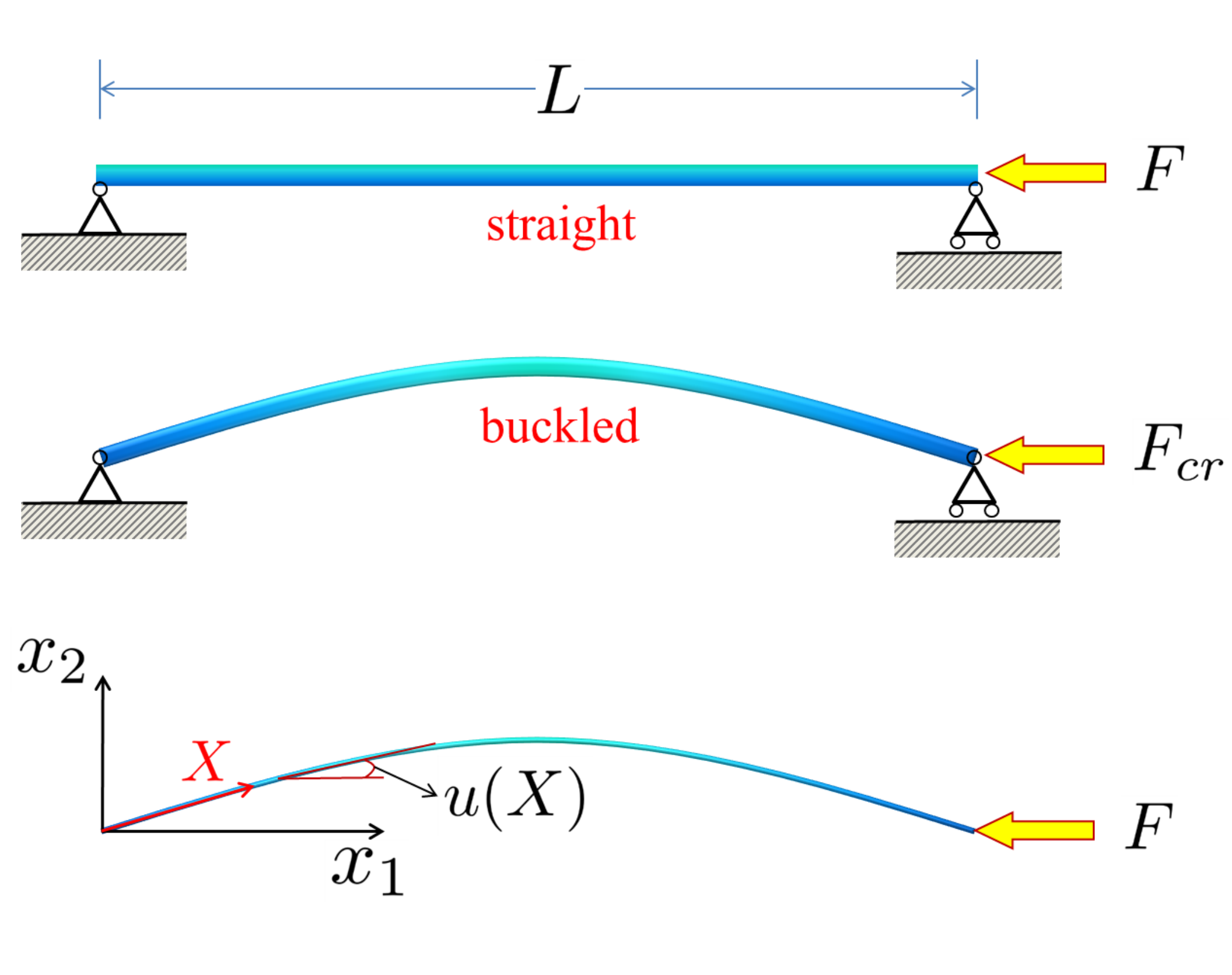}
\label{bifur-4a}}
\subfigure[]{%
\includegraphics[width=2.9in]{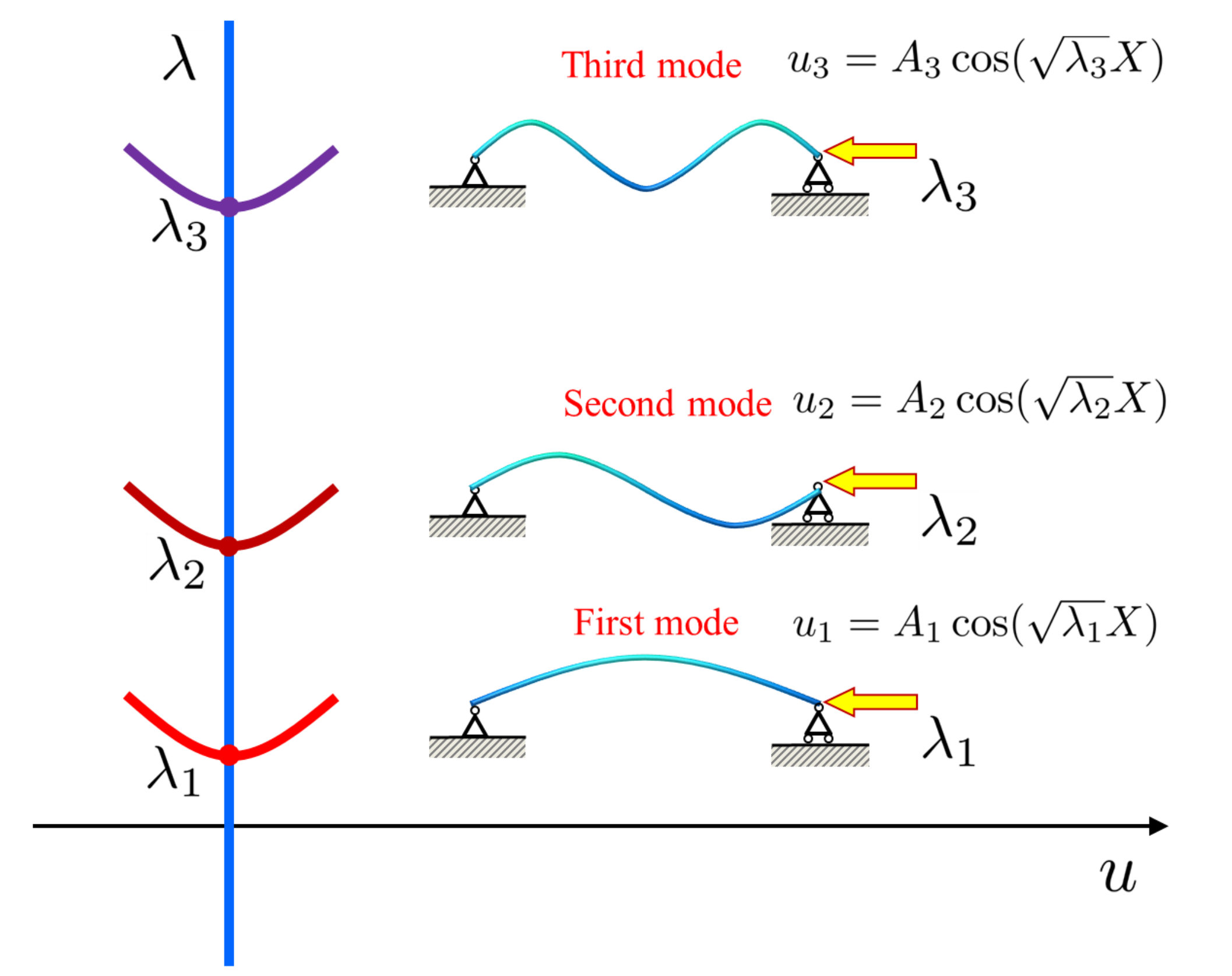}
\label{bifur-4b}}
\caption{Buckling of Euler's column with pinned-pinned ends. (a) Schematic of one model of buckling and the coordinates used for analysis. $u(X)$ is the angle between the tangent of the buckled column and the horizontal axis. (b) Bifurcation diagram and the first three modes of buckling. The straight column corresponds to the principal branch that is denoted by the vertical axis, $u=0$. The first three bifurcation points are $(0, \lambda_1)$, $(0, \lambda_2)$, and $(0, \lambda_3)$. Here the notation is $\lambda=\frac{F}{EI}$, and $\lambda_1 = \frac{\pi^2}{L^2}$, $\lambda_2 = \frac{4 \pi^2}{L^2}$, and $\lambda_3 = \frac{9 \pi^2}{L^2}$.}
\label{bifur-4}
\end{figure}

{\it Example}: Consider Euler's buckling in continuum media (see Fig.~\ref{bifur-4a}). The equilibrium equation of the incompressible Euler's column is \cite{antman2013nonlinear, chen2001singularity}
\begin{equation} \label{bifur-con-4a}
f (u(X), \lambda) = \frac{d^2 u}{d X^2} (X) + \lambda \sin u(X) = 0, \quad 0 < X < L,
\end{equation}
and the boundary conditions are 
\begin{equation} \label{bifur-con-4b}
\frac{d u}{d X} (0) = \frac{d u}{d X} (L) = 0.
\end{equation}

The unbuckled column admits a principal solution $u_0(X) =0$ (see Fig.~\ref{bifur-4a}). We now linearize the above boundary-value problem \eqref{bifur-con-4a} and \eqref{bifur-con-4b}. Consider a small variation $\delta u = u^{\star}$. By using Eq.~\eqref{bifur-con-3}, the resulting incremental forms of Eqs.~\eqref{bifur-con-4a} and \eqref{bifur-con-4b} are
\begin{equation} \label{bifur-con-5a}
\frac{d^2 u^{\star}}{d X^2} + \lambda u^{\star} = 0, \quad 0 < X < L,
\end{equation}
and 
\begin{equation} \label{bifur-con-5b}
\frac{d u^{\star}}{d X} (0) = \frac{d u^{\star}}{d X} (L) = 0.
\end{equation}

Nonzero solutions of Eqs.~\eqref{bifur-con-5a} and \eqref{bifur-con-5b} are
\begin{equation} \label{bifur-con-5c}
u_n (X) = A_n \cos (\sqrt{\lambda_n} X) \qquad {\rm at} \ \lambda_n = \frac{n^2 \pi^2}{L^2},
\end{equation}
where $n \in \mathbf Z^*$ is a non-zero integer and $A_n$ is the amplitude. The bifurcation diagram and the first three modes of buckling are shown in Fig.~\ref{bifur-4b}.

\subsection{Closure on the basic concepts underlying stability and bifurcation} %
The issues of bifurcation in the equilibrium solutions of nonlinear solids and their stability are two important but different topics \cite{ericksen1956implications, hill1957uniqueness, ogden1997non, chen2003stability, chen2014stability}. Generally speaking, stability theory mostly studies the response of deformed bodies or structures under disturbance, while bifurcation theory mainly focuses on the non-uniqueness of solutions in solid mechanics. \\

These two different concepts and the underlying theories are linked. For example, the buckling of Euler's column corresponds to a bifurcation point at which the unbuckled deformation becomes unstable. Other examples of the relation between the stability and the bifurcation include an inflated elastic cylinder \cite{chen2003stability}, a soap film spanning a flexible loop \cite{chen2014stability}, and surface instability of a compressed half-space of elastic materials \cite{biot1963surface, chen2018surface}, etc. In the examples mentioned above, the bifurcation points and the unstable critical points coincide with each other. Such straightforward examples, however, are rather few, making a complete understanding of the connection between the two issues difficult. \\

Two landmark works in the mechanics community were carried out by Ericksen and Toupin \cite{ericksen1956implications} and Hill \cite{hill1957uniqueness} for the relation between stability and bifurcation. They showed that there was no bifurcation from the trivial solution if the corresponding state of the trivial solution was stable, i.e., stability implies uniqueness, but the converse does not follow.

\section{One-dimensional electrostatics of deformable dielectrics} \label{section-1D-continuum}
In this section, before proceeding to a 3D formulation, we present a 1D theory of electrostatics of deformable media. The objective is to establish the key ideas of electroelasticity in a simple manner without contending with the tedium of tensor algebra.

\subsection{Kinematics}
Kinematics in mechanics of deformable solids just describes the motion of a body {\it without} considering the force or the electric field or other fields that caused the motion \cite{gurtin1982introduction, gurtin2010mechanics, holzapfel2000nonlinear}. In this tutorial, we only focus on static systems. 

\subsubsection{Deformation of a body}
Consider a 1D body occupying the domain $\Omega_R^1 \subseteq \mathbb R$ in 1D Euclidean space. The 1D body can be identified with the occupied
domain in some fixed configuration, which we term the {\it reference configuration}. The domain in a reference configuration is denoted by $\Omega_R^1 \subseteq \mathbb R$, and the 1D body identified with $\Omega_R^1 = \{ X\in \mathbb R: X_a < X < X_b \}$ is called the reference
body. A point $X \in \Omega_R^1$ is called a {\it material point}. The boundary of $\Omega_R^1$ is
denoted by $\partial \Omega_R^1 = \{X=X_a \ \& \ X_b\}$. \\

Consider a smooth function $\chi: \Omega_R^1 \to \mathbb R $ that assigns to each material point $X \in \Omega_R^1$ a point
\begin{equation} \label{1d-spa-point}
x = \chi (X) = X + u (X) \in \Omega^1, 
\end{equation}
where $x$ is called the {\it spatial point} and $u(X)$ is the {\it displacement} that relates the material point $X \in \Omega_R^1$ and the spatial point $x \in \Omega^1$. \\

The function $\chi$ in Eq.~\eqref{1d-spa-point} is  {\it invertible} and there exists a unique function $\chi^{-1}: \Omega^1 \to \mathbb R$ that assigns to each spatial point $x \in \Omega^1$ a material point
\begin{equation} \label{1d-mater-point}
X = \chi^{-1} (x) \in \Omega_R^1.
\end{equation}

Mathematically, with the properties of inverse functions, each spatial point $x \in \Omega^1$ must correspond to no more than one material point $X \in \Omega_R$ and vice versa. Physically, a particle, denoted by $X \in \Omega_R^1$, cannot break into two particles, and two different particles, for example, $X_1, X_2 \in \Omega_R^1$, cannot occupy the same spatial position during deformation.

\subsubsection{Deformation gradient and strain}

The (scalar) deformation gradient of 1D deformation is defined by 
\begin{equation} \label{1d-def-grad}
F = \frac{\partial \chi (X)}{\partial X} = 1 + \frac{\partial u(X)}{\partial X} > 0,
\end{equation}
where ${\partial u(X)} / {\partial X}$ is the displacement gradient. The {\it inverse} of the deformation gradient is 
\begin{equation} \label{1d-def-inver}
F^{-1} = \frac{1}{F} = \frac{\partial \chi^{-1} (x)}{\partial x}.
\end{equation}

The inverse $F^{-1}$ in Eq.~\eqref{1d-def-inver} is related to the derivative of the inverse function that can be given by the inverse function theorem, such that $\big(f^{-1} \big)' (x) = 1 / f '(X)$, where the prime denotes the derivative. \\

By using Eqs.~\eqref{1d-spa-point}, \eqref{1d-def-grad} and \eqref{1d-def-inver}, the deformation gradient $F$ and the inverse $F^{-1}$ relate to the reference length element $d X$ and the current length element $d x$ as
\begin{equation} \label{dx-dX-rel}
dx = F dX \quad {\rm and} \quad dX = F^{-1} dx.
\end{equation}

In addition, the strain is often used to describe the deformation. Consider the change in the squared lengths $dx^2 - dX^2$. Here $dx^2 $ (or the expression $(d x)^2$) is the squared distance of two spatial points while $dX^2$ is the squared original distance of two material points. We may introduce two scalars $E_s$ and $e_s$,
namely
\begin{equation} \label{1d-dx2-dX2}
dx^2 - dX^2 = 2 E_s \, dX^2 = 2 e_s \, dx^2,
\end{equation}
where $E_s$ is the so-called Green strain and $e_s$ is the Euler strain. Substituting Eq.~\eqref{dx-dX-rel} into Eq.~\eqref{1d-dx2-dX2}, we have
\begin{equation} \label{1d-g-e-s}
E_s = \frac{1}{2} (F^2 -1) \quad {\rm and}  \quad  e_s = \frac{1}{2} (1 - F^{-2}).
\end{equation}
Alternatively, the strains are expressed in terms of the displacement gradient in Eq.~\eqref{1d-def-grad}:
\begin{equation} \label{1d-g-e-u}
E_s = \frac{\partial u}{\partial X} + \frac{1}{2} \Big ( \frac{\partial u}{\partial X} \Big)^2 \quad {\rm and} \quad e_s = \frac{1}{2} (1 - (1+ \frac{\partial u}{\partial X})^{-2}).
\end{equation}

In particular, for small deformation $\|\frac{\partial u}{\partial X}\| \ll 1$, the high order terms in Eq.~\eqref{1d-g-e-u} are omitted, then the Green and the Euler strains are equivalent, such that
\begin{equation} \label{1d-u-small-d}
E_s = e_s = \frac{\partial u}{\partial X}.
\end{equation}

\subsubsection{Differentiation and integration}
Differentiations in the reference and current configurations are also called the material and spatial derivatives. They can be connected by the deformation gradient $F$ in Eq.~\eqref{1d-def-grad}. The material and spatial derivatives are
\begin{equation} \label{1d-m-s-deri}
\frac{\partial }{\partial X} \quad {\rm and} \quad \frac{\partial }{\partial x}.
\end{equation}

Consider a smooth scalar field, which is represented by $\phi (x)$ in the current configuration ($\Omega^1$) and by $\phi_R (X)$ in the reference configuration ($\Omega_R^1$). With the chain-rule, we have the following relation
\begin{equation} \label{1d-G-g-rela}
\frac{\partial \phi_R (X)}{\partial X} = \frac{\partial \phi (x)}{\partial x} \frac{\partial \chi (X)}{\partial X} \Big |_{x=\chi(X)} = F \frac{\partial \phi(x)}{\partial x} \Big |_{x=\chi(X)} \quad {\rm or} \quad \frac{\partial \phi}{\partial X} = \frac{\partial \phi}{\partial x} \frac{\partial \chi}{\partial X} = F \frac{\partial \phi}{\partial x}.
\end{equation}
The latter one $\eqref{1d-G-g-rela}_2$ is more concise, where we have dropped the arguments and the subscript. \\

In addition, with the length element relation $dx = F dX$ in Eq.~\eqref{dx-dX-rel},
we have the following relation between the spatial and material integrals
\begin{equation} \label{1d-line-int}
\int_{\Omega^1} \phi(x) d x = \int_{\Omega_R^1} \phi_R (X) F(X) d X \quad {\rm or} \quad \int_{\Omega^1} \phi d x = \int_{\Omega_R^1} \phi F d X.
\end{equation}

\subsection{Maxwell's equations}

\subsubsection{Spatial representation}
Maxwell's equations of electrostatics govern the behavior of materials when subject to the presence of charges or potential difference. In 1D electrostatics, the Maxwell equations in the current configuration are (SI units convention)
\begin{equation} \label{1d-MW-eq-true}
E^1 = - \frac{\partial \xi}{\partial x}    \quad {\rm and} \quad \frac{\partial D^1}{\partial x} = \rho_{\rm f}^1,
\end{equation}
where $E^1$ is the (true) electric field, $\xi$ is the electric potential, $D^1$ is the (true) electric displacement field, and $\rho_{\rm f}^1$ is the (free) charge density (free charge per unit length in the current configuration). Free charges are those which exist and remain in the body regardless of any applied electrostatic boundary conditions. We remark that Maxwell's equations, as written in most physics and non-mechanics books, are always in the current configuration.  \\

The electric boundary conditions (voltage-controlled) on $x=x_a$ and $x=x_b$ are given by 
\begin{equation} \label{1d-eBCs-eq-true}
\xi (x_a) = \xi_a    \quad {\rm and} \quad \xi (x_b) = \xi_b.
\end{equation}

For physical reasons, we may introduce another field, the polarization of a body and is defined as: 
\begin{equation} \label{1d-DEP-rela}
D^1 = {\varepsilon_0} E^1 + P^1,
\end{equation}
where $\varepsilon_0$ is the vacuum permittivity and $P^1$ is the (true) polarization field. The vacuum does not polarize and therefore the polarization $P^1$ is zero in the vacuum and the relation \eqref{1d-DEP-rela} reduces to $D^1 = {\varepsilon_0} E^1$.\\

The electric constitutive relation in the current configuration that is most frequently used is the one that linearly relates the electric displacement to the electric field:

\begin{equation} \label{1d-e-cons-true}
D^1 = \varepsilon E^1,
\end{equation}
where $\varepsilon$ is the permittivity of the material. While the relation in vacuum is expected to be exact, the above equation is a constitutive \emph{choice} and is applicable to only linear dielectrics.\\

A direct consequence of Eqs.~\eqref{1d-DEP-rela} and \eqref{1d-e-cons-true} for linear dielectrics is
\begin{equation} \label{1d-PE-r-rela}
P^1 = (\varepsilon - \varepsilon_0) E^1.
\end{equation}


\subsubsection{Material representation}

Since we have to couple elasticity with electrostatics, we often need to invoke Maxwell's equations in the reference configuration also. They are (SI units convention)
\begin{equation} \label{1d-MW-eq-refs}
\tilde{E}^1 = - \frac{\partial \xi}{\partial X}    \quad {\rm and} \quad    \frac{\partial \tilde{D}^1}{\partial X} =\tilde\rho_{\rm f}^1,
\end{equation}
where $\tilde E^1$ is the (nominal) electric field, $\xi$ is the electric potential, $\tilde{D}^1$ is the (nominal) electric displacement, and $\tilde \rho_{\rm f}^1$ is the (free) charge density (free charge per unit length in the reference configuration). \\

The electric boundary conditions (voltage-controlled) on $X=X_a$ and $X=X_b$ are given by 
\begin{equation} \label{1d-eBCs-eq-ref}
\xi (X_a) =\xi_a  \quad {\rm and} \quad \xi (X_b) = \xi_b.
\end{equation}

Recalling the relation \eqref{1d-G-g-rela}, the nominal electric field $\tilde{E}^1$ is related to the true electric field $E^1$ by
\begin{equation} \label{1d-MW-rela-ref-cur}
\tilde{E}^1 = - \frac{\partial \xi}{\partial X} = - \frac{\partial \xi}{\partial x} \frac{\partial x}{\partial X} = E^1 F.
\end{equation}

The nominal electric displacement field in the reference configuration is equal to the true electric displacement field, namely\footnote{Actually, the definition \eqref{1d-MW-D=d} comes from the conservation of the charge during the deformation, and a detailed derivation will be given in the general 3D formulation later on. Here we can show that $\int_{x_a}^{x_b}\frac{\partial D^1}{\partial x} dx = \int_{x_a}^{x_b} \rho^1 dx = \int_{X_a}^{X_b} \tilde\rho^1 dX = \int_{X_a}^{X_b} \frac{\partial \tilde{D}^1}{\partial X} dX = \int_{X_a}^{X_b} \frac{\partial \tilde{D}^1}{\partial x} \frac{\partial x}{\partial X} dX$. For linear dielectrics, a direct consequence of the definition \eqref{1d-MW-D=d} is $\tilde D^1 = D^1 = \varepsilon E^1 = \varepsilon F^{-1} \tilde{E}^1$. }
\begin{equation} \label{1d-MW-D=d}
\tilde D^1 = D^1,
\end{equation}
which leads to the relation 
\begin{equation} \label{1d-MW-D=E-refs}
\tilde D^1 = \varepsilon F^{-1} \tilde{E}^1.
\end{equation}

\subsection{Balance of forces}%
The contact forces on the boundaries $x=x_a$ and $x=x_b$ in 1D formulation are
\begin{equation} \label{1d-force-r-c}
t^1 (x_a) = t_{x_a}^1 \quad {\rm and} \quad  t^1 (x_b) = t_{x_b}^1.
\end{equation}
In 1D statics, balance of forces in the deformed body $\Omega^1 = \{ x \in \mathbb R: x_a < x < x_b \}$ is
\begin{equation} \label{1d-balance-f}
t_{x_b}^1 - t_{x_a}^1 + \int_{x_a}^{x_b} b^1 (x) dx = 0,
\end{equation}
where $b^1 (x)$ is the body force density (per unit length) in the current configuration. The smooth function $t^1: \Omega^1 \to \mathbb R$ can be interpreted as the force (or the stress) on the 1D body. We define that $t^1(x) > 0$ corresponds to an extended (spatial) point $x$ while $t^1(x) < 0$ corresponds to a compressed (spatial) point $x$. Using integration by parts, an alternative form of Eq.~\eqref{1d-balance-f} is
\begin{equation} \label{1d-balance-f-1}
 \int_{x_a}^{x_b} \Big( \frac{\partial t^1(x)}{\partial x}+ b^1 (x) \Big) dx = 0.
\end{equation}

Since Eq.~\eqref{1d-balance-f-1} holds for any $\Omega^1 = \{ x \in \mathbb R: x_a < x < x_b \}$, differential form or local form of balance of forces is given by
\begin{equation} \label{1d-bal-f-local}
\frac{\partial t^1(x)}{\partial x}+ b^1 (x) = 0.
\end{equation}

With the relation \eqref{1d-line-int}, the integral of the body force in Eq.~\eqref{1d-balance-f} can be written in the reference configuration as
\begin{equation} \label{1d-body-f-rel}
\int_{X_a}^{X_b} b_R^1 (X) F(X) dX = \int_{x_a}^{x_b} b^1 (x) dx \quad {\rm or} \quad \int_{X_a}^{X_b} b_R^1 F dX = \int_{x_a}^{x_b} b^1 dx.
\end{equation}

With the traction boundary conditions
\begin{equation} \label{1d-force-r-c-refs}
t_R^1 (X_a) = t_{X_a}^1 \quad {\rm and} \quad  t_R^1 (X_b) = t_{X_b}^1,
\end{equation}
together with Eqs.~\eqref{1d-balance-f} and \eqref{1d-body-f-rel}, the global form of balance of forces in the reference configuration becomes
\begin{equation} \label{1d-bal-ref}
t_{X_b}^1 - t_{X_a}^1 + \int_{X_a}^{X_b} b_R^1 (X) F dX = 0.
\end{equation}
By using integration by parts, similar to the procedure used in Eqs.~\eqref{1d-balance-f-1} and \eqref{1d-bal-f-local}, we have the local form of balance of forces in the reference configuration
\begin{equation} \label{1d-bal-local-ref}
\frac{\partial t^1(X)}{\partial X}+ b_0^1 (X) = 0,
\end{equation}
where the subscript `$R$' is dropped and $b_0^1 (X) = b^1 (X) F$ is the body force (density) in the reference configuration. \\

It is important to emphasize that the forces (or stresses) $t_R^1$ and $t^1$ in the two configurations are the same. In contrast, the body forces $b_0^1$ and $b_1$ are related by $b_0^1 = b^1 F$. In the way we have presented the 1D formulation, there is no change of the cross-sectional area, but the length elements are related by $dx = F dX$.

\subsection{Constitutive equations}

In electroelasticity, the stress $t^1$ can be decomposed into two parts:
\begin{equation} \label{1d-total-stress}
t^1 = t^{e} + t^{M},
\end{equation}
where $t^e$ is the elastic stress and $t^M$ is the so-called Maxwell stress (to be defined shortly). Note that the nominal and true stresses are the same in 1D continuum mechanics. The elastic stress is a function of the deformation gradient:
\begin{equation} \label{1d-elastic-stress}
t^e = f^1 (F).
\end{equation}
In one widely used constitutive choice (linear Hooke's law), the elastic stress can be represented by
\begin{equation} \label{1d-elastic-stress-102601}
t^e = C (F-1) = C \frac{\partial u} {\partial X},
\end{equation}
where $C$ is the elastic modulus for small deformation. \\

The divergence of the Maxwell stress can be regarded as the electric body force. The readers can refer to the works \cite{mcmeeking2005electrostatic, bustamante2009electric, dorfmann2017nonlinear} for further discussion of other definitions of the Maxwell stress. The electric body force comes from the Lorentz force. For a charge distribution $\rho_{\rm f}^1$ in an electric field $E^1$, the Lorentz force per unit volume is equal to $\rho_{\rm f}^1 E^1$ in the electrostatics. By the identification,  ${\partial t^M}/{\partial x} = \rho_{\rm f}^1 E^1$, i.e., the divergence of the Maxwell stress can be related to the electric body force, the 1D Maxwell stress of linear soft dielectrics in the current configuration as\footnote{While it is common to introduce the Maxwell stress in this manner (as evident from many works), we believe that its emergence is most natural and clear by first constructing a suitable energy functional and then using a variational principle. The reader is referred to the works \cite{liu2013energy, liu2014energy} of Liu where this is elaborated further.}
\begin{equation} \label{1d-MW-stress}
t^M = \frac{\varepsilon}{2} (E^1)^2.
\end{equation}

It follows from Eqs.~\eqref{1d-MW-eq-true}, \eqref{1d-e-cons-true} and \eqref{1d-MW-stress} that ${\partial t^M}/{\partial x} = \varepsilon E^1 {\partial E^1}/{\partial x} = E^1 {\partial D^1}/{\partial x} = \rho_{\rm f}^1 E^1$. A more general derivation of the Maxwell stress in 3D from the balance of force is shown in Sec.~\ref{sec-maxwell stress-4.5.3}. By Eqs.~\eqref{1d-MW-rela-ref-cur} and \eqref{1d-MW-D=E-refs}, the 1D Maxwell stress \eqref{1d-MW-stress} can also be written as either $t^M = \frac{\varepsilon}{2} (F^{-1} \tilde E^1)^2$ or $t^M = \frac{1}{2 \varepsilon} (\tilde D^1)^2$. Thus, the total stress $t^1$ in Eq.~\eqref{1d-total-stress} can be written as
\begin{equation} \label{1d-total-stress-expli}
t^1 = f^1 (F) + \frac{\varepsilon}{2} (F^{-1} \tilde E^1)^2.
\end{equation}

\subsection{Summary of the boundary-value problem} \label{section3.5-solution}
The boundary-value problem of the electrostatics of deformable dielectrics in the reference configuration can be compactly summarized as:
\begin{equation} \label{1d-total-bvp-refs}
\left.
\begin{aligned}
& \underbrace{ \frac{\partial t^1(X)}{\partial X}+ b_0^1 (X) = 0}_{\text{equilibrium equation}}, \quad  \underbrace{\tilde{E}^1 = - \frac{\partial \xi}{\partial X},  \ \frac{\partial \tilde{D}^1}{\partial X} =\tilde\rho_{\rm f}^1}_{\text{Maxwell's equations}}, \quad \underbrace{\tilde D^1 = \varepsilon F^{-1} \tilde{E}^1}_{\text{relation (linear dielectrics)}}, \\
& \underbrace{ t_R^1 (X_a) = t_{X_a}^1, \ t_R^1 (X_b) = t_{X_b}^1}_{\text{mechanical BCs}}, \quad \underbrace{\xi (X_a) = \xi_a, \ \xi (X_b) = \xi_b}_{\text{electric BCs}}, \\
& \underbrace{t^1 = f^1 (F) + \frac{\varepsilon}{2} (F^{-1} \tilde E^1)^2}_{\text{constitutive equation}}, \quad \underbrace{F = 1 + \frac{\partial u(X)}{\partial X}}_{\text{kinematics}}.
\end{aligned}
\right\}
\end{equation}

In the above representation, we have chosen the constitutive law representative of a linear dielectric, and the nominal electric field is our state variable (in addition to the deformation). Other state variables can also be used in the formulation, and, based on the review by Liu \cite{liu2014energy}, we will elaborate that later on in the paper when we discuss the 3D formulation. For the sake of brevity, we have only emphasized the force-traction and voltage-controlled boundary conditions in Eq.~\eqref{1d-total-bvp-refs}. \\

In contrast to Eq.~\eqref{1d-total-bvp-refs}, the boundary-value problem in the current configuration may be summarized as:
\begin{equation} \label{1d-total-bvp-current}
\left.
\begin{aligned}
& \underbrace{ \frac{\partial t^1(x)}{\partial x}+ b^1 (x) = 0}_{\text{equilibrium equation}}, \quad  \underbrace{E^1 = - \frac{\partial \xi}{\partial x},  \ \frac{\partial D^1}{\partial x} =\rho_{\rm f}^1}_{\text{Maxwell's equations}}, \quad \underbrace{D^1 = \varepsilon E^1}_{\text{relation (linear dielectrics)}}, \\
& \underbrace{ t^1 (x_a) = t_{x_a}^1, \ t^1 (x_b) = t_{x_b}^1}_{\text{mechanical BCs}}, \quad \underbrace{\xi (x_a) = \xi_a, \ \xi (x_b) = \xi_b}_{\text{electric BCs}}, \\
& \underbrace{t^1 = f^1 (e_s) + \frac{\varepsilon}{2} (E^1)^2}_{\text{constitutive equation}}, \quad \underbrace{e_s = \frac{1}{2} (1 - (1+ \frac{\partial u}{\partial X})^{-2})\big|_{X=\chi^{-1}(x)}}_{\text{Euler strain}}.
\end{aligned}
\right\}
\end{equation}

\subsection{Incremental formulation and bifurcation analysis} \label{1d-incremental-section}
Of interest here is the bifurcation analysis of the electromechanical behaviour of 1D deformable dielectrics. We focus on the onset of bifurcation from a given trivial solution to other solutions of the boundary-value problem in Sec.~\ref{section3.5-solution}. Based on the implicit function theorem in Sec.~\ref{subsection-bifurcation}, the equations of equilibrium have a non-trivial solution bifurcating from the trivial solution only if their linearized forms possess a non-zero solution. The linearized equations describe the response of 1D deformable dielectrics, in a state of equilibrium, to infinitesimal increments of the deformation and the electric field (i.e., either the electric field or the electric displacement). We now linearize the above boundary-value problem \eqref{1d-total-bvp-refs}. \\

The infinitesimal increment of the deformation $\chi$ is denoted by $u^{\star} (X)$, i.e., $\chi (X) \to \chi (X) + u^{\star} (X)$, with sufficiently small $\Vert u^{\star} (X) \Vert$ and $\Vert \frac{\partial}{\partial X} u^{\star} (X) \Vert$. Consider the variation of the deformation gradient $F= \frac{\partial \chi}{\partial X}$. By the definition \eqref{bifur-con-2a}, the infinitesimal increment of the deformation gradient is
\begin{equation} \label{1d-incre-form-000}
F^{\star} = \left. \frac{d}{d \tau} \left( \frac{\partial}{\partial X} (\chi(X) + \tau u^\star (X))\right) \right|_{\tau=0} = \frac{\partial u^{\star}}{\partial X}.
\end{equation}

It can be understood that the consequence of the change $\chi (X) \to \chi (X) + u^{\star} (X)$ is $F \to F + F^\star$. Similarly, the infinitesimal increment of the inverse $F^{-1}$ is
\begin{equation} \label{1d-incre-form-001}
(F^{-1})^{\star} = \left. \frac{d}{d \tau} \left( \frac{\partial}{\partial X} (\chi(X) + \tau u^\star (X))\right)^{-1} \right|_{\tau=0} = - \left( \frac{\partial \chi(X)}{\partial X}\right)^{-2} \frac{\partial u^{\star}}{\partial X} = - F^{-2} F^{\star}.
\end{equation}

By the variation $F \to F + F^\star$ with sufficiently small $\| F^\star \|$, the infinitesimal increment $(F^{-1})^{\star}$ can be obtained as follows: $(F^{-1})^{\star} = \left. \frac{d}{d \tau} \left( F + \tau F^\star\right)^{-1} \right|_{\tau=0} = - F^{-2} F^{\star}$. One can also take the following derivation. Consider the identity $F F^{-1} = 1$. Suppose that the deformation is changed from $\chi (X)$ to $\chi(X) + \tau u^\star (X)$. Differentiating both sides of the identity by $\tau$ at $\tau=0$, we obtain $F^\star F^{-1} + F (F^{-1})^\star =0$, and, hence, that $(F^{-1})^\star = - F^{-2} F^\star$.\\

The infinitesimal increment of the electric potential $\xi$ is denoted by $\xi^{\star} (X)$, i.e., $\xi (X) \to \xi (X) + \xi^{\star} (X)$, with sufficiently small $\Vert \xi^{\star} (X) \Vert$ and $\Vert \frac{\partial}{\partial X} \xi^{\star} (X) \Vert$. The infinitesimal increment of the nominal electric field $\tilde E^1$ in Eq.~\eqref{1d-MW-eq-refs} is expressed as
\begin{equation} \label{1d-incre-form-1001}
\tilde{E}^{\star} = \left. \frac{d}{d \tau} \left( - \frac{\partial}{\partial X} (\xi + \tau \xi^{\star}) \right)\right|_{\tau=0} = - \frac{\partial \xi^{\star}}{\partial X}.
\end{equation}

The infinitesimal increment $\tilde{D}^{\star}$ of the nominal electric displacement $\tilde D^1$ in Eq.~\eqref{1d-MW-D=E-refs} is\footnote{The detailed derivation is as follows: $\tilde{D}^{\star} = (\varepsilon F^{-1} \tilde{E}^1 )^\star = \left.\frac{d}{d\tau} \left\{\varepsilon \left( \frac{\partial}{\partial X} (\chi + \tau u^\star)\right)^{-1} \cdot \left( - \frac{\partial}{\partial X} (\xi + \tau \xi^{\star}) \right) \right\}\right|_{\tau=0} = \left. \varepsilon \left\{\left[\frac{d}{d\tau} \left( \frac{\partial}{\partial X} (\chi + \tau u^\star)\right)^{-1} \right] \cdot \left( - \frac{\partial}{\partial X} (\xi + \tau \xi^{\star}) \right) \right\}\right|_{\tau=0} + \left. \varepsilon \left\{ \left( \frac{\partial}{\partial X} (\chi + \tau u^\star)\right)^{-1} \cdot \frac{d}{d\tau} \left( - \frac{\partial}{\partial X} (\xi + \tau \xi^{\star}) \right) \right\}\right|_{\tau=0} = \varepsilon \left[-\left( \frac{\partial \chi}{\partial X} \right)^{-2} \frac{\partial u^\star}{\partial X}\right] \left( - \frac{\partial \xi}{\partial X} \right)   +   \varepsilon \left( \frac{\partial \chi}{\partial X}\right)^{-1} \left( - \frac{\partial \xi^\star}{\partial X} \right)  =   \varepsilon (F^{-1})^{\star} \tilde{E}^1 + \varepsilon F^{-1} \tilde{E}^{\star}$. One can also take the following derivation. By the variations $F^{-1} \to F^{-1} + (F^{-1})^\star$ and $\tilde{E}^1 \to \tilde{E}^1 + \tilde{E}^{\star}$, we have $\tilde{D}^{\star} = \left. \frac{d}{d \tau} [\varepsilon (F^{-1} + \tau (F^{-1})^\star) (\tilde{E}^1 + \tau\tilde{E}^{\star})] \right|_{\tau=0} = \varepsilon (F^{-1})^{\star} \tilde{E}^1 + \varepsilon F^{-1} \tilde{E}^{\star}$.}
\begin{equation} \label{1d-incre-form-2}
\tilde{D}^{\star} = (\varepsilon F^{-1} \tilde{E}^1 )^\star = \varepsilon (F^{-1})^{\star} \tilde{E}^1 + \varepsilon F^{-1} \tilde{E}^{\star} = \varepsilon F^{-1} (- F^{-1} \tilde{E}^1 F^{\star} + \tilde{E}^{\star}).
\end{equation}

The infinitesimal increment $t^\star$ of the total stress $t^1$ in \eqref{1d-total-stress-expli} is
\begin{equation}  \label{1d-incre-form-3}
t^\star = \frac{\partial f^1 (F)}{\partial F} F^\star + (F^{-1} \tilde E^1) \tilde {D}^\star.
\end{equation}

Now the incremental version of the boundary-value problem \eqref{1d-total-bvp-refs} can be represented by
\begin{equation} \label{1d-incre-form-4}
\frac{\partial t^\star (X)}{\partial X} = 0, \quad \frac{\partial \tilde{D}^\star}{\partial X} =0, \quad \underbrace{ t^\star (X_a) = t^\star (X_b) = 0, \ \xi^\star (X_a) = \xi^\star (X_b) = 0}_{\text{incremental BCs}},
\end{equation}
along with Eqs.~\eqref{1d-incre-form-000}$-$\eqref{1d-incre-form-3}.

\subsection{An energy formulation of the electrostatics of deformable dielectrics}
As opposed to a ``stress centric" approach described in the preceding section, as done in purely mechanical continuum mechanics, we may also start by formulating free energy and then invoke a suitable variational principle to arrive at the theoretical formulation. In the energy formulation, the deformation gradient $F$ can be chosen as one state variable related to the mechanical field. The other state variable related to the electric fields can be chosen to be either the nominal electric field ${\tilde E}^1$, the nominal electric displacement ${\tilde D}^1$, or the nominal polarization ${\tilde P}^1$. For a rigorous formulation that presents a relative comparison of the different formulations may be found in Liu \cite{liu2014energy}. To proceed in the 1D formulation and simplify the discussion, we will assume the cross-section of the 1D structure as a constant (unit area) that is independent of the deformation.

\subsubsection{Formulation in terms of the nominal electric field ${\tilde E}^1$}
Consider a smooth mapping $\chi: \Omega_R^1 \to \mathbb R $. We use $(\chi, {\tilde E}^1)$ as the independent variables to formulate the 1D theory of electrostatics. By the identity ${\tilde E}^1 = - \partial \xi/\partial X$ and the electric and mechanical boundary conditions
\begin{equation} \label{1d-energy-form-d-e-2a}
\xi (X_a)= \xi_a, \quad \xi (X_b)= \xi_b, \quad t_R^1 (X_a) = t_{X_a}^1, \quad t_R^1 (X_b) = t_{X_b}^1,
\end{equation}
we define the energy functional of the 1D system is defined as
\begin{equation} \label{1d-energy-form-d-e-1}
F_e [\chi, \xi] = \int_{X_a}^{X_b} W^1 (F, {\tilde E}^1) - \int_{X_a}^{X_b} b_0^1 \cdot \chi - (t_R^1 \cdot \chi) \Big|_{X_a}^{X_b},
\end{equation}
where $b_0^1$ is the body force and $t_R^1$ is the traction force. The equilibrium state is then determined by the variational principle \cite{liu2014energy}
\begin{equation} \label{1d-energy-form-d-e-2}
\min_{\chi} \max_{\xi} F_e [\chi, \xi].
\end{equation}

\noindent {$\bullet$ \it Variation of the electric field}\\
Let us suppose that ${\tilde E}^1$ is the function required to make the free energy $F_e$ stationary. Consider the variation ${\tilde E}^1 \to {\tilde E}^1 + \tau {\tilde E}_1^1$. Here $\tau$ is a sufficiently small parameter and ${\tilde E}_1^1$ is an arbitrary function that satisfies all the kinematically admissible deformations. The variation of the electric potential $\xi$ related to the electric reads $\xi \to \xi + \tau \xi_1$, where the arbitrary potential function $\xi_1$ equals zero at both end-points, i.e., $\xi_1 (X_a) = \xi_1 (X_b)= 0$. \\

The first variation of the energy functional \eqref{1d-energy-form-d-e-1} with respect to the electric potential admits
\begin{equation} \label{1d-energy-form-d-e-3}
\left.\frac{\partial F_e [\chi, \xi + \tau \xi_1]}{\partial \tau}\right|_{\tau=0} = \int_{X_a}^{X_b} \frac{\partial W^1}{\partial {\tilde E}^1} \frac{\partial \xi_1}{\partial X} = \left. \left(\xi_1 \frac{\partial W^1}{\partial {\tilde E}^1}\right) \right|_{X_a}^{X_b} - \int_{X_a}^{X_b} \xi_1 \frac{\partial}{\partial X} \left(\frac{\partial W^1}{\partial {\tilde E}^1}\right) = 0.
\end{equation}
By $\xi_1 (X_a)= 0$ and $\xi_1 (X_b)= 0$ and the standard lemma of the calculus of variations, Eq.~\eqref{1d-energy-form-d-e-3} gives
\begin{equation} \label{1d-energy-form-d-e-4}
\frac{\partial}{\partial X} \left(\frac{\partial W^1}{\partial {\tilde E}^1}\right) = 0.
\end{equation}

\noindent {$\bullet$ \it Variation of the deformation}\\
Let us suppose that the deformation $\chi$ makes the free energy $F_e$ stationary. Consider the variation $\chi \to \chi + \tau \chi_1$. Here $\chi_1$ is arbitrary and satisfies all the kinematically admissible deformations. The electric quantities are assumed to be independent of the variation of the deformation. Then the zero first variation of the energy functional \eqref{1d-energy-form-d-e-1} with respect to the deformation reads
\begin{equation} \label{1d-energy-form-d-e-5}
\left.\frac{\partial F_e [\chi + \tau \chi_1, \xi]}{\partial \tau}\right|_{\tau=0} = \int_{X_a}^{X_b} \frac{\partial W^1}{\partial F} \frac{\partial \chi_1}{\partial X} - \int_{X_a}^{X_b} b_0^1 \cdot \chi_1 - (t_R^1 \cdot \chi_1) \Big|_{X_a}^{X_b} = 0,
\end{equation}
which, with integration by parts and the lemma of the calculus of variations, gives the equilibrium equation
\begin{equation} \label{1d-energy-form-d-e-6}
\frac{\partial}{\partial X} \left(\frac{\partial W^1}{\partial F}\right) + b_0^1= 0
\end{equation}
and the natural (traction) boundary conditions
\begin{equation} \label{1d-energy-form-d-e-7}
\frac{\partial W^1}{\partial F} = t_R^1 \quad {\rm at} \ X= X_a \, \& \, X_b.
\end{equation}
We may further partition the energy function $W^1 (F, {\tilde E}^1)$ of a dielectric elastomer as: 
\begin{equation} \label{1d-energy-form-d-e-8}
W^1 (F, {\tilde E}^1) = W_{\rm elast}^1 (F) - \frac{\varepsilon}{2} F^{-1}({\tilde E}^1)^2.
\end{equation}
It follows from Eqs.~\eqref{1d-energy-form-d-e-4} and \eqref{1d-energy-form-d-e-8} that
\begin{equation} \label{1d-energy-form-d-e-9}
\frac{\partial}{\partial X} \left(\varepsilon F^{-1} {\tilde E}^1\right) = 0,
\end{equation}
which, by Eq.~\eqref{1d-MW-D=E-refs}, is the Maxwell equation $\eqref{1d-MW-eq-refs}_2$ in the absence of free charges. By Eq.~\eqref{1d-energy-form-d-e-8}, the total stress is
\begin{equation} \label{1d-energy-form-d-e-10}
\frac{\partial W^1}{\partial F} = \frac{\partial W_{\rm elast}^1}{\partial F} + \frac{\varepsilon}{2} (F^{-1}{\tilde E}^1)^2 = f^1 (F) +  \frac{\varepsilon}{2} (F^{-1}{\tilde E}^1)^2,
\end{equation}
which is exactly the total stress shown in Eq.~\eqref{1d-total-stress-expli}. A suitable mechanical constitutive law may be chosen by specifying $W_{\rm elast}^1$.

\subsubsection{Formulation in terms of the nominal polarization ${\tilde P}^1$}
We once again follow Liu \cite{liu2014energy}. Consider a smooth mapping $\chi: \Omega_R^1 \to \mathbb R $. We use $(\chi, {\tilde P}^1)$ as the independent variables to formulate the 1D theory of electrostatics. By using the identity ${\tilde E}^1 = - \partial \xi/\partial X$, the 1D Maxwell equation $\eqref{1d-MW-eq-refs}_2$ in the reference configuration in the absence of free charges is
\begin{equation} \label{1d-energy-form-d-p-1}
\frac{\partial}{\partial X} \left( - \varepsilon_0 F^{-1} \frac{\partial \xi}{\partial X} + F^{-1} {\tilde P}^1\right) = 0.
\end{equation}
The electric and mechanical boundary conditions are
\begin{equation}  \label{1d-energy-form-d-p-2}
\xi (X_a)= \xi_a, \quad \xi (X_b)= \xi_b, \quad t_R^1 (X_a) = t_{X_a}^1, \quad t_R^1 (X_b) = t_{X_b}^1.
\end{equation}

The free energy of the electrostatic system is
\begin{equation}  \label{1d-energy-form-d-p-3}
F_p [\chi, {\tilde P}^1]=  F^{\rm mech} [\chi] + F^{\rm elct} [\chi, {\tilde P}^1].
\end{equation}
We claim that the equilibrium state is determined by the variational principle \cite{liu2014energy}
\begin{equation} \label{1d-energy-form-d-p-3a}
\min_{\chi} \min_{{\tilde P}^1} F_p [\chi, {\tilde P}^1]
\end{equation}
subjected to the Maxwell equation \eqref{1d-energy-form-d-p-1} (which can be treated as a constraint) and the boundary conditions \eqref{1d-energy-form-d-p-2}. The mechanical part $F^{\rm mech} [\chi]$ in Eq.~\eqref{1d-energy-form-d-p-3} can be written as
\begin{equation}  \label{1d-energy-form-d-p-4}
F^{\rm mech} [\chi] = \int_{X_a}^{X_b} W_{\rm elast}^1 (F) - \int_{X_a}^{X_b} b_0^1 \cdot \chi - (t_R^1 \cdot \chi) \Big|_{X_a}^{X_b}
\end{equation}
while the electric part $F^{\rm elct} [\chi, {\tilde P}^1]$ for linear dielectric elastomers is 
\begin{equation}  \label{1d-energy-form-d-p-5}
F^{\rm elct} [\chi, {\tilde P}^1] = \frac{\varepsilon_0}{2} \int_{X_a}^{X_b} F^{-1} \left(\frac{\partial \xi}{\partial X}\right)^2 + \left.\left[ \xi \left( - \varepsilon_0 F^{-1} \frac{\partial \xi}{\partial X} + F^{-1} {\tilde P}^1\right) \right] \right|_{X_a}^{X_b} + \int_{X_a}^{X_b} \frac{F^{-1} ({\tilde P}^1)^2}{2 (\varepsilon - \varepsilon_0)}.
\end{equation}

\noindent {$\bullet$ \it Variation of the polarization}
Suppose that ${\tilde P}^1$ is the function required to make the free energy $F_p$ stationary. Consider the variation ${\tilde P}^1 \to {\tilde P}^1 + \tau {\tilde P}_1^1$. Here $\tau$ is a sufficiently small parameter and ${\tilde P}_1^1$ is an arbitrary function that satisfies all the kinematically admissible deformations. The variation of the electric potential $\xi$ related to the polarization reads $\xi \to \xi + \tau \xi_1$, where the arbitrary potential function $\xi_1$ equals zero at both end-points. The variation of the 1D Maxwell equation \eqref{1d-energy-form-d-p-1} reads
\begin{equation} \label{1d-energy-form-d-p-9-20-1}
\frac{\partial}{\partial X} \left( - \varepsilon_0 F^{-1} \frac{\partial \xi_1}{\partial X} + F^{-1} {\tilde P}_1^1\right) = 0.
\end{equation}

The first variation of the free energy \eqref{1d-energy-form-d-p-3} with respect to the polarization is
\begin{equation}  \label{1d-energy-form-d-p-6}
\begin{aligned}
\left.\frac{\partial F_p [\chi, {\tilde P}^1+ \tau {\tilde P}_1^1]}{\partial \tau}\right|_{\tau=0} & = \varepsilon_0 \int_{X_a}^{X_b} F^{-1} \left(\frac{\partial \xi}{\partial X}\right) \left(\frac{\partial \xi_1}{\partial X}\right) + \int_{X_a}^{X_b} \frac{F^{-1}{\tilde P}^1 {\tilde P}_1^1}{\varepsilon - \varepsilon_0} \\
& \quad + \left.\left[ \xi_1 \left( - \varepsilon_0 F^{-1} \frac{\partial \xi}{\partial X} + F^{-1} {\tilde P}^1\right) + \xi \left( - \varepsilon_0 F^{-1} \frac{\partial \xi_1}{\partial X} + F^{-1} {\tilde P}_1^1\right)\right]\right|_{X_a}^{X_b}.
\end{aligned}
\end{equation}
By $\xi_1 (X_a) = \xi_1 (X_b)= 0$ and Eq.~\eqref{1d-energy-form-d-p-9-20-1}, the third term on the RHS of Eq.~\eqref{1d-energy-form-d-p-6} can be written as
\begin{equation}  \label{1d-energy-form-d-p-9-20-2}
\begin{aligned}
\left. \left[ \xi \left( - \varepsilon_0 F^{-1} \frac{\partial \xi_1}{\partial X} + F^{-1} {\tilde P}_1^1\right) \right] \right|_{X_a}^{X_b}
& = \int_{X_a}^{X_b} \frac{\partial}{\partial X} \left[ \xi \left( - \varepsilon_0 F^{-1} \frac{\partial \xi_1}{\partial X} + F^{-1} {\tilde P}_1^1\right) \right] \\
&= \int_{X_a}^{X_b} \left( - \varepsilon_0 F^{-1}  \frac{\partial \xi}{\partial X}  \frac{\partial \xi_1}{\partial X} + F^{-1}  \frac{\partial \xi}{\partial X} {\tilde P}_1^1\right).
\end{aligned}
\end{equation}
Thus, the zero first variation \eqref{1d-energy-form-d-p-6} is reformulated as
\begin{equation}  \label{1d-energy-form-d-p-9-20-3}
\left.\frac{\partial F_p [\chi, {\tilde P}^1+ \tau {\tilde P}_1^1]}{\partial \tau}\right|_{\tau=0} = \int_{X_a}^{X_b}\left( \frac{\partial \xi}{\partial X} + \frac{{\tilde P}^1}{\varepsilon - \varepsilon_0} \right) F^{-1} {\tilde P}_1^1 = 0.
\end{equation}
Since Eq.~\eqref{1d-energy-form-d-p-9-20-3} must be satisfied for arbitrary ${\tilde P}_1^1$, it is easy to see that we require
\begin{equation}  \label{1d-energy-form-d-p-7}
{\tilde P}^1 = -(\varepsilon -\varepsilon_0) \frac{\partial \xi}{\partial X}.
\end{equation}
A consequence of Eqs.~\eqref{1d-energy-form-d-p-1} and \eqref{1d-energy-form-d-p-7} is
\begin{equation} \label{1d-energy-form-d-p-8}
\frac{\partial}{\partial X} \left( - \varepsilon F^{-1} \frac{\partial \xi}{\partial X}\right) = 0,
\end{equation}
which is the same as Eq.~\eqref{1d-energy-form-d-e-9}. By substituting Eq.~\eqref{1d-energy-form-d-p-7} into Eq.~\eqref{1d-energy-form-d-p-5} and using integration by parts and Eq.~\eqref{1d-energy-form-d-p-8}, we can show that
\begin{equation}  \label{1d-energy-form-d-p-9}
F^{\rm elct} [\chi, {\tilde P}^1] = - \int_{X_a}^{X_b} \frac{\varepsilon}{2} F^{-1} \left(\frac{\partial \xi}{\partial X}\right)^2.
\end{equation}
It follows from Eqs.~\eqref{1d-energy-form-d-p-4} and \eqref{1d-energy-form-d-p-9} that the free energy \eqref{1d-energy-form-d-p-3} can then be written as
\begin{equation}  \label{1d-energy-form-d-p-10}
F_p [\chi, {\tilde P}^1] = \int_{X_a}^{X_b} \left( W_{\rm elast}^1 (F) - \frac{\varepsilon}{2} F^{-1} \left(\frac{\partial \xi}{\partial X}\right)^2 \right) - \int_{X_a}^{X_b} b_0^1 \cdot \chi - (t_R^1 \cdot \chi) \Big|_{X_a}^{X_b}.
\end{equation}

\noindent {$\bullet$ \it Variation of the deformation}
Suppose that the deformation $\chi$ makes the free energy $F_p$ stationary. Consider the variation $\chi \to \chi + \tau \chi_1$. Here $\chi_1$ is arbitrary and satisfies all the kinematically admissible deformations. The electric quantities are assumed to be independent of the variation of the deformation. Then the zero first variation of the free energy \eqref{1d-energy-form-d-p-10} with respect to the deformation is
\begin{equation}  \label{1d-energy-form-d-p-11}
\begin{aligned}
0 & =\left.\frac{\partial F_p [\chi + \tau \chi_1, {\tilde P}^1]}{\partial \tau}\right|_{\tau=0} \\
& = \int_{X_a}^{X_b} \left( \frac{\partial W_{\rm elast}^1}{\partial F} + t^M \right)\frac{\partial \chi_1}{\partial X} - \int_{X_a}^{X_b} b_0^1 \cdot \chi_1 - (t_R^1 \cdot \chi_1) \Big|_{X_a}^{X_b}\\
& =\left. \left[\left( \frac{\partial W_{\rm elast}^1}{\partial F} + t^M - t_R^1\right) \cdot \chi_1 \right] \right|_{X_a}^{X_b} - \int_{X_a}^{X_b} \left[ \frac{\partial}{\partial X} \left( \frac{\partial W_{\rm elast}^1}{\partial F} + t^M \right)+ b_0^1 \right] \cdot \chi_1,
\end{aligned}
\end{equation}
where $t^M$ is the Maxwell stress
\begin{equation} \label{1d-energy-form-d-p-14}
t^M = \frac{\varepsilon}{2} \left(F^{-1} \frac{\partial \xi}{\partial X}\right)^2 = \frac{\varepsilon}{2} (F^{-1}{\tilde E}^1)^2
\end{equation}
that is the same as the one in Eq.~\eqref{1d-energy-form-d-e-10} (or Eq.~\eqref{1d-total-stress-expli}). Since $\chi_1$ is an arbitrary function, the zero first variation \eqref{1d-energy-form-d-p-11} gives the equilibrium equation
\begin{equation} \label{1d-energy-form-d-p-12}
\frac{\partial}{\partial X} \left(\frac{\partial W_{\rm elast}^1}{\partial F} + t^M \right) + b_0^1= 0
\end{equation}
and the natural (traction) boundary conditions
\begin{equation} \label{1d-energy-form-d-p-13}
\frac{\partial W_{\rm elast}^1}{\partial F} + t^M = t_R^1 \quad {\rm at} \ X= X_a \, \& \, X_b.
\end{equation}

A third type of the energy formulation is also possible---in terms of the deformation $\chi$ and the nominal electric displacement ${\tilde D}^1$---and is omitted here. The reader can refer to the work by Liu \cite{liu2014energy}. 

\subsection{Examples of large deformation and bifurcation analysis in soft dielectrics}

\begin{figure}[h] %
    \centering
    \includegraphics[width=4in]{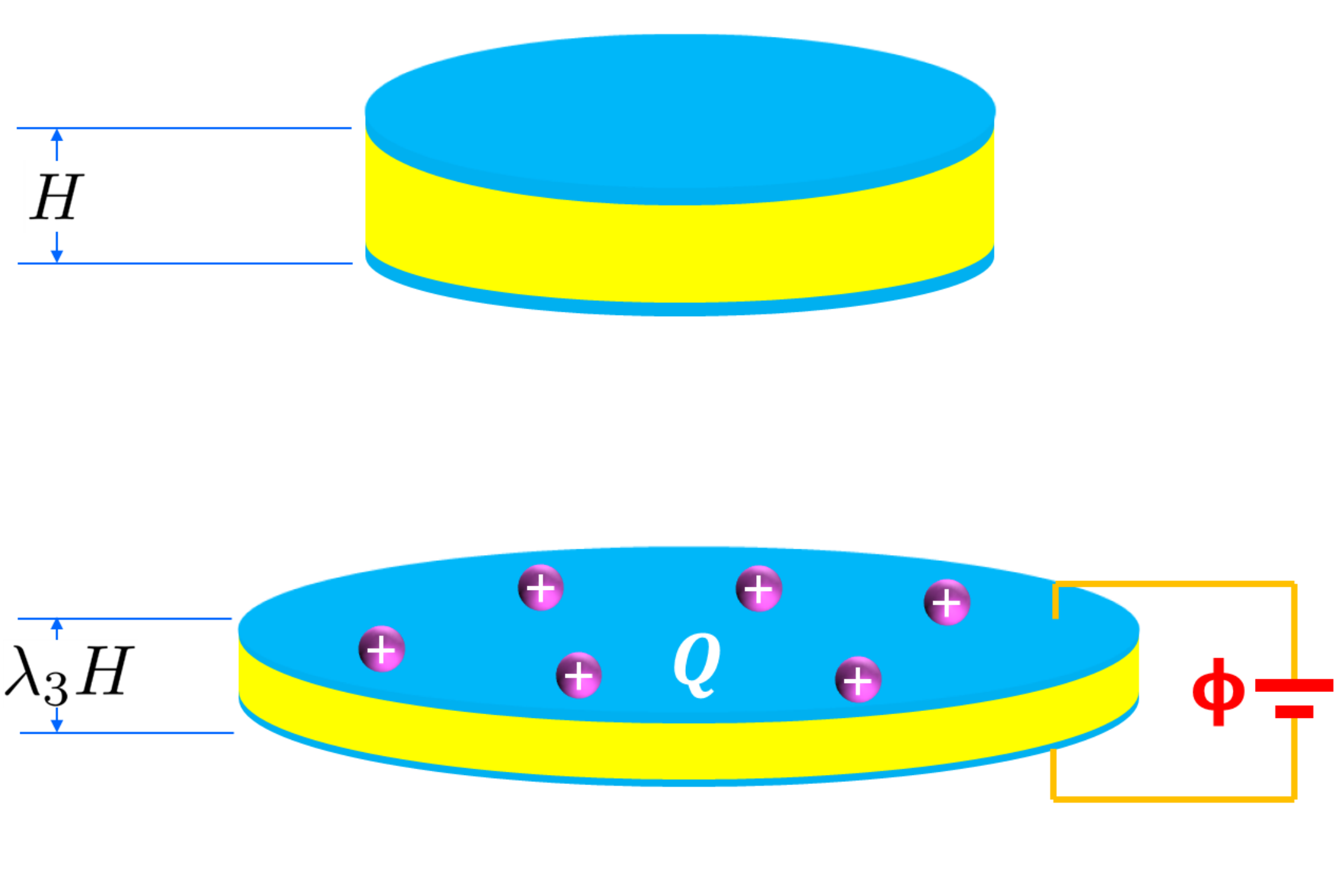}
    \caption{Schematic of the deformation of a circular disc of soft dielectrics subjected to an applied voltage $\Phi$. Two layers of compliant electrodes are coated on the upper and bottom surfaces. The disc deforms from thickness $H$ to $\lambda_3 H$, and both the upper and bottom electrodes gain an electric charge of magnitude $Q$.}
    \label{stability-2}
\end{figure}

\subsubsection{Large deformation} 
Consider a circular disc of soft dielectrics subjected to an applied voltage $\Phi$ (see Fig.~\ref{stability-2}). Recall the BVP \eqref{1d-total-bvp-refs}, the electric and mechanical boundary conditions are
\begin{equation}  \label{section-3-7-1}
\xi (0)= 0, \quad \xi (H)= \Phi, \quad t_R^1 (0) = t_R^1 (H) = 0.
\end{equation}

We specialize to homogeneous deformations and the deformation gradient $F= \frac{\partial \chi}{\partial X} = \frac{\lambda_3 H}{H} = \lambda_3$ is a constant. The Maxwell equation in the BVP \eqref{1d-total-bvp-refs}, together with the electric boundary conditions \eqref{section-3-7-1}, gives the solution of the potential $\xi (X) = \frac{X}{H} \Phi$. Then the nominal electric field is
\begin{equation}  \label{section-3-7-2}
{\tilde E}^1 = -\frac{\partial \xi}{\partial X} = - \frac{\Phi}{H}.
\end{equation}
It follows from Eqs.~\eqref{section-3-7-2} that the Maxwell stress in the BVP \eqref{1d-total-bvp-refs} is
\begin{equation}  \label{section-3-7-3}
t^M = \frac{\varepsilon}{2 \lambda_3^2} \left(\frac{\Phi}{H}\right)^2.
\end{equation}
The total stress in \eqref{1d-total-bvp-refs} then is
\begin{equation}  \label{section-3-7-3a}
t^1 = f^1 (\lambda_3) + \frac{\varepsilon}{2 \lambda_3^2} \left(\frac{\Phi}{H}\right)^2,
\end{equation}
which is independent of the coordinate $X$ due to a constant $\lambda_3$. Thus, the mechanical boundary conditions in Eq.~\eqref{section-3-7-1} finally give the governing equation
\begin{equation}  \label{section-3-7-3b}
f^1 (\lambda_3) + \frac{\varepsilon}{2 \lambda_3^2} \left(\frac{\Phi}{H}\right)^2 = 0,
\end{equation}
which governs the deformation $\lambda_3$ of the disc subjected to the applied voltage $\Phi$.

\subsubsection{Incremental boundary-value problem} 
In contrast to Eq.~\eqref{section-3-7-1}, the incremental electric and mechanical boundary conditions are
\begin{equation}  \label{10-24-19-1}
\xi^\star (0)=\xi^\star (H)= 0 \quad {\rm and} \quad t^\star (0) = t^\star (H) = 0.
\end{equation}

The infinitesimal increments of the potential and displacement are assumed to be $\xi^\star$ and $u^\star = \lambda_3 ^\star X$, where $\lambda_3 ^\star$ is sufficiently small and independent of the coordinates. It follows from Eqs.~\eqref{1d-incre-form-000} and \eqref{1d-incre-form-1001} that 
\begin{equation}  \label{10-24-19-2}
F^\star = \lambda_3 ^\star \quad {\rm and} \quad \tilde{E}^{\star} = - \frac{\partial \xi^{\star}}{\partial X}.
\end{equation}

Since $F=\lambda_3$ and $F^\star=\lambda_3^\star$ are independent of the coordinates, the increment $\tilde{D}^{\star}$ in Eq.~\eqref{1d-incre-form-2} and the incremental Maxwell equation $\frac{\partial \tilde{D}^\star}{\partial X} =0$ in Eq.~\eqref{1d-incre-form-4} implies $\frac{\partial \tilde{E}^\star}{\partial X} =0$. By Eq.~\eqref{10-24-19-2}, we have $\frac{\partial^2 \xi^{\star}}{\partial X^2} = 0$, which, together with the incremental electric boundary conditions $\eqref{10-24-19-1}_1$, gives $\xi^{\star} (X) = 0$. Thus, 
\begin{equation}  \label{10-24-19-3}
\tilde{E}^{\star} = 0 \quad {\rm and} \quad \tilde{D}^{\star} = \varepsilon \lambda_3 ^{-2} \frac{\Phi}{H} \lambda_3 ^\star,
\end{equation}
where $\lambda_3$ is the solution to the governing equation \eqref{section-3-7-3b}. The infinitesimal increment $t^\star$ in \eqref{1d-incre-form-3} then is
\begin{equation}  \label{10-24-19-4}
t^\star = \frac{\partial f^1 (\lambda_3)}{\partial \lambda_3} \lambda_3 ^\star - \varepsilon \lambda_3 ^{-3} \left(\frac{\Phi}{H}\right)^2 \lambda_3 ^\star.
\end{equation}

Thus, the incremental mechanical boundary conditions $\eqref{10-24-19-1}_2$ give $
 \left[ \frac{\partial f^1 (\lambda_3)}{\partial \lambda_3} - \varepsilon \lambda_3 ^{-3} \left(\frac{\Phi}{H}\right)^2 \right] \lambda_3 ^\star =0$, which requires 
\begin{equation}  \label{10-24-19-5}
\frac{\partial f^1 (\lambda_3)}{\partial \lambda_3} - \varepsilon \lambda_3 ^{-3} \left(\frac{\Phi}{H}\right)^2  =0,
\end{equation}
for nonzero $\lambda_3 ^\star$ and Eq.~\eqref{10-24-19-5} is the necessary condition for bifurcation. One can also obtain Eq.~\eqref{10-24-19-5} directly from the governing equation \eqref{section-3-7-3b} by using the implicit function theorem in Sec.~\ref{section-bifurcation-points}. The procedure is the same as that of the first two bifurcation examples in Sec.~\ref{subsection-bifurcation-exs}. Let $x=\lambda_3$ and $\lambda = \frac{\varepsilon}{2} \left(\frac{\Phi}{H}\right)^2$ in Eq.~\eqref{section-3-7-3b} for notation simplicity, we have the reformulated governing equation $f_3 (x, \lambda) = f^1 (x) + x^{-2} \lambda = 0$. The condition of bifurcation is $\frac{\partial f_3 (x, \lambda)}{\partial x} = \frac{\partial f^1 (x)}{\partial x} -2 x^{-3} \lambda =0$, which is exactly the same as Eq.~\eqref{10-24-19-5}. In the following, we will consider two kinds of the constitutive laws. \\

\noindent {$\bullet$ \it Linear constitutive law} Consider the constitutive law \eqref{1d-elastic-stress-102601}. The governing equation \eqref{section-3-7-3b} becomes
\begin{equation}  \label{10-24-19-6}
C (\lambda_3 -1) + \frac{\varepsilon}{2 \lambda_3^2} \left(\frac{\Phi}{H}\right)^2 = 0
\end{equation}
while the necessary condition \eqref{10-24-19-5} for bifurcation is
\begin{equation}  \label{10-24-19-7}
C - \varepsilon \lambda_3 ^{-3} \left(\frac{\Phi}{H}\right)^2  =0.
\end{equation}

\noindent {$\bullet$ \it Nonlinear constitutive law} Assuming the neo-Hookean constitutive law to describe the mechanical behavior of the solid, the strain-energy function $W_{\rm elast}^1$ in Eqs.~\eqref{1d-energy-form-d-p-12} and \eqref{1d-energy-form-d-p-13} can be written as $W_{\rm elast}^1 (\lambda_3) =  \frac{\mu}{2} (\lambda_3^2 + 2\lambda_3^{-1} - 3)$, where $\mu$ is the shear modulus at small deformation. By the elastic stress $t^e = f^1 (\lambda_3) = \partial W_{\rm elast}^1/ \partial \lambda_3$, the governing equation \eqref{section-3-7-3b} becomes
\begin{equation}  \label{section-3-7-5}
\mu (\lambda_3 - \lambda_3^{-2}) + \frac{\varepsilon}{2 \lambda_3^2} \left(\frac{\Phi}{H}\right)^2 = 0
\end{equation}
and while the necessary condition \eqref{10-24-19-5} for bifurcation is
\begin{equation}  \label{10-24-19-8}
\mu (1 +2 \lambda_3^{-3}) - \varepsilon \lambda_3 ^{-3} \left(\frac{\Phi}{H}\right)^2  =0.
\end{equation}
Regarding the governing equation \eqref{section-3-7-5}, a similar equation can be found, see eq.(28) in the work \cite{zhao2014harnessing}:
\begin{equation}  \label{section-3-7-6}
\mu (\lambda_3 - \lambda_3^{-2}) + \frac{\varepsilon}{\lambda_3^3} \left(\frac{\Phi}{H}\right)^2 = 0.
\end{equation}

The difference between Eq.~\eqref{section-3-7-5} and Eq.~\eqref{section-3-7-6} is due to the simplified 1D model in our tutorial which does not consider the deformation of the cross-section. \\

\begin{figure}[h] %
    \centering
    \includegraphics[width=4in]{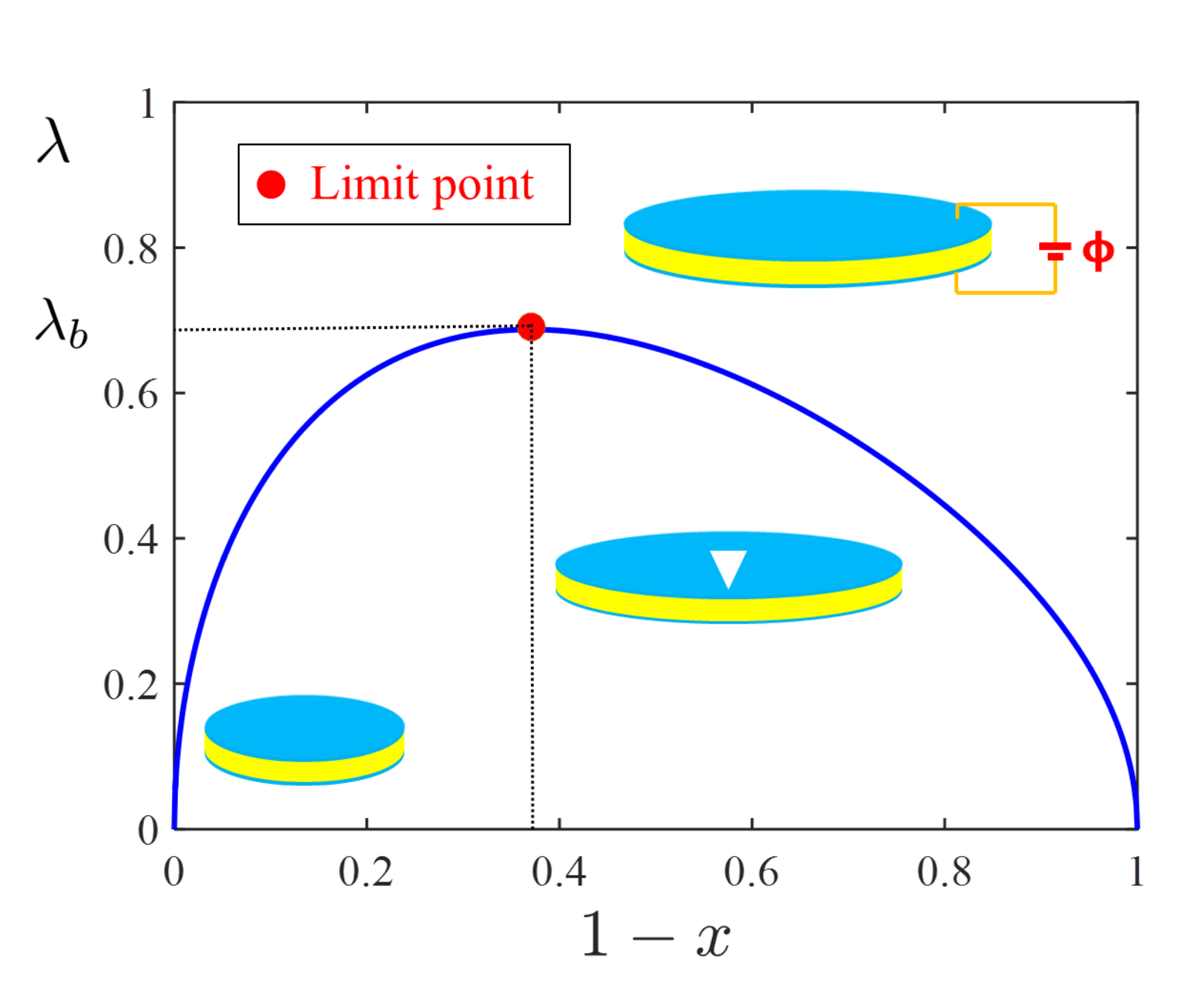}
    \caption{Bifurcation of the equilibrium state, $\sqrt{x - x^4} - \lambda = 0$ in Eq.~\eqref{bifur-ex-3}, of a film of neo-Hookean ideal dielectrics subjected to an applied electric field. The load parameter $\lambda$ here is the nominal electric field $\tilde E_3 \sqrt{\varepsilon/\mu}$ in equation (29) by Zhao and Wang \cite{zhao2014harnessing}, and the state variable $x$ is the stretch $\lambda_3$ in the thickness direction. The limit point, $(1-x_b, \lambda_b)=(0.37, 0.69)$, is found by the bifurcation analysis.}
    \label{bifur-3}
\end{figure}

To make a direct comparison to the stability analysis in the work \cite{zhao2014harnessing}, we consider the equilibrium equation \eqref{section-3-7-6}, which is reformulated as
\begin{equation} \label{bifur-ex-3}
f_4 (x, \lambda) = \sqrt{x - x^4} - \lambda = 0,
\end{equation}
where $x=\lambda_3$ and $\lambda = \frac{\Phi}{H} \sqrt{{\varepsilon}/{\mu}}$ for notation simplicity. By using the implicit function theorem in Sec.~\ref{section-bifurcation-points} or considering the statement below Eq.~\eqref{10-24-19-5}, the existence of the bifurcation point requires 
\begin{equation} \label{bifur-ex-3a}
\frac{\partial f_4}{\partial x} =\frac{1}{2} \frac{1-4x^3}{\sqrt{x-x^4}} = 0,
\end{equation}
giving $x=(\frac{1}{4})^{1/3} = 0.63$ that is exactly the critical point for the onset of pull-in instability in the work \cite{zhao2014harnessing}. The bifurcation diagram is plotted in Fig.~\ref{bifur-3}.

\section{Three-dimensional electrostatics of soft deformable dielectrics} \label{section-3D-continuum}

In this section, we summarize nonlinear electroelasticity in the context of three-dimensional solids. Several books may be consulted for foundations of mechanics \cite{gurtin1982introduction, ogden1997non, holzapfel2000nonlinear, truesdell2004non, reddy2007introduction, gurtin2010mechanics, fung2001classical, huang2003fundamentals}. For the nonlinear field theory of deformable dielectrics,  the following articles are good references: \cite{toupin1956elastic, toupin1960stress, ericksen2002electromagnetic, dorfmann2005nonlinear, mcmeeking2005electrostatic, ericksen2007theory, fosdick2007electrodynamics, suo2008nonlinear, suo2010theory, dorfmann2014nonlinear, liu2013energy, liu2014energy, dorfmann2017nonlinear, dorfmann2019instabilities}.

\subsection{Mathematical preliminaries}

\subsubsection{Basic tensor algebra}

A (second-order) tensor can be seen as a linear transformation from ${\mathbb R}^3$ space to ${\mathbb R}^3$ space. Thus a tensor ${\bf S}$ is a linear mapping of vectors to vectors; that is, for a vector ${\bf a} \in {\mathbb R}^3$, 
\begin{equation} \label{tensor-def}
{\bf b} = {\bf S}{\bf a} \ \in \ {\mathbb R}^3.
\end{equation}
The (Cartesian) components $S_{ij}$ of the tensor ${\bf S}$ are
\begin{equation} \label{tensor-comp}
S_{ij} = {\bf e}_i \cdot {\bf S}{\bf e}_j,
\end{equation}
where the dot in Eq.~\eqref{tensor-comp} denotes the inner (dot) product between vectors. An alternative (component) form of ${\bf b} = {\bf S}{\bf a}$ in Eq.~\eqref{tensor-def} is given by
\begin{equation} \label{ten-com-def}
b_i = S_{ij} a_j.
\end{equation}

In this tutorial, the vector form (i.e., Eq.~\eqref{tensor-def}) rather than the component form (i.e., Eq.~\eqref{ten-com-def}) of tensor algebra is often used. In the following, we just summarize some commonly used properties of second-order tensors in Table \ref{table-tensor}.

{
\setlength{\arrayrulewidth}{0.5 mm}
\setlength{\tabcolsep}{10 pt}
\renewcommand{\arraystretch}{2}
\begin{table}
\centering
\caption{A brief summary of properties of a second-order tensor ${\bf S}$ \cite{gurtin1982introduction, gurtin2010mechanics, holzapfel2000nonlinear}.}
\label{table-tensor}
{\rowcolors{1}{green!80!yellow!10}{green!70!yellow!20}
\begin{tabular}{ |p{0.8 in}| p{0.8 in}|p{2 in}|p{1.2 in}|   }
\hline
%
Name &   Symbol   &   How obtained   &   Notes  \\
\hline
 ${\color{blue} {\rm Components}}$          & $S_{ij}=({\bf S})_{ij}$        &    $S_{ij} = {\bf e}_i \cdot {\bf S}{\bf e}_j $   for basis $\{{\bf e}_i \}$              & ${\bf S} {\bf e}_j = S_{ij} {\bf e}_i$  \\
 ${\color{blue} {\rm Transpose}}$          & ${\bf S}^T$        & ${\bf b} \cdot {\bf S}{\bf a}= {\bf a} \cdot {\bf S}^T {\bf b},$  \ $\forall \ {\bf a, b} \in \mathbb R^3$                                   & ${\bf S}^T = S_{ji} {\bf e}_i \otimes {\bf e}_j$  \\
 ${\color{blue} {\rm Trace}}$                 & tr ${\bf S}$         & tr ${\bf S} = S_{ii}$ \ (in sum) for $\{{\bf e}_i \}$          &         tr ${\bf S}$ $=$ tr ${\bf S}^T = {\bf S}: {\bf I}$  \\  
  ${\color{blue} {\rm Inverse}}$                 & ${\bf S}^{-1}$         &  ${\bf S} {\bf S}^{-1} = {\bf S}^{-1} {\bf S} = {\bf I}$  &         $({\bf S}^T)^{-1}$ $=$ $({\bf S}^{-1})^{T}$  \\  
${\color{blue} {\rm Matrix}}$ \ ${\color{blue}{\rm representation}}$    & $[{\bf S}]$          & $[{\bf S}] = \left[
  \begin{bmatrix} 
S_{11} & S_{12} & S_{13} \\
S_{21} & S_{22} & S_{23} \\
S_{31} & S_{32} & S_{33}
  \end{bmatrix} \right.$                &   ~  \\
 ${\color{blue} {\rm Determinant}}$     & $\det {\bf S}$ or $|{\bf S}|$         & $\det {\bf S} = \frac{{\bf Su}\cdot({\bf Sv} \times {\bf Sw})}{{\bf u}\cdot({\bf v} \times {\bf w})},$ for any basis \{{\bf u}, {\bf v}, {\bf w}\}     &      $\det{\bf S} = \det{\bf S}^T$, $\det{\bf S}^{-1} = (\det{\bf S})^{-1}$   \\
  ${\color{blue} {\rm Principal}}$ \ \ \ ${\color{blue}{\rm invariants}}$         & $I_1 ({\bf S})$, \qquad $I_2 ({\bf S})$, \qquad \ $I_3 ({\bf S})$         & $I_1 ({\bf S})$$=$tr ${\bf S}$, \ $I_2 ({\bf S}) = \frac{1}{2} [{\rm tr}^2 ({\bf S}) - {\rm tr}({\bf S}^2)]$, \ $I_3 ({\bf S}) = \det {\bf S}$                                   & Cayley-Hamilton equation  \\
\hline
\end{tabular}
}
\end{table}
}

\subsubsection{Tensor analysis: differentiation and integration}

For a scalar function $\phi({\bf x})$ and a vector function ${\bf v}({\bf x})$ of vector ${\bf x}$, their gradients are
\begin{equation} \label{grad-f-v}
{\rm grad} \ \phi({\bf x}) = \frac{\partial\phi({\bf x})}{\partial x_i} {\bf e}_i \quad {\rm and} \quad  {\rm grad} \ {\bf v}({\bf x}) = \frac{\partial v_i ({\bf x})}{\partial x_j} {\bf e}_i \otimes {\bf e}_j,
\end{equation}
respectively. In addition, the divergence of a vector field ${\bf v}({\bf x})$ and a tensor field ${\bf S}({\bf x})$ may be defined by:
\begin{equation} \label{div-v-T}
{\rm div} \ {\bf v}({\bf x}) = \frac{\partial v_i ({\bf x})}{\partial x_i} \quad {\rm and} \quad  {\rm div} \ {\bf S}({\bf x}) = \frac{\partial S_{ij} ({\bf x})}{\partial x_j} {\bf e}_i.
\end{equation}

Deformation is a mapping from material points ${\bf X}$ to spatial points ${\bf x}$. Thus, it is necessary to distinguish between the gradient (or the divergence) with respect to ${\bf X}$ and ${\bf x}$. In this tutorial, we use ``$\nabla$'' for the gradient and ``${\rm Div}$'' for the divergence with respect to material points ${\bf X}$, for example, 
\begin{equation} \label{Grad-Div-f-v}
\nabla \phi({\bf X}) = \frac{\partial\phi({\bf X})}{\partial X_i} {\bf e}_i, \quad \nabla {\bf v}({\bf X}) = \frac{\partial v_i ({\bf X})}{\partial X_j} {\bf e}_i \otimes {\bf e}_j, \quad  {\rm Div} \ {\bf v}({\bf X}) = \frac{\partial v_i ({\bf X})}{\partial X_i}.
\end{equation}

The expressions \eqref{grad-f-v}$-$\eqref{Grad-Div-f-v} may be simply regarded as the operations in the Cartesian coordinates. Actually, if $x_i$(or $X_i$) and $\{{\bf e}_i\}$ are taken as the general coordinates and the corresponding basis, the gradient and the divergence in the general coordinates also have similar forms of Eqs.~\eqref{grad-f-v}$-$\eqref{Grad-Div-f-v}. Specifically, vector operators in orthogonal curvilinear coordinates, such as cylindrical polar coordinates and spherical polar coordinates, as well as other concepts in tensor analysis, can be found in textbooks \cite{reddy2007introduction, riley2011essential, truesdell2004non, fung2001classical, huang2003fundamentals}.

Consider a scalar function $\phi({\bf S})$ of a tensor variable ${\bf S}$. The derivative $\partial \phi({\bf S})/ \partial {\bf S}$ is the tensor function, namely
\begin{equation} \label{ten-derive}
\frac{\partial \phi({\bf S})}{\partial {\bf S}} = \frac{\partial \phi({\bf S})}{\partial S_{ij}} {\bf e}_i \otimes {\bf e}_j.
\end{equation}
The chain-rule reads 
\begin{equation} \label{chain-rule}
\dot{\overline{ \phi({\bf S}) }}= \frac{\partial \phi({\bf S})}{\partial {\bf S}} : \dot{\overline{{\bf S}}},
\end{equation}
where ${\bf S}$ is a function of $t$ and the dot denotes the derivative with respect to $t$. The chain-rule is a simple and efficient way of calculating the derivative $\partial \phi({\bf S})/ \partial {\bf S}$. 

One form of the divergence theorem reads 
\begin{equation} \label{div-theo}
\int_{\partial \Omega_R} {\bf S} {\bf N} \ dA = \int_{\Omega_R} {\rm Div} \, {\bf S} \ dV, 
\end{equation}
where $\Omega_R \in \mathbb R^3$ is a bounded region with smooth boundary $\partial \Omega_R$, ${\bf S}$ is a differentiable tensor field over $\Omega_R$, and ${\bf N}$ is the outward unit normal to $\partial\Omega_R$. An important identity from the divergence theorem \eqref{div-theo} is 
\begin{equation} \label{div-conse}
\int_{\partial \Omega_R} {\bf u} \cdot {\bf S} {\bf N} \ dA = \int_{\Omega_R} ({\bf u} \cdot {\rm Div} \, {\bf S} + {\bf S}: \nabla {\bf u}) \ dV, 
\end{equation}
where ${\bf u}$ is a differentiable vector field over $\Omega_R$.

\subsection{Kinematics}%

\subsubsection{Deformation of a body}%

\begin{figure}[h] %
    \centering
    \includegraphics[width=4in]{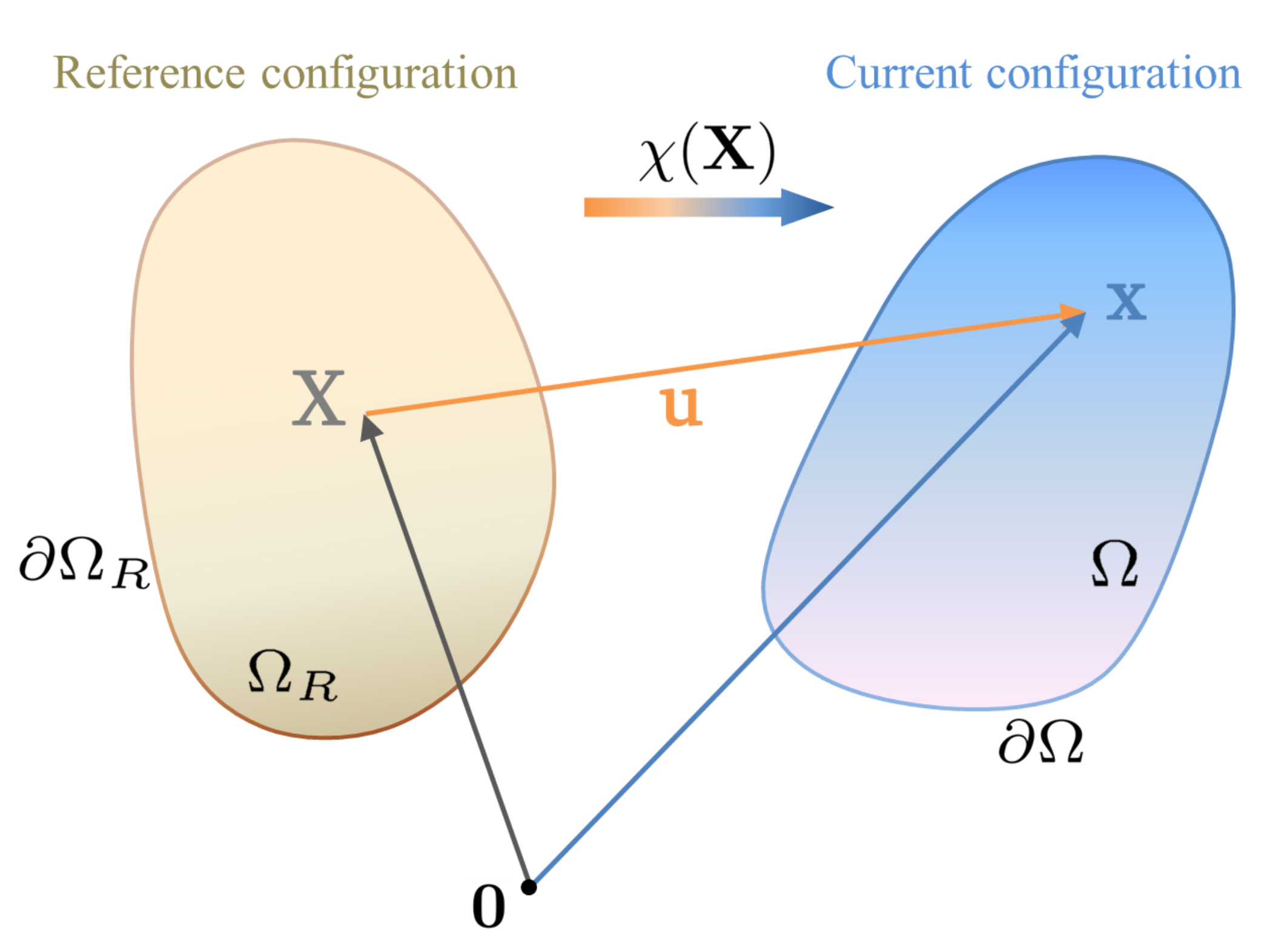}
    \caption{Reference and current configurations of a body.}
    \label{2-kinematic}
\end{figure}

A body may occupy a domain in Euclidean space. We may identify the body with the occupied domain in a (or some) fixed configuration, which is called the {\it reference configuration}. The domain in a reference configuration is denoted by $\Omega_R$, and the body identified with $\Omega_R$ is called the {\it reference body}. A point ${\bf X} \in \Omega_R$ is called a {\it material point}. The (sufficiently regular) boundary of $\Omega_R$ is denoted by $\partial \Omega_R$ (see Fig.~\ref{2-kinematic}).  \\

Consider a smooth function $\chi: \Omega_R \to \Omega$ that assigns to each material point ${\bf X} \in \Omega_R$ a point
\begin{equation} \label{spatial-point}
{\bf x} = \chi ({\bf X}) = {\bf u}({\bf X}) + {\bf X} \in \Omega,
\end{equation}
where ${\bf x}$ is called the {\it spatial point} and ${\bf u}({\bf X})$ is the {\it displacement vector} that relates the material point ${\bf X} \in \Omega_R$ and the spatial point ${\bf x} \in \Omega$ (see Fig.~\ref{2-kinematic}). The smooth function $\chi$ is the deformation and $\Omega = \chi (\Omega_R)$ is the domain occupied by the body at the {\it current configuration}. 

\subsubsection{Deformation gradient, stretch and strain tensors}%
The deformation gradient is defined by
\begin{equation} \label{2-Def-Grad}
{\bf F} = \nabla \chi ({\bf X}) = \nabla{\bf u} + {\bf I}, \quad F_{ij} = \frac{\partial x_i}{\partial X_j} = \frac{\partial u_i}{\partial X_j} + \delta_{ij},
\end{equation}
where $\delta_{ij}$ is the Kronecker delta. The deformation gradient ${\bf F}$ is a key measure of the deformation and other related quantities (the various forms of strain tensors) are derived from it. \\

The (volumetric) {\it Jacobian}, $J ({\bf X})$, relates to the change in volume between the reference, $d V$, and the current volume element, $d v$, as
\begin{equation} \label{2-Jacobian-1}
d v = J ({\bf X}) d V, 
\end{equation}
where the Jacobian $J ({\bf X})$ is equal to the determinant of the deformation gradient:
\begin{equation} \label{2-Jacobian-2}
J ({\bf X}) = \det {\bf F}({\bf X}) = \det \nabla \chi ({\bf X}) > 0. 
\end{equation}
In particular, the Jacobian for incompressible materials equals unity, i.e.,
\begin{equation} \label{2-J-unit}
J ({\bf X}) = 1. 
\end{equation}
The {\it polar decomposition} reads
\begin{equation} \label{polar-dec}
 {\bf F} = {\bf RU} = {\bf VR},
\end{equation}
where ${\bf R}$ is a rotation, ${\bf U}$ and ${\bf V}$ are positive-definite symmetric tensors. The explicit representations of ${\bf U}$ and ${\bf V}$ are
\begin{equation} \label{right-left-str}
{\bf U} = \sqrt{{\bf F}^T {\bf F}} \quad {\rm and} \quad {\bf V} = \sqrt{{\bf F} {\bf F}^T}.
\end{equation}
Note that ${\bf U}$ and ${\bf V}$ are the so-called right and left stretch tensors, namely
\begin{equation} \label{stretch-s-stre}
\lambda = |{\bf U}({\bf X}) {\bf e}|, \quad \lambda^2 = {\bf e} \cdot {\bf U}^2({\bf X}) {\bf e} = {\bf e} \cdot {\bf C}({\bf X}) {\bf e},
\end{equation}
where $\lambda$ is the {\it stretch} at material point ${\bf X}$ in the material direction ${\bf e}$, and ${\bf C} = {\bf U}^2 = {\bf F}^T {\bf F}$ is the right Cauchy-Green tensor. In contrast, ${\bf B} = {\bf V}^2 = {\bf F} {\bf F}^T$ is the left Cauchy-Green tensor. \\

It is known that the eigenvalues of a positive-definite symmetric matrix (tensor) are real and positive, and the eigenvectors corresponding to different eigenvalues are orthogonal \cite{meyer2000matrix, riley2011essential}. Therefore, spectral representations of the symmetric and positive-definite tensors ${\bf U}$, ${\bf V}$, ${\bf C}$, and ${\bf B}$ are \cite{gurtin2010mechanics, gurtin1982introduction}
\begin{equation} \label{4-spectral}
{\bf U} = \sum_{i=1}^{3} \lambda_i {\bf r}_i \otimes {\bf r}_i, \quad {\bf V} = \sum_{i=1}^{3} \lambda_i {\bf l}_i \otimes {\bf l}_i, \quad {\bf C} = \sum_{i=1}^{3} \lambda_i^2 {\bf r}_i \otimes {\bf r}_i, \quad {\rm and} \quad {\bf B} = \sum_{i=1}^{3} \lambda_i^2 {\bf l}_i \otimes {\bf l}_i,
\end{equation}
where $\lambda_i > 0$ are the principle stretches, ${\bf r}_i$ and ${\bf l}_i$ are the right and left principal directions. \\

In addition to the stretch, the strain tensors are often used to describe the deformation. Consider the change in the squared lengths $|d {\bf x}|^2 - |d {\bf X}|^2$. Here $|d {\bf x}|^2 = d {\bf x} \cdot d {\bf x}$ is the squared distance of two spatial points in the current configuration while $|d {\bf X}|^2$ is the squared original distance of two material points in the reference configuration. We introduce two second-order tensors ${\bf E}^s$ and ${\bf e}^s$, namely
\begin{equation}
|d {\bf x}|^2 - |d {\bf X}|^2 = 2 d {\bf X} \cdot {\bf E}^s d {\bf X}= 2 d {\bf x} \cdot {\bf e}^s d {\bf x}, 
\end{equation}
where ${\bf E}^s$ is called the Green-St. Venant (Lagrangian) strain tensor or simply the Green strain tensor
\begin{equation} \label{green-strain}
{\bf E}^s = \frac{1}{2} ({\bf F}^T {\bf F} - {\bf I}) = \frac{1}{2} ({\bf C} - {\bf I}) = \frac{1}{2} (\nabla {\bf u} + \nabla {\bf u}^T + \nabla {\bf u}^T \nabla {\bf u})
\end{equation}
and ${\bf e}^s$ is called the Almansi-Hamel (Eulerian)  strain tensor or simply the Euler
strain tensor
\begin{equation} \label{euler-strain}
{\bf e}^s = \frac{1}{2} ({\bf I} - {\bf F}^{-T} {\bf F}^{-1} ) = \frac{1}{2} ({\bf I} - {\bf B}^{-1} ).
\end{equation}

In linear elasticity, for small deformation $\Vert \nabla {\bf u} \Vert \ll 1$, the Green strain tensor ${\bf E}^s$ and the Euler strain tensor ${\bf e}^s$ are equivalent.

\subsubsection{Material and spatial gradient, divergence, and curl}%


In the reference configuration, the (material) gradient, the (material) divergence, and the (material) curl are denoted, respectively, by 
\begin{equation} \label{M-G-D-C}
\nabla, \quad {\rm Div}, \quad {\rm and} \quad {\rm Curl},
\end{equation}
while in the current configuration, the (spatial) gradient, the (spatial) divergence, and the (spatial) curl are denoted, respectively, by 
\begin{equation} \label{s-g-d-c}
{\rm grad}, \quad {\rm div}, \quad {\rm and} \quad {\rm curl}.
\end{equation} 

For a scalar field $\varphi$ and a vector field ${\bf g}$, with the chain-rule, we have the following relation between the material and the spatial gradient:
\begin{equation} \label{G-g-relation_c}
\frac{\partial \varphi}{\partial X_i} = \frac{\partial \varphi}{\partial x_j} \frac{\partial \chi_j}{\partial X_i} = F_{ji} \frac{\partial \varphi}{\partial x_j}   \quad  {\rm and}    \quad   \frac{\partial g_i}{\partial X_j} = \frac{\partial g_i}{\partial x_k} \frac{\partial \chi_k}{\partial X_j} =  \frac{\partial g_i}{\partial x_k} F_{kj}
\end{equation}
or, the vector forms
\begin{equation} \label{G-g-relation}
\nabla \varphi = {\bf F}^T {\rm grad}\, \varphi  \quad {\rm and} \quad \nabla {\bf g} = ({\rm grad} \, {\bf g}) {\bf F}.
\end{equation}
An equivalent form of Eq.~\eqref{G-g-relation} is
\begin{equation} \label{g-G-relation}
 {\rm grad} \, \varphi = {\bf F}^{-T} \nabla \varphi  \quad {\rm and} \quad {\rm grad}\, {\bf g} = (\nabla {\bf g})  {\bf F}^{-1}.
\end{equation}
In addition, the relation between the material and the spatial divergence reads:
\begin{equation} \label{D-d-relation}
{\rm Div} \, {\bf g} = {\bf F}^{T} : {\rm grad} \, {\bf g}  \quad {\rm and} \quad {\rm div} \, {\bf g} = {\bf F}^{-T} : \nabla {\bf g}.
\end{equation}

\subsubsection{Material and spatial integration}%

Recall Eq.~\eqref{2-Jacobian-1}, the spacial volume element $dv$ is related to the material volume element $dV$ through the Jacobian $J = \det {\bf F} $, that is,
$dv = J \, dV$. Consider a scalar field which is represented by $\phi ({\bf x})$ in the current configuration ($\Omega$) and by $\phi_R ({\bf X})$ in the reference configuration ($\Omega_R$). With the volume element relation \eqref{2-Jacobian-1}, $dv = J \, dV$, we have the following volume integral 
\begin{equation} \label{M-S-V-int}
\int_{\Omega} \phi ({\bf x}) d v ({\bf x}) = \int_{\Omega_R} \phi_R ({\bf X}) J({\bf X}) d V ({\bf X}) \quad {\rm or} \quad \int_{\Omega} \phi d v = \int_{\Omega_R} \phi J d V.
\end{equation}
The latter equation $\eqref{M-S-V-int}_2$ admits a more concise form in which the arguments are suppressed and the subscript of $\phi$ is dropped. \\

In contrast to the volumetric Jacobian $J = {\rm det} {\bf F}$ in Eq.~\eqref{2-Jacobian-1}, the areal Jacobian $J_a$ is defined by
\begin{equation} \label{area-Jacobian}
J_a = J |{\bf F}^{-T} {\bf N}|.
\end{equation}
Thus the spatial area element $da = |{\bf n} \, da|$ and the material area element $dA = |{\bf N} \, dA|$ have the relation
\begin{equation} \label{area-Jacobian-1}
da = J_a dA, \quad {\bf n} \, da = J {\bf F}^{-T}{\bf N} \, dA = {\bf F}^{C}{\bf N} \, dA,
\end{equation}
where ${\bf F}^{C} = J {\bf F}^{-T}$ is called the {\it cofactor} of ${\bf F}$. \\

Consider a vector field ${\bf u}$ and a tensor field ${\bf G}$. With Eq.~\eqref{area-Jacobian-1}, we have the following surface integrals 
\begin{equation} \label{M-S-S-int}
\int_{\partial \Omega} {\bf u} \cdot {\bf n}  d a = \int_{\partial \Omega_R} {\bf u} \cdot {\bf F}^{C} {\bf N} \, dA \quad {\rm and} \quad \int_{\partial \Omega} {\bf G}{\bf n}  d a = \int_{\partial \Omega_R} {\bf G}{\bf F}^{C} {\bf N} \, dA.
\end{equation}

\subsection{Maxwell's equations}

\subsubsection{Spatial representation}%
In the absence of currents and magnetic fields, the Maxwell equations in the current configuration become (SI units convention)
\begin{equation} \label{MW-eq-true}
{\rm curl} \, {\bf E} = {\bf 0} \quad {\rm and} \quad {\rm div} \, \bf D = \rho_{\rm f},
\end{equation}
where ${\bf E}$ is the (true) electric field, ${\bf D}$ is the (true) electric displacement, and $\rho_{\rm f}$ is the (free) charge density per unit volume in the current configuration.  \\

Equation $\eqref{MW-eq-true}_1$ implies that the electric field ${\bf E}$ in electrostatics is {\it irrotational} \cite{riley2011essential}, that is, there exists a scalar potential field $\xi$, such that
\begin{equation} \label{EgV-true}
{\bf E} = - {\rm grad} \, \xi.
\end{equation}
Equation $\eqref{MW-eq-true}_2$ means that the electric flux leaving a domain is proportional to the charge inside. \\

The electric boundary conditions on the boundary $\partial \Omega = \partial \Omega^\xi \cup \partial \Omega^d$ in the current configuration are given by
\begin{equation} \label{y-current-e-BCs}
\xi = \xi_0 \quad {\rm on} \ \partial \Omega^\xi  \quad {\rm and} \quad  {\bf D} \cdot {\bf n} = q_{\rm f} \quad {\rm on} \ \partial \Omega^d,
\end{equation}
where $\xi_0$ is the prescribed voltage on the surface $\partial \Omega^\xi$ and $q_{\rm f}$ is the (free) charge density per unit area on the surface $\partial \Omega^d$ in the current configuration.\\

The electric displacement ${\bf D}$ in \eqref{MW-eq-true} is defined as
\begin{equation} \label{DEP-relation}
{\bf D} = \varepsilon_0 {\bf E} + {\bf P},
\end{equation}
where ${\bf P}$ is the (true) polarization field in the current configuration. The relation \eqref{DEP-relation} becomes ${\bf D} = \varepsilon_0 {\bf E}$ in vacuum. The constitutive relation for dielectrics can be written as ${\bf D} = {\bm \varepsilon} (\bf x) {\bf E}$, where ${\bm \varepsilon} (\bf x)$ is the second-order dielectric permittivity tensor. We will mostly study isotropic systems so ${\bm \varepsilon} (\bf x) = \varepsilon {\bf I}$, where $\varepsilon$ is a constant and ${\bf I}$ is the second-order identity tensor, the constitutive relation for \emph{linear} dielectrics in the current configuration is simply:
\begin{equation} \label{elec-cons-true}
{\bf D} = \varepsilon {\bf E}.
\end{equation}

A direct consequence of Eqs.~\eqref{DEP-relation} and \eqref{elec-cons-true} for linear dielectrics is
\begin{equation} \label{PE-rela-true}
{\bf P} = (\varepsilon - \varepsilon_0) {\bf E}.
\end{equation}

\subsubsection{Material representation} \label{sec-4.3.2}

In contrast to Eq.~\eqref{MW-eq-true}, the Maxwell equations in the reference configuration are
\begin{equation} \label{MW-eq-refer}
{\rm Curl} \, \tilde {\bf E} = {\bf 0} \quad {\rm and} \quad {\rm Div} \, \tilde {\bf D} = \tilde\rho_{\rm f},
\end{equation}
where $\tilde {\bf E}$ is the nominal electric field, $\tilde {\bf D}$ is the nominal electric displacement, and $\tilde \rho_{\rm f}$ is the (free) charge density per unit volume in the reference configuration. \\

The irrotational field $\tilde {\bf E}$ in Eq.~$\eqref{MW-eq-refer}_1$ indicates a scalar potential field $\xi$, together with Eqs.~\eqref{G-g-relation} and \eqref{EgV-true}, we have
\begin{equation} \label{EgV-nominal-e}
\tilde {\bf E} = - \nabla \xi = - {\bf F}^T {\rm grad} \, \xi = {\bf F}^T {\bf E}.
\end{equation}

Recall the surface integral \eqref{M-S-S-int} of the spatial and material representations. If we replace the vector field ${\bf u}$ in Eq.~\eqref{M-S-S-int} by the (true) electric displacement ${\bf D}$, we have
\begin{equation} \label{d-D-int}
\int_{\partial \Omega} {\bf D} \cdot {\bf n}  d a = \int_{\partial \Omega_R} J {\bf D} \cdot {\bf F}^{-T} {\bf N} \, dA = \int_{\partial \Omega_R} J {\bf F}^{-1}{\bf D} \cdot {\bf N} \, dA = \int_{\partial \Omega_R} {\tilde {\bf D}} \cdot {\bf N} \, dA,
\end{equation}
which indicates the relation between the nominal and true electric displacements, namely
\begin{equation} \label{def-ND-d}
{\tilde {\bf D}} = J {\bf F}^{-1} {\bf D}.
\end{equation}
The 3D relation \eqref{def-ND-d} can be directly reduced to the 1D relation \eqref{1d-MW-D=d} since the Jacobian $J$ and deformation gradient ${\bf F}$ in 1D formulation admit $J=F$. \\

The electric boundary conditions on the boundary $\partial \Omega_R = \partial \Omega_R^\xi \cup \partial \Omega_R^d$ in the reference configuration are defined as
\begin{equation} \label{y-ref-e-BCs}
\xi = \xi_0 \quad {\rm on} \ \partial \Omega_R^\xi  \quad {\rm and} \quad  \tilde {\bf D} \cdot {\bf N} = \tilde q_{\rm f} \quad {\rm on} \ \partial \Omega_R^d,
\end{equation}
where $\xi_0$ is the prescribed voltage and $\tilde q_{\rm f}$ is the (free) charge density per unit area. 
\\

The definition of the relation between the nominal and true polarizations is not unique. For example, the definition in the work \cite{liu2014energy} is
\begin{equation} \label{def-NP-p}
{\tilde {\bf P}} = J {\bf P}.
\end{equation}
By Eqs.~\eqref{DEP-relation}, \eqref{def-ND-d} and \eqref{def-NP-p}, the auxiliary field in the reference configuration can be written as
\begin{equation} \label{DEP-r-refer}
\tilde {\bf D} = \varepsilon_0 J {\bf C}^{-1} \tilde {\bf E} + {\bf F}^{-1} \tilde {\bf P}.
\end{equation}
The constitutive relation of linear dielectrics in the reference configuration, for example, is
\begin{equation} \label{elec-cons-ref}
\tilde {\bf D} = \varepsilon J {\bf C}^{-1} \tilde {\bf E}.
\end{equation}
Combining Eqs.~\eqref{DEP-r-refer} and \eqref{elec-cons-ref} for linear dielectrics, we have the relation
\begin{equation} \label{PE-rela-refer}
 \tilde {\bf P} = (\varepsilon - \varepsilon_0) J {\bf F}^{-T} \tilde {\bf E}.
\end{equation}

For convenience, several important relations related to electrostatics of a linear isotropic dielectric deformable media are summarized in Table \ref{table-DEP-nominal-true}.

{
\setlength{\arrayrulewidth}{0.5 mm}
\setlength{\tabcolsep}{10 pt}
\renewcommand{\arraystretch}{3}
\begin{table}
\centering
\caption{Conversion formulas}
\label{table-DEP-nominal-true}
{\rowcolors{1}{green!80!yellow!10}{green!70!yellow!20}
\begin{tabular}{ |p{0.3 in}|| p{0.5 in}|p{0.5 in}|p{0.75 in}|p{0.5 in}|p{0.5 in}|p{0.7 in}|   }
\hline
\multicolumn{7}{|c|}{Nominal/true electric displacement $(\tilde {\bf D} / {\bf D})$, electric field $(\tilde {\bf E} / {\bf E})$, polarization $(\tilde {\bf P} / {\bf P})$} \\
\hline
\hline
 ~ &   $\tilde {\bf D}=$   &   $\tilde {\bf E}=$   &   $\tilde {\bf P}=$   &   ${\bf D}=$   &   ${\bf E}=$   & ${\bf P}=$  \\
\hline
\hline
 ${\color{blue}(\tilde {\bf D}, {\bf F})}$ & ${\color{red} \tilde {\bf D} }$ & $\displaystyle\frac{{\bf C}{\tilde {\bf D}}}{\varepsilon J}$ & $\displaystyle\frac{\varepsilon-\varepsilon_0}{\varepsilon} {\bf F} \tilde {\bf D}$  & $\displaystyle\frac{{\bf F} \tilde {\bf D}}{J}$ & $\displaystyle\frac{{\bf F} \tilde {\bf D}}{\varepsilon J}$ & $\displaystyle\frac{\varepsilon - \varepsilon_0}{\varepsilon J} {\bf F} \tilde {\bf D}$ \\
 ${\color{blue}(\tilde {\bf E}, {\bf F})}$   &    $\varepsilon J{\bf C}^{-1} \tilde {\bf E}$    &    ${\color{red} \tilde {\bf E}}$    &   \scriptsize$(\varepsilon-\varepsilon_0) J {\bf F}^{-T} \tilde {\bf E}$     &    $\varepsilon {\bf F}^{-T} \tilde {\bf E}$     &    ${\bf F}^{-T} \tilde {\bf E}$     &     \scriptsize$(\varepsilon-\varepsilon_0) {\bf F}^{-T} \tilde {\bf E}$ \\
 ${\color{blue}(\tilde {\bf P}, {\bf F})}$ &  $\displaystyle\frac{\varepsilon {\bf F}^{-1} \tilde{\bf P}}{\varepsilon-\varepsilon_0}$  & $\displaystyle \frac{{\bf F}^T \tilde {\bf P}}{(\varepsilon-\varepsilon_0)J}$ & ${\color{red} \tilde {\bf P}} $ & $\displaystyle \frac{\varepsilon \tilde {\bf P}}{(\varepsilon-\varepsilon_0)J}$ & $\displaystyle \frac{\tilde {\bf P}}{(\varepsilon-\varepsilon_0)J}$ & $\displaystyle\frac{\tilde {\bf P}}{J}$ \\
 ${\color{blue}({\bf D}, {\bf F})}$             &          $J {\bf F}^{-1} {\bf D}$         &       $\displaystyle\frac{{\bf F}^T {\bf D}}{\varepsilon}$     &       $\displaystyle\frac{(\varepsilon-\varepsilon_0)J}{\varepsilon} {\bf D}$      & ${\color{red} {\bf D}}$ &  $\displaystyle\frac{{\bf D}}{\varepsilon}$      &        $\displaystyle\frac{\varepsilon-\varepsilon_0}{\varepsilon} {\bf D}$ \\
 ${\color{blue}({\bf E}, {\bf F})}$ &    $\varepsilon J {\bf F}^{-1} {\bf E}$         &       ${\bf F}^T {\bf E}$     &       $(\varepsilon-\varepsilon_0) J {\bf E}$      &    $\varepsilon {\bf E}$   &  ${\color{red} {\bf E}}$      &        $\displaystyle(\varepsilon-\varepsilon_0) {\bf E}$ \\

 ${\color{blue}({\bf P}, {\bf F})}$ &  $\displaystyle\frac{\varepsilon J {\bf F}^{-1} {\bf P}}{\varepsilon-\varepsilon_0}$ & $\displaystyle\frac{{\bf F}^{T} {\bf P}}{\varepsilon-\varepsilon_0}$ & $J {\bf P}$ & $\displaystyle\frac{\varepsilon {\bf P}}{\varepsilon-\varepsilon_0}$ & $\displaystyle\frac{\bf P}{\varepsilon-\varepsilon_0}$ & ${\color{red}{\bf P}}$ \\
\hline
\end{tabular}
}
\end{table}
}

\subsection{Balances of forces and moments}%
The mechanical laws, including balances of forces and moments, can be represented respectively in the current and reference configurations. We first discuss the balance laws and the stress tensor field in the current configuration.
\subsubsection{Spatial representation}%
The balances of forces and moments for the static equilibrium of a deformed body $\Omega$ are
\begin{equation} \label{balance-f-true}
\int_{\partial \Omega} {\bf t} ({\bf n}) \, da + \int_{\Omega} {\bf b_0} \, dv = {\bf 0}, \quad {\rm (global) \ balance \ of \ forces}
\end{equation}
and
\begin{equation} \label{balance-m-true}
\int_{\partial \Omega} {\bf r} \times {\bf t} ({\bf n}) \, da + \int_{\Omega} {\bf r} \times {\bf b_0} \, dv = {\bf 0}, \quad {\rm (global) \ balance \ of \ moments},
\end{equation}
where ${\bf t} ({\bf n}, {\bf x})$ is the (true) traction force on the boundary ${\partial \Omega}$, ${\bf n}({\bf x})$ is the outward unit normal to the boundary ${\partial \Omega}$, ${\bf b}_0 ({\bf x})$ is the (true) body force in the domain $\Omega$, and ${\bf r}({\bf x})$ is the position vector. \\

A well-known theorem in continuum mechanics is {\it Cauchy's theorem} for the existence of stress, that is, there exists a spatial tensor field ${\boldsymbol \sigma}$, known as the {\it Cauchy stress} (field), due to the balance of forces Eq.~\eqref{balance-f-true}, namely
\begin{equation} \label{def-Cauchy-s}
{\bf t} ({\bf n}) = {\boldsymbol \sigma}{\bf n}, \quad t_i = \sigma_{ij} n_j. 
\end{equation}

In contrast to Eqs.~\eqref{balance-f-true} and \eqref{balance-m-true}, we have local forms of the force and moment balances by using Cauchy's theorem \eqref{def-Cauchy-s}, the divergence theorem \eqref{div-theo}, the identity \eqref{div-conse}, and the property of skew tensors, such that
\begin{equation} \label{local-b-f-true}
{\rm div} \, {\boldsymbol \sigma} + {\bf b}_0 = {\bf 0}, \quad \sigma_{ij,j} + b_{0i} = 0, \quad {\rm (local) \ balance \ of \ forces}
\end{equation}
and
\begin{equation} \label{local-b-m-true}
{\boldsymbol \sigma} = {\boldsymbol \sigma}^T, \quad \sigma_{ij} = \sigma_{ji}, \quad {\rm (local) \ balance \ of \ moments}.
\end{equation}

Clearly, the local form \eqref{local-b-m-true} is that the Cauchy stress tensor ${\boldsymbol \sigma}$ is symmetric. It should be noted that Eqs.~\eqref{def-Cauchy-s}$-$\eqref{local-b-m-true} are consequences of Eqs.~\eqref{balance-f-true} and \eqref{balance-m-true}. \\

The traction boundary conditions are given by
\begin{equation} \label{y-current-traction-BCs}
{\boldsymbol \sigma}{\bf n} = {\bf t}_0,
\end{equation}
where ${\bf t}_0$ is the traction force on the surface $\partial \Omega^t$ in the current configuration.

\subsubsection{Material representation}%
Recall the surface integral \eqref{M-S-S-int} of the spatial and material representations. If we replace the tensor field ${\bf G}$ in Eq.~\eqref{M-S-S-int} by the Cauchy stress tensor $\boldsymbol \sigma$, we have
\begin{equation} \label{sigma-S-int}
\int_{\partial \Omega} {\boldsymbol \sigma}{\bf n}  d a = \int_{\partial \Omega_R} {\boldsymbol \sigma}{\bf F}^{C} {\bf N} \, dA = \int_{\partial \Omega_R} {\bf T}{\bf N} \, dA,
\end{equation}
where 
\begin{equation} \label{def-PK-s}
{\bf T} = {\boldsymbol \sigma}{\bf F}^{C} = J{\boldsymbol \sigma}{\bf F}^{-T}
\end{equation}
representing the stress measured per unit area in the reference configuration. In contrast to the Cauchy stress $\boldsymbol \sigma$ in Eq.~\eqref{def-Cauchy-s}, ${\bf T}$ in Eq.~\eqref{def-PK-s} is referred to as the (first) Piola-Kirchhoff stress. \\

With the volume integral \eqref{M-S-V-int}, we obtain
\begin{equation} \label{body-f-int}
\int_{\Omega} {\bf b}_0 \, d v = \int_{\Omega_R} J \, {\bf b}_0 \, d V = \int_{\Omega_R} {\bf b}_{0_R} \, d V,
\end{equation}
where
\begin{equation} \label{body-f-ref}
{\bf b}_{0_R} = J \, {\bf b}_{0},
\end{equation}
is the body force measured per unit volume in the reference configuration. \\

Combining Eqs.~\eqref{sigma-S-int} and \eqref{body-f-int} and Cauchy's theorem \eqref{def-Cauchy-s}, the spatial form of balance of forces \eqref{balance-f-true} is equivalent to
\begin{equation} \label{balance-f-ref}
\int_{\partial \Omega_R} {\bf T}{\bf N} \, dA + \int_{\Omega_R} {\bf b}_{0_R} \, d V = {\bf 0},  \quad {\rm (global) \ balance \ of \ forces}.
\end{equation}
With the divergence theorem \eqref{div-theo} in an arbitrary chosen domain $\Omega_R$, we have the local form
\begin{equation} \label{local-b-f-ref}
{\rm Div} \, {\bf T} + {\bf b}_{0_R} = {\bf 0}, \quad T_{ij,j} + b_{0_Ri} = 0, \quad {\rm (local) \ balance \ of \ forces}.
\end{equation}
An alternative form of the stress relation \eqref{def-PK-s} is
\begin{equation} \label{Cauchy-PK-rel}
{\boldsymbol \sigma} = J^{-1} {\bf T} {\bf F}^{T}.
\end{equation}
Since the Cauchy stress is symmetric, ${\boldsymbol \sigma} = {\boldsymbol \sigma}^T$ in Eq.~\eqref{local-b-m-true}, we have the material form of balance of moments
\begin{equation} \label{local-b-m-ref}
{\bf T} {\bf F}^{T} = {\bf F} {\bf T}^T, \quad T_{ik} F_{jk} = F_{ik} T_{jk}, \quad {\rm (local) \ balance \ of \ moments}.
\end{equation}

The traction boundary conditions are given by
\begin{equation} \label{y-ref-traction-BCs}
{\bf T}{\bf N} = {\bf t}_{0_R},
\end{equation}
where ${\bf t}_{0_R}$ is the prescribed traction force on the surface $\partial \Omega_R^t$ in the reference configuration. In addition to the traction boundary conditions, the displacement boundary conditions are ${\bf u} = {\bf u}_0$ on the surface $\partial \Omega_R^u = \partial \Omega_R \setminus \partial \Omega_R^t$, where ${\bf u}_0$ is the prescribed displacement.
\\

\subsection{Constitutive equations of soft dielectrics}%
\subsubsection{Decomposition of the stress tensors}%
In the electrostatic of soft dielectrics, both the (first) Piola-Kirchhoff stress ${\bf T} $ in Eq.~\eqref{def-PK-s} and the Cauchy stress ${\boldsymbol \sigma} $ in Eq.~\eqref{Cauchy-PK-rel} can be decomposed into two parts:
\begin{equation} \label{Piola-s-decom}
{\bf T} = {\bf T}^e + {\bf T}^M
\end{equation}
and
\begin{equation} \label{Cauchy-s-decom}
{\boldsymbol \sigma} = {\boldsymbol \sigma}^e + {\boldsymbol \sigma}^M.
\end{equation}
Here ${\bf T}^e$ and ${\boldsymbol \sigma}^e$ are elastic stresses. In contrast, ${\bf T}^M$ in Eq.~\eqref{Piola-s-decom} is the Piola-Maxwell stress and ${\boldsymbol \sigma}^M$ in Eq.~\eqref{Cauchy-s-decom} is the true Maxwell stress

\subsubsection{Elastic stress tensors}%
Consider a strain-energy function $W^e = W^e ({\bf F})$. Evidently, $W^e = W^e ({\bf F})$ is a scalar-valued function of one tensor variable ${\bf F}$. The elastic part ${\bf T}^e$ of the Piola-Kirchhoff stress in Eq.~\eqref{Piola-s-decom} is
\begin{equation} \label{elastic-PK-s}
{\bf T}^e = \left \{
                           \begin{aligned}
                           & \frac{\partial W^e ({\bf F})}{\partial {\bf F}} \\
                           & \frac{\partial W^e ({\bf F})}{\partial {\bf F}} - \mathcal L_a {\bf F}^{-T}
                           \end{aligned} \: ,
                           \right.
\quad 
T_{ij}^e = \left \{
                           \begin{aligned}
                           & \frac{\partial W^e (F_{ij})}{\partial F_{ij}}, & {\rm compressible \ solids}, \\
                           & \frac{\partial W^e (F_{ij})}{\partial F_{ij}} - \mathcal L_a {F}_{ji}^{-1}, &  {\rm incompressible \ solids},
                           \end{aligned}
                           \right.
\end{equation}
where $\mathcal L_a$ is the Lagrange multiplier related to the constraint of incompressibility $J= \det {\bf F} =1$. Also, $\mathcal L_a$ in Eq.~\eqref{elastic-PK-s} may be interpreted as the hydrostatic pressure.  With Eqs.~\eqref{Cauchy-PK-rel} and \eqref{elastic-PK-s}, the elastic part ${\boldsymbol \sigma}^e$ of the Cauchy stress in Eq.~\eqref{Cauchy-s-decom} is 
\begin{equation}  \label{elastic-Cau-s}
{\boldsymbol \sigma}^e = J^{-1} {\bf T}^e {\bf F}^{T}, \quad \sigma_{ij}^e = J^{-1} T_{ik}^e F_{jk}.
\end{equation}

In the following, we will briefly discuss the consequences of {\it frame-indifference} in the strain-energy functions and stresses. Consider a rotation ${\bf Q} \in$ Orth$^+ = $ \{all rotations\}. The frame-indifference of the strain-energy functions reads \cite{gurtin2010mechanics}
\begin{equation} \label{frame-indiff}
W^e ({\bf F}) = W^e ({\bf QF}).
\end{equation}
With the polar decomposition ${\bf F} = {\bf RU}$, ${\bf R} \in$ Orth$^+ $, in Eq.~\eqref{polar-dec}, and letting ${\bf Q} = {\bf R}^T$ in Eq.~\eqref{frame-indiff}, we have
\begin{equation} \label{frame-indiff-1}
W^e ({\bf F}) = W^e ({\bf R}^T {\bf RU}) = W^e ({\bf U}).
\end{equation}

Since ${\bf U} =\sqrt{ {\bf U}^T {\bf U}} = \sqrt{ {\bf U}^T {\bf R}^T {\bf R} {\bf U}} = \sqrt{ {\bf F}^T {\bf F}} = \sqrt{{\bf C}}$, we can introduce a strain-energy function $\bar W^e ({\bf C})$, such that
\begin{equation} \label{frame-ind-2}
\bar W^e ({\bf C}) = W^e (\sqrt{\bf C}) = W^e ({\bf U}) = W^e ({\bf F}).
\end{equation}

Similarly, with the Green strain tensor ${\bf E}^s = \frac{1}{2} ({\bf C} - {\bf I})$, Eqs.~\eqref{frame-indiff-1} and \eqref{frame-ind-2}, we can have a strain-energy function ${\tilde W^e} ({\bf E}^e)$, namely
\begin{equation} \label{frame-ind-3}
{\tilde W^e} ({\bf E}^e) = \bar W^e ({\bf C}) = W^e ({\bf U}) = W^e ({\bf F}).
\end{equation}
The derivatives of the strain-energy functions in Eq.~\eqref{frame-ind-3} have the following relations
\begin{equation}
\frac{\partial W^e ({\bf F})}{\partial {\bf F}} = 2 {\bf F} \frac{\partial \bar W^e ({\bf C})}{\partial {\bf C}} = {\bf F} \frac{\partial {\tilde W^e} ({\bf E}^e)}{\partial {\bf E}^e}.
\end{equation}

In addition, assuming isotropy, the strain-energy function $W^e$ depends on ${\bf F}$ through the principal stretches $\lambda_1$, $\lambda_2$ and $\lambda_3$:
\begin{equation}  \label{elastic-F-stretch-9-11-1}
W^e ({\bf F}) = W^e (\lambda_1, \lambda_2, \lambda_3).
\end{equation}
In contrast to Eqs.~\eqref{elastic-PK-s} and \eqref{elastic-Cau-s}, the elastic part of the first Piola-Kirchhoff stress and the Cauchy stress can be written as
\begin{equation}  \label{elastic-PK-s-stretch}
{\bf T}^e = \sum_{i=1}^3 \Big( \frac{\partial W^e}{\partial \lambda_i} - \frac{1}{\lambda_i} \mathcal L_a \Big) {\hat {\bf n}}_i \otimes {\hat {\bf N}}_i
\end{equation}
and
\begin{equation} \label{elastic-Cau-s-stretch}
{\boldsymbol \sigma}^e = \sum_{i=1}^3 \Big( J^{-1} \lambda_i \frac{\partial W^e}{\partial \lambda_i} - \mathcal L_a \Big) {\hat {\bf n}}_i \otimes {\hat {\bf n}}_i,
\end{equation}
where the orthonormal vectors ${\hat {\bf n}}_i$ and ${\hat {\bf N}}_i$ are the spatial and referential directions, respectively; and $J=\lambda_1 \lambda_2 \lambda_3$ and $\mathcal L_a$ is zero for compressible solids while $J=1$ and $\mathcal L_a$ is the hydrostatic pressure for incompressible solids. \\

For rubber-like elastic materials, the commonly used models are the neo-Hookean model, the Mooney-Rivlin model, the Ogden model, and the Gent model, etc \cite{mooney1940theory, rivlin1948large, ogden1972large, truesdell2004non, ogden1997non, ARRUDA1993389, gent1996new, holzapfel2000nonlinear, Boyce2000Constitutive, XIANG2018110, Destrade2022Ogden}. The strain-energy function of incompressible neo-Hookean materials, for example, is defined by
\begin{equation} \label{ex-neo-H-strain-energy}
W^e ({\bf F}) = \frac{\mu}{2} ({\rm tr} ({\bf F}^T {\bf F}) -3) \quad {\rm or} \quad W^e (\lambda_1, \lambda_2, \lambda_3) = \frac{\mu}{2} (\lambda_1^2 + \lambda_2^2 + \lambda_3^2 -3),
\end{equation}
where $\mu$ is the shear modulus at small deformation. The constraint of incompressibility reads
\begin{equation}
\det{\bf F} = 1 \quad {\rm or} \quad \lambda_1 \lambda_2 \lambda_3 = 1.
\end{equation}
The first Piola-Kirchhoff stress Eq.~\eqref{elastic-PK-s} or Eq.~\eqref{elastic-PK-s-stretch} is 
\begin{equation}\label{ex-neo-H-PK}
{\bf T}^e = \mu {\bf F} - \mathcal L_a {\bf F}^{-T} \quad {\rm or} \quad {\bf T}^e = \sum_{i=1}^3 \Big( \mu \lambda_i - \frac{1}{\lambda_i} \mathcal L_a \Big) {\hat {\bf n}}_i \otimes {\hat {\bf N}}_i
\end{equation}
while the Cauchy stress Eq.~\eqref{elastic-Cau-s} or Eq.~\eqref{elastic-Cau-s-stretch} is 
\begin{equation}\label{ex-neo-H-Cauchy}
{\boldsymbol \sigma}^e = \mu {\bf F}{\bf F}^T - \mathcal L_a {\bf I} \quad {\rm or} \quad {\boldsymbol \sigma}^e = \sum_{i=1}^3 \Big( \mu \lambda_i^2 - \mathcal L_a \Big) {\hat {\bf n}}_i \otimes {\hat {\bf n}}_i.
\end{equation}

The strain-energy function of the Gent model \cite{gent1996new}, for example, is 
\begin{equation} \label{sy-film-14-gent}
W^e (\lambda_1, \lambda_2, \lambda_3) = - \frac{\mu J_m}{2} \ln \Big( 1 - \frac{I_1 - 3}{J_m} \Big),
\end{equation}
where $I_1 = \lambda_1^2 + \lambda_2^2 + \lambda_3^2$ and $J_m$ is a material constant.

\subsubsection{Maxwell stress tensor for an ideal dielectric elastomer} \label{sec-maxwell stress-4.5.3}
In this section, we consider linear dielectrics that admit the relation \eqref{elec-cons-true} between the electric field ${\bf E}$ and the electric displacement ${\bf D}$ in the current configuration. \\

 Consider a charge density $\rho_{\rm f}$ in an electric field ${\bf E}$. The electric body force is equal to the Lorentz force $\rho_{\rm f} \bf E$ in the electrostatics. By the definition, ${\rm div} \, {\boldsymbol \sigma}^M = \rho_{\rm f} \bf E$, i.e., the divergence of the Maxwell stress ${\boldsymbol \sigma}^M$ equals to the electric body force $\rho_{\rm f} \bf E$, we have the Maxwell stress in the deformed linear dielectrics
\begin{equation}\label{def-MW-E-true}
{\boldsymbol \sigma}^M = \varepsilon {\bf E} \otimes {\bf E} - \frac{\varepsilon |{\bf E}|^2}{2} {\bf I}, \quad \sigma_{ij}^M = \varepsilon E_{i} E_{j} - \frac{\varepsilon (E_{1}^2 + E_2^2 + E_3^2)}{2} \delta_{ij}.
\end{equation}
Note that the divergence, ${\rm div} \, {\boldsymbol \sigma}^M$, of the Maxwell stress is equal to the electric body force $\rho_{\rm f} \bf E$ in linear dielectrics.\footnote{By the identities ${\rm div} \, ({\bf E} \otimes {\bf E}) = (\rm div\, {\bf E}) {\bf E} + ({\rm grad} \, {\bf E}) {\bf E}$ and ${\rm div} \, (|{\bf E}|^2 {\bf I}) = {\rm grad} \, |{\bf E}|^2 = 2({\rm grad} \, {\bf E})^T {\bf E}$, together with Eqs.~$\eqref{MW-eq-true}_2$ and \eqref{elec-cons-true}, we can obtain the divergence ${\rm div} \, {\boldsymbol \sigma}^M = \rho_{\rm f} \bf E +\varepsilon [ ({\rm grad} \, {\bf E}) {\bf E} - ({\rm grad} \, {\bf E})^T {\bf E}]$. Since the electric field $\bf E$ in Eq.~\eqref{EgV-true} is irrotational, we have ${\rm grad} \, {\bf E} = ({\rm grad} \, {\bf E})^T$. Then, $({\rm grad} \, {\bf E}) {\bf E} = ({\rm grad} \, {\bf E})^T {\bf E}$ and the divergence finally reduces to ${\rm div} \, {\boldsymbol \sigma}^M = \rho_{\rm f} \bf E$.}\\

It follows from Eqs.~\eqref{def-PK-s} and \eqref{def-MW-E-true} that the Piola-Maxwell stress ${\bf T}^M$ in Eq.~\eqref{Piola-s-decom} is
\begin{equation}\label{def-MW-E-nominal}
{\bf T}^M = \varepsilon J {\bf E} \otimes {\bf F}^{-1}{\bf E} - \frac{\varepsilon J |{\bf E}|^2}{2} {\bf F}^{-T}.
\end{equation}
By Eqs.~\eqref{elec-cons-true} and \eqref{def-ND-d}, ${\boldsymbol \sigma}^M$ in Eq.~\eqref{def-MW-E-true} and ${\bf T}^M$ in Eq.~\eqref{def-MW-E-nominal} can be, respectively, reformulated as \cite{zhao2014harnessing}
\begin{equation}\label{def-MW-D-true}
{\boldsymbol \sigma}^M = \frac{1}{\varepsilon} {\bf D} \otimes {\bf D} - \frac{|{\bf D}|^2}{2 \varepsilon} {\bf I}, \quad \sigma_{ij}^M = \frac{1}{\varepsilon} D_{i} D_{j} - \frac{(D_1^2+D_2^2+D_3^2)}{2 \varepsilon} \delta_{ij},
\end{equation}
and
\begin{equation}\label{def-MW-E-nominal-zhao}
{\bf T}^M =  \frac{1}{\varepsilon J} {\bf F}{\tilde {\bf D}} \otimes {\tilde {\bf D}} - \frac{1}{2 \varepsilon J} |{\bf F}{\tilde {\bf D}}|^2{\bf F}^{-T}.
\end{equation}

By the conversion formulas in Table \ref{table-DEP-nominal-true}, the Piola-Maxwell stress can also be written as either ${\bf T}^M =  \frac{1}{\varepsilon} {\bf D} \otimes {\tilde {\bf D}} - \frac{J}{2 \varepsilon} |{\bf D}|^2{\bf F}^{-T}$ or ${\bf T}^M =  {\bf E} \otimes {\tilde {\bf D}} - \frac{\varepsilon J}{2} |{\bf E}|^2{\bf F}^{-T}$.

\subsection{Constitutive law for soft dielectrics}%

The choice of a constitutive model depends on the material being modeled and, of course, the independent state variables chosen. As an example, in several works, Suo et al. \cite{suo2008nonlinear} assumed that the free energy density is a function of the deformation gradient ${\bf F}$ and the nominal electric displacement $\tilde{\bf D}$, i.e., $W({\bf F}, \tilde{\bf D})$. Thus the nominal stress and the nominal electric field are
\begin{equation}\label{suo-08-T-E-1}
{\bf T} = \frac{\partial W ({\bf F}, \tilde {\bf D})}{\partial {\bf F}} \quad  {\rm and} \quad \tilde {\bf E} = \frac{\partial W ({\bf F}, \tilde {\bf D})}{\partial \tilde {\bf D}}.
\end{equation}

For the model of ideal soft dielectrics \cite{zhao2007electromechanical, zhao2014harnessing}, the free energy density can be written as
\begin{equation}\label{suo-08-T-E-2}
W ({\bf F}, \tilde {\bf D}) = W^e ({\bf F}) + \frac{1}{2 \varepsilon J} |{\bf F} \tilde {\bf D}|^2 - \mathcal L_a (J -1),
\end{equation}
where $W^e ({\bf F})$ is the strain-energy function due to the mechanical deformation, and $\mathcal L_a$ serves as the Lagrange multiplier related to the constraint of incompressibility, i.e., $J=1$. Typically, $\mathcal L_a$ is equal to zero in the case of compressible dielectrics.\\

Substitution of Eq.~\eqref{suo-08-T-E-2} into Eq.~\eqref{suo-08-T-E-1} gives
\begin{subequations}\label{suo-08-T-E-3}
\begin{align}
\label{suo-08-T-E-3a}
{\bf T} & = \frac{\partial W ({\bf F}, \tilde {\bf D})}{\partial {\bf F}} = \frac{\partial W^e ({\bf F})}{\partial {\bf F}} + {\bf T}^M - \mathcal L_a {\bf F}^{-T}, \\
\label{suo-08-T-E-3b}
{\tilde {\bf E}} & = \frac{\partial W ({\bf F}, \tilde {\bf D})}{\partial \tilde {\bf D}} = \frac{1}{\varepsilon J} {\bf F}^T ({\bf F} {\tilde {\bf D}}) = \frac{1}{\varepsilon J} {\bf C} {\tilde {\bf D}},
\end{align}
\end{subequations}
where the Maxwell stress ${\bf T}^M$ is given by Eq.~\eqref{def-MW-E-nominal-zhao}. In contrast, the constitutive law can also be formulated \cite{liu2014energy}, for example, by using the deformation gradient ${\bf F}$ and the nominal polarization $\tilde {\bf P}$, the deformation gradient ${\bf F}$ and the nominal electric field $\tilde {\bf E}$. 

\subsection{Summary of the boundary-value problem} \label{section4.7-summary}
In contrast to the 1D BVP Eq.~\eqref{1d-total-bvp-refs}, the 3D BVP of electrostatics of deformable dielectrics in the reference configuration are summarized as:
\begin{equation} \label{3d-total-bvp-refs}
\left.
\begin{aligned}
& \underbrace{ {\rm Div} \, {\bf T} + {\bf b}_{0_R} = {\bf 0}}_{\text{equilibrium equation}}, \quad  \underbrace{{\rm Curl} \, \tilde {\bf E} = {\bf 0}, \ {\rm Div} \, \tilde {\bf D} = \tilde\rho_{\rm f}}_{\text{Maxwell's equations}}, \quad \underbrace{\tilde {\bf D} = \varepsilon J {\bf C}^{-1} \tilde {\bf E}}_{\text{relation (linear dielectrics)}}, \\
& \underbrace{{\bf u} = {\bf u}_0 \ {\rm on} \ \partial \Omega_R^u, \ {\bf T}{\bf N} = {\bf t}_{0_R} \ {\rm on} \ \partial \Omega_R^t}_{\text{mechanical BCs}}, \quad \underbrace{\xi = \xi_0 \ {\rm on} \ \partial \Omega_R^{\xi}, \ \tilde {\bf D} \cdot {\bf N} = \tilde q_{\rm f} \ {\rm on} \ \partial \Omega_R^d}_{\text{electric BCs}}, \\
& {\bf T} = {\bf T}^e + {\bf T}^M, \quad \underbrace{{\bf F} = \nabla \chi ({\bf X}) = \nabla{\bf u} + {\bf I}}_{\text{kinematics}},\\
&{\bf T}^e =
\left\{ 
\begin{aligned}
& \frac{\partial W^e ({\bf F})}{\partial {\bf F}}, & {\text{compressible solids}},\\
&  \frac{\partial W^e ({\bf F})}{\partial {\bf F}} - \mathcal L_a {\bf F}^{-T}, & J=1, \ {\text{incompressible solids}},
\end{aligned}\right. \\
& \underbrace{{\bf T}^M = \frac{1}{\varepsilon J} {\bf F} {\tilde {\bf D}} \otimes {\tilde {\bf D}} - \frac{1}{2 \varepsilon J} |{\bf F} {\tilde {\bf D}}|^2 {\bf F}^{-T}}_{\text {Maxwell\rq{s} stress tensor}}.
\end{aligned}
\right\}
\end{equation}

In the above representation, we have chosen the constitutive law representative of a linear dielectric, and the nominal electric field is our state variable (in addition to the deformation).

\subsection{Incremental formulation and bifurcation analysis} \label{section4.8-incremental}

In contrast to Sec.~\ref{1d-incremental-section}, we now linearize the 3D BVP Eq.~\eqref{3d-total-bvp-refs}. The infinitesimal increment of the deformation $\chi$ is denoted by ${\bf u}^{\star} ({\bf X})$, i.e., $\chi ({\bf X}) \to \chi ({\bf X}) + {\bf u}^{\star} ({\bf X})$, with sufficiently small $\Vert {\bf u}^{\star} ({\bf X}) \Vert$ and $\Vert \nabla {\bf u}^{\star} ({\bf X}) \Vert$. In contrast to Eq.~\eqref{1d-incre-form-000}, the infinitesimal increment of the deformation gradient is
\begin{equation} \label{3d-incre-form-1}
{\bf F}^{\star} = \left. \frac{d}{d \tau} \Big( \nabla \big(\chi({\bf X}) + \tau {\bf u}^\star ({\bf X}) \big)\Big) \right|_{\tau=0} = \nabla {\bf u}^{\star}.
\end{equation}

Similarly, it is easy to show that $({\bf F}^T)^{\star} = ({\bf F}^{\star})^T$. By the variation ${\bf F} \to {\bf F} + {\bf F}^\star$, $\Vert {\bf F}^\star \Vert \ll 1$, one can also get the following infinitesimal increments
\begin{equation} \label{3d-incre-form-2}
\left.
\begin{aligned}
 ({\bf F}^{-1})^{\star} & = - {\bf F}^{-1} {\bf F}^\star {\bf F}^{-1}, \qquad \  \,  ({\bf F}^{-T})^{\star} = - {\bf F}^{-T} ({\bf F}^\star)^T {\bf F}^{-T} = [({\bf F}^{-1})^{\star}]^T, \\
 {\bf C}^\star & = {\bf F}^T {\bf F}^\star + ({\bf F}^T)^\star {\bf F}, \quad ({\bf C}^{-1})^{\star} = - {\bf C}^{-1} {\bf C}^\star {\bf C}^{-1}, \\
 J^\star & = J {\bf F}^{-T} : {\bf F}^\star, \qquad \quad   \   \  (J^{-1})^\star = - J^{-2} J^\star.
\end{aligned}
\right\}
\end{equation}

By the definition \eqref{bifur-con-2a}, the detailed derivations of the infinitesimal increments in Eq.~\eqref{3d-incre-form-2} are listed as follows: 
\[ \begin{aligned}
&\begin{aligned}
({\bf F}^{-1})^{\star} & = \left. \frac{d}{d \tau} \left( {\bf F} + \tau {\bf F}^\star\right)^{-1} \right|_{\tau=0} = \lim_{\tau \to 0} \frac{\left( {\bf F} + \tau {\bf F}^\star\right)^{-1} - {\bf F}^{-1}}{\tau} \\
&= \lim_{\tau \to 0} \frac{\left( {\bf F} + \tau {\bf F}^\star\right)^{-1} \left\{ {\bf F} - \left( {\bf F} + \tau {\bf F}^\star\right) \right\} {\bf F}^{-1}}{\tau}  = - {\bf F}^{-1} {\bf F}^\star {\bf F}^{-1},
\end{aligned}\\
&\begin{aligned}
({\bf F}^{-T})^{\star} & = \lim_{\tau \to 0} \frac{\left( {\bf F} + \tau {\bf F}^\star\right)^{-T} - {\bf F}^{-T}}{\tau} \\
& = \lim_{\tau \to 0} \frac{\left( {\bf F} + \tau {\bf F}^\star\right)^{-T} \left\{ {\bf F}^T - \left( {\bf F} + \tau {\bf F}^\star\right)^T \right\} {\bf F}^{-T}}{\tau} = - {\bf F}^{-T} ({\bf F}^\star)^T {\bf F}^{-T},
\end{aligned}\\
&\begin{aligned}
{\bf C}^\star & = ({\bf F}^T {\bf F})^\star = \left. \frac{d}{d \tau} \left( {\bf F} + \tau {\bf F}^\star\right)^T \left( {\bf F} + \tau {\bf F}^\star\right)\right|_{\tau=0} \\
& = \lim_{\tau \to 0} \frac{\left( {\bf F} + \tau {\bf F}^\star\right)^T \left( {\bf F} + \tau {\bf F}^\star\right) - {\bf F}^T {\bf F}}{\tau} \\
& = {\bf F}^T {\bf F}^\star + ({\bf F}^\star)^T {\bf F}  = {\bf F}^T {\bf F}^\star + ({\bf F}^T)^\star {\bf F},
\end{aligned}\\
&\begin{aligned}
({\bf C}^{-1})^{\star} & = ({\bf F}^{-1} {\bf F}^{-T})^{\star} = \lim_{\tau \to 0} \frac{\left( {\bf F} + \tau {\bf F}^\star\right)^{-1}\left( {\bf F} + \tau {\bf F}^\star\right)^{-T} - {\bf F}^{-1} {\bf F}^{-T}}{\tau} \\
& = \lim_{\tau \to 0} \frac{\left( {\bf F} + \tau {\bf F}^\star\right)^{-1}\left( {\bf F} + \tau {\bf F}^\star\right)^{-T} \left\{{\bf F}^{T} {\bf F} - \left( {\bf F} + \tau {\bf F}^\star\right)^{T}\left( {\bf F} + \tau {\bf F}^\star\right) \right\}{\bf F}^{-1} {\bf F}^{-T}}{\tau} \\
& = - {\bf C}^{-1} {\bf C}^\star {\bf C}^{-1},
\end{aligned}\\
& J^\star = (\det {\bf F})^\star = \lim_{\tau \to 0} \frac{\det \left( {\bf F} + \tau {\bf F}^\star\right) - \det {\bf F}}{\tau} = J {\bf F}^{-T} : {\bf F}^\star.
\end{aligned}
\]

Here we use the relation $\det \left( {\bf F} + \tau {\bf F}^\star\right) = (\det {\bf F}) \det \left( {\bf I} + \tau {\bf F}^{-1} {\bf F}^\star\right) = J [1 + \tau ({\bf F}^{-T}:{\bf F}^\star) + o (\tau)]$, which comes from the Cayley-Hamilton equation, cf., e.g., page 35 in \cite{gurtin2010mechanics}. From the forms of the above increments, we have $\Vert {\bf C}^\star \Vert \sim \Vert {\bf F}^\star \Vert \ll 1$, $\Vert ({\bf F}^{-1})^\star \Vert \sim \Vert {\bf F}^\star \Vert \ll 1$, $\Vert ({\bf C}^{-1})^\star \Vert \sim \Vert {\bf C}^\star \Vert \sim \Vert {\bf F}^\star \Vert \ll 1$, and $\Vert {J}^\star \Vert \sim \Vert {\bf F}^\star \Vert \ll 1$. By the variation $J \to J + J^\star$, $\Vert J^\star \Vert \ll 1$, we have $(J^{-1})^\star = \lim_{\tau \to 0} \frac{(J + \tau J^\star)^{-1} - J^{-1}}{\tau} = \lim_{\tau \to 0} \frac{J - (J + \tau J^\star)}{\tau J (J+\tau J^\star)} = - J^{-2} J ^\star$, and, hence, that $\Vert (J^{-1})^\star \Vert \sim \Vert {J}^\star \Vert \sim \Vert {\bf F}^\star \Vert \ll 1$.\\

The infinitesimal increment of the electric potential $\xi$ is denoted by $\xi^{\star} ({\bf X})$, i.e., $\xi ({\bf X}) \to \xi ({\bf X}) + \xi^{\star} ({\bf X})$, with sufficiently small $\Vert \xi^{\star} ({\bf X}) \Vert$ and $\Vert \nabla \xi^{\star} ({\bf X}) \Vert$. The infinitesimal increment of the nominal electric field $\tilde {\bf E}$ is expressed as
\begin{equation} \label{3d-incre-form-3}
\tilde {\bf E}^{\star} = \left. \frac{d}{d \tau} \Big( - \nabla \big(\xi ({\bf X}) + \tau \xi^{\star} ({\bf X}) \big) \Big)\right|_{\tau=0} = - \nabla \xi^{\star}.
\end{equation}

Consider the second row with variables $(\tilde {\bf E}, {\bf F})$ in Table \ref{table-DEP-nominal-true}. The infinitesimal increment $\tilde {\bf D}^\star$ of the nominal electric displacement $\tilde {\bf D} = \varepsilon J {\bf C}^{-1} \tilde {\bf E}$ is 
\begin{equation} \label{3d-incre-form-4}
\tilde {\bf D}^\star = \varepsilon J^\star {\bf C}^{-1} \tilde {\bf E} + \varepsilon J ({\bf C}^{-1})^\star \tilde {\bf E} + \varepsilon J {\bf C}^{-1} \tilde {\bf E}^\star,
\end{equation}
and the infinitesimal increment ${\bf E}^\star$ of the true electric field ${\bf E} = {\bf F}^{-T} \tilde {\bf E}$ is 
\begin{equation} \label{3d-incre-form-5}
{\bf E}^\star = ({\bf F}^{-T})^\star \tilde {\bf E} + {\bf F}^{-T} \tilde {\bf E}^\star,
\end{equation}
where the infinitesimal increments $J^\star$, $({\bf C}^{-1})^\star$, $({\bf F}^{-T})^\star$, and $\tilde {\bf E}^\star$ are given in Eqs.~\eqref{3d-incre-form-2} and \eqref{3d-incre-form-3}. The detailed derivations are $\qquad \tilde {\bf D}^\star = \left. \frac{d}{d \tau}\Big(\varepsilon (J + \tau J^\star) ({\bf C}^{-1} + \tau ({\bf C}^{-1})^\star) (\tilde {\bf E} + \tau \tilde {\bf E}^\star) \Big) \right|_{\tau=0} \qquad$ and $\qquad {\bf E}^\star = \left. \frac{d}{d \tau} \Big( ({\bf F}^{-T} + \tau ({\bf F}^{-T})^\star) (\tilde {\bf E} +\tau \tilde {\bf E}^\star) \Big)\right|_{\tau=0}$.\\

The infinitesimal increment of the Maxwell stress, i.e.,  ${\bf T}^M =  {\bf E} \otimes {\tilde {\bf D}} - \frac{\varepsilon J}{2 } |{\bf E}|^2{\bf F}^{-T}$, then is
\begin{equation}  \label{3d-incre-form-6}
({\bf T}^M)^\star =  {\bf E}^\star \otimes {\tilde {\bf D}} + {\bf E} \otimes {\tilde {\bf D}}^\star - \frac{\varepsilon}{2 } \left\{ J^\star |{\bf E}|^2{\bf F}^{-T} + 2 J ({\bf E} \cdot {\bf E}^\star){\bf F}^{-T} + J |{\bf E}|^2 ({\bf F}^{-T})^\star \right\}.
\end{equation}

Now the incremental version of the boundary-value problem \eqref{3d-total-bvp-refs} can be represented by
\begin{equation}  \label{3d-incre-form-7}
\left.
\begin{aligned}
& {\rm Div} \, {\bf T}^\star = {\bf 0}, \quad {\rm Curl} \, {\tilde {\bf E}}^\star = {\bf 0}, \quad {\rm Div} \, {\tilde {\bf D}}^\star = 0, \\
& \underbrace{{\bf u}^\star = {\bf 0} \ {\rm on} \ \partial \Omega_R^u, \ {\bf T}^\star {\bf N} = {\bf 0} \ {\rm on} \ \partial \Omega_R^t}_{\text{incremental mechanical BCs}},\quad \underbrace{\xi^\star = 0 \ {\rm on} \ \partial \Omega_R^{\xi}, \ {\tilde {\bf D}}^\star \cdot {\bf N} = 0 \ {\rm on} \ \partial \Omega_R^d}_{\text{incremental electric BCs}}, \\
& {\bf T}^\star = ({\bf T}^e)^\star + ({\bf T}^M)^\star, \\
& ({\bf T}^e)^\star =
\left\{ 
\begin{aligned}
& \frac{\partial^2 W^e ({\bf F})}{\partial {\bf F}^2} {\bf F}^\star, & {\text{compressible solids}},\\
&  \frac{\partial^2 W^e ({\bf F})}{\partial {\bf F}^2} {\bf F}^\star - \mathcal L_a^\star {\bf F}^{-T} - \mathcal L_a ({\bf F}^{-T})^\star, & J^\star=0 \ {\text{for incompressible solids}},
\end{aligned}\right.
\end{aligned}
\right\}
\end{equation}
along with Eqs.~\eqref{3d-incre-form-1}$-$\eqref{3d-incre-form-6}.

\subsection{An energy formulation of the electrostatics of deformable dielectrics}
In this section, we revisit the energy formulation of the electrostatic system of dielectric elastomer in three-dimensional space. The two independent variables in the energy formulation are the deformation gradient $\bf F$ and the nominal polarization $\tilde {\bf P}$. We take the variation of the energy functional to obtain the governing equations of the electrostatic system in equilibrium. Detailed energy formulation and the first variation of the energy functional can refer to the two works \cite{liu2014energy, yang2017revisiting}.

\subsubsection{Preliminary considerations}
Consider an electrostatic system that occupies a domain $\Omega_R$ with the boundary $\partial \Omega_R$ in the reference configuration. The mechanical boundary conditions on the boundary $\partial \Omega_R = \partial \Omega_R^u \cup \partial \Omega_R^t$ are 
\begin{equation} \label{y-3D-energyF-ME-BCs-1}
{\bf x} = {\bf x}_0 \quad {\rm on} \ \partial \Omega_R^u \quad {\rm and} \quad {\bf t} = {\bf t}_{0_R} \quad {\rm on} \ \partial \Omega_R^t,
\end{equation}
where ${\bf x}_0$ is the prescribed displacement and ${\bf t}_{0_R}$ is the prescribed traction force. \\

The electric boundary conditions on the boundary $\partial \Omega_R = \partial \Omega_R^\xi \cup \partial \Omega_R^d$, see Eq.~\eqref{y-ref-e-BCs}, are rewritten here 
\begin{equation} \label{y-3D-energyF-ME-BCs-2}
\xi = \xi_0 \quad {\rm on} \ \partial \Omega_R^\xi \quad {\rm and} \quad \tilde {\bf D} \cdot {\bf N} = {\tilde q}_{\rm f} \quad {\rm on} \ \partial \Omega_R^d,
\end{equation}
where $\xi_0$ is the prescribed voltage and ${\tilde q}_{\rm f}$ is the free charge density per unit area in the reference configuration.

Subjected to the electrostatic loads, the domain $\Omega_R$ of the system deforms into $\Omega$. Recall Sec. \ref{sec-4.3.2}, the relations of the true fields $(\mathbf{E}, \mathbf{D}, \mathbf{P})$ in the deformed domain $\Omega$ and the nominal fields $(\tilde{\mathbf{E}}, \tilde{\mathbf{D}}, \tilde{\mathbf{P}})$ in the undeformed domain $\Omega_R$ are listed as follows:
\begin{equation} \label{y-3D-energyF-relations}
\tilde{\mathbf{E}} = \mathbf{F}^T \mathbf{E}, \quad \tilde{\mathbf{D}} = J \mathbf{F}^{-1} \mathbf{D}, \quad \tilde{\mathbf{P}} = J \mathbf{P}.
\end{equation}
In terms of the above fields in Eq.~\eqref{y-3D-energyF-relations}, the Maxwell equations and the relations can be written as
\begin{equation} \label{y-3D-energyF-MW-relations}
\mathbf{F}^T \mathbf{E} = - \nabla \xi, \quad \text{Div}\, \tilde{\mathbf{D}} = 0, \quad \mathbf{F}\tilde{\mathbf{D}} = \varepsilon_0 J \mathbf{E} + \tilde{\mathbf{P}}.
\end{equation}

\subsubsection{Free energy of the electrostatic system}
Subjected to the electromechanical loads, the total free energy (see, for example, \cite{liu2014energy, deng2014flexoelectricity}) of the electrostatic system consists of three parts:
\begin{equation}\label{y-3D-energyF-free_E1}
\mathcal{F} [\mathbf{x},\tilde{\mathbf{P}}] = \mathcal{U} [\mathbf{x},\tilde{\mathbf{P}}] + \mathcal{E}^{\text{elect}} [\mathbf{x},\tilde{\mathbf{P}}] - \int_{\partial \Omega_R^t} {\bf t}_{0_R} \cdot {\bf x} dA.
\end{equation}
The last term on the RHS of Eq.~\eqref{y-3D-energyF-free_E1} represents the work done by the prescribed traction force ${\bf t}_{0_R}$. The first term $\mathcal{U} [\mathbf{x}, \tilde{\mathbf{P}}]$ is the internal energy, namely
\begin{equation}\label{y-3D-energyF-inter_E1}
\mathcal{U} [\mathbf{x}, \tilde{\mathbf{P}}] = \int_{\Omega_R} \Psi (\mathbf{F}, \tilde{\textbf{P}}) dV,
\end{equation}
where $\Psi (\mathbf{F}, \tilde{\textbf{P}})$ is the internal energy density. The electric energy, $\mathcal{E}^{\text{elect}} [\mathbf{x}, \tilde{\mathbf{P}}]$ in Eq.~\eqref{y-3D-energyF-free_E1}, is
\begin{equation}\label{y-3D-energyF-elec_E1}
\mathcal{E}^{\text{elect}} [\mathbf{x}, \tilde{\mathbf{P}}] = \frac{\varepsilon_0}{2}\int_{\Omega_R} J |\mathbf{E}|^2 dV + \int_{\partial \Omega_R^\xi} \xi \tilde{\mathbf{D}} \cdot \mathbf{N} dA,
\end{equation}
where the relations between $\mathbf{E}$, $\tilde{\mathbf{D}}$, and $\tilde{\mathbf{P}}$ are given by Eq.~\eqref{y-3D-energyF-relations}. The equilibrium state $(\mathbf{x}, \tilde{\mathbf{P}})$ is then determined by the variational principle of the total free energy, such that
\begin{equation}
\min_{(\mathbf{x},\tilde{\mathbf{P}})} \mathcal{F} [\mathbf{x},\tilde{\mathbf{P}}].
\end{equation}

An equilibrium state $(\mathbf{x}, \tilde{\mathbf{P}})$ is stable if its free energy $\mathcal{F} [\mathbf{x},\tilde{\mathbf{P}}]$ is lower than that of all the neighborhood states $\mathcal{F} [\mathbf{x} + \delta \mathbf{x},\tilde{\mathbf{P}} + \delta \tilde{\mathbf{P}}]$, such that
\begin{equation}
\mathcal{F} [\mathbf{x},\tilde{\mathbf{P}}] \le \mathcal{F} [\mathbf{x} + \delta \mathbf{x},\tilde{\mathbf{P}} + \delta \tilde{\mathbf{P}}].
\end{equation}

Here $\max\{||\delta \mathbf{x}||, ||\delta \tilde{\mathbf{P}}||\}$ is sufficiently small and $||\cdot||$ is a properly defined norm for the space of deformation functions. In the following, we take the variation of the energy functional to find the governing equations of the equilibrium state $(\mathbf{x}, \tilde{\mathbf{P}})$.

\subsubsection{Variation of the fields}
The infinitesimal variations of the deformation $\bf x$ and
the polarization $\tilde{\bf P}$ are defined as
\begin{equation}\label{y-3D-energyF-AA1} 
 \delta \mathbf{x} = \tau_1 {\mathbf{x}}_{1}, \quad \delta \tilde{\mathbf{P}} = \tau_2 \tilde{\mathbf{P}}_{1},
\end{equation}
where $\tau_1$ and $\tau_2$ are scalars with $\max \{|\tau_1|, |\tau_2|\} \ll 1$, and ${\mathbf{x}}_{1}$ and $\tilde{\mathbf{P}}_{1}$ are two smooth variations. \\

Based on $\delta \mathbf{x}$ and $\delta \tilde{\mathbf{P}}$ in Eq.~\eqref{y-3D-energyF-AA1}, the first variations of other fields, omitting the higher-order terms, are
\begin{equation}\label{y-3D-energyF-AA2} 
\mathbf{F} \to \mathbf{F} + \tau_1 {\mathbf{F}}_{d}, \  J \to J + \tau_1 J_{d}, \  \xi \to \xi + \tau_1 \xi_{d} + \tau_2 \xi_{p}, \  \tilde{\mathbf{D}} \to \tilde{\mathbf{D}} + \tau_1 \tilde{\mathbf{D}}_{d} +\tau_2 \tilde{\mathbf{D}}_{p}, \  \mathbf{E} \to \mathbf{E} + \tau_1 \mathbf{E}_{d} + \tau_2 \mathbf{E}_{p},
\end{equation}
where the subscripts \lq{}$d$\rq{} and \lq{}$p$\rq{} denote the variations with the deformation and the polarization, respectively. For example, the relations between the variations can be obtained as follows:
    \begin{equation} \label{y-3D-energyF-AA3} 
       {\mathbf{F}}_{d} = \frac{\textrm{d}}{\textrm{d} \tau_1} \nabla \big( {\mathbf{x} + \tau_1 \mathbf{x}_{1}} \big)\Big|_{\tau_1 =0} = \nabla {\mathbf{x}}_{1}, \quad
                        J_{d} = \frac{\textrm{d}}{\textrm{d} \tau_1} \textrm{det} \big( \mathbf{F} + \tau_1 {\mathbf{F}}_{d}\big) \Big|_{\tau_1 =0} = J \mathbf{F}^{-T} \cdot \nabla {\mathbf{x}}_{1}.
     \end{equation}

Moreover, taking the partial derivatives of the relations \eqref{y-3D-energyF-MW-relations} with respect to $\tau_1$ and $\tau_2$ at $\tau_1=\tau_2=0$, we have the following identities
\begin{equation} \label{y-3D-energyF-AA4} 
\begin{aligned} 
\mathbf{F}^T \mathbf{E}_{d} + (\mathbf{F}^T)_d \mathbf{E} = - \nabla \xi_d,& \quad \mathbf{F}^T \mathbf{E}_p = - \nabla \xi_p, \\
\mathbf{F}\tilde{\mathbf{D}}_{d} + \mathbf{F}_{d} \tilde{\mathbf{D}} = \varepsilon_0 J \mathbf{E}_{d} + \varepsilon_0 J_{d} \mathbf{E},& \quad \mathbf{F}\tilde{\mathbf{D}}_{p} = \varepsilon_0 J \mathbf{E}_{p} + \tilde{\mathbf{P}}_{1} 
\end{aligned}
 \end{equation}
and the variations of the Maxwell equations
\begin{equation} \label{y-3D-energyF-AA5} 
\text{Div} \, \tilde{\mathbf{D}}_{d} = \text{Div} \, \tilde{\mathbf{D}}_{p} = 0.
\end{equation}

Similarly, the variations of the boundary conditions \eqref{y-3D-energyF-ME-BCs-1} and \eqref{y-3D-energyF-ME-BCs-2} are
\begin{equation} \label{y-3D-energyF-var-ME-BCs-1}
{\bf x}_1 = {\bf 0}  \  {\rm on} \ \partial \Omega_R^u, \quad
\xi_d = \xi_p = 0 \ {\rm on} \ \partial \Omega_R^\xi,  \quad \tilde {\bf D}_d \cdot {\bf N}  = \tilde {\bf D}_p \cdot {\bf N} = 0 \ {\rm on} \ \partial \Omega_R^d.
\end{equation}

\subsubsection{First variation with respect to the polarization}
The first variation of the energy functional \eqref{y-3D-energyF-free_E1} with respect to the polarization $\tilde{\mathbf{P}}$ is defined as
\begin{equation} \label{y-3D-energyF-1st-p} 
\begin{aligned} 
& \frac{\textrm{d}}{\textrm{d} \tau_2} \mathcal{F} [\mathbf{x},\tilde{\mathbf{P}} + \tau_2 \tilde{\mathbf{P}}_1] \Big|_{\tau_2 =0} = \frac{\textrm{d}}{\textrm{d} \tau_2} \mathcal{U} [\mathbf{x},\tilde{\mathbf{P}} + \tau_2 \tilde{\mathbf{P}}_1] \Big|_{\tau_2 =0} + \frac{\textrm{d}}{\textrm{d} \tau_2} \mathcal{E}^{\text{elect}} [\mathbf{x},\tilde{\mathbf{P}} + \tau_2 \tilde{\mathbf{P}}_1] \Big|_{\tau_2 =0} \\
& = \int_{\Omega_R} \frac{\partial \Psi}{\partial \tilde{\mathbf{P}}}\cdot \tilde{\mathbf{P}}_{1} dV + \varepsilon_0 \int_{\Omega_R} J \mathbf{E} \cdot \mathbf{E}_{p} dV +  \int_{\partial \Omega_R^\xi} \big( \xi_p \tilde {\mathbf{D}} \cdot \mathbf{N} + \xi \tilde {\mathbf{D}}_{p} \cdot \mathbf{N} \big) dA.
\end{aligned}
 \end{equation}

By the divergence theorem and the relations $\eqref{y-3D-energyF-AA4}-\eqref{y-3D-energyF-var-ME-BCs-1}$, the vanishing of the first variation \eqref{y-3D-energyF-1st-p} can be further reduced to
\begin{equation} \label{y-3D-energyF-1st-p1} 
\begin{aligned} 
0 & = \frac{\textrm{d}}{\textrm{d} \tau_2} \mathcal{F} [\mathbf{x},\tilde{\mathbf{P}} + \tau_2 \tilde{\mathbf{P}}_1] \Big|_{\tau_2 =0} \\
& = \int_{\Omega_R} \Big( \frac{\partial \Psi}{\partial \tilde{\mathbf{P}}}\cdot \tilde{\mathbf{P}}_{1} + \varepsilon_0 J \mathbf{E} \cdot \mathbf{E}_{p} \Big) dV + \int_{\partial \Omega_R} \xi \tilde {\mathbf{D}}_p \cdot \mathbf{N} dA \\
& =  \int_{\Omega_R} \Big( \frac{\partial \Psi}{\partial \tilde{\mathbf{P}}}\cdot \tilde{\mathbf{P}}_{1} + \varepsilon_0 J \mathbf{E} \cdot \mathbf{E}_{p} \Big) dV + \int_{\Omega_R} \big( \xi \textrm{Div} \, \tilde {\mathbf{D}}_p + \tilde {\mathbf{D}}_p \cdot \nabla \xi \big) dV \\
& =  \int_{\Omega_R} \Big( \frac{\partial \Psi}{\partial \tilde{\mathbf{P}}}\cdot \tilde{\mathbf{P}}_{1} + \varepsilon_0 J \mathbf{E} \cdot \mathbf{E}_{p} \Big) dV - \int_{\Omega_R} \mathbf{E} \cdot \mathbf{F}\tilde {\mathbf{D}}_p dV \\
& = \int_{\Omega_R} \Big( \frac{\partial \Psi}{\partial \tilde{\mathbf{P}}}\cdot \tilde{\mathbf{P}}_{1} + \mathbf{E} \cdot \big( \varepsilon_0 J \mathbf{E}_{p} - \mathbf{F}\tilde {\mathbf{D}}_{p} \big) \Big) dV \\
& = \int_{\Omega_R} \big( \frac{\partial \Psi}{\partial \tilde{\mathbf{P}}} - \mathbf{E} \big) \cdot \tilde{\mathbf{P}}_{1} dV,
\end{aligned}
 \end{equation}
where $\tilde{\mathbf{P}}_{1}$ is arbitrary. Based on the basic lemma of calculus of variations, the identity \eqref{y-3D-energyF-1st-p1} holds if and only if
\begin{equation} \label{y-3D-energyF-1st-p-EP} 
\mathbf{E} = \frac{\partial \Psi}{\partial \tilde{\mathbf{P}}} \quad {\rm in} \ \Omega_R.
\end{equation}

Eq.~\eqref{y-3D-energyF-1st-p-EP} is the relation between the true electric field $\bf E$ and the nominal polarization $\tilde{\bf P}$. The relation can also be found in Eq.~(3.22) in the work \cite{liu2014energy} or Eq.~(14) in the work \cite{yang2017revisiting}.

\subsubsection{First variation with respect to the deformation}
By incorporating the constraint of incompressibility $J=1$, the energy functional \eqref{y-3D-energyF-free_E1} is reformulated as follows:
\begin{equation}\label{y-3D-energyF-free_E1-J}
\hat {\mathcal{F}} [\mathbf{x},\tilde{\mathbf{P}}] = \int_{\Omega_R} \left(\Psi (\mathbf{F}, \tilde{\textbf{P}}) + \frac{\varepsilon_0}{2} J |\mathbf{E}|^2 - \mathcal L_a (J-1)\right)dV + \int_{\partial \Omega_R^\xi} \xi \tilde{\mathbf{D}} \cdot \mathbf{N} dA - \int_{\partial \Omega_R^t} {\bf t}_{0_R} \cdot {\bf x} dA,
\end{equation}
where $\mathcal L_a$ is the Lagrange multiplier. The first variation of Eq.~\eqref{y-3D-energyF-free_E1-J} with respect to the deformation $\mathbf{x}$ is
\begin{equation} \label{y-3D-energyF-1st-x} 
\begin{aligned}
\frac{\textrm{d}}{\textrm{d} \tau_1} \hat{\mathcal{F}} [\mathbf{x}+ \tau_1 \mathbf{x}_1,\tilde{\mathbf{P}}] \Big|_{\tau_1 =0} & = \int_{\Omega_R} \Big( \frac{\partial \Psi}{\partial \mathbf{F}}\cdot \nabla\textbf{x}_1 + \frac{\varepsilon_0}{2}J_d |\mathbf{E}|^2 + \varepsilon_0 J \mathbf{E} \cdot \mathbf{E}_{d} - \mathcal L_a J_d \Big) dV \\
& \quad + \int_{\partial \Omega_R^\xi} \big( \xi_d \tilde {\mathbf{D}} \cdot \mathbf{N} + \xi \tilde {\mathbf{D}}_{d} \cdot \mathbf{N} \big) dA - \int_{\partial \Omega_R^t} {\bf t}_{0_R} \cdot {\bf x}_1 dA.
\end{aligned}
\end{equation}

By the relations $\eqref{y-3D-energyF-AA4}-\eqref{y-3D-energyF-var-ME-BCs-1}$ and the divergence theorem, the variation \eqref{y-3D-energyF-1st-x} can be rewritten as
\begin{equation}  \label{y-3D-energyF-1st-x0} 
\begin{aligned}
& \frac{\textrm{d}}{\textrm{d} \tau_1} \hat{\mathcal{F}} [\mathbf{x}+ \tau_1 \mathbf{x}_1,\tilde{\mathbf{P}}] \Big|_{\tau_1 =0} \\
& = \int_{\Omega_R} \Big( \frac{\partial \Psi}{\partial \mathbf{F}}\cdot \nabla\textbf{x}_1 + \frac{\varepsilon_0}{2}J_d |\mathbf{E}|^2 + \varepsilon_0 J \mathbf{E} \cdot \mathbf{E}_{d} - \mathcal L_a J_d \Big) dV + \int_{\partial \Omega_R} \xi \tilde {\mathbf{D}}_{d} \cdot \mathbf{N} dA - \int_{\partial \Omega_R^t} {\bf t}_{0_R} \cdot {\bf x}_1 dA \\
& = \int_{\Omega_R} \Big( \frac{\partial \Psi}{\partial \mathbf{F}}\cdot \nabla\textbf{x}_1 + \frac{\varepsilon_0}{2}J_d |\mathbf{E}|^2 + \varepsilon_0 J \mathbf{E} \cdot \mathbf{E}_{d} - \mathcal L_a J_d \Big) dV + \int_{\Omega_R} (\xi \textrm{Div} \, \tilde {\mathbf{D}}_{d} + \tilde {\mathbf{D}}_{d} \cdot \nabla \xi) dV - \int_{\partial \Omega_R^t} {\bf t}_{0_R} \cdot {\bf x}_1 dA \\
& = \int_{\Omega_R} \Big( \frac{\partial \Psi}{\partial \mathbf{F}}\cdot \nabla\textbf{x}_1 + \frac{\varepsilon_0}{2}J_d |\mathbf{E}|^2 - \mathcal L_a J_d + \mathbf{E} \cdot \big( \varepsilon_0 J \mathbf{E}_{d} - \mathbf{F}\tilde {\mathbf{D}}_{d} \big) \Big) dV - \int_{\partial \Omega_R^t} {\bf t}_{0_R} \cdot {\bf x}_1 dA \\
& = \int_{\Omega_R} \Big( \frac{\partial \Psi}{\partial \mathbf{F}}\cdot \nabla\textbf{x}_1 + \frac{\varepsilon_0}{2}J_d |\mathbf{E}|^2 - \mathcal L_a J_d + \mathbf{E} \cdot \big( \mathbf{F}_{d}\tilde {\mathbf{D}} - \varepsilon_0 J_d \mathbf{E} \big) \Big) dV - \int_{\partial \Omega_R^t} {\bf t}_{0_R} \cdot {\bf x}_1 dA \\
& = \int_{\Omega_R} \Big( \frac{\partial \Psi}{\partial \mathbf{F}} + \mathbf{E} \otimes \tilde {\mathbf{D}} - \frac{\varepsilon_0 J}{2} |\mathbf{E}|^2 \mathbf{F}^{-T}- \mathcal L_a J \mathbf{F}^{-T}  \Big) \cdot \nabla \textbf{x}_1 dV - \int_{\partial \Omega_R^t} {\bf t}_{0_R} \cdot {\bf x}_1 dA.
\end{aligned}
\end{equation}

By introducing the Piola-Maxwell stress tensor
\begin{equation}  \label{y-3D-energyF-pMS} 
\tilde{\bm\Sigma} = \mathbf{E} \otimes \tilde {\mathbf{D}} - \frac{\varepsilon_0 J}{2} |\mathbf{E}|^2 \mathbf{F}^{-T},
\end{equation}
the variation \eqref{y-3D-energyF-1st-x0} further becomes
\begin{equation}  \label{y-3D-energyF-1st-x1} 
\begin{aligned}
& \frac{\textrm{d}}{\textrm{d} \tau_1} \hat{\mathcal{F}} [\mathbf{x}+ \tau_1 \mathbf{x}_1,\tilde{\mathbf{P}}] \Big|_{\tau_1 =0} = \int_{\Omega_R} \Big( \frac{\partial \Psi}{\partial \mathbf{F}} + \tilde{\bm\Sigma} - \mathcal L_a J \mathbf{F}^{-T}  \Big) \cdot \nabla \textbf{x}_1 dV - \int_{\partial \Omega_R^t} {\bf t}_{0_R} \cdot {\bf x}_1 dA \\
& = \int_{\partial \Omega_R} \textbf{x}_1 \cdot \Big( \frac{\partial \Psi}{\partial \mathbf{F}} + \tilde{\bm\Sigma} - \mathcal L_a J \mathbf{F}^{-T}  \Big) \textbf{N} dA - \int_{\Omega_R} \textbf{x}_1 \cdot \textrm{Div} \Big( \frac{\partial \Psi}{\partial \mathbf{F}} + \tilde{\bm\Sigma} - \mathcal L_a J \mathbf{F}^{-T}  \Big) dV - \int_{\partial \Omega_R^t} {\bf t}_{0_R} \cdot {\bf x}_1 dA.
\end{aligned}
\end{equation}

Based on the basic lemma of calculus of variations, the vanishing of the variation \eqref{y-3D-energyF-1st-x1} gives the equilibrium equations
\begin{equation}  \label{y-3D-energyF-1st-x2} 
\textrm{Div} \Big( \frac{\partial \Psi}{\partial \mathbf{F}} + \tilde{\bm\Sigma} - \mathcal L_a J \mathbf{F}^{-T}  \Big) = {\bf 0}  \quad {\rm in} \ \Omega_R
\end{equation}
and the natural boundary conditions
\begin{equation}  \label{y-3D-energyF-1st-x3} 
\Big( \frac{\partial \Psi}{\partial \mathbf{F}} + \tilde{\bm\Sigma} - \mathcal L_a J \mathbf{F}^{-T}  \Big) \textbf{N} = {\bf t}_{0_R}  \quad {\rm on} \ \partial \Omega_R^t.
\end{equation}

\subsubsection{Summary of the boundary-value problem}
Based on the first variation of the energy functional, the 3D BVP of electrostatics of deformable dielectrics in the reference configuration are summarized as:
\begin{equation}  \label{y-3D-energyF-BVP} 
\left.
\begin{aligned}
& \underbrace{ \textrm{Div} \Big( \frac{\partial \Psi}{\partial \mathbf{F}} + \tilde{\bm\Sigma} - \mathcal L_a J \mathbf{F}^{-T}  \Big) = {\bf 0}}_{\text{equilibrium equation}}, \quad  \underbrace{\mathbf{F}^T \mathbf{E} = - \nabla \xi, \ \text{Div}\, \tilde{\mathbf{D}} = 0}_{\text{Maxwell's equations}}, \\
& \underbrace{\mathbf{E} = \frac{\partial \Psi}{\partial \tilde{\mathbf{P}}}, \quad \mathbf{F}\tilde{\mathbf{D}} = \varepsilon_0 J \mathbf{E} + \tilde{\mathbf{P}}}_{\text{relations}}, \quad \underbrace{\tilde{\bm\Sigma} = \mathbf{E} \otimes \tilde {\mathbf{D}} - \frac{\varepsilon_0 J}{2} |\mathbf{E}|^2 \mathbf{F}^{-T}}_{\text {Maxwell\rq{s} stress tensor}}, \\
& \underbrace{{\bf x} = {\bf x}_0 \ {\rm on} \ \partial \Omega_R^u, \quad \Big( \frac{\partial \Psi}{\partial \mathbf{F}} + \tilde{\bm\Sigma} - \mathcal L_a J \mathbf{F}^{-T}  \Big) \textbf{N} = {\bf t}_{0_R}}_{\text{mechanical BCs}}, \\
& \underbrace{\xi = \xi_0 \ {\rm on} \ \partial \Omega_R^{\xi}, \quad \tilde {\bf D} \cdot {\bf N} = \tilde q_{\rm f} \ {\rm on} \ \partial \Omega_R^d}_{\text{electric BCs}}, \quad \underbrace{{\bf F} = \nabla {\bf x}}_{\text{kinematics}}.
\end{aligned}
\right\}
\end{equation}

In the above representation, we have chosen the deformation gradient $\bf F$ and the nominal polarization $\tilde{\bf P}$ as our independent state variables. In addition, the body force is omitted. 

\subsection{Example 1: Pull-in instability of a dielectric elastomer film by using the incremental method}

 \begin{figure}[h] %
    \centering
    \includegraphics[width=5in]{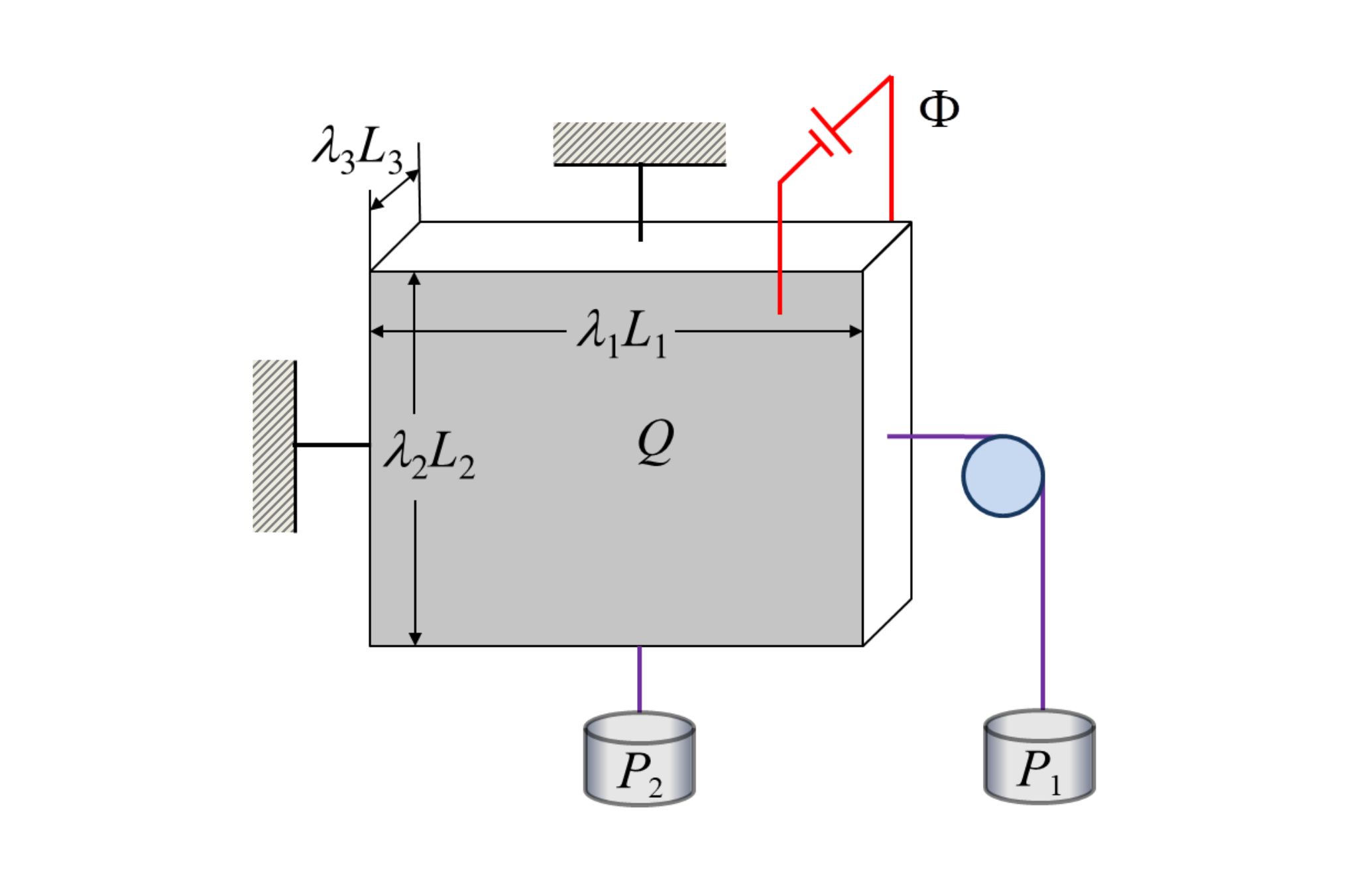}
    \caption{Schematic of a deformed film of dielectric elastomers coated with two compliant electrodes. The dead loads $P_1$ and $P_2$, and the voltage $\Phi$ deform the film from $L_1$, $L_2$, and $L_3$ to $\lambda_1 L_1$, $\lambda_2 L_2$, and $\lambda_3 L_3$, as well as induce an electric charge of magnitude $Q$ on either electrode \cite{zhao2007method}.}
    \label{C-example-1}
\end{figure}

Consider an undeformed film of dielectric elastomers with the dimensions $L_1$, $L_2$, and $L_3$. The upper and bottom surfaces of the film are both coated with compliant electrodes. The film deforms from $L_1 \times L_2 \times L_3$ to $\lambda_1 L_1 \times \lambda_2 L_2 \times \lambda_3 L_3$ by applying the in-plane dead loads $P_1$ and $P_2$ and in-thickness voltage $\Phi$ (see Fig.~\ref{C-example-1}).\\

Taking the Cartesian coordinates $(X_1, X_2, X_3)$ with an orthonormal basis $(\bf e_1, \bf e_2, \bf e_3)$, the domain occupied by the undeformed film is represented by
\begin{equation} \label{2019-10-17-1}
\Omega_R = \left\{ (X_1, X_2, X_3) \in \mathbb{R}^3: 0 \le X_1 \le L_1, 0 \le X_2 \le L_2, 0 \le X_3 \le L_3 \right\}.
\end{equation}
The boundary $\partial \Omega_R$ of $\Omega_R$ consists of six parts:
\begin{equation} \label{2019-10-17-2}
\left.
\begin{aligned}
\mathcal S_{1-} & = \left\{ {\bf X} \in \mathbb{R}^3: X_1 =0 \right\}, \quad \mathcal S_{1+} = \left\{ {\bf X} \in \mathbb{R}^3: X_1 =L_1 \right\}, \\
\mathcal S_{2-} & = \left\{ {\bf X} \in \mathbb{R}^3: X_2 =0 \right\}, \quad \mathcal S_{2+} = \left\{ {\bf X} \in \mathbb{R}^3: X_2 =L_2 \right\}, \\
\mathcal S_{3-} & = \left\{ {\bf X} \in \mathbb{R}^3: X_3 =0 \right\}, \quad \mathcal S_{3+} = \left\{ {\bf X} \in \mathbb{R}^3: X_3 =L_3 \right\}.
\end{aligned}
\right\}
\end{equation}

The mechanical and electric boundary conditions are
\begin{equation} \label{2019-10-17-3}
\left.
\begin{aligned}
{\bf T} (- {\bf e}_1) & = {\bf t}_R = - s_1 {\bf e}_1, \ \tilde {\bf D} \cdot (-{\bf e}_1) = 0 \ {\rm on} \ \mathcal S_{1-} , \quad {\bf T} {\bf e}_1 = {\bf t}_R = s_1 {\bf e}_1, \ \tilde {\bf D} \cdot {\bf e}_1 = 0  \ {\rm on} \ \mathcal S_{1+}, \\
{\bf T} (- {\bf e}_2) & = {\bf t}_R = - s_2 {\bf e}_2, \ \tilde {\bf D} \cdot (-{\bf e}_2) = 0 \ {\rm on} \ \mathcal S_{2-} , \quad {\bf T} {\bf e}_2 = {\bf t}_R = s_2 {\bf e}_2, \ \tilde {\bf D} \cdot {\bf e}_2 = 0 \ {\rm on} \ \mathcal S_{2+}, \\
{\bf T} (- {\bf e}_3) & = {\bf t}_R =  {\bf 0}, \ \xi = \Phi \quad {\rm on} \ \mathcal S_{3-} , \qquad \qquad \quad \ \ \,  {\bf T} {\bf e}_3 = {\bf t}_R = {\bf 0}, \ \xi = 0 \quad {\rm on} \ \mathcal S_{3+}, 
\end{aligned}
\right\}
\end{equation}
where $s_1 = P_1 / (L_2 L_3)$ and $s_2 = P_2 / (L_1 L_3)$.

\subsubsection{Large deformation}
We assume that the dielectric film admits the following homogeneous deformation
\begin{equation} \label{2019-10-17-4}
x_1 = \lambda_1 X_1, \ x_2 = \lambda_2 X_2, \ x_3 = \lambda_3 X_3,
\end{equation}
where $\lambda_1$, $\lambda_2$, and $\lambda_3$ are constant stretches. Then the matrix presentation of the deformation gradient is a $3 \times 3$ diagonal matrix
\begin{equation}\label{2019-10-17-5}
{\mathbf F} := 
\begin{pmatrix}
\lambda_1 & 0 & 0 \\
  0 &\lambda_2 & 0 \\
   0 & 0 & \lambda_3
\end{pmatrix}
= {\rm diag} \, (\lambda_1, \lambda_2, \lambda_3).
\end{equation}
Consequently, 
\begin{equation}\label{2019-10-17-6}
{\mathbf F}^{-1} = {\mathbf F}^{-T} := {\rm diag} \, (\lambda_1^{-1}, \lambda_2^{-1}, \lambda_3^{-1}), \quad
{\mathbf C}^{-1} := {\rm diag} \, (\lambda_1^{-2}, \lambda_2^{-2}, \lambda_3^{-2}), \quad J = \det \, {\mathbf F} = \lambda_1 \lambda_2 \lambda_3 = 1.
\end{equation}

Note that ${\mathbf F}$, ${\mathbf F}^{-1}$, ${\mathbf F}^{-T}$ and ${\mathbf C}^{-1}$ are independent of the coordinates. For linear dielectric elastomers without considering the free change, the Maxwell equations in the BVP \eqref{3d-total-bvp-refs} give the three-dimensional Laplace equation
\begin{equation} \label{2019-10-17-7}
\frac{\partial^2 \xi}{\partial X_1^2} + \frac{\partial^2 \xi}{\partial X_2^2} + \frac{\partial^2 \xi}{\partial X_3^2} = 0.
\end{equation}
By the electric boundary conditions in Eq.~\eqref{2019-10-17-3}, the solution of $\xi$ is given by
\begin{equation} \label{2019-10-17-8}
\xi ({\bf X}) = \Phi \frac{L_3 - X_3}{L_3},
\end{equation}
and the nominal electric field and displacement are
\begin{equation} \label{2019-10-17-9}
\tilde {\bf E} = - \nabla \xi= \frac{\Phi}{L_3} {\bf e}_3 = \tilde E {\bf e}_3 \quad {\rm and} \quad  \tilde {\bf D} = \varepsilon J {\bf C}^{-1} \tilde {\bf E} = \varepsilon \lambda_3^{-2} \tilde E {\bf e}_3 = {\tilde D} {\bf e}_3
\end{equation}
that are independent of the coordinates. By Eqs.~\eqref{2019-10-17-5} and \eqref{2019-10-17-8}, the Maxwell stress is obtained as
\begin{equation} \label{2019-10-17-10}
{\bf T}^M := \frac{\varepsilon {\tilde E}^2}{2} {\rm diag} \, ( -\lambda_1^{-1} \lambda_3^{-2}, -\lambda_2^{-1} \lambda_3^{-2}, \lambda_3^{-3} ).
\end{equation}

For incompressible neo-Hookean materials, the elastic stress is given by Eq.~\eqref{ex-neo-H-PK}. By Eqs.~\eqref{2019-10-17-5}, \eqref{2019-10-17-6} and \eqref{2019-10-17-10}, the total stress tensor is 
\begin{equation} \label{2019-10-17-11}
{\bf T} := {\rm diag} \, (\mu \lambda_1 - \mathcal L_a \lambda_1^{-1} - \frac{1}{2} \varepsilon \lambda_1^{-1} \lambda_3^{-2} {\tilde E}^2, \mu \lambda_2 - \mathcal L_a \lambda_2^{-1} - \frac{1}{2} \varepsilon \lambda_2^{-1} \lambda_3^{-2} {\tilde E}^2, \mu \lambda_3 - \mathcal L_a \lambda_3^{-1} + \frac{1}{2} \varepsilon \lambda_3^{-3} {\tilde E}^2).
\end{equation}
By the mechanical boundary conditions in Eq.~\eqref{2019-10-17-3}, we finally have
\begin{subequations} \label{2019-10-17-12}
\begin{align}
\label{2019-10-17-12a}
\mu \lambda_1 - \mathcal L_a \lambda_1^{-1} - \frac{1}{2} \varepsilon \lambda_1^{-1} \lambda_3^{-2} {\tilde E}^2 & = s_1, \\
\label{2019-10-17-12b}
\mu \lambda_2 - \mathcal L_a \lambda_2^{-1} - \frac{1}{2} \varepsilon \lambda_2^{-1} \lambda_3^{-2} {\tilde E}^2 & = s_2, \\
\label{2019-10-17-12c}
\mu \lambda_3 - \mathcal L_a \lambda_3^{-1} + \frac{1}{2} \varepsilon \lambda_3^{-3} {\tilde E}^2 & = 0,
\end{align}
\end{subequations}
which, together with the constraint of incompressibility, i.e., $\lambda_1 \lambda_2 \lambda_3 = 1$, can give the solutions of $(\lambda_1, \lambda_2, \lambda_3, \mathcal L_a)$. One can also reduce the number of the algebraic equations from four to two. By substituting $\lambda_3 = \lambda_1^{-1} \lambda_2^{-1}$ and $\mathcal L_a = \mu \lambda_3^2 + \frac{1}{2} \varepsilon \lambda_3^{-2} {\tilde E}^2$ into Eqs.~\eqref{2019-10-17-12a} and \eqref{2019-10-17-12a}, we have
\begin{subequations} \label{2019-10-17-13}
\begin{align}
\label{2019-10-17-13a}
\mu (\lambda_1 - \lambda_1^{-3} \lambda_2^{-2}) - \varepsilon \lambda_1 \lambda_2^{2} {\tilde E}^2 & = s_1, \\
\label{2019-10-17-13b}
\mu (\lambda_2 - \lambda_2^{-3} \lambda_1^{-2}) - \varepsilon \lambda_2 \lambda_1^{2} {\tilde E}^2 & = s_2,
\end{align}
\end{subequations}
which, by the relation $\tilde E = \varepsilon^{-1} \lambda_1^{-2} \lambda_2^{-2} \tilde D$ in Eq.~$\eqref{2019-10-17-9}_2$, are the same as Eqs. (6a) and (6b) in \cite{zhao2007method}.\\

A trivial solution to the boundary-value problem is summarized here: the nominal electric field is $\tilde {\bf E} = \frac{\Phi}{L_3} {\bf e}_3 = \tilde E {\bf e}_3$ in Eq.~\eqref{2019-10-17-9}, stretches $\lambda_1$ and $\lambda_2$ are the solutions of Eqs.~\eqref{2019-10-17-13a} and \eqref{2019-10-17-13b}, $\lambda_3 = \lambda_1^{-1} \lambda_2^{-1}$ and the Lagrange multiplier is $\mathcal L_a = \mu \lambda_3^2 + \frac{1}{2} \varepsilon \lambda_3^{-2} {\tilde E}^2$.

\subsubsection{Incremental boundary-value problem}
In contrast to Eq.~\eqref{2019-10-17-3}, the incremental mechanical and electric boundary conditions are
\begin{equation} \label{2019-10-18-0}
\left.
\begin{aligned}
{\bf T}^\star (- {\bf e}_1) & = {\bf 0}, \ \tilde {\bf D}^\star \cdot (-{\bf e}_1) = 0 \ {\rm on} \ \mathcal S_{1-} , \quad {\bf T}^\star {\bf e}_1 = {\bf 0}, \ \tilde {\bf D}^\star \cdot {\bf e}_1 = 0  \ {\rm on} \ \mathcal S_{1+}, \\
{\bf T}^\star (- {\bf e}_2) & = {\bf 0}, \ \tilde {\bf D}^\star \cdot (-{\bf e}_2) = 0 \ {\rm on} \ \mathcal S_{2-} , \quad {\bf T}^\star {\bf e}_2 = {\bf 0}, \ \tilde {\bf D}^\star \cdot {\bf e}_2 = 0 \ {\rm on} \ \mathcal S_{2+}, \\
{\bf T}^\star (- {\bf e}_3) & = {\bf 0}, \ \xi^\star = 0 \quad {\rm on} \ \mathcal S_{3-} , \qquad \quad \quad {\bf T}^\star {\bf e}_3 = {\bf 0}, \ \xi^\star = 0 \quad {\rm on} \ \mathcal S_{3+}.
\end{aligned}
\right\}
\end{equation}
At the trivial solution, the infinitesimal increments of the displacement are assumed to be
\begin{equation} \label{2019-10-18-1}
u_1^\star = \lambda_1^\star X_1, \ u_2^\star = \lambda_2^\star X_2, \ u_3^\star = \lambda_3^\star X_3,
\end{equation}
where $\lambda_1^\star$, $\lambda_2^\star$, and $\lambda_3^\star$ are sufficiently small and independent of the coordinates. By Eqs.~\eqref{3d-incre-form-1} and \eqref{2019-10-18-1}, the infinitesimal increment of the deformation gradient is
\begin{equation} \label{2019-10-18-2}
{\bf F}^\star := {\rm diag} \, (\lambda_1^\star, \lambda_2^\star, \lambda_3^\star).
\end{equation}
With Eqs.~\eqref{2019-10-17-5} and \eqref{2019-10-18-2}, other increments in Eq.~\eqref{3d-incre-form-2} at the trivial solution can be obtained as
\begin{equation} \label{2019-10-18-3}
\left.
\begin{aligned}
& ({\bf F}^{-1})^{\star}  = ({\bf F}^{-T})^{\star} := -{\rm diag} \, (\lambda_1^{-2} \lambda_1^\star, \lambda_2^{-2} \lambda_2^\star, \lambda_3^{-2} \lambda_3^\star), \quad  J^\star = \lambda_2 \lambda_3 \lambda_1^\star + \lambda_3 \lambda_1 \lambda_2^\star +\lambda_1 \lambda_2 \lambda_3^\star, \\
& {\bf C}^\star := 2 \, {\rm diag} \, (\lambda_1 \lambda_1^\star, \lambda_2 \lambda_2^\star, \lambda_3 \lambda_3^\star), \quad ({\bf C}^{-1})^{\star} := -2 \, {\rm diag} \, (\lambda_1^{-3}\lambda_1^\star, \lambda_2^{-3}\lambda_2^\star, \lambda_3^{-3}\lambda_3^\star).
\end{aligned}
\right\}
\end{equation}
For incompressible solids, the infinitesimal increment of the Jacobian is zero, i.e.,
\begin{equation} \label{2019-10-18-4}
J^\star = \lambda_2 \lambda_3 \lambda_1^\star + \lambda_3 \lambda_1  \lambda_2^\star +\lambda_1 \lambda_2 \lambda_3^\star = 0.
\end{equation}

Consider incompressible neo-Hookean solids \eqref{ex-neo-H-strain-energy}. By Eqs.~\eqref{2019-10-17-5} and \eqref{2019-10-18-3}, the incremental elastic stress tensor in the incremental BVP \eqref{3d-incre-form-7} is obtained as
\begin{equation} \label{2019-10-18-4a}
({\bf T}^e)^\star := {\rm diag} \, (\mu \lambda_1^\star - \lambda_1^{-1} \mathcal L_a^\star + \mathcal L_a \lambda_1^{-2} \lambda_1^\star, \mu \lambda_2^\star - \lambda_2^{-1} \mathcal L_a^\star + \mathcal L_a \lambda_2^{-2} \lambda_2^\star, \mu \lambda_3^\star - \lambda_3^{-1} \mathcal L_a^\star + \mathcal L_a \lambda_3^{-2} \lambda_3^\star),
\end{equation}
where $\mathcal L_a = \mu \lambda_3^2 + \frac{1}{2} \varepsilon \lambda_3^{-2} {\tilde E}^2$ and $\mathcal L_a^\star$ is the infinitesimal increment of the Lagrange multiplier.\\

The infinitesimal increment of the electric potential is $\xi^\star$, then the incremental nominal electric field \eqref{3d-incre-form-3} is $\tilde {\bf E}^\star = - \nabla \xi^\star$. Also, the incremental nominal electric displacement \eqref{3d-incre-form-4} is $\tilde {\bf D}^\star = \varepsilon ({\bf C}^{-1})^\star \tilde {\bf E} + \varepsilon {\bf C}^{-1} \tilde {\bf E}^\star$ by using $J=1$ and $J^\star = 0$. Since  ${\bf C}^{-1}$ in Eq.~\eqref{2019-10-17-6}, $({\bf C}^{-1})^\star$ in Eq.~\eqref{2019-10-18-3}, $\tilde {\bf E}$ in Eq.~\eqref{2019-10-17-9} are independent of the coordinates, the incremental Maxwell equation ${\rm Div} \, {\tilde {\bf D}}^\star = 0$ in Eq.~\eqref{3d-incre-form-7} implies ${\rm Div} \, {\tilde {\bf E}}^\star = 0$, namely
\begin{equation} \label{2019-10-18-5}
\frac{\partial^2 \xi^\star}{\partial X_1^2} + \frac{\partial^2 \xi^\star}{\partial X_2^2} + \frac{\partial^2 \xi^\star}{\partial X_3^2} = 0,
\end{equation}
which, together with the incremental electric boundary conditions in Eq.~\eqref{2019-10-18-0}, gives the solution
\begin{equation} \label{2019-10-18-6}
\xi^\star ({\bf X}) = 0.
\end{equation}
Then the increments at the trivial solution become
\begin{equation} \label{2019-10-18-7}
\tilde {\bf E}^{\star} = {\bf 0} \quad {\rm and} \quad \tilde {\bf D}^\star = -2\varepsilon \lambda_3^{-3} {\tilde E} \lambda_3^\star {\bf e}_3 .
\end{equation}

By Eqs.~$\eqref{2019-10-17-6}_1$ and $\eqref{2019-10-17-9}_1$, the true electric field is ${\bf E} = {\bf F}^{-T} \tilde {\bf E} = \lambda_3^{-1} \tilde E {\bf e}_3$. It follows from Eqs.~$\eqref{2019-10-18-3}_1$, $\eqref{2019-10-17-9}_1$ and $\eqref{2019-10-18-7}_1$ that the increment ${\bf E}^\star$ in Eq.~\eqref{3d-incre-form-5} is obtained as ${\bf E}^\star = - \lambda_3^{-2} {\tilde E} \lambda_3^\star {\bf e}_3$. Thus, the incremental Maxwell stress \eqref{3d-incre-form-6} is 
\begin{equation} \label{2019-10-18-8}
({\bf T}^M)^\star := \frac{\varepsilon \tilde E^2}{2} {\rm diag} \, (\lambda_1^{-2} \lambda_3^{-2} \lambda_1^\star + 2 \lambda_1^{-1} \lambda_3^{-3} \lambda_3^\star, \lambda_2^{-2} \lambda_3^{-2} \lambda_2^\star + 2 \lambda_2^{-1} \lambda_3^{-3} \lambda_3^\star, -3 \lambda_3^{-4} \lambda_3^\star).
\end{equation}

By Eqs.~\eqref{2019-10-18-4a} and \eqref{2019-10-18-8}, we have the increment ${\bf T}^\star$ of the total stress in the incremental BVP \eqref{3d-incre-form-7}. Thus, the incremental mechanical boundary conditions in Eq.~\eqref{2019-10-18-0} give
\begin{subequations} \label{2019-10-18-9}
\begin{align}
 \label{2019-10-18-9a}
  (\mu + \mu \lambda_1^{-2} \lambda_3^2 +\varepsilon \lambda_1^{-2} \lambda_3^{-2} \tilde E^2)\lambda_1^\star + \varepsilon \lambda_1^{-1} \lambda_3^{-3} \tilde E^2 \lambda_3^\star - \lambda_1^{-1} \mathcal L_a^\star & = 0, \\
  \label{2019-10-18-9b}
 (\mu + \mu \lambda_2^{-2} \lambda_3^2 + \varepsilon \lambda_2^{-2} \lambda_3^{-2}\tilde E^2) \lambda_2^\star + \varepsilon \lambda_2^{-1} \lambda_3^{-3} \tilde E^2 \lambda_3^\star - \lambda_2^{-1} \mathcal L_a^\star & = 0,      \\
   \label{2019-10-18-9c}
  (2\mu - \varepsilon \lambda_3^{-4}\tilde E^2)\lambda_3^\star - \lambda_3^{-1} \mathcal L_a^\star & = 0,
\end{align}
\end{subequations}
which, together with Eq.~\eqref{2019-10-18-4}, can give the solutions of $(\lambda_1^\star, \lambda_2^\star, \lambda_3^\star, \mathcal L_a^\star)$. 

\subsubsection{Solution of the incremental boundary-value problem}
Here we have 4 equations in 4 unknowns, i.e., Eqs.~\eqref{2019-10-18-4}, \eqref{2019-10-18-9a}$-$\eqref{2019-10-18-9c} with variables $(\lambda_1^\star, \lambda_2^\star, \lambda_3^\star, \mathcal L_a^\star)$. It follows from Eq.~\eqref{2019-10-18-9c} that $\mathcal L_a^\star = (2\mu \lambda_3 - \varepsilon \lambda_3^{-3} \tilde E^2)\lambda_3^\star$. By substituting $\mathcal L_a^\star$ into Eqs.~\eqref{2019-10-18-9a} and \eqref{2019-10-18-9b}, we can reduce the set of simultaneous equations to 3 equations in 3 unknowns. The set of equations may be expressed as
\begin{equation} \label{2019-10-18-10}
\begin{pmatrix}
\vspace{0.1in}
\mu + \mu \lambda_1^{-2} \lambda_3^2 +\varepsilon \lambda_1^{-2} \lambda_3^{-2} \tilde E^2 & 0  & 2\varepsilon \lambda_1^{-1} \lambda_3^{-3} \tilde E^2 - 2\mu \lambda_1^{-1} \lambda_3 \\
\vspace{0.1in}
0  & \mu + \mu \lambda_2^{-2} \lambda_3^2 + \varepsilon \lambda_2^{-2} \lambda_3^{-2} \tilde E^2 & 2\varepsilon \lambda_2^{-1} \lambda_3^{-3} \tilde E^2 - 2\mu \lambda_2^{-1} \lambda_3 \\
\vspace{0.1in}
\lambda_2 \lambda_3 & \lambda_3 \lambda_1 & \lambda_1 \lambda_2
\end{pmatrix} 
\begin{pmatrix}
\vspace{0.1in}
\lambda_1^\star\\
\vspace{0.1in}
\lambda_2^\star\\
\vspace{0.1in}
\lambda_3^\star
\end{pmatrix}
= {\bf 0}.
\end{equation}

If the determinant of the $3 \times 3$ matrix is non-zero, the simultaneous linear equations only have the zero solution, i.e., $\lambda_1^\star = \lambda_2^\star = \lambda_3^\star = 0$. If the determinant is zero, there exists an infinite number of solutions. Thus, the necessary condition for the existence of incremental solutions is a zero determinant. \\

In the following, we will show that the condition for a zero determinant of the $3 \times 3$ matrix in Eq.~\eqref{2019-10-18-10} is the same as that of the Hessian matrix in \cite{zhao2007method}. \\

Consider the $3 \times 3$ matrix in Eq.~\eqref{2019-10-18-10} as Matrix(I). Taking $\lambda_3 = \lambda_1^{-1} \lambda_2^{-1}$ and replacing row (1) by row (1) + $2\mu \lambda_1^{-3} \lambda_2^{-2} \times$  row (3), row (2) by row (2) + $2\mu \lambda_1^{-2} \lambda_2^{-3} \times$  row (3) in Matrix(I), we obtain Matrix(II). Subsequently, replacing column (1) by column (1) + $\lambda_1^{-2} \lambda_2^{-1} \times$  column (3), column (2) by column (2) + $\lambda_1^{-1} \lambda_2^{-2} \times$  column (3) in Matrix(II), we get Matrix(III). We multiply column (3) in Matrix(III) by $-\varepsilon^{-1} {\tilde E}^{-1} \lambda_1^{-3} \lambda_2^{-3}$ and obtain Matrix(IV). Multiplying row (3) in Matrix(IV) by $-{\tilde E}$, we finally get Matrix(V), which, by the relation $\tilde E = \varepsilon^{-1} \lambda_1^{-2} \lambda_2^{-2} \tilde D$, is exact the Hessian matrix in \cite{zhao2007method}. \\

By the effects of row/column operations of determinants \cite{meyer2000matrix}, we have the following relations: $\det \, (\text{Matrix(I)}) = \det \, (\text{Matrix(II)}) = \det \, (\text{Matrix(III)}) = \det \, (\text{Matrix(IV)}) / (-\varepsilon^{-1} {\tilde E}^{-1} \lambda_1^{-3} \lambda_2^{-3})$ and $\det \, (\text{Matrix(IV)}) = \det \, (\text{Matrix(V)}) / (-{\tilde E})$. Thus, the condition for a zero determinant of the $3 \times 3$ matrix in Eq.~\eqref{2019-10-18-10} is equivalent to that of the Hessian matrix in \cite{zhao2007method}.

\subsection{Example 2: Wrinkle surface instability of a dielectric elastomer by using the incremental method} \label{surface-wrinkling-ex}

In addition to pull-in instability in the homogeneous deformation, surface instabilities, which lead to inhomogeneous deformation (wrinkling), are also common. Given the existence of an extensive body of work on the electromechanical instabilities related to inhomogeneous deformations, we simply point to a small sample of the works that the reader may consult (and the references therein): \cite{dorfmann2010nonlinear, katia2011instabilities, wang2013creasing, park2013electromechanical, dorfmann2014instabilities, lu2015electro, liang2017new, seifi2017electro, bortot2017tuning, bortot2018prismatic, mao2017morphology, zurlo2017catastrophic, mao2018voltage, su2018wrinkles, su2019finite, greaney2018out, fu2018reduced, YANG2022111306, huang2011electromechanical, he2009dielectric}. \\

\begin{figure}[h] %
\centering
\subfigure[]{%
\includegraphics[width=2.9in]{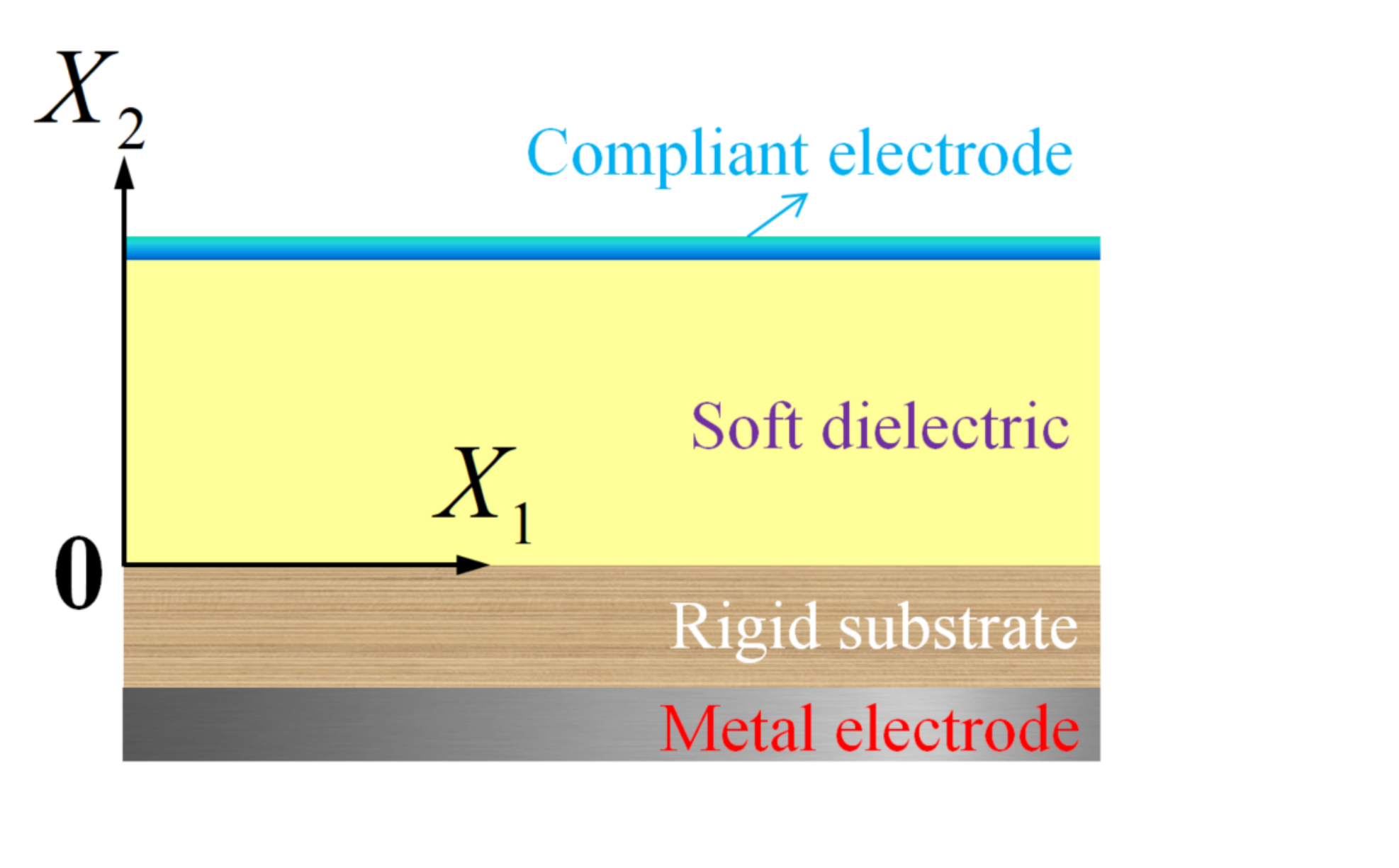}
\label{wrinkle-1a}}
\subfigure[]{%
\includegraphics[width=2.9in]{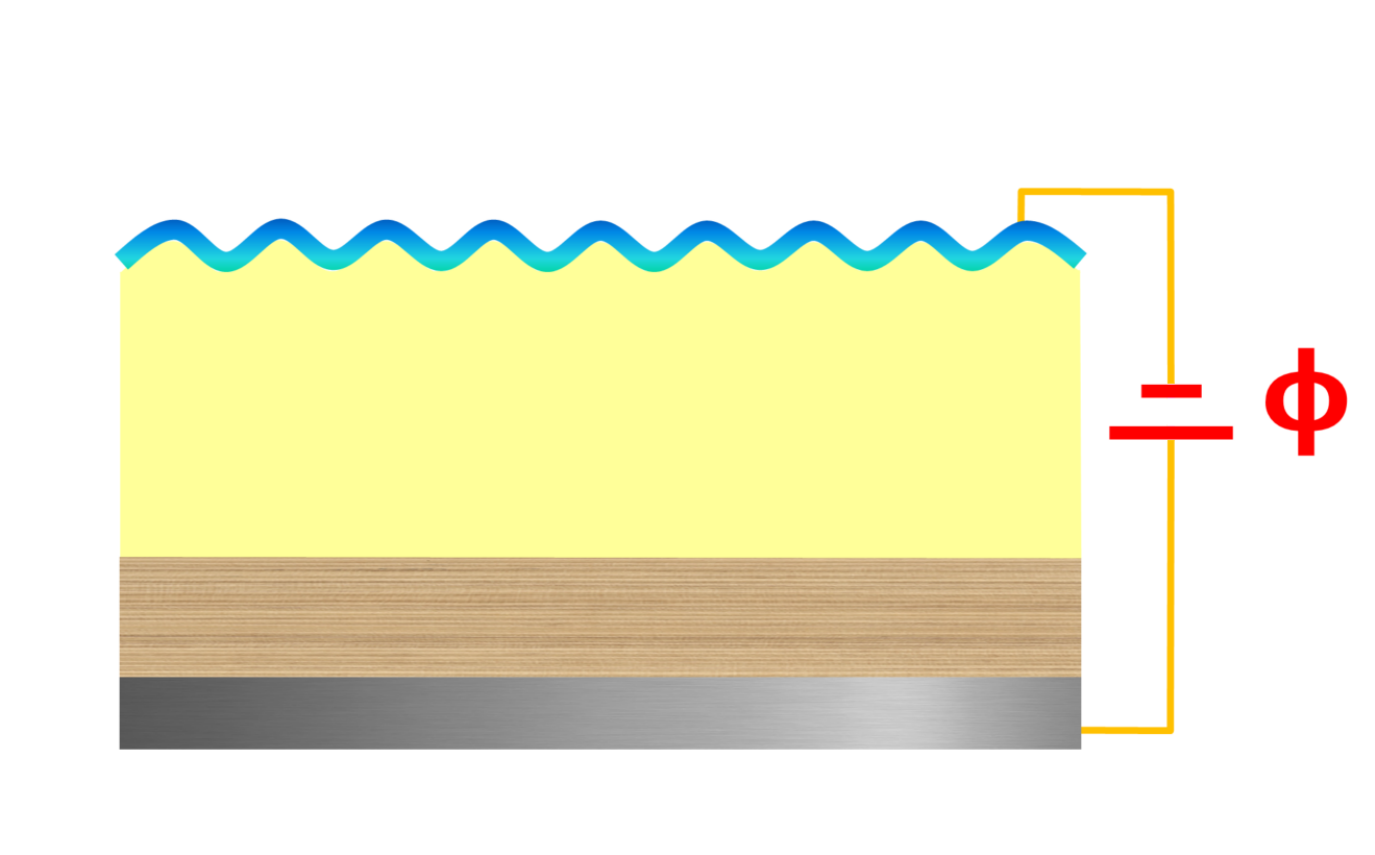}
\label{wrinkle-1b}}
\caption{Schematic of the surface wrinkling of a dielectric block bonded on a rigid substrate. The upper surface is bonded with a compliant electrode while the rigid substrate is coated with a metal electrode. A voltage is applied to the two electrodes. The corresponding experiment can refer to the work \cite{wang2013creasing}. (a) Undeformed state with a flat surface. (b) Wrinkled state by increasing the voltage at the threshold.}
\label{wrinkle-1}
\end{figure}

The schematic of the onset of surface wrinkling induced by an applied electric field is shown in Fig.~\ref{wrinkle-1}. Due to inhomogeneous deformations, the electro-wrinkling problem is more involved. We will use linear bifurcation analysis to find the necessary condition for the existence of a non-trivial solution (surface wrinkling). We employ a similar procedure as what can be found in the appendix of the work by Wang and Zhao \cite{wang2013creasing}.

\subsubsection{Large deformation} 
For the electromechanical system in Fig.~\ref{wrinkle-1}, the displacement of the bottom surface is fixed. We just consider a 2D formulation for the plane strain assumption. Thus the kinematical boundary conditions at the bottom surface are:
\begin{equation} \label{w-z-13-1}
{\bf u}({\bf X}) = {\bf 0} \quad {\rm or} \quad u_1 (X_1, X_2) = u_2 (X_1, X_2) = 0 \quad {\rm at} \ X_2 = 0.
\end{equation}
The deformation gradient is 
\begin{equation}  \label{w-z-13-2}
{\bf F} = \nabla ({\bf X} + {\bf u}) = {\bf I} + \nabla {\bf u} := 
\begin{pmatrix}
1+ u_{1,1} & u_{1,2} \\
u_{2,1} & 1+ u_{2,2}
\end{pmatrix}.
\end{equation}
The constraint of incompressibility reads
\begin{equation}  \label{w-z-13-3}
J = \det {\bf F} = 
\begin{vmatrix}
1+ u_{1,1} & u_{1,2} \\
u_{2,1} & 1+ u_{2,2}
\end{vmatrix} = 1.
\end{equation}
Alternatively, calculation of the determinant \eqref{w-z-13-3} gives
\begin{equation}  \label{w-z-13-4}
u_{1,1} + u_{2,2} + u_{1,1} u_{2,2} - u_{1,2} u_{2,1} = 0.
\end{equation}
The traction-free boundary conditions at the upper surface are\footnote{${\bf T} {\bf e}_2 = (T_{11} {\bf e}_1 \otimes {\bf e}_1 + T_{12} {\bf e}_1 \otimes {\bf e}_2 + T_{21}{\bf e}_2 \otimes {\bf e}_1 + T_{22}{\bf e}_2 \otimes {\bf e}_2) {\bf e}_2 = T_{12} {\bf e}_1 + T_{22} {\bf e}_2 = {\bf 0}$.}
\begin{equation} \label{w-z-13-5}
{\bf T} {\bf e}_2 = {\bf 0} \quad {\rm or} \quad T_{12} (X_1, X_2) = T_{22} (X_1, X_2) = 0 \quad {\rm at} \ X_2 = H.
\end{equation}
The true electric field induced by an applied voltage $\Phi$ is approximately written as
\begin{equation} \label{w-z-13-6}
{\bf E} = E_{\delta} \, {\bf e}_2 = (0, E_{\delta})^T,
\end{equation}
where
\begin{equation} \label{w-z-13-7}
E_{\delta} = \frac{\Phi}{H + \delta H} = \frac{\Phi}{H + u_2(X_1, H)}.
\end{equation}
The total Cauchy stress $\boldsymbol{\sigma}$ in Eq.~\eqref{Cauchy-s-decom} is rewritten here as
\begin{equation} \label{w-z-13-8}
\boldsymbol{\sigma} = \boldsymbol{\sigma}^e + \boldsymbol{\sigma}^M,
\end{equation}
where the elastic stress and the Maxwell stress for neo-Hookean ideal dielectrics are
\begin{equation} \label{w-z-13-9}
{\boldsymbol \sigma}^e = \mu {\bf F}{\bf F}^T - \mathcal L_a {\bf I} = \mu ({\bf I} + \nabla {\bf u} + \nabla{\bf u}^T  + \nabla {\bf u} \nabla{\bf u}^T) - \mathcal L_a {\bf I},
\end{equation}
and
\begin{equation} \label{w-z-13-10}
\boldsymbol{\sigma}^M = \varepsilon {\bf E} \otimes {\bf E} - \frac{\varepsilon |{\bf E}|^2}{2} {\bf I} := \frac{\varepsilon}{2} E_{\delta}^2
\begin{pmatrix}
- 1  & 0 \\
0    &  1
\end{pmatrix}.
\end{equation}
In contrast to Eq.~\eqref{w-z-13-8}, the (first) Piola-Kirchhoff stress, with the relation \eqref{Cauchy-PK-rel}, is given by
\begin{equation} \label{w-z-13-11}
{\bf T} = J {\boldsymbol \sigma} {\bf F}^{-T} = (\boldsymbol{\sigma}^e + \boldsymbol{\sigma}^M){\bf F}^{-T},
\end{equation}
where $J=1$ for incompressible materials. Without the body force, the equilibrium equation is
\begin{equation} \label{w-z-13-12}
{\rm Div} \,  {\bf T} = {\bf 0}.
\end{equation}
Now we have constructed a boundary-value problem:
\begin{equation} \label{w-z-13-13}
\left. 
\begin{aligned}
& {\rm Div} \,  {\bf T} = {\bf 0}, \quad \det {\bf F} =1, \quad {\bf E} = E_{\delta} {\bf e}_2, \\
& u_1 (X_1, X_2) = u_2 (X_1, X_2) = 0 \qquad \, {\rm at} \ X_2 = 0, \\
& T_{12} (X_1, X_2) = T_{22} (X_1, X_2) = 0 \quad \ {\rm at} \ X_2 = H,
\end{aligned}
\right\}
\end{equation}
with
\begin{equation} \label{w-z-13-14}
\left.
\begin{aligned}
& {\bf F} = {\bf I} + \nabla {\bf u}, \quad {\bf T} = (\boldsymbol{\sigma}^e + \boldsymbol{\sigma}^M){\bf F}^{-T}, \quad E_{\delta} = \frac{\Phi}{H + u_2(X_1, H)}, \\
& {\boldsymbol \sigma}^e = \mu ({\bf I} + \nabla {\bf u} + \nabla{\bf u}^T  + \nabla {\bf u} \nabla{\bf u}^T) - \mathcal L_a {\bf I}, \quad \boldsymbol{\sigma}^M := \frac{\varepsilon}{2} E_{\delta}^2
\begin{pmatrix}
- 1  & 0 \\
0    &  1
\end{pmatrix}.
\end{aligned}
\right\}
\end{equation}

A trivial solution to the boundary-value problem is
\begin{equation} \label{w-z-13-15}
{\bf u}_0 = {\bf 0}, \quad \mathcal L_{a0} = \mu + \frac{\varepsilon}{2} \Big( \frac{\Phi}{H}\Big)^2,
\end{equation}
implying
\begin{equation} \label{w-z-13-16}
{\bf F}_0 = {\bf I}, \quad \boldsymbol{\sigma}_0^{e} := \frac{\varepsilon}{2} \Big( \frac{\Phi}{H}\Big)^2
\begin{pmatrix}
-1  & 0 \\
0    &  -1
\end{pmatrix}, \quad
\boldsymbol{\sigma}_0^{M} := \frac{\varepsilon}{2} \Big( \frac{\Phi}{H}\Big)^2
\begin{pmatrix}
- 1  & 0 \\
0    &  1
\end{pmatrix}.
\end{equation}

\subsubsection{Small deformation simplification} 

Prior to the following analysis, we would like to simplify the problem in small deformation. Subsequently, the constraint of incompressibility \eqref{w-z-13-4} reduces to
\begin{equation} \label{w-z-13-s1}
u_{1,1} + u_{2,2} = 0 \quad {\rm or} \quad {\rm Div} \,  {\bf u} = 0,
\end{equation}
the magnitude of the true electric field \eqref{w-z-13-7} becomes
\begin{equation} \label{w-z-13-s2}
E_{\delta} = \frac{\Phi}{H} \left(1 -\frac{u_2(x_1, H)}{H}\right),
\end{equation}
the elastic stress \eqref{w-z-13-9} and the Maxwell stress \eqref{w-z-13-10} are reduced to
\begin{align}
\label{w-z-13-s4}
& \boldsymbol{\sigma}^e = \mu ({\bf I} + \nabla {\bf u} + \nabla {\bf u}^T) - \mathcal L_a {\bf I}, \\
\label{w-z-13-s5}
& \boldsymbol{\sigma}^M := \frac{\varepsilon E_0^2}{2} \left(1 -2\frac{u_2(x_1, H)}{H}\right)
\begin{pmatrix}
- 1  & 0 \\
0    &  1
\end{pmatrix}.
\end{align}
Now the boundary-value problem \eqref{w-z-13-13} and \eqref{w-z-13-14} reduces to
\begin{equation} \label{w-z-13-s6}
\left. 
\begin{aligned}
& {\rm Div} \, {\boldsymbol \sigma} = {\bf 0}, \quad {\rm Div} \, {\bf u} =0, \quad E_{\delta} = E_0 \left( 1 -\frac{u_2(x_1, H)}{H} \right), \\
& u_1 (x_1, x_2) = u_2 (x_1, x_2) = 0 \qquad \, {\rm at} \ x_2 = 0, \\
& \sigma_{12} (x_1, x_2) = \sigma_{22} (x_1, x_2) = 0 \quad \ {\rm at} \ x_2 = H,
\end{aligned}
\right\}
\end{equation}
with
\begin{equation}  \label{w-z-13-s7}
\left.
\begin{aligned}
& \boldsymbol{\sigma} = \boldsymbol{\sigma}^e + \boldsymbol{\sigma}^M, \quad {\boldsymbol \sigma}^e = \mu ({\bf I} + \nabla {\bf u} + \nabla {\bf u}^T) - \mathcal L_a {\bf I}, \\
& \boldsymbol{\sigma}^M := \frac{\varepsilon E_0^2}{2} \left (1  -2 \frac{u_2(x_1, H)}{H}\right)
\begin{pmatrix}
- 1  & 0 \\
0    &  1
\end{pmatrix}.
\end{aligned}
\right\}
\end{equation}
\subsubsection{Incremental boundary-value problem} 
At the trivial solution \eqref{w-z-13-15}, the incremental forms of the displacement and the Lagrange multiplier are
\begin{equation} \label{w-z-13-s8}
{\bf u} = {\bf u}_0 + {\bf u}^\star = {\bf u}^\star, \quad \mathcal L_a = \mathcal L_{a0} + \mathcal L_a^\star,
\end{equation}
where ${\bf u}^\star$ and $\mathcal L_a^\star$ are increments with sufficiently small $\Vert {\bf u}^\star \Vert$, $\Vert \nabla {\bf u}^\star \Vert$, $\Vert \mathcal L_a^\star \Vert$, and $\Vert \nabla \mathcal L_a^\star \Vert$. Then the incremental form of the constraint ${\rm Div} \, {\bf u} =0$ in Eq.~\eqref{w-z-13-s6} is
\begin{equation} \label{w-z-13-s9}
{\rm Div} \, {\bf u}^\star = u_{1,1}^\star + u_{2,2}^\star = 0.
\end{equation}
The incremental form of the elastic stress ${\boldsymbol \sigma}^e$ in Eq.~\eqref{w-z-13-s7} is
\begin{equation} \label{w-z-13-s10}
{\boldsymbol \sigma}^e = {\boldsymbol \sigma}_0^{e} + ({\boldsymbol \sigma}^e)^\star = {\boldsymbol \sigma}_0^{e} + \mu (\nabla {\bf u}^\star + (\nabla{\bf u}^\star)^T) - \mathcal L_a^\star {\bf I}.
\end{equation}
The magnitude of the electric field in Eq.~\eqref{w-z-13-s6} is
\begin{equation} \label{w-z-13-s11}
E_{\delta} = E_0 \left( 1 - \frac{u_2^\star(x_1, H)}{H} \right)
\end{equation}
and the incremental form of the Maxwell stress is
\begin{equation} \label{w-z-13-s12} 
\boldsymbol{\sigma}^{M} = \boldsymbol{\sigma}_0^{M} + (\boldsymbol{\sigma}^{M})^\star := \frac{\varepsilon}{2} E_0^2
\begin{pmatrix}
- 1  & 0 \\
0    &  1
\end{pmatrix} - \varepsilon E_0^2 \frac{u_2^\star (x_1, H)}{H}
\begin{pmatrix}
- 1  & 0 \\
0    &  1
\end{pmatrix}.
\end{equation}
Therefore, the incremental form of the boundary-value problem \eqref{w-z-13-s6} and \eqref{w-z-13-s7} is
\begin{equation} \label{w-z-13-s13}
\left. 
\begin{aligned}
&{\rm Div} \, {\boldsymbol \sigma}^\star = {\bf 0}, \quad u_{1,1}^\star + u_{2,2}^\star =0, \quad E_{\delta}^\star = - E_0 \frac{u_2^\star (x_1, H)}{H}, \\
& u_1^\star (x_1, x_2) = u_2^\star (x_1, x_2) = 0 \qquad \, {\rm at} \ x_2 = 0, \\
& \sigma_{12}^\star (x_1, x_2) = \sigma_{22}^\star (x_1, x_2) = 0 \quad \ {\rm at} \ x_2 = H,
\end{aligned}
\right\}
\end{equation}
with
\begin{equation}  \label{w-z-13-s14}
\left.
\begin{aligned}
& \boldsymbol{\sigma}^\star = (\boldsymbol{\sigma}^e)^\star + (\boldsymbol{\sigma}^M)^\star, \quad ({\boldsymbol \sigma}^e)^\star = \mu (\nabla {\bf u}^\star + (\nabla{\bf u}^T)^\star) - \mathcal L_a^\star {\bf I}, \\
& (\boldsymbol{\sigma}^M)^\star := -\varepsilon E_0^2 \frac{u_2^\star (x_1, H)}{H}
\begin{pmatrix}
- 1  & 0 \\
0    &  1
\end{pmatrix}.
\end{aligned}
\right\}
\end{equation}
\subsubsection{Solution of the incremental boundary-value problem} 
An approximation used here is that ${\rm Div} \, (\boldsymbol{\sigma}^M)^\star = {\bf 0}$ \cite{wang2013creasing, seifi2017electro}. With the constraint, $u_{1,1}^\star + u_{2,2}^\star = 0$, we have ${\rm Div} \, (\nabla{\bf u}^T)^\star = {\bf 0}$.\footnote{Since $u_{1,1}^\star + u_{2,2}^\star =u_{i,i}^\star = u_{j,j}^\star=0$, $i,j = 1,2$, then ${\rm Div} \, (\nabla{\bf u}^T)^\star = {\rm Div} \, (u_{j,i}^\star {\bf e}_i \otimes {\bf e}_j) = u_{j,ij}^\star {\bf e}_i = \frac{\partial}{\partial x_i} \Big( u_{j,j}^\star {\bf e}_i \Big) = {\bf 0}.$} Thus, the equilibrium equation in Eq.~\eqref{w-z-13-s13} reduces to
\begin{equation}  \label{w-z-13-s15}
{\rm Div} \, (\mu \nabla {\bf u}^\star - \mathcal L_a^\star {\bf I}) = \mu \nabla^2 {\bf u}^\star - \nabla \mathcal L_a^\star = {\bf 0}.
\end{equation}
With the constraint $u_{1,1}^\star + u_{2,2}^\star = 0$, the general solution of ${\bf u}^\star (x_1, x_2)$ can be written as 
\begin{equation}  \label{w-z-13-s16} 
u_1^\star (x_1, x_2) = \frac{\partial \phi}{\partial x_2} \quad {\rm and} \quad u_2^\star (x_1, x_2) = - \frac{\partial \phi}{\partial x_1},
\end{equation}
where $\phi (x_1, x_2)$ is an arbitrary function. Substitution of Eq.~\eqref{w-z-13-s16} into Eq.~\eqref{w-z-13-s14} gives
\begin{equation}  \label{w-z-13-s17} 
\boldsymbol \sigma ^\star := 
\begin{pmatrix}
\displaystyle 2 \mu \frac{\partial^2 \phi}{\partial x_2 \partial x_1} - \mathcal L_a^\star & \displaystyle \mu \Big( \frac{\partial^2 \phi}{\partial x_2^2} - \frac{\partial^2 \phi}{\partial x_1^2} \Big) \\
& \\
\displaystyle \mu \Big( \frac{\partial^2 \phi}{\partial x_2^2} - \frac{\partial^2 \phi}{\partial x_1^2} \Big)       & \displaystyle -2 \mu \frac{\partial^2 \phi}{\partial x_1\partial x_2} - \mathcal L_a^\star
\end{pmatrix}
+ \frac{\varepsilon E_0^2}{H} \frac{\partial \phi (x_1, H)}{\partial x_1}
\begin{pmatrix}
-1 & 0 \\
0  & 1
\end{pmatrix}.
\end{equation}
By using the general solution \eqref{w-z-13-s16}, the equilibrium equation \eqref{w-z-13-s15} becomes
\begin{equation}  \label{w-z-13-s18} 
\mu \Big( \frac{\partial^3 \phi}{\partial x_2 \partial x_1^2} + \frac{\partial^3 \phi}{\partial x_2^3} \Big) - \frac{\partial \mathcal L_a^\star}{\partial x_1} = 0, \quad \mu \Big( \frac{\partial^3 \phi}{\partial x_1^3} + \frac{\partial^3 \phi}{\partial x_1 \partial x_2^2} \Big) + \frac{\partial \mathcal L_a^\star}{\partial x_2} = 0,
\end{equation}
the kinematical boundary conditions in Eq.~\eqref{w-z-13-s13} read
\begin{equation}  \label{w-z-13-s19} 
\frac{\partial \phi}{\partial x_2} = \frac{\partial \phi}{\partial x_1} = 0 \quad {\rm at} \ x_2 =0,
\end{equation}
and the traction-free boundary conditions in Eq.~\eqref{w-z-13-s13} are
\begin{equation}  \label{w-z-13-s20} 
\frac{\partial^2 \phi}{\partial x_2^2} - \frac{\partial^2 \phi}{\partial x_1^2} = 0, \quad -2 \mu \frac{\partial^2 \phi}{\partial x_1\partial x_2} - \mathcal L_a^\star+\frac{\varepsilon E_0^2}{H} \frac{\partial \phi (x_1, H)}{\partial x_1} = 0 \quad {\rm at} \ x_2 = H.
\end{equation}

The boundary-value problem \eqref{w-z-13-s18}$ - $\eqref{w-z-13-s20} is traditional and the solution procedure can refer to, for example, the work \cite{wang2013creasing, seifi2017electro, yang2017revisiting, yang2017wrinkle}. Consider the forms of the solution 
\begin{equation}  \label{w-z-13-s21} 
\phi (x_1, x_2) = \bar\phi(x_2) \cos (k x_1)   \quad    {\rm and}    \quad    \mathcal L_a^\star (x_1, x_2) = \bar{ \mathcal L} (x_2) \sin (k x_1).
\end{equation}
Then the boundary-value problem \eqref{w-z-13-s18}$ - $\eqref{w-z-13-s20} is converted to
\begin{subequations}
\begin{equation}  \label{w-z-13-s22a} 
\mu \Big( - k^2 \frac{\partial \bar\phi}{\partial x_2} + \frac{\partial^3 \bar\phi}{\partial x_2^3} \Big) - k \bar {\mathcal L} = 0, \quad \mu \Big( k^3 \bar \phi - k \frac{\partial^2 \bar\phi}{\partial x_2^2} \Big) + \frac{\partial \bar{\mathcal L}}{\partial x_2} = 0, \quad x_2 \in (0, H),
\end{equation}
\begin{equation}  \label{w-z-13-s22b} 
\frac{\partial \bar \phi}{\partial x_2} = 0, \quad \bar \phi = 0 \quad {\rm at} \ x_2 =0,
\end{equation}
\begin{equation}  \label{w-z-13-s22c} 
\frac{\partial^2 \bar \phi}{\partial x_2^2} + k^2 \bar\phi  = 0, \quad 2 \mu k \frac{\partial \bar\phi}{\partial x_2} - \bar{\mathcal L} - \frac{\varepsilon E_0^2}{H} k \bar{\phi} = 0 \quad {\rm at} \ x_2 = H.
\end{equation}
\end{subequations}
By eliminating $\bar{\mathcal L}$ in Eq.~\eqref{w-z-13-s22a}, we have a fourth-order differential equation
\begin{equation}  \label{w-z-13-s23a} 
\frac{\partial^4 \bar \phi}{\partial x_2^4} - 2k^2 \frac{\partial^2 \bar \phi}{\partial x_2^2} + k^4 \bar \phi = 0
\end{equation}
with the general solution
\begin{equation}  \label{w-z-13-s23b} 
\bar \phi (x_2) = (C_1 +C_2 x_2) e^{k x_2} + (C_3 +C_4 x_2) e^{-k x_2},
\end{equation}
where $C_i$, $i=1,2,3,4$, are constant. Substituting Eq.~\eqref{w-z-13-s23b} into Eq.~$\eqref{w-z-13-s22a}_1$, we have
\begin{equation}  \label{w-z-13-s23c} 
\bar{\mathcal L} (x_2) = 2 \mu k (C_2 e^{k x_2} + C_4 e^{-k x_2}).
\end{equation}

Consider the boundary conditions \eqref{w-z-13-s22b} and \eqref{w-z-13-s22c}. We encounter a set of simultaneous linear equations, that is, we have four equations in four unknowns $C_1$, $C_2$, $C_3$, $C_4$ of the form
\begin{equation}  \label{w-z-13-s23d} 
\begin{pmatrix}
k                           &                       1                    &               -k             &                      1 \\
1                           &                       0                    &               1              &                       0 \\
2 k^2 e^{k H}       &       2 (1+kH) k e^{k H}          &    2 k^2 e^{-k H}     & 2 (-1+kH) k e^{-k H} \\
(\displaystyle\frac{\varepsilon E_0^2}{H} - 2k\mu ) k e^{kH}   & (\varepsilon E_0^2 - 2kH\mu ) k e^{kH} &               (\displaystyle\frac{\varepsilon E_0^2}{H} + 2k\mu ) k e^{-kH}             &      (\varepsilon E_0^2 + 2kH\mu ) k e^{-kH} 
\end{pmatrix}
 \begin{pmatrix}
 C_1 \\
 C_2 \\
 C_3 \\
 C_4
 \end{pmatrix} ={\bf 0}.
\end{equation}

The above equations \eqref{w-z-13-s23d} only have non-trivial solutions $C_i$, $i=1,2,3,4$, if the determinant of the coefficient matrix vanishes, which gives
\begin{equation}  \label{w-z-13-s23e} 
\frac{\varepsilon E_0^2}{\mu} =\frac{ 2kH \left (e^{4kH} + (2+ 4k^2 H^2) e^{2kH} +1\right)}{e^{4kH} - 4kH e^{2kH} - 1}.
\end{equation}

 \begin{figure}[h] %
    \centering
    \includegraphics[width=5in]{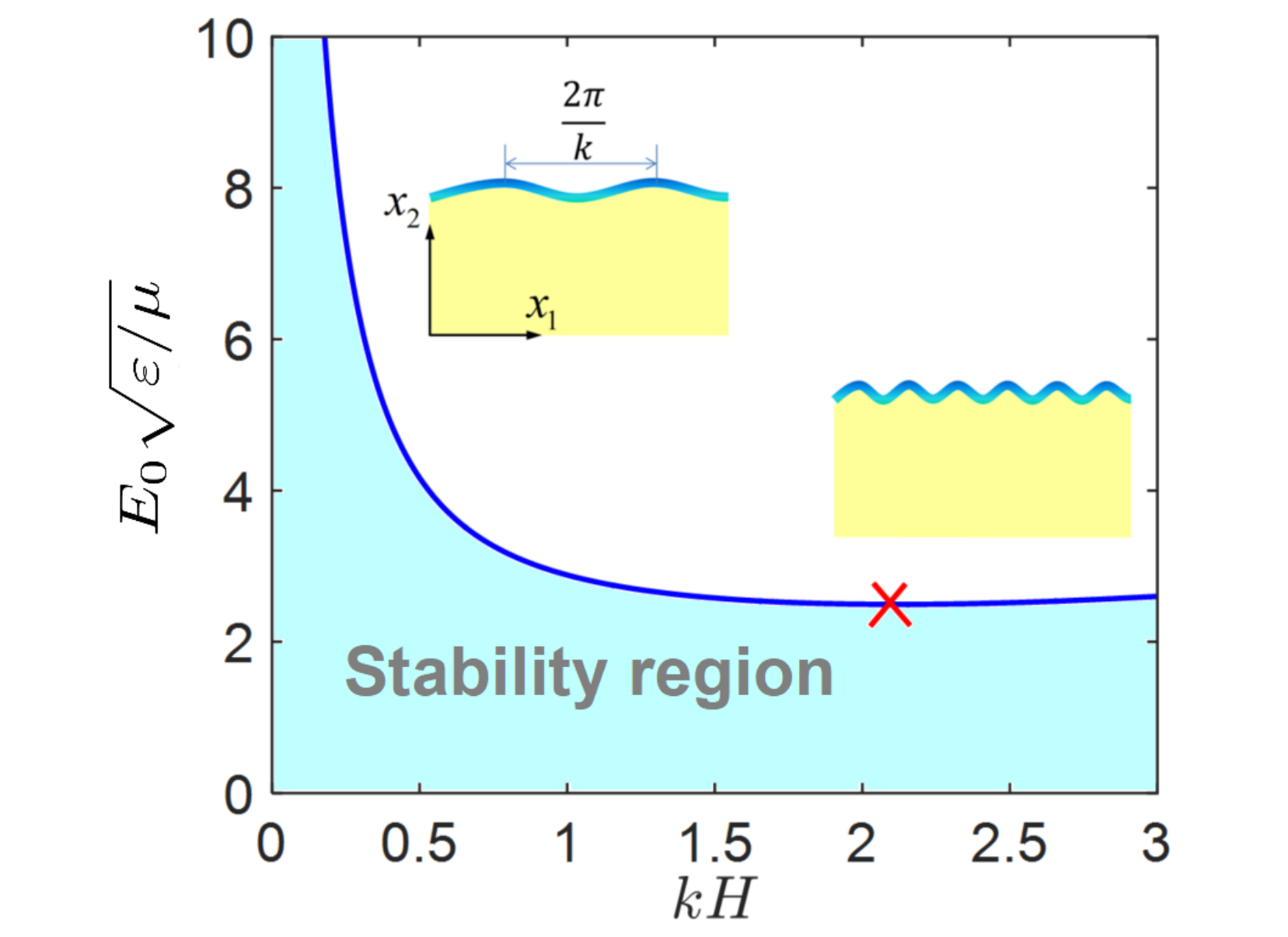}
    \caption{The electric field for the onset of surface wrinkling of neo-Hookean ideal dielectrics. The (blue) curve comes from the numerical plot of Eq.~\eqref{w-z-13-s23e}. The lowest point on the curve is denoted by a cross mark at $E_0 \sqrt{\varepsilon/\mu} = 2.49$ and $kH=2.12$. The cross mark corresponds to the lowest electric field and can be regarded as the threshold of the surface wrinkling instability. Below the lowest point of the curve, the elastomer has a flat surface, which means that the system shown in Fig.~\ref{wrinkle-1} is stable and there is no electromechanical wrinkling.}
    \label{wrinkle2}
\end{figure}

Equation \eqref{w-z-13-s23e} is exactly equation (1) in the work by Wang and Zhao \cite{wang2013creasing} in the absence of surface energy. In Fig.~\ref{wrinkle2}, we plot how the electric field $E_0 \sqrt{\varepsilon/\mu}$ varies with respect to the ratio $kH$. The lowest electric field (the threshold) for the onset of electromechanical surface wrinkling is $E_0 \sqrt{\varepsilon/\mu} = 2.49$ at $kH=2.12$. If the applied electric field is less than the threshold, there is no electromechanical wrinkling and the elastomer admits a flat surface. \\

The block dimensions would indeed affect the critical electric field and the wavelength $2 \pi / k$ (or the wavenumber $k$) of surface wrinkling. In the analysis, we regard a continuous wavenumber $k$ to carry out our plot. This treatment is accurate enough if the length of the elastomer block is sufficiently large, which admits relatively dense discrete values of $k$. In addition to the length, the height $H$ of the elastomer block affects the wrinkling phenomenon in terms of the product $k H$. For the onset of surface wrinkling in Eq.~\eqref{w-z-13-s23e}, a larger electric field corresponds to a smaller product $k H$, giving a larger wavelength $2 \pi / k$ of wrinkles for an elastomer block with a given geometry. A detailed discussion of the effects of block dimensions and side boundary conditions on the surface wrinkling is out of the scope of this tutorial.

\section{Electromechanical instability and bifurcation of soft dielectrics: pull-in instability} \label{section-e-stability-zhao}
In this section, we mainly revisit pull-in instability of soft dielectrics by using the energy method. Other types of instability can refer to two good reviews \cite{zhao2014harnessing, dorfmann2019instabilities}. There is an extensive body of work on pull-in instability of a dielectric elastomer film in the homogeneous deformation \cite{zhao2007method, norris2008comment, diaz2008electromechanical, xu2010electromechanical, li2011electromechanical, yang2017avoiding, alameh2018emergent, ChenLL2020nonlinear, LI2021106507}. Here we focus on two works \cite{zhao2007method, yang2017avoiding} and discuss pull-in instability by both the principle of minimum energy (i.e., the energy method) and the linear bifurcation analysis.

\subsection{Pull-in instability of a dielectric elastomer film}
We review the example of electromechanical instability of a film of a dielectric elastomer by Zhao and Suo \cite{zhao2007method}. The stability is examined by using the principle of minimum energy. Since, in this example, the deformation is homogeneous, this is the same as the so-called Hessian approach.

\subsubsection{Stability analysis by using the Hessian approach}
Consider the electrostatic system shown in Fig.~\ref{C-example-1}. Subject to in-plane dead loads $P_1$ and $P_2$ and in-thickness voltage $\Phi$, the film deforms from $L_1 \times L_2 \times L_3$ to $\lambda_1 L_1 \times \lambda_2 L_2 \times \lambda_3 L_3$. The constraint of incompressibility gives the relation $\lambda_3 = 1/ (\lambda_1 \lambda_2)$ between the stretches. The free energy $G$ of the system is given by \cite{zhao2007method, suo2008nonlinear}
\begin{equation} \label{zhao-07-1}
\frac{G }{L_1 L_2 L_3}=  W (\lambda_1, \lambda_2, \tilde{D}) - \frac{P_1}{L_2 L_3} \lambda_1 - \frac{P_2}{L_1 L_3} \lambda_2 - \frac{\Phi}{L_3} \tilde{D},
\end{equation}
where $W (\lambda_1, \lambda_2, \tilde{D})$ is the free energy function. With the small variations of the generalized coordinates, $\delta \lambda_1$, $\delta \lambda_2$, $\delta \tilde {D}$, the variation of the free energy is 
\begin{align} \label{zhao-07-2}
\frac{\Delta G }{L_1 L_2 L_3} &=  \Big( \frac{\partial W}{\partial \lambda_1} - \frac{P_1}{L_2 L_3} \Big) \delta \lambda_1 + \Big( \frac{\partial W}{\partial \lambda_2} - \frac{P_2}{L_1 L_3} \Big) \delta \lambda_2 + \Big( \frac{\partial W}{\partial \tilde D}- \frac{\Phi}{L_3} \Big) \delta \tilde{D} \nonumber \\
                                           & \quad + \frac{1}{2} \frac{\partial^2 W}{\partial \lambda_1^2} (\delta \lambda_1)^2 + \frac{1}{2} \frac{\partial^2 W}{\partial \lambda_2^2} (\delta \lambda_2)^2 + \frac{1}{2} \frac{\partial^2 W}{\partial {\tilde D}^2} (\delta \tilde D)^2  \nonumber \\
                                           & \quad + \frac{\partial^2 W}{\partial \lambda_1 \partial \lambda_2} \delta \lambda_1 \delta \lambda_2 + \frac{\partial^2 W}{\partial \lambda_1 \partial {\tilde D}} \delta \lambda_1 \delta {\tilde D} + \frac{\partial^2 W}{\partial \lambda_2 \partial \tilde D} \delta \lambda_2 \delta \tilde D,
\end{align}
where only the first and second variations are retained and all the high-order terms are omitted. \\

In equilibrium, the first variation is zero, we have
\begin{equation} \label{zhao-07-3}
 \frac{\partial W}{\partial \lambda_1} - \frac{P_1}{L_2 L_3} = 0, \quad \frac{\partial W}{\partial \lambda_2} - \frac{P_2}{L_1 L_3} =0, \quad \frac{\partial W}{\partial \tilde D}- \frac{\Phi}{L_3} = 0.
\end{equation}

In Eq.~\eqref{zhao-07-2}, the second variation is related through the real-symmetric Hessian matrix
\begin{equation} \label{zhao-07-4}
{\bf H} = 
\begin{pmatrix}
\vspace{0.1in}
\displaystyle \frac{\partial^2 W}{\partial \lambda_1^2}                          & \displaystyle \frac{\partial^2 W}{\partial \lambda_1 \partial \lambda_2}  & \displaystyle \frac{\partial^2 W}{\partial \lambda_1 \partial \tilde D} \\
\vspace{0.1in}
\displaystyle \frac{\partial^2 W}{\partial \lambda_2 \partial \lambda_1}   & \displaystyle \frac{\partial^2 W}{\partial \lambda_2^2}                         & \displaystyle \frac{\partial^2 W}{\partial \lambda_2 \partial \tilde D} \\
\displaystyle \frac{\partial^2 W}{\partial \tilde D \partial \lambda_1}       & \displaystyle \frac{\partial^2 W}{\partial \tilde D \partial \lambda_2}      & \displaystyle \frac{\partial^2 W}{\partial \tilde D ^2}
\end{pmatrix},
\end{equation}
which must be positive definite if the equilibrium state is stable. A real-symmetric positive definite matrix requires that all the principal minors of the matrix are positive \cite{meyer2000matrix}. The $1 \times 1$ principal minors are the diagonal entries
\begin{equation} \label{zhao-07-5}
\frac{\partial^2 W}{\partial \lambda_1^2} , \quad \frac{\partial^2 W}{\partial \lambda_2^2}, \quad \frac{\partial^2 W}{\partial \tilde D ^2}.
\end{equation}

The $2 \times 2$ principal minors are

\begin{equation} \label{zhao-07-6}
\begin{vmatrix}
\vspace{0.1in}
\displaystyle \frac{\partial^2 W}{\partial \lambda_1^2}                          & \displaystyle \frac{\partial^2 W}{\partial \lambda_1 \partial \lambda_2}  \\
\vspace{0.1in}
\displaystyle \frac{\partial^2 W}{\partial \lambda_2 \partial \lambda_1}   & \displaystyle \frac{\partial^2 W}{\partial \lambda_2^2}
\end{vmatrix}, 
\quad 
\begin{vmatrix}
\vspace{0.1in}
\displaystyle \frac{\partial^2 W}{\partial \lambda_1^2} & \displaystyle \frac{\partial^2 W}{\partial \lambda_1 \partial \tilde D} \\
\vspace{0.1in}
\displaystyle \frac{\partial^2 W}{\partial \tilde D \partial \lambda_1}       & \displaystyle \frac{\partial^2 W}{\partial \tilde D ^2}
\end{vmatrix},
\quad
\begin{vmatrix}
\vspace{0.1in}
\displaystyle \frac{\partial^2 W}{\partial \lambda_2^2}                         & \displaystyle \frac{\partial^2 W}{\partial \lambda_2 \partial \tilde D} \\
\displaystyle \frac{\partial^2 W}{\partial \tilde D \partial \lambda_2}      & \displaystyle \frac{\partial^2 W}{\partial \tilde D ^2}
\end{vmatrix},
\end{equation}
and the only $3 \times 3$ principal minor is 
\begin{equation} \label{zhao-07-7}
\det \, {\bf H} = 
\begin{vmatrix}
\vspace{0.1in}
\displaystyle \frac{\partial^2 W}{\partial \lambda_1^2}                          & \displaystyle \frac{\partial^2 W}{\partial \lambda_1 \partial \lambda_2}  & \displaystyle \frac{\partial^2 W}{\partial \lambda_1 \partial \tilde D} \\
\vspace{0.1in}
\displaystyle \frac{\partial^2 W}{\partial \lambda_2 \partial \lambda_1}   & \displaystyle \frac{\partial^2 W}{\partial \lambda_2^2}                         & \displaystyle \frac{\partial^2 W}{\partial \lambda_2 \partial \tilde D} \\
\displaystyle \frac{\partial^2 W}{\partial \tilde D \partial \lambda_1}       & \displaystyle \frac{\partial^2 W}{\partial \tilde D \partial \lambda_2}      & \displaystyle \frac{\partial^2 W}{\partial \tilde D ^2}
\end{vmatrix}.
\end{equation}

For an ideal dielectric elastomer \eqref{suo-08-T-E-2}, the free energy function is given by
\begin{equation} \label{zhao-07-8}
W (\lambda_1, \lambda_2, \tilde{D}) =  \frac{\mu}{2} (\lambda_1^2 + \lambda_2^2 + \lambda_1^{-2} \lambda_2^{-2} - 3) + \frac{{\tilde D}^2}{2 \varepsilon}\lambda_1^{-2} \lambda_2^{-2}.
\end{equation}

Substitution of Eq.~\eqref{zhao-07-8} into the equilibrium equations \eqref{zhao-07-3} and the Hessian matrix \eqref{zhao-07-4} yields the necessary equations for the stability analysis as can be found in the works \cite{zhao2007method, diaz2008electromechanical}.\\

For notational expediency, we summarize the stability analysis in vector form. The free energy in Eq.~\eqref{zhao-07-1} and the variation in Eq.~\eqref{zhao-07-2} can be written as
\begin{equation} \label{zhao-07-8-r1}
G ({\bf x}; {\bf L}) \quad {\rm and} \quad G ({\bf x} + \delta {\bf x}; {\bf L}) = G ({\bf x}; {\bf L}) + \frac{\partial G}{\partial {\bf x}} \cdot \delta {\bf x} + \delta {\bf x} \cdot \frac{\partial^2 G}{\partial {\bf x}^2} [\delta {\bf x}] + o(\delta{\bf x}^2),
\end{equation}
where the state is ${\bf x} = (\lambda_1, \lambda_2, \tilde D)^T$, the increments are $\delta {\bf x} = (\delta \lambda_1, \delta\lambda_2, \delta\tilde D)^T$, and the load parameters are ${\bf L} = (P_1, P_2, \Phi)^T$. The function $o(\delta{\bf x}^2)$ has the limiting behavior, i.e., the ratio $\frac{\|o(\delta{\bf x}^2)\|}{\| \delta{\bf x}^2 \|}$ approaches zero as $\| \delta{\bf x}^2 \| \to 0$.

The stability of the state ${\bf x}$ subjected to the load parameter ${\bf L}$ requires a local energy minimization, such that $G ({\bf x}; {\bf L}) \le G ({\bf x} + \delta {\bf x}; {\bf L})$ for sufficiently small $\Vert \delta {\bf x} \Vert$. Thus we have the equilibrium equations
\begin{equation} \label{zhao-07-8-r2}
\frac{\partial G ({\bf x}; {\bf L})}{\partial {\bf x}} \cdot \delta {\bf x} = 0 \quad \implies \quad g_j ({\bf x}; {\bf L})= \frac{\partial G ({\bf x}; {\bf L})}{\partial x_j} = 0,
\end{equation}
which correspond to Eq.~\eqref{zhao-07-3}, and the stability conditions
\begin{equation} \label{zhao-07-8-r3}
\delta {\bf x} \cdot \frac{\partial^2 G}{\partial {\bf x}^2} [\delta {\bf x}] \ge 0  \quad \implies \quad 
\left\{
\begin{aligned}
& \text{a Hessian matrix ${\bf H} = \frac{\partial^2 G}{\partial {\bf x}^2}$ is (semi) positive definite}\\
& \text {and the matrix elements are} \ H_{ij} = \frac{\partial^2 G ({\bf x}; {\bf L})}{\partial x_i \partial x_j},
\end{aligned}
\right.
\end{equation}
where $i,j,k=1,2,3$.

It follows from Eqs.~\eqref{zhao-07-8-r2}, \eqref{zhao-07-8}, and \eqref{zhao-07-1} that the equilibrium equations are
\begin{subequations}  \label{zhao-07-r3-add1}
\begin{align}
\label{zhao-07-r3-add1-a}
 \mu (\lambda_1 - \lambda_1^{-3}\lambda_2^{-2}) - \frac{{\tilde D}^2}{\varepsilon}\lambda_1^{-3} \lambda_2^{-2}- \frac{P_1}{L_2 L_3} & = 0, \\
\label{zhao-07-r3-add1-b}
 \mu (\lambda_2 - \lambda_1^{-2}\lambda_2^{-3}) - \frac{{\tilde D}^2}{\varepsilon}\lambda_1^{-2} \lambda_2^{-3}- \frac{P_2}{L_1 L_3} & = 0, \\
\label{zhao-07-r3-add1-c}
\frac{\tilde D}{\varepsilon}\lambda_1^{-2} \lambda_2^{-2}- \frac{\Phi}{L_3} & = 0.
\end{align}
\end{subequations}

It follows from Eqs.~\eqref{zhao-07-8-r3}, \eqref{zhao-07-8}, and \eqref{zhao-07-1} that the Hessian matrix is
\begin{equation} \label{zhao-07-4-add2}
{\bf H} = 
\begin{pmatrix}
\vspace{0.1in}
\mu (1 + 3\lambda_1^{-4}\lambda_2^{-2}) + \displaystyle\frac{3{\tilde D}^2}{\varepsilon}\lambda_1^{-4} \lambda_2^{-2}                     & 2\mu \lambda_1^{-3}\lambda_2^{-3} + \displaystyle\frac{2{\tilde D}^2}{\varepsilon}\lambda_1^{-3} \lambda_2^{-3}  & - \displaystyle\frac{2{\tilde D}}{\varepsilon}\lambda_1^{-3} \lambda_2^{-2} \\
\vspace{0.1in}
2\mu \lambda_1^{-3}\lambda_2^{-3} + \displaystyle\frac{2{\tilde D}^2}{\varepsilon}\lambda_1^{-3} \lambda_2^{-3}    & \mu (1 + 3\lambda_1^{-2}\lambda_2^{-4}) + \displaystyle\frac{3{\tilde D}^2}{\varepsilon}\lambda_1^{-2} \lambda_2^{-4} & - \displaystyle\frac{2{\tilde D}}{\varepsilon}\lambda_1^{-2} \lambda_2^{-3} \\
- \displaystyle\frac{2{\tilde D}}{\varepsilon}\lambda_1^{-3} \lambda_2^{-2}      & - \displaystyle\frac{2{\tilde D}}{\varepsilon}\lambda_1^{-2} \lambda_2^{-3}      & \displaystyle\frac{1}{\varepsilon}\lambda_1^{-2} \lambda_2^{-2}
\end{pmatrix}.
\end{equation}

As stated in Eq.~\eqref{zhao-07-8-r3}, the Hessian matrix \eqref{zhao-07-4-add2} is positive definite when the electrostatic system shown in Fig.~\ref{C-example-1} at equilibrium \eqref{zhao-07-r3-add1} is stable.

\subsubsection{Bifurcation analysis} \label{bifurcation-ex-1}
We now discuss the bifurcation problem pertaining to the electromechanical system shown in Fig.~\ref{C-example-1}. Based on the equilibrium equations \eqref{zhao-07-3}, we will formulate the incremental equilibrium equations and then find the incremental solutions.\footnote{The method of incremental solutions is based on the adjacent equilibrium stability criterion. This stability criterion asserts that a primary equilibrium state becomes unstable at a threshold where other nearby equilibrium states exist. In the bifurcation theory, the existence of an incremental solution is just a {\it necessary} condition for the primary equilibrium state at the threshold to be a bifurcation point. However, this condition is not sufficient. An extensive nonlinear bifurcation analysis \cite{golubitsky2012singularities, chen2001singularity} is needed to verify the existence of the bifurcation solution branches, and therefore the adjacent equilibrium states.} The construction of the incremental equilibrium equations be found in the following books\cite{ogden1997non, dorfmann2014nonlinear}. \\

Let us describe the idea of linear bifurcation analysis generally at first and then we discuss the bifurcation of the system shown in Fig.~\ref{C-example-1} in detail. Consider the equilibrium equations \eqref{zhao-07-8-r2}. A small perturbation of the equilibrium state ${\bf x} = (\lambda_1, \lambda_2, \tilde D)^T$ is given by $\delta {\bf x} = (\delta \lambda_1, \delta\lambda_2, \delta\tilde D)^T$ at the load parameters ${\bf L} = (P_1, P_2, \Phi)^T$. The new state is ${\bf x} +\delta {\bf x} = (\lambda_1 +\delta \lambda_1, \lambda_2 + \delta \lambda_2, \tilde D + \delta \tilde D)^T$ with sufficiently small $\Vert \delta {\bf x} \Vert$. Then the expansion of the function $g_j$ in Eq.~\eqref{zhao-07-8-r2} for the new state is 
\begin{equation} \label{zhao-07-8-r4}
g_j ({\bf x} +\delta {\bf x}; {\bf L}) = g_j ({\bf x}; {\bf L}) + \frac{\partial g_j}{\partial {\bf x}}\Big|_{({\bf x}; {\bf L})} \cdot \delta {\bf x} + o(\delta {\bf x}).
\end{equation}

By the implicit function theorem in Sec.~\ref{section-bifurcation-points}, a necessary condition for $({\bf x}; {\bf L})$ to be a bifurcation point is that the Fr\'{e}chet derivative
\begin{equation} \label{zhao-07-8-r5}
\frac{\partial g_j}{\partial {\bf x}}\Big|_{({\bf x}; {\bf L})}
\end{equation}
 be not invertible. Otherwise, there exists a function ${\bf h}$, such that $g_j ({\bf h}({\bf L}); {\bf L}) =0$. \\

In contrast to the implicit form, we consider the explicit forms of the bifurcation analysis of the system shown in Fig.~\ref{C-example-1}. Consider the equilibrium equations \eqref{zhao-07-3}. The infinitesimal increments of the variables $\lambda_1$, $\lambda_2$, and $\tilde D$ are $\delta \lambda_1$, $\delta \lambda_2$, and $\delta \tilde D$, respectively. With the Taylor series expansion at the equilibrium solutions $(\lambda_1, \lambda_2, \tilde D)$, we have
\begin{subequations} \label{zhao-07-8-0}
\begin{align}
 \frac{\partial W}{\partial \lambda_1} & =  \left. \frac{\partial W}{\partial \lambda_1} \right|_e + \left. \frac{\partial^2 W}{\partial \lambda_1^2} \right|_e \delta \lambda_1 + \left. \frac{\partial^2 W}{\partial \lambda_1 \partial \lambda_2} \right|_e \delta \lambda_2 + \left. \frac{\partial^2 W}{\partial \lambda_1 \partial \tilde D} \right|_e \delta \tilde D + o (\delta \lambda_1, \delta \lambda_2, \delta \tilde D), \\
 \frac{\partial W}{\partial \lambda_2} & =  \left. \frac{\partial W}{\partial \lambda_2} \right|_e + \left. \frac{\partial^2 W}{\partial \lambda_2 \partial \lambda_1} \right|_e \delta \lambda_1 + \left. \frac{\partial^2 W}{\partial \lambda_2^2} \right|_e \delta \lambda_2 + \left. \frac{\partial^2 W}{\partial \lambda_2 \partial \tilde D} \right|_e \delta \tilde D + o (\delta \lambda_1, \delta \lambda_2, \delta \tilde D), \\
 \frac{\partial W}{\partial \tilde D} & =  \left. \frac{\partial W}{\partial \tilde D} \right|_e + \left. \frac{\partial^2 W}{\partial \tilde D \partial \lambda_1} \right|_e \delta \lambda_1 + \left. \frac{\partial^2 W}{\partial \tilde D \partial \lambda_2} \right|_e \delta \lambda_2 + \left. \frac{\partial^2 W}{\partial {\tilde D}^2} \right|_e \delta \tilde D + o (\delta \lambda_1, \delta \lambda_2, \delta \tilde D).
\end{align}
\end{subequations}
Substituting Eq.~\eqref{zhao-07-8} into the equilibrium equations \eqref{zhao-07-3} and ignoring the higher order terms, we have 
\begin{subequations} \label{zhao-07-8-1}
\begin{align}
\left. \frac{\partial W}{\partial \lambda_1} \right|_e + \left. \frac{\partial^2 W}{\partial \lambda_1^2} \right|_e \delta \lambda_1 + \left. \frac{\partial^2 W}{\partial \lambda_1 \partial \lambda_2} \right|_e \delta \lambda_2 + \left. \frac{\partial^2 W}{\partial \lambda_1 \partial \tilde D} \right|_e \delta \tilde D - \frac{P_1}{L_2 L_3} & = 0, \\
\left. \frac{\partial W}{\partial \lambda_2} \right|_e + \left. \frac{\partial^2 W}{\partial \lambda_2 \partial \lambda_1} \right|_e \delta \lambda_1 + \left. \frac{\partial^2 W}{\partial \lambda_2^2} \right|_e \delta \lambda_2 + \left. \frac{\partial^2 W}{\partial \lambda_2 \partial \tilde D} \right|_e \delta \tilde D - \frac{P_2}{L_1 L_3} & = 0, \\
\left. \frac{\partial W}{\partial \tilde D} \right|_e + \left. \frac{\partial^2 W}{\partial \tilde D \partial \lambda_1} \right|_e \delta \lambda_1 + \left. \frac{\partial^2 W}{\partial \tilde D \partial \lambda_2} \right|_e \delta \lambda_2 + \left. \frac{\partial^2 W}{\partial {\tilde D}^2} \right|_e \delta \tilde D - \frac{\Phi}{L_3} & = 0.
\end{align}
\end{subequations}
Since the equilibrium solutions of $\lambda_1$, $\lambda_2$, and $\tilde D$ satisfy the equilibrium equations \eqref{zhao-07-3}, that is
\begin{equation} \label{zhao-07-9}
 \left.\frac{\partial W}{\partial \lambda_1}\right|_e - \frac{P_1}{L_2 L_3} = 0, \quad \left.\frac{\partial W}{\partial \lambda_2}\right|_e - \frac{P_2}{L_1 L_3} =0, \quad \left.\frac{\partial W}{\partial \tilde D}\right|_e - \frac{\Phi}{L_3} = 0,
\end{equation}
then Eq.~\eqref{zhao-07-8-1} reduces to a set of simultaneous linear equations that can be expressed as a single matrix equation, or, written
out in full, as
\begin{equation} \label{zhao-07-11}
\begin{pmatrix}
\vspace{0.1in}
\left.\displaystyle \frac{\partial^2 W}{\partial \lambda_1^2}\right|_e                        & \left.\displaystyle \frac{\partial^2 W}{\partial \lambda_1 \partial \lambda_2}\right|_e  & \left.\displaystyle \frac{\partial^2 W}{\partial \lambda_1 \partial \tilde D}\right|_e \\
\vspace{0.1in}
\left.\displaystyle \frac{\partial^2 W}{\partial \lambda_2 \partial \lambda_1}\right|_e   & \left.\displaystyle \frac{\partial^2 W}{\partial \lambda_2^2}\right|_e                         & \left.\displaystyle \frac{\partial^2 W}{\partial \lambda_2 \partial \tilde D}\right|_e \\
\left.\displaystyle \frac{\partial^2 W}{\partial \tilde D \partial \lambda_1}\right|_e       & \left.\displaystyle \frac{\partial^2 W}{\partial \tilde D \partial \lambda_2}\right|_e      & \left.\displaystyle \frac{\partial^2 W}{\partial \tilde D ^2}\right|_e
\end{pmatrix} 
\begin{pmatrix}
\vspace{0.22in}
\delta \lambda_1\\
\vspace{0.22in}
\delta \lambda_2\\
\delta \tilde D
\end{pmatrix}
= {\bf 0}.
\end{equation}

If the determinant of the $3 \times 3$ matrix in Eq.~\eqref{zhao-07-11} is non-zero, the simultaneous linear equations only have the trivial solution $\delta \lambda_1 = \delta \lambda_2 = \delta \tilde D =0$. If the determinant is zero, there exists an infinite number of solutions of $\delta \lambda_1$, $\delta \lambda_2$, and $\delta \tilde D$. Thus, the necessary condition for the existence of incremental solutions is a zero determinant of the $3 \times 3$ matrix in Eq.~\eqref{zhao-07-11}. Consider the free energy function \eqref{zhao-07-8} of an ideal dielectric elastomer. By Eq.~\eqref{zhao-07-11}, the bifurcation condition of the electrostatic system shown in Fig.~\ref{C-example-1} at equilibrium \eqref{zhao-07-9} is given by
\begin{equation} \label{zhao-07-4-section-5-1-add1}
\begin{vmatrix}
\vspace{0.1in}
\mu (1 + 3\lambda_1^{-4}\lambda_2^{-2}) + \displaystyle\frac{3{\tilde D}^2}{\varepsilon}\lambda_1^{-4} \lambda_2^{-2}                     & 2\mu \lambda_1^{-3}\lambda_2^{-3} + \displaystyle\frac{2{\tilde D}^2}{\varepsilon}\lambda_1^{-3} \lambda_2^{-3}  & - \displaystyle\frac{2{\tilde D}}{\varepsilon}\lambda_1^{-3} \lambda_2^{-2} \\
\vspace{0.1in}
2\mu \lambda_1^{-3}\lambda_2^{-3} + \displaystyle\frac{2{\tilde D}^2}{\varepsilon}\lambda_1^{-3} \lambda_2^{-3}    & \mu (1 + 3\lambda_1^{-2}\lambda_2^{-4}) + \displaystyle\frac{3{\tilde D}^2}{\varepsilon}\lambda_1^{-2} \lambda_2^{-4} & - \displaystyle\frac{2{\tilde D}}{\varepsilon}\lambda_1^{-2} \lambda_2^{-3} \\
- \displaystyle\frac{2{\tilde D}}{\varepsilon}\lambda_1^{-3} \lambda_2^{-2}      & - \displaystyle\frac{2{\tilde D}}{\varepsilon}\lambda_1^{-2} \lambda_2^{-3}      & \displaystyle\frac{1}{\varepsilon}\lambda_1^{-2} \lambda_2^{-2}
\end{vmatrix}
=0.
\end{equation}

\subsection{Avoiding the pull-in instability of a dielectric elastomer film}

Inspired by the work by Zhao and Suo \cite{zhao2007method}, Yang et al. \cite{yang2017avoiding} concocted a set of simple, experimentally implementable, conditions that render the dielectric elastomer film impervious to pull-in instability for all practical loading conditions. Here we take the example in the work \cite{yang2017avoiding} to discuss electromechanical instability and bifurcation. 

\subsubsection{Stability analysis by using the energy method}

 \begin{figure}[h] %
    \centering
    \includegraphics[width=5in]{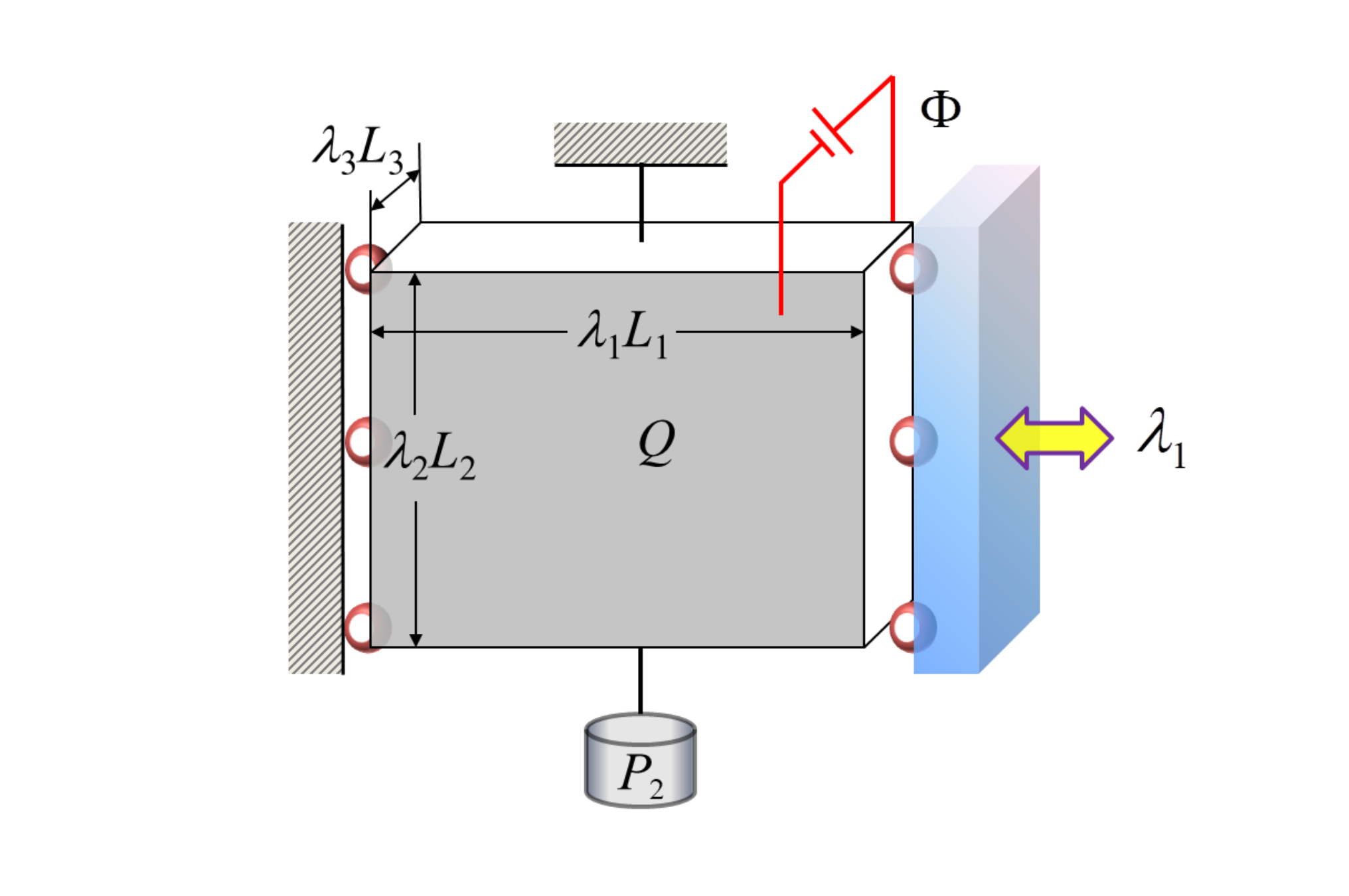}
    \caption{Schematic of a deformed film of a dielectric elastomer. The film can be extended/compressed between two well-lubricated (with rollers), rigid plates by means of a controlled displacement (or the stretch $\lambda_1$) in the $X_1$ direction. A dead load $P_2$ is applied in the $X_2$ direction and a voltage $\Phi$ is applied in the thickness direction of the film that is coated with two compliant electrodes. The controlled stretch $\lambda_1$, the dead load $P_2$ and the voltage $\Phi$ deform the film from $L_1$, $L_2$, and $L_3$ to $\lambda_1 L_1$, $\lambda_2 L_2$, and $\lambda_3 L_3$, as well as induce an electric charge of magnitude $Q$ on either electrode \cite{yang2017avoiding}.}
    \label{C-example-2}
\end{figure}

The free energy $G$ of the system shown in Fig.~\ref{C-example-2} is \cite{yang2017avoiding}
\begin{equation} \label{yang-17-1}
\frac{G }{L_1 L_2 L_3}=  W (\lambda_1, \lambda_2, \tilde{D}) - \frac{P_2}{L_1 L_3} \lambda_2 - \frac{\Phi}{L_3} \tilde{D},
\end{equation}
where $W (\lambda_1, \lambda_2, \tilde{D})$ is the free energy function. In contrast to three variables in Eq.~\eqref{zhao-07-1}, there are only two variables $\lambda_2$ and $\tilde{D}$ in Eq.~\eqref{yang-17-1}. Here the reduction of variables from three to two is due to the displacement-controlled boundary condition introduced. \\ 

With the small variations $\delta \lambda_2$ and $\delta \tilde {D}$, the variation of the free energy \eqref{yang-17-1} is 
\begin{equation} \label{yang-17-2}
\frac{\Delta G }{L_1 L_2 L_3} = \Big( \frac{\partial W}{\partial \lambda_2} - \frac{P_2}{L_1 L_3} \Big) \delta \lambda_2 + \Big( \frac{\partial W}{\partial \tilde D}- \frac{\Phi}{L_3} \Big) \delta \tilde{D}+ \frac{1}{2} \frac{\partial^2 W}{\partial \lambda_2^2} (\delta \lambda_2)^2 + \frac{1}{2} \frac{\partial^2 W}{\partial {\tilde D}^2} (\delta \tilde D)^2 + \frac{\partial^2 W}{\partial \lambda_2 \partial \tilde D} \delta \lambda_2 \delta \tilde D.
\end{equation}
In equilibrium, the first variation of the free energy is zero, we have
\begin{equation} \label{yang-17-3}
\frac{\partial W}{\partial \lambda_2} - \frac{P_2}{L_1 L_3} =0 \quad {\rm and} \quad \frac{\partial W}{\partial \tilde D} - \frac{\Phi}{L_3} = 0.
\end{equation}
From the principle of minimum energy, the stability analysis is related to the Hessian matrix
\begin{equation} \label{yang-17-4}
{\bf H} = 
\begin{pmatrix}
\vspace{0.1in}
\displaystyle \frac{\partial^2 W}{\partial \lambda_2^2}                         & \displaystyle \frac{\partial^2 W}{\partial \lambda_2 \partial \tilde D} \\
\displaystyle \frac{\partial^2 W}{\partial \tilde D \partial \lambda_2}      & \displaystyle \frac{\partial^2 W}{\partial \tilde D ^2}
\end{pmatrix}.
\end{equation}

A stable equilibrium state requires that the Hessian matrix \eqref{yang-17-4} must be positive definite, that is, all the principal minors of ${\bf H}$ are positive in equilibrium. The $1 \times 1$ principal minors are
\begin{equation} \label{yang-17-5}
\frac{\partial^2 W}{\partial \lambda_2^2}, \quad \frac{\partial^2 W}{\partial \tilde D ^2},
\end{equation}
and the only $2 \times 2$ principal minor is 
\begin{equation} \label{yang-17-6}
\det \, \bf H = \begin{vmatrix}
\vspace{0.1in}
\displaystyle \frac{\partial^2 W}{\partial \lambda_2^2}                         & \displaystyle \frac{\partial^2 W}{\partial \lambda_2 \partial \tilde D} \\
\displaystyle \frac{\partial^2 W}{\partial \tilde D \partial \lambda_2}      & \displaystyle \frac{\partial^2 W}{\partial \tilde D ^2}
\end{vmatrix}.
\end{equation}

Again, if all the principal minors in Eqs.~\eqref{yang-17-5} and \eqref{yang-17-6} are positive, the equilibrium state \eqref{yang-17-3} is stable. Consider the free energy function \eqref{zhao-07-8}. The equilibrium equations \eqref{yang-17-3} are
\begin{equation} \label{re-yang-17-7}
{\mu} \Big[ \lambda_2 - \Big( 1 + \frac{\tilde{D}^2}{\varepsilon \mu} \Big) \lambda_1^{-2}\lambda_2^{-3} \Big] = \frac{P_2}{L_1 L_3} \quad {\rm and} \quad \frac{\tilde D}{\varepsilon}\lambda_1^{-2} \lambda_2^{-2} = \frac{\Phi}{L_3}.
\end{equation}
The Hessian matrix \eqref{yang-17-4} is
\begin{equation} \label{re-yang-17-8}
{\bf H} = 
\begin{pmatrix}
{\mu} \Big[ 1 + 3\Big( 1 + \displaystyle \frac{\tilde{D}^2}{\varepsilon \mu} \Big) \lambda_1^{-2}\lambda_2^{-4} \Big] & - \displaystyle \frac{2 \tilde{D}}{\varepsilon} \lambda_1^{-2}\lambda_2^{-3} \\
        & \\
       - \displaystyle \frac{2 \tilde{D}}{\varepsilon} \lambda_1^{-2}\lambda_2^{-3} & \displaystyle \frac{1}{\varepsilon} \lambda_1^{-2}\lambda_2^{-2}
\end{pmatrix}.
\end{equation}

The $1 \times 1$ principle minors in Eq.~\eqref{re-yang-17-8} are the diagonal entries and they are always positive no matter the state is in equilibrium or not. The determinant of the Hessian in Eq.~\eqref{re-yang-17-8} is
\begin{equation} \label{re-yang-17-9}
\det \, {\bf H} = \displaystyle \frac{\mu}{\varepsilon} \lambda_1^{-2}\lambda_2^{-3} \Big[ 4\lambda_1^{-2}\lambda_2^{-3} + \lambda_2 - \Big(1 + \displaystyle \frac{\tilde{D}^2}{\varepsilon \mu} \Big) \lambda_1^{-2}\lambda_2^{-3} \Big].
\end{equation}
By Eq.~$\eqref{re-yang-17-7}_1$, we have
\begin{equation} \label{re-yang-17-10}
\lambda_2 - \Big( 1 + \frac{\tilde{D}^2}{\varepsilon \mu} \Big) \lambda_1^{-2}\lambda_2^{-3} = \frac{1}{\mu} \frac{P_2}{L_1 L_3}.
\end{equation}
Then we find that the determinant at the equilibrium state is always positive
\begin{equation} \label{re-yang-17-11}
\det \, {\bf H} = \displaystyle \frac{\mu}{\varepsilon} \lambda_1^{-4}\lambda_2^{-6} \Big( 4 + \frac{1}{\mu} \frac{P_2}{L_1 L_3} \lambda_1^{2}\lambda_2^{3} \Big) > 0 
\end{equation}
due to the positive stretches $\lambda_1$ and $\lambda_2$ and the positive applied dead load $P_2$. Thus, a positive definite Hessian matrix in this example indicates that the equilibrium state is stable.

\subsubsection{Bifurcation analysis}  \label{bifurcation-ex-2}
The bifurcation analysis here is similar to that of the example in Sec.~\ref{bifurcation-ex-1}. Consider the equilibrium equations \eqref{yang-17-3}. The infinitesimal increments of the variables $\lambda_2$ and $\tilde D$ are $\delta \lambda_2$ and $\delta \tilde D$, respectively. The Taylor series expansion gives
\begin{subequations} \label{yang-17-7}
\begin{align}
 \frac{\partial W}{\partial \lambda_2} & =  \left. \frac{\partial W}{\partial \lambda_2} \right|_e + \left. \frac{\partial^2 W}{\partial \lambda_2^2} \right|_e \delta \lambda_2 + \left. \frac{\partial^2 W}{\partial \lambda_2 \partial \tilde D} \right|_e \delta \tilde D + o (\delta \lambda_2, \delta \tilde D), \\
 \frac{\partial W}{\partial \tilde D} & =  \left. \frac{\partial W}{\partial \tilde D} \right|_e + \left. \frac{\partial^2 W}{\partial \tilde D \partial \lambda_2} \right|_e \delta \lambda_2 + \left. \frac{\partial^2 W}{\partial {\tilde D}^2} \right|_e \delta \tilde D + o (\delta \lambda_2, \delta \tilde D).
\end{align}
\end{subequations}
Substituting Eq.~\eqref{yang-17-7} into the equilibrium equations \eqref{yang-17-3} and ignoring the higher order terms, we have
\begin{subequations} \label{yang-17-8-1}
\begin{align}
\left. \frac{\partial W}{\partial \lambda_2} \right|_e + \left. \frac{\partial^2 W}{\partial \lambda_2^2} \right|_e \delta \lambda_2 + \left. \frac{\partial^2 W}{\partial \lambda_2 \partial \tilde D} \right|_e \delta \tilde D  - \frac{P_2}{L_1 L_3} & = 0, \\
\left. \frac{\partial W}{\partial \tilde D} \right|_e + \left. \frac{\partial^2 W}{\partial \tilde D \partial \lambda_2} \right|_e \delta \lambda_2 + \left. \frac{\partial^2 W}{\partial {\tilde D}^2} \right|_e \delta \tilde D - \frac{\Phi}{L_3} & = 0.
\end{align}
\end{subequations}
Together with the equilibrium equations
\begin{equation} \label{yang-17-8}
\left.\frac{\partial W}{\partial \lambda_2}\right|_e - \frac{P_2}{L_1 L_3} =0 \quad {\rm and} \quad \left.\frac{\partial W}{\partial \tilde D}\right|_e - \frac{\Phi}{L_3} = 0,
\end{equation}
we encounter a set of simultaneous linear equations that is expressed as a single matrix equation
\begin{equation} \label{yang-17-10}
\begin{pmatrix}
\vspace{0.1in}
\left.\displaystyle \frac{\partial^2 W}{\partial \lambda_2^2}\right|_e                         & \left.\displaystyle \frac{\partial^2 W}{\partial \lambda_2 \partial \tilde D}\right|_e \\
\left.\displaystyle \frac{\partial^2 W}{\partial \tilde D \partial \lambda_2}\right|_e      & \left.\displaystyle \frac{\partial^2 W}{\partial \tilde D ^2}\right|_e
\end{pmatrix} 
\begin{pmatrix}
\vspace{0.25in}
\delta \lambda_2\\
\delta \tilde D
\end{pmatrix}
= {\bf 0}.
\end{equation}

If the determinant of the $2 \times 2$ matrix in Eq.~\eqref{yang-17-10} is non-zero, namely
\begin{equation} \label{yang-17-11}
\begin{vmatrix}
\vspace{0.1in}
\left.\displaystyle \frac{\partial^2 W}{\partial \lambda_2^2}\right|_e                         & \left.\displaystyle \frac{\partial^2 W}{\partial \lambda_2 \partial \tilde D}\right|_e \\
\left.\displaystyle \frac{\partial^2 W}{\partial \tilde D \partial \lambda_2}\right|_e      & \left.\displaystyle \frac{\partial^2 W}{\partial \tilde D ^2}\right|_e
\end{vmatrix} 
\neq 0,
\end{equation}
the simultaneous linear equations only have the trivial solution $\delta \lambda_2 = \delta \tilde D =0$. This means that there is no bifurcation from the primary equilibrium state (solutions of Eq.~\eqref{yang-17-8}). Consider the example of an ideal dielectric elastomer in Eqs.~\eqref{re-yang-17-7} and \eqref{re-yang-17-8}. The determinant of the Hessian matrix at equilibrium is
\begin{equation} \label{yang-17-12-section-5-2-1}
\begin{vmatrix}
{\mu} \Big[ 1 + 3\Big( 1 + \displaystyle \frac{\tilde{D}^2}{\varepsilon \mu} \Big) \lambda_1^{-2}\lambda_2^{-4} \Big] & - \displaystyle \frac{2 \tilde{D}}{\varepsilon} \lambda_1^{-2}\lambda_2^{-3} \\
        & \\
       - \displaystyle \frac{2 \tilde{D}}{\varepsilon} \lambda_1^{-2}\lambda_2^{-3} & \displaystyle \frac{1}{\varepsilon} \lambda_1^{-2}\lambda_2^{-2}
\end{vmatrix} 
= \displaystyle \frac{\mu}{\varepsilon} \lambda_1^{-4}\lambda_2^{-6} \Big( 4 + \frac{1}{\mu}\frac{P_2}{L_1 L_3} \lambda_1^{2}\lambda_2^{3} \Big) \ne 0,
\end{equation}
thus there is no bifurcation.

\section{Stability of the homogeneous deformation of soft dielectrics: an alternative energy formulation} \label{section-e-stability-yang}

In this section, we revisit electromechanical instability manifested in the homogeneous deformation of soft dielectrics as presented in the work \cite{zhao2007method}. Here, we will use the energy formulation of nonlinear electroelasticity in terms of the deformation and the polarization \cite{liu2013energy, liu2014energy}. The procedure of the stability analysis can also be found in a recent paper by Alameh et al. \cite{alameh2018emergent} about the magneto-electro-mechanical instability.

\subsection{A dielectric film}
\begin{figure}[h] 
\centering
\subfigure[]{%
\includegraphics[width=2.9in]{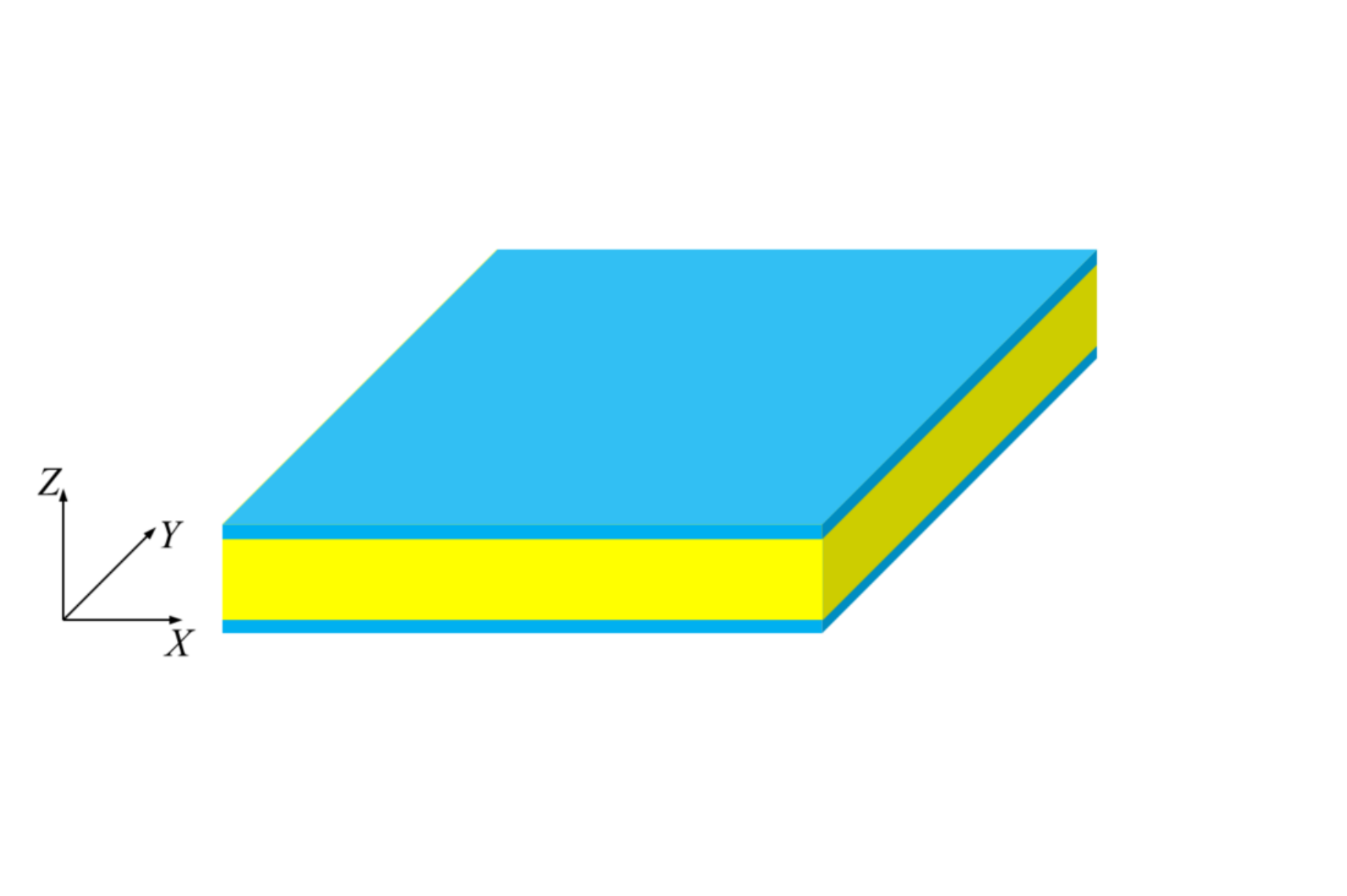}
\label{film-1a}}
\subfigure[]{%
\includegraphics[width=2.9in]{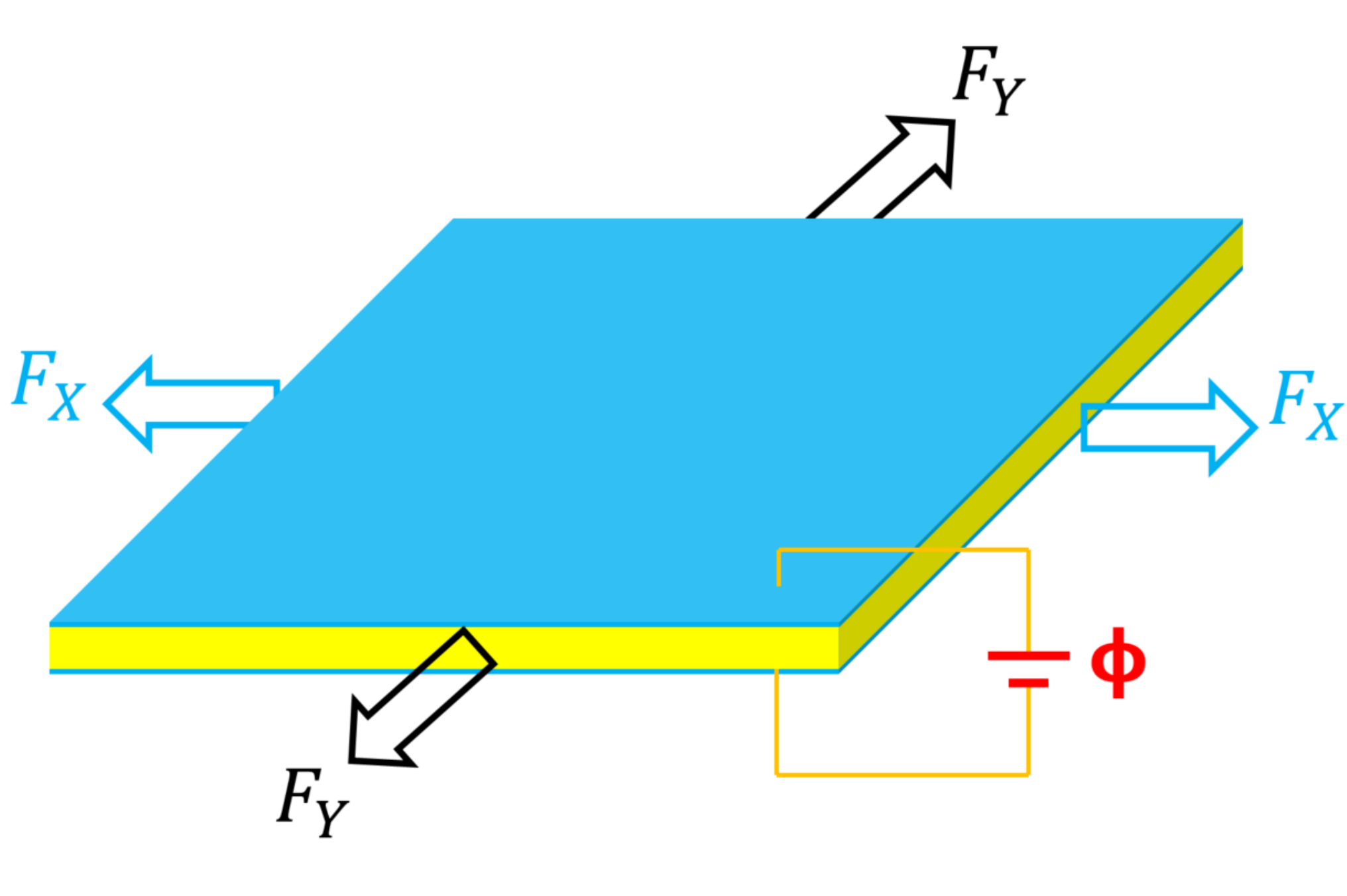}
\label{film-1b}}
\caption{Schematic of the deformation of a soft dielectric film subjected to an electric voltage $\Phi$ and two pairs of dead loads $F_X$ and $F_Y$. The dielectric film is coated with two compliant electrodes on the upper and bottom surfaces. (a) Undeformed with dimension $L_X \times L_Y \times L_Z$. (b) Deformed with dimension $\lambda_X L_X \times \lambda_Y L_Y \times \lambda_Z L_Z$.}
\label{film-1}
\end{figure}

Consider the homogeneous deformation of a thin film of soft dielectrics subject to two pairs of equal in-plane forces (dead loads) and an electric voltage in the thickness direction. The dielectric film deforms from the original dimension $L_X \times L_Y \times L_Z$ to $\lambda_X L_X \times \lambda_Y L_Y \times \lambda_Z L_Z$ (see Fig.~\ref{film-1}), such that
\begin{equation} \label{film-displace}
x = \lambda_X X, \quad y = \lambda_Y Y, \quad z = \lambda_Z Z,
\end{equation}
and the deformation gradient is 
\begin{equation} \label{film-dg-F}
{\bf F} := {\rm diag} \, (\lambda_X, \lambda_Y, \lambda_Z),
\end{equation}
where $\lambda_X$, $\lambda_Y$, and $\lambda_Z$ are {\it constant} stretches. Then, the Jacobian is 
\begin{equation} \label{film-Jaco}
J = \det {\bf F} = \lambda_X \lambda_Y \lambda_Z
\end{equation}
and the inverses of ${\bf F}$ and ${\bf F}^T$ are
\begin{equation} \label{sy-film-4}
{\bf F}^{-1} = {\bf F}^{-T} := {\rm diag} \, (\lambda_X^{-1}, \lambda_Y^{-1}, \lambda_Z^{-1}).
\end{equation}

In the homogeneous deformation \eqref{film-displace}, the nominal polarization ${\tilde {\bf P}}$ and the nominal electric field ${\tilde {\bf E}}$ in Eq.~\eqref{EgV-nominal-e} are uniform and only in the $Z-$direction, such that
\begin{equation} \label{sy-film-5}
{\tilde {\bf P}} = - \tilde P {\bf e}_Z, \quad {\tilde {\bf E}} = -\nabla \xi = -\frac{\partial \xi}{\partial Z} {\bf e}_Z= -\frac{\Phi}{L_Z} {\bf e}_Z = -{\tilde E}_0 {\bf e}_Z.
\end{equation}
With Eqs.~\eqref{sy-film-4} and \eqref{sy-film-5}, the true electric field in Eq.~\eqref{EgV-nominal-e} is
\begin{equation} \label{sy-film-6}
{\bf E} = - {\bf F}^{-T} \nabla \xi = - \lambda_Z^{-1} {\tilde E}_0 {\bf e}_Z,
\end{equation}
and then the nominal electric displacement in Eq.~\eqref{DEP-r-refer} reads
\begin{equation} \label{sy-film-7}
{\tilde {\bf D}} = {\bf F}^{-1} (\varepsilon_0 J {\bf E} + {\tilde {\bf P}}) = - \lambda_Z^{-1} (\varepsilon_0 \lambda_X \lambda_Y {\tilde E}_0 +  \tilde P) {\bf e}_Z.
\end{equation}
The total free energy of the system shown in Fig.~\ref{film-1} is \cite{deng2014flexoelectricity, liu2014energy}
\begin{align} \label{film-free-en}
\mathcal F (\lambda_X, \lambda_Y, \lambda_Z, \tilde P) & = L_X L_Y L_Z \Psi (\lambda_X, \lambda_Y, \lambda_Z, \tilde P) - (F_X \lambda_X L_X + F_Y \lambda_Y L_Y) \nonumber \\
                                                                   & \quad +  L_X L_Y L_Z \frac{\varepsilon_0}{2} \lambda_Z^{-1}\lambda_X \lambda_Y {\tilde E}_0^2 - L_X L_Y \Phi \lambda_Z^{-1} (\varepsilon_0 \lambda_X \lambda_Y {\tilde E}_0 +  \tilde P).
\end{align}
Divided by the volume $L_X L_Y L_Z$, the free energy density reads
\begin{equation} \label{film-free-den}
f (\lambda_X, \lambda_Y, \lambda_Z, \tilde P) = \Psi (\lambda_X, \lambda_Y, \lambda_Z, \tilde P) - S_X \lambda_X - S_Y \lambda_Y - \frac{\varepsilon_0}{2} \lambda_Z^{-1} \lambda_X \lambda_Y {\tilde E}_0^2 - \lambda_Z^{-1} {\tilde E}_0 \tilde P,
\end{equation}
where $S_X = F_X / (L_Y L_Z)$ and $S_Y = F_Y / (L_Z L_X)$. The internal energy function $\Psi$ in Eq.~\eqref{film-free-den} consists of two parts:
\begin{equation} \label{inter-energy}
\Psi (\lambda_X, \lambda_Y, \lambda_Z, \tilde P) = W^e (\lambda_X, \lambda_Y, \lambda_Z)  + \frac{\tilde {P}^2}{2 J (\varepsilon - \varepsilon_0)}.
\end{equation}

The first term $W^e$ on the RHS of Eq.~\eqref{inter-energy} represents the elastic strain-energy function, whilst the second term reflects the usual linear dielectric behavior, i.e., the permittivity $\varepsilon$ of the dielectric elastomer is independent of the deformation. In this section, we discuss the strain-energy functions for both compressible and incompressible materials.

\subsubsection{Compressible soft dielectrics}

The zero first variation of Eq.~\eqref{film-free-den} gives the equilibrium equations
\begin{equation} \label{film-com-1st}
\frac{\partial f}{\partial \lambda_X} = 0, \quad \frac{\partial f}{\partial \lambda_Y} = 0, \quad \frac{\partial f}{\partial \lambda_Z} = 0, \quad \frac{\partial f}{\partial \tilde P} = 0.
\end{equation}
The Hessian matrix of Eq.~\eqref{film-free-den} is
\begin{equation} \label{film-com-Hessian}
{\bf H} = 
\begin{pmatrix}
\vspace{0.1in}
\displaystyle \frac{\partial^2 f}{\partial \lambda_X^2}                             & \displaystyle \frac{\partial^2 f}{\partial \lambda_X \partial \lambda_Y} & \displaystyle \frac{\partial^2 f}{\partial \lambda_X \partial \lambda_Z} & \displaystyle \frac{\partial^2 f}{\partial \lambda_X \partial \tilde P} \\
\vspace{0.1in}
\displaystyle \frac{\partial^2 f}{\partial \lambda_Y \partial \lambda_X }     & \displaystyle \frac{\partial^2 f}{\partial \lambda_Y^2}                        & \displaystyle \frac{\partial^2 f}{\partial \lambda_Y \partial \lambda_Z} & \displaystyle \frac{\partial^2 f}{\partial \lambda_Y \partial \tilde P} \\
\vspace{0.1in}
\displaystyle \frac{\partial^2 f}{\partial \lambda_Z \partial \lambda_X }     & \displaystyle \frac{\partial^2 f}{\partial \lambda_Z \partial \lambda_Y}  & \displaystyle \frac{\partial^2 f}{\partial \lambda_Z^2}                        & \displaystyle \frac{\partial^2 f}{\partial \lambda_Z \partial \tilde P} \\
\displaystyle \frac{\partial^2 f}{\partial \tilde P \partial \lambda_X }          & \displaystyle \frac{\partial^2 f}{\partial \tilde P \partial \lambda_Y }       & \displaystyle \frac{\partial^2 f}{\partial \tilde P \partial \lambda_Z }     & \displaystyle \frac{\partial^2 f}{\partial \tilde P ^2}
\end{pmatrix}.
\end{equation}

For isotropic compressible materials, the strain-energy function, for example, can be written as \cite{deng2014flexoelectricity, deng2014electrets, liu2014energy}:
\begin{equation} \label{sy-film-13}
W^e (\lambda_X, \lambda_Y, \lambda_Z) = \frac{\mu}{2}\Big[ J^{-2/3} (\lambda_X^2 + \lambda_Y^2 + \lambda_Z^2) - 3 \Big] + \frac{\kappa}{2} (J-1)^2,
\end{equation}
where $\mu$ and $\kappa$ are the shear and bulk moduli, respectively. \\

Substituting the free energy density \eqref{film-free-den}, with Eqs.~\eqref{inter-energy} and \eqref{sy-film-13}, into \eqref{film-com-1st}, we can obtain four algebraic equations of four variables $\lambda_X$, $\lambda_Y$, $\lambda_Z$, $\tilde P$. Solutions of these four algebraic equations give the equilibrium states $(\lambda_X, \lambda_Y, \lambda_Z, \tilde P)$ of the dielectric film shown in Fig.~\ref{film-1}. Then the stability of the equilibrium states can be verified by evaluating the positive definiteness of the Hessian matrix \eqref{film-com-Hessian} at equilibrium. 

\subsubsection{Incompressible soft dielectrics}

In contrast to compressible materials, the constraint of incompressibility requires that
\begin{equation} \label{film-in-Jaco}
J = \det {\bf F} = \lambda_X \lambda_Y \lambda_Z =1.
\end{equation}

i): {\it Lagrange multiplier method}

The Lagrange multiplier approach is often used to address the extremum problem subjected to a constraint, for example, the constraint of incompressibility \cite{ogden1997non, holzapfel2000nonlinear, reddy2007introduction}. Consider a Lagrange multiplier ${\mathcal L}_a$. The internal energy function $\Psi$ in Eq.~\eqref{inter-energy} is modified as
\begin{equation} \label{film-in-inter}
\bar{\Psi} (\lambda_X, \lambda_Y, \lambda_Z, \tilde P, {\mathcal L}_a) = \Psi (\lambda_X, \lambda_Y, \lambda_Z, \tilde P) - {\mathcal L}_a (\lambda_X \lambda_Y \lambda_Z - 1)
\end{equation}
and the free energy density $f (\lambda_X, \lambda_Y, \lambda_Z, \tilde P) $ in Eq.~\eqref{film-free-den} becomes 
\begin{equation} \label{film-m-f-den}
\bar{f} (\lambda_X, \lambda_Y, \lambda_Z, \tilde P, {\mathcal L}_a) = f (\lambda_X, \lambda_Y, \lambda_Z, \tilde P) - {\mathcal L}_a (\lambda_X \lambda_Y \lambda_Z - 1).
\end{equation}
Then, the governing equations are
\begin{equation} \label{film-in-1st}
\frac{\partial \bar{f}}{\partial \lambda_X} = 0, \quad \frac{\partial \bar{f}}{\partial \lambda_Y} = 0, \quad \frac{\partial \bar{f}}{\partial \lambda_Z} = 0, \quad \frac{\partial \bar{f}}{\partial \tilde P} = 0, \quad \frac{\partial \bar{f}}{\partial \mathcal L_a} = 0.
\end{equation}

\noindent ii): {\it Reduced energy function method}

The constraint of incompressibility \eqref{film-in-Jaco} gives the following relation between the stretches
\begin{equation} \label{film-re-Lz}
\lambda_Z = \frac{1}{\lambda_X \lambda_Y}.
\end{equation}
Then, the internal energy function \eqref{inter-energy} can be defined by a reduced form
\begin{equation} \label{film-re-inter}
\hat{\Psi} (\lambda_X, \lambda_Y, \tilde P) = \Psi (\lambda_X, \lambda_Y, \frac{1}{\lambda_X \lambda_Y}, \tilde P)
\end{equation}
and the free energy density \eqref{film-free-den} becomes
\begin{equation} \label{film-re-free-den}
\hat f (\lambda_X, \lambda_Y, \tilde P) = \hat \Psi (\lambda_X, \lambda_Y, \tilde P) - S_X \lambda_X - S_Y \lambda_Y - \frac{\varepsilon_0}{2} \lambda_X^2 \lambda_Y^2 {\tilde E}_0^2 - \lambda_X \lambda_Y {\tilde E}_0 \tilde P.
\end{equation}
The zero first variation of Eq.~\eqref{film-re-free-den} gives the equilibrium equations
\begin{equation} \label{film-re-1st}
\frac{\partial \hat f}{\partial \lambda_X} (\lambda_X, \lambda_Y, \tilde P) = 0, \quad \frac{\partial \hat f}{\partial \lambda_Y} (\lambda_X, \lambda_Y, \tilde P) = 0, \quad \frac{\partial \hat f}{\partial \tilde P} (\lambda_X, \lambda_Y, \tilde P) = 0.
\end{equation}
The related Hessian matrix of Eq.~\eqref{film-re-free-den} is
\begin{equation} \label{film-re-Hessian}
\hat {\bf H} = 
\begin{pmatrix}
\vspace{0.1in}
\displaystyle \frac{\partial^2 \hat f}{\partial \lambda_X^2}                          & \displaystyle \frac{\partial^2 \hat f}{\partial \lambda_X \partial \lambda_Y}  & \displaystyle \frac{\partial^2 \hat f}{\partial \lambda_X \partial \tilde P} \\
\vspace{0.1in}
\displaystyle \frac{\partial^2 \hat f}{\partial \lambda_Y \partial \lambda_X}   & \displaystyle \frac{\partial^2 \hat f}{\partial \lambda_Y^2}                         & \displaystyle \frac{\partial^2 \hat f}{\partial \lambda_Y \partial \tilde P} \\
\displaystyle \frac{\partial^2 \hat f}{\partial \tilde P \partial \lambda_X}       & \displaystyle \frac{\partial^2 \hat f}{\partial \tilde P \partial \lambda_Y}      & \displaystyle \frac{\partial^2 \hat f}{\partial \tilde P ^2}
\end{pmatrix}.
\end{equation}

\subsubsection{Example: a film of incompressible neo-Hookean dielectrics}
We take the incompressible neo-Hookean dielectrics as an example to illustrate the above analysis. Thus the internal energy function \eqref{inter-energy}, by the relation $\lambda_Z = 1/ (\lambda_X \lambda_Y)$ and the reduced energy function \eqref{film-re-inter}, may be rewritten as
\begin{equation} \label{film-re-neo}
\hat \Psi (\lambda_X, \lambda_Y, \tilde {P}) = \frac{\mu}{2} (\lambda_X^2 + \lambda_Y^2 + \frac{1}{\lambda_X^2 \lambda_Y^2} -3) + \frac{\tilde {P}^2}{2 (\varepsilon - \varepsilon_0)}.
\end{equation}
Then the reduced free energy density \eqref{film-re-free-den} becomes
\begin{equation} \label{film-re-neo-free}
\hat f (\lambda_X, \lambda_Y, \tilde P) = \frac{\mu}{2} (\lambda_X^2 + \lambda_Y^2 + \frac{1}{\lambda_X^2 \lambda_Y^2} -3) + \frac{\tilde {P}^2}{2 (\varepsilon - \varepsilon_0)} - S_X \lambda_X - S_Y \lambda_Y - \frac{\varepsilon_0}{2} \lambda_X^2 \lambda_Y^2 {\tilde E}_0^2 - \lambda_X \lambda_Y {\tilde E}_0 \tilde P.
\end{equation}
By Eq.~\eqref{film-re-neo-free}, the governing equations \eqref{film-re-1st} are
\begin{subequations} \label{film-re-neo-1stabc}
\begin{align}
\label{film-re-neo-1sta}
\mu (\lambda_X - \frac{1}{\lambda_X^3 \lambda_Y^2}) - S_X - \varepsilon_0 \lambda_X \lambda_Y^2 {\tilde E}_0^2 - \lambda_Y {\tilde E}_0 \tilde P & = 0, \\
\label{film-re-neo-1stb}
\mu(\lambda_Y - \frac{1}{\lambda_X^2 \lambda_Y^3}) - S_Y - \varepsilon_0 \lambda_X^2 \lambda_Y {\tilde E}_0^2 - \lambda_X {\tilde E}_0 \tilde P & = 0, \\ 
\label{film-re-neo-1stc}
\frac{\tilde {P}}{\varepsilon - \varepsilon_0} - \lambda_X \lambda_Y {\tilde E}_0 & = 0,
\end{align}
\end{subequations}
and the Hessian matrix \eqref{film-re-Hessian} reads
\begin{equation} \label{film-re-neo-H}
\hat {\bf H} = 
\begin{pmatrix}
\vspace{0.2in}
\displaystyle \mu (1 + \frac{3}{\lambda_X^4 \lambda_Y^2}) - \varepsilon_0 \lambda_Y^2 \tilde E_0^2  & \displaystyle \mu \frac{2}{\lambda_X^3 \lambda_Y^3} - 2 \varepsilon_0 \lambda_X \lambda_Y \tilde E_0^2 - \tilde E_0 \tilde P  & \displaystyle - \lambda_Y \tilde E_0 \\
\vspace{0.2in}
~                                                                                                                                                 & \displaystyle \mu (1 + \frac{3}{\lambda_X^2 \lambda_Y^4}) - \varepsilon_0 \lambda_X^2 \tilde E_0^2                                    & \displaystyle - \lambda_X \tilde E_0 \\
\displaystyle {sym}       & ~      & \displaystyle \frac{1}{\varepsilon - \varepsilon_0}
\end{pmatrix}.
\end{equation}

In the following, we will discuss the solutions of Eqs.~\eqref{film-re-neo-1sta}$-$\eqref{film-re-neo-1stc}, and then study the stability of the dielectric film by checking the positive definiteness of the Hessian matrix \eqref{film-re-neo-H} at equilibrium. \\

The equilibrium equation \eqref{film-re-neo-1stc} directly gives the relation
\begin{subequations}
\begin{equation} \label{film-re-neo-P}
{\tilde P} = (\varepsilon - \varepsilon_0) \lambda_X \lambda_Y \tilde E_0.
\end{equation}
By Eq.~\eqref{film-re-neo-P}, Eqs.~\eqref{film-re-neo-1sta} and \eqref{film-re-neo-1stb} become
\begin{equation} \label{film-re-neo-eq1}
 \mu (\lambda_X - \frac{1}{\lambda_X^3 \lambda_Y^2}) - S_X - \varepsilon \lambda_X \lambda_Y^2 {\tilde E}_0^2 = 0,
\end{equation}
\begin{equation} \label{film-re-neo-eq2}
 \mu(\lambda_Y - \frac{1}{\lambda_X^2 \lambda_Y^3}) - S_Y - \varepsilon \lambda_X^2 \lambda_Y {\tilde E}_0^2 =0,
\end{equation}
\end{subequations}
and the Hessian matrix \eqref{film-re-neo-H} is recast as
\begin{equation} \label{film-re-neo-H1}
\hat {\bf H} = 
\begin{pmatrix}
\vspace{0.2in}
\displaystyle \mu (1 + \frac{3}{\lambda_X^4 \lambda_Y^2}) - \varepsilon_0 \lambda_Y^2 \tilde E_0^2  & \displaystyle \mu \frac{2}{\lambda_X^3 \lambda_Y^3} - (\varepsilon_0+\varepsilon) \lambda_X \lambda_Y \tilde E_0^2  & \displaystyle - \lambda_Y \tilde E_0 \\
\vspace{0.2in}
~                                                                                                                                                 & \displaystyle \mu (1 + \frac{3}{\lambda_X^2 \lambda_Y^4}) - \varepsilon_0 \lambda_X^2 \tilde E_0^2                                    & \displaystyle - \lambda_X \tilde E_0 \\
\displaystyle {sym}       & ~      & \displaystyle \frac{1}{\varepsilon - \varepsilon_0}
\end{pmatrix}.
\end{equation}

Note that Eqs.~\eqref{film-re-neo-eq1} and \eqref{film-re-neo-eq2} are the same as equations (6a) and (6b) in the work of Zhao and Suo \cite{zhao2007method} with appropriate notation changes. For example, the subscripts `1' and `2' in the work \cite{zhao2007method} correspond to the subscripts `$X$' and `$Y$' in this tutorial. If we replace $\tilde D$ in (6a) and (6b) by $\tilde E$ in (6c) in their work, then we get exactly the same two algebraic equations as Eqs.~\eqref{film-re-neo-eq1} and \eqref{film-re-neo-eq2}. This equivalence serves as a direct validation of the alternative formulation presented in this tutorial. In the following, we will use some special but easy cases to illustrate the behavior of a dielectric film. \\

Consider the case of a uniaxial prestress, i.e., $S_Y=0$. The equilibrium equation \eqref{film-re-neo-eq2} can be recast as
\begin{equation} \label{Ly-Lx-E0}
\lambda_Y = [ \lambda_X^2 - \lambda_X^4 ({\tilde E}_0 \sqrt{\varepsilon/\mu})^2 ]^{-1/4} .
\end{equation}
Substituting Eq.~\eqref{Ly-Lx-E0} into the equilibrium equation \eqref{film-re-neo-eq1}, we get the following algebraic equation for $\lambda_X$:
\begin{equation} \label{Lx-Sx-E0}
\lambda_X - \frac{\sqrt{1 - \lambda_X^2 ({\tilde E}_0 \sqrt{\varepsilon/\mu})^2}}{\lambda_X^2} - \frac{({\tilde E}_0 \sqrt{\varepsilon/\mu})^2}{\sqrt{1 - \lambda_X^2 ({\tilde E}_0 \sqrt{\varepsilon/\mu})^2}} = \frac{S_X}{\mu}.
\end{equation}

\begin{figure}[h] 
\centering
\subfigure[]{%
\includegraphics[width=2.8in]{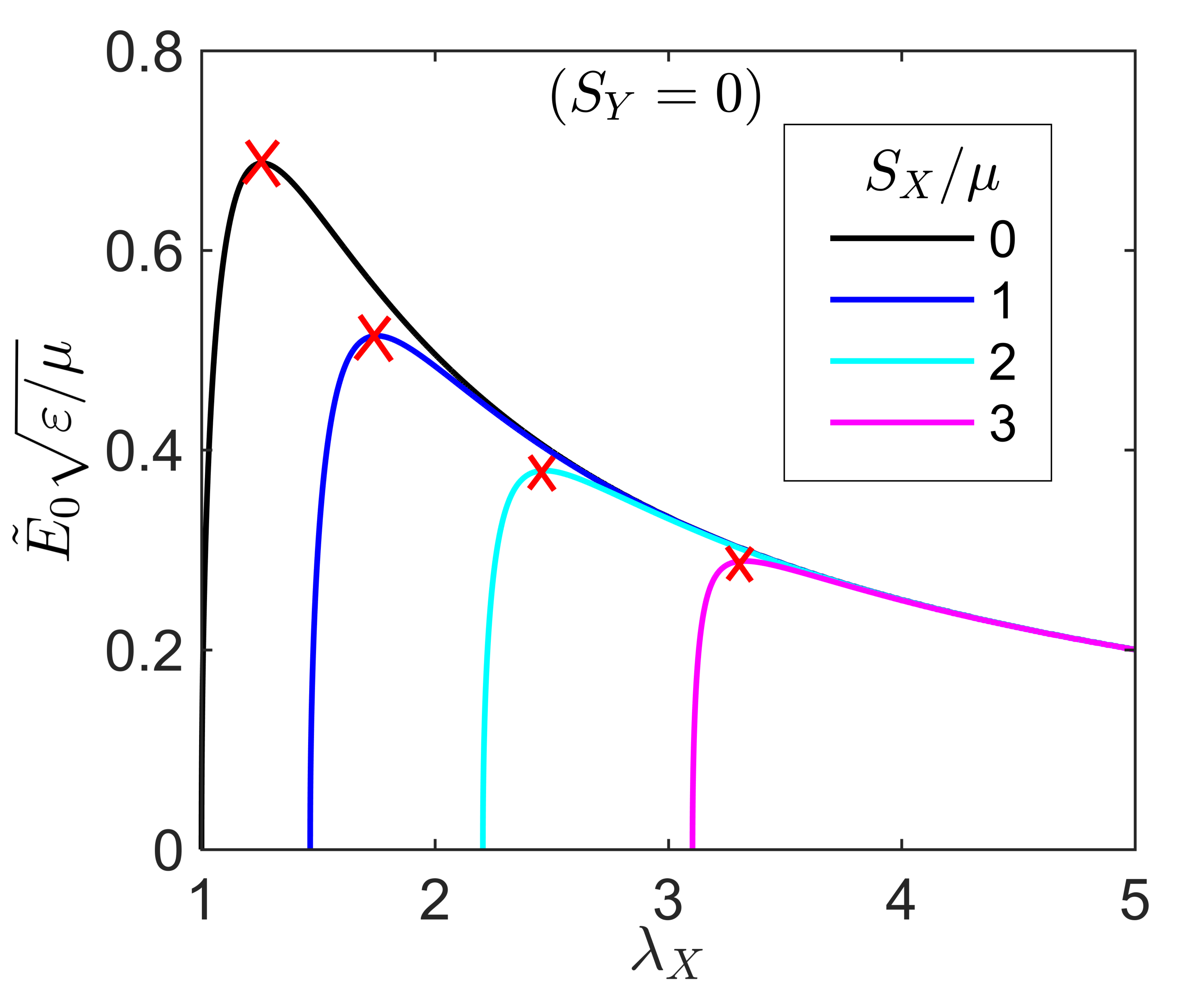}
\label{film-2a}}
\subfigure[]{%
\includegraphics[width=2.8in]{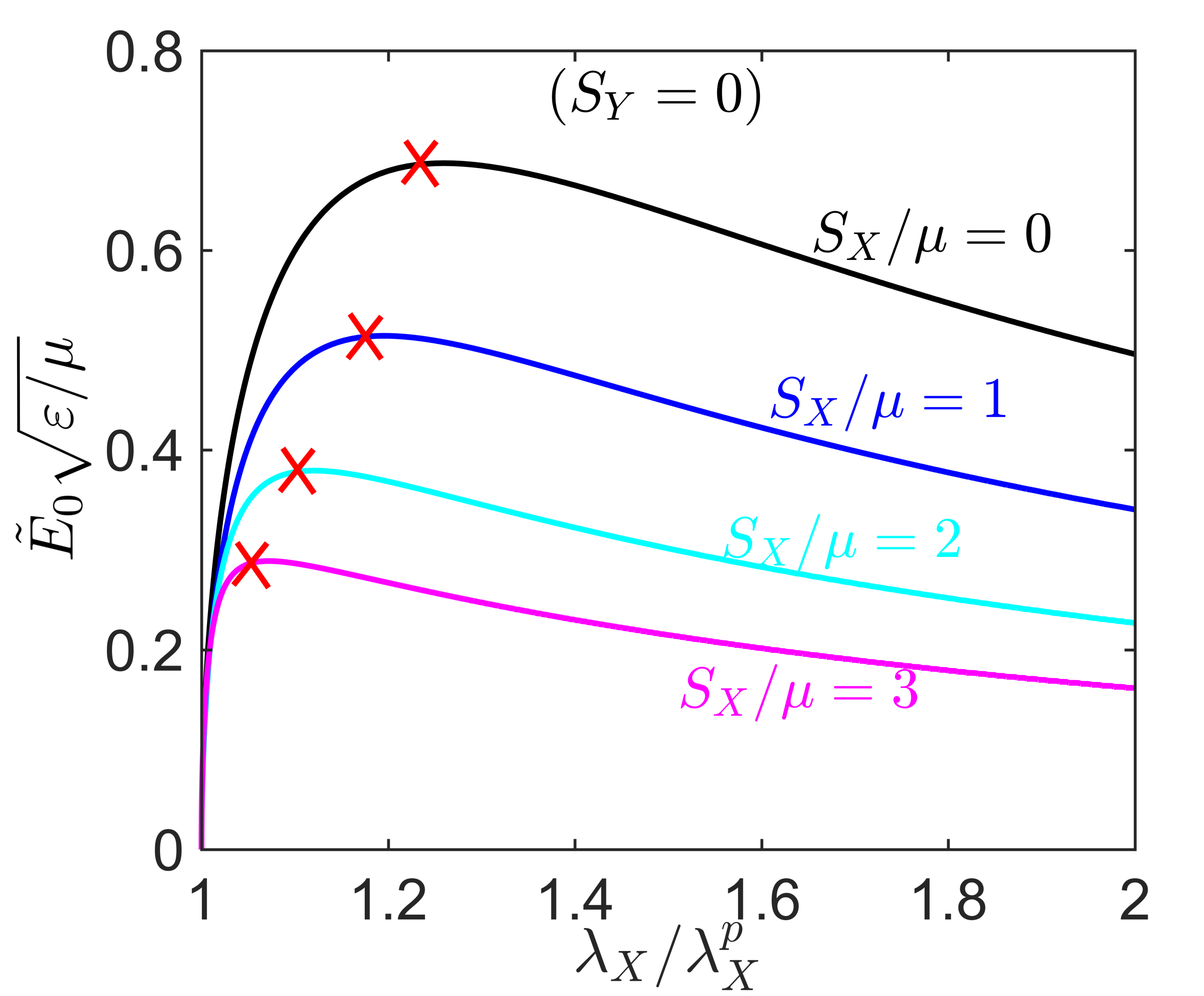}
\label{film-2b}}
\caption{Deformation of the dielectric film subject to an electric voltage and a uniaxial prestress ($S_Y=0$). (a) Stretch $\lambda_X$ {\it vs.} the nominal electric field under several forces $S_X/\mu$. (b) Actuation stretch $\lambda_X/\lambda_X^p$ {\it vs.} the nominal electric field under several forces $S_X/\mu$. The critical point for the onset of pull-in instability is denoted by a cross in each curve.}
\label{film-2}
\end{figure}

The equilibrium solution of $\lambda_X$ in Eq.~\eqref{Lx-Sx-E0} for a given pair of $(S_X, \tilde E_0)$ is shown in Fig.~\ref{film-2a}. Without the applied dead loads, i.e., $S_X = S_Y = 0$, the nominal electric field (prior to the onset of instability) induces the film to expand but not significantly. When the nominal electric field increases to the threshold (i.e., the cross in each curve), pull-in instability occurs and the film expands dramatically without an appreciable increase in the electric field. Finally, the thinned film is damaged by electric breakdown. In addition, the dead load $S_X$ expands the film significantly and the electric field assists the expansion, and pull-in instability occurs when the electric field increases to the threshold. In contrast, the threshold of the nominal electric field $\tilde E_0$ decreases as the dead load $S_X$ increases. \\

To investigate the electric actuation, we define the prestretch in the $X-$direction as $\lambda_X^p$, which is the stretch due to the dead load $S_X$ in the absence of the voltage. Explicitly, $\lambda_X^p$ is the solution of the algebraic equation \eqref{Lx-Sx-E0} with a zero electric field $\tilde E_0 =0$. In contrast to the prestretch $\lambda_X^p$, the actuation stretch in the $X-$direction is defined as $\lambda_X/\lambda_X^p$. Figure \ref{film-2b} shows the variation of the actuation stretch $\lambda_X/\lambda_X^p$ with respect to the nominal electric field $\tilde E_0$ under several uniaxial dead loads $S_X$. The maximum actuation stretch in each curve is the critical stretch at the occurrence of pull-in instability (marked by a cross). Without the dead load $S_X=0$, the film can attain the largest maximum actuation stretch. By increasing the dead load $S_X$, the maximum actuation stretch decreases. Note that even the maximum actuation stretch decreases, the total stretch $\lambda_X$ does increase as the increasing of the dead load before the onset of pull-in instability. The difference between the total stretch $\lambda_X$ and the actuation stretch $\lambda_X/\lambda_X^p$ can be observed by comparing Figs.~\ref{film-2a} and \ref{film-2b}.

\subsection{A dielectric disc}%
Another tractable example that we consider is that of a dielectric disc. To that end, we consider a circular dielectric plate with radius $R_a$ and thickness $H_a$ in the undeformed state (see Fig.~\ref{disk-1}). Taking the cylindrical coordinates $(R, \varTheta, Z)$ with an {\it orthonormal} basis $({\bf e}_R, {\bf e}_{\varTheta}, {\bf e}_Z)$, the domain of the dielectric disc in the reference configuration is represented by
\begin{equation} \label{disk-domain}
\Omega_R = \Big\{ (R, \varTheta, Z) \in {\mathbb R}^3: 0 \le R \le R_a, \ 0 \le \varTheta < 2 \pi, \ 0 \le Z \le H_a \Big\}.
\end{equation}

\subsubsection{Large deformation}%
Subject to both an applied voltage $\Phi$ in the thickness direction (the $Z-$direction) and a surrounding uniform traction force $F_R$, the dielectric disc expands its in-plane area and decreases its thickness from $H_a$ to $\lambda_Z H_a$. Here the stretch $\lambda_Z$ in the thickness direction is a constant that is independent of the position. We assume that the dielectric disc admits the cylindrical deformation
\begin{equation} \label{disk-displace}
r = r (R), \quad \theta = \varTheta, \quad z = \lambda_Z Z.
\end{equation}
The deformation gradient in the cylindrical coordinates is
\begin{equation} \label{disk-dg-F}
{\bf F} := {\rm diag} \, ( \frac{d r}{d R}, \frac{r}{R}, \lambda_Z ).
\end{equation}

\begin{figure}[h] %
\centering
\subfigure[]{%
\includegraphics[width=2.9in]{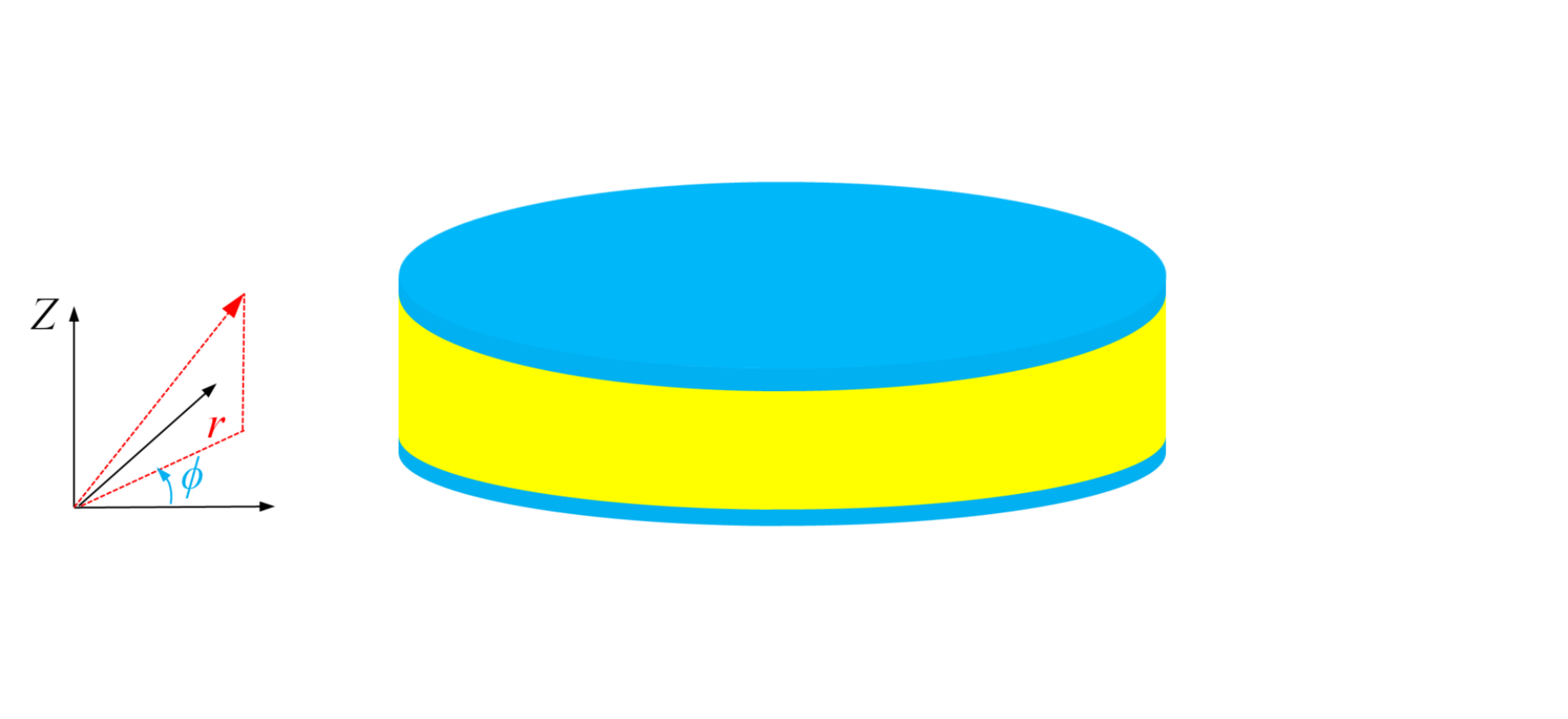}
\label{disk-1a}}
\subfigure[]{%
\includegraphics[width=2.9in]{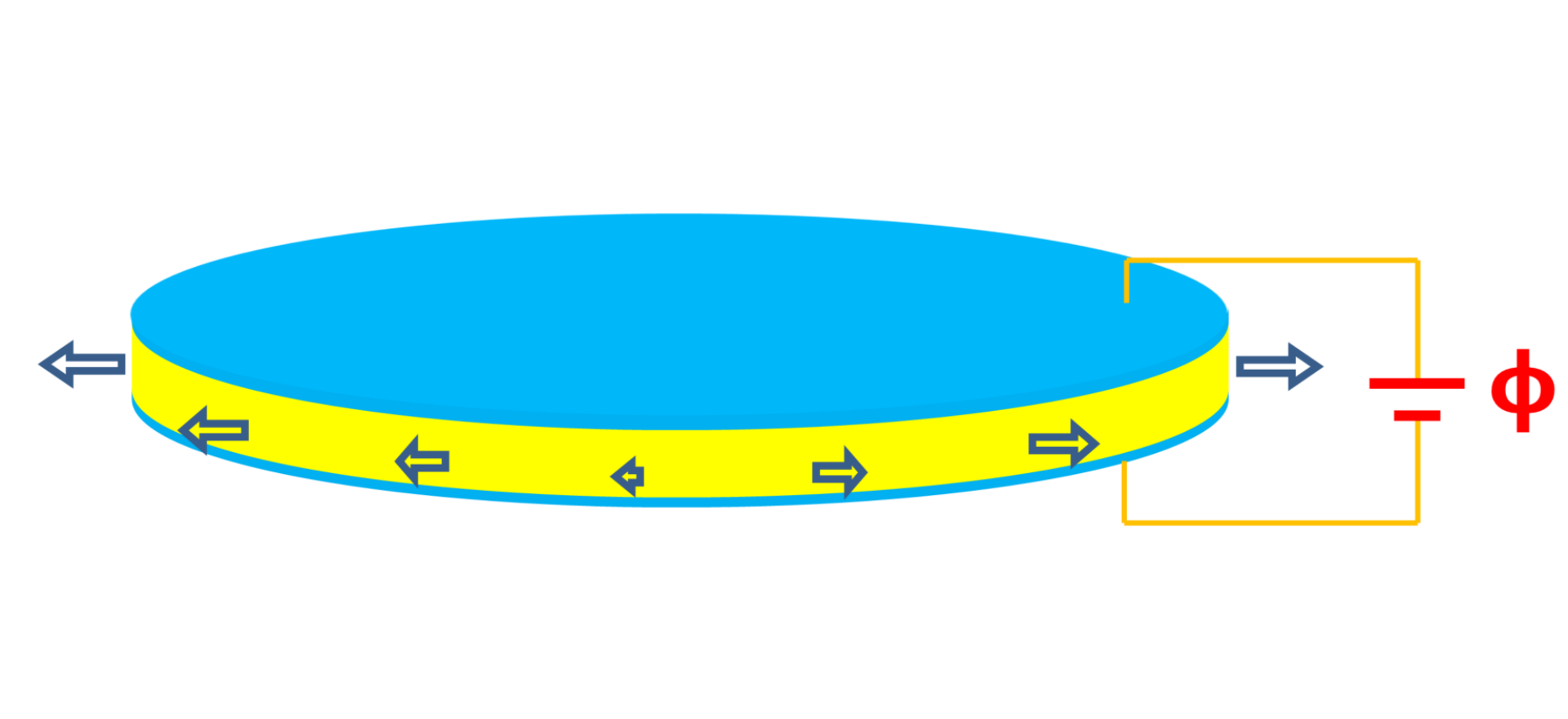}
\label{disk-1b}}
\caption{Schematic of the deformation of a dielectric disc subjected to an electric voltage $\Phi$ and a surrounding dead load $F_R$. The disc is coated with two compliant electrodes on the upper and bottom surfaces. (a) Undeformed disc with radius $R_a$ and thickness $H_a$. (b) Deformed disc with thickness $\lambda_Z H_a$.}
\label{disk-1}
\end{figure}

\noindent Incompressibility requires that 
\begin{equation} \label{disk-Jaco}
J = \det {\bf F} = \frac{d r}{d R} \frac{r}{R} \lambda_Z = 1.
\end{equation}
Integrating Eq.~\eqref{disk-Jaco} with respect to $R$ and with $r(0) =0$, we have 
\begin{equation}
r (R) = \frac{R}{\sqrt{\lambda_Z}}. 
\end{equation}
Thus, the deformation gradient \eqref{disk-dg-F} becomes
\begin{equation} \label{disk-dg-F1}
{\bf F} := {\rm diag} \, ( \frac{1}{\sqrt{\lambda_Z}}, \frac{1}{\sqrt{\lambda_Z}}, \lambda_Z ).
\end{equation}

The thinning of the disc shown in Fig.~\ref{disk-1} is similar to the deformation in Fig.~\ref{film-1}. By the cylindrical deformation \eqref{disk-displace}, the equations of the polarization, the electric field, and the electric displacement are the same as Eqs.~\eqref{sy-film-5}$-$\eqref{sy-film-7}. \\

With the energy formulation in the work \cite{deng2014flexoelectricity, liu2014energy} and the deformation gradient \eqref{disk-dg-F1}, the free energy of the dielectric disc in Fig.~\ref{disk-1} is written as
\begin{align} \label{disk-free-en}
\mathcal F (\lambda_Z, \tilde P) & = \pi R_a^2 H_a \Psi (\lambda_Z, \tilde P) -  \frac{F_R R_a}{\sqrt{\lambda_Z}} +  \pi R_a^2 H_a \frac{\varepsilon_0}{2} \lambda_Z^{-2}{\tilde E}_0^2 - \pi R_a^2 \Phi \lambda_Z^{-1} (\varepsilon_0 \lambda_Z^{-1} {\tilde E}_0 +  \tilde P).
\end{align}
Divided by the volume $\pi R_a^2 H_a$, the free energy density reads
\begin{align} \label{disk-free-den}
f (\lambda_Z, \tilde P) = \Psi (\lambda_Z, \tilde P) -  \frac{S_R}{\sqrt{\lambda_Z}} - \frac{\varepsilon_0}{2} \lambda_Z^{-2}{\tilde E}_0^2 - \lambda_Z^{-1} {\tilde E}_0 \tilde P,
\end{align}
where $S_R = F_R / (\pi R_a H_a)$. Consider incompressible neo-Hookean dielectrics. By Eq.~\eqref{disk-dg-F1}, the internal energy function reads
\begin{equation} \label{disk-neo-en}
\Psi (\lambda_Z, \tilde P) = \frac{\mu}{2} (\lambda_Z^2 + 2 \lambda_Z^{-1} - 3) + \frac{{\tilde P}^2}{2(\varepsilon - \varepsilon_0)}
\end{equation}
and the free energy density \eqref{disk-free-den} becomes
\begin{align} \label{disk-free-den1}
f (\lambda_Z, \tilde P) = \frac{\mu}{2} (\lambda_Z^2 + 2 \lambda_Z^{-1} - 3) + \frac{{\tilde P}^2}{2(\varepsilon - \varepsilon_0)} - \frac{S_R}{\sqrt{\lambda_Z}} - \frac{\varepsilon_0}{2} \lambda_Z^{-2}{\tilde E}_0^2 - \lambda_Z^{-1} {\tilde E}_0 \tilde P.
\end{align}
The zero first variation of Eq.~\eqref{disk-free-den1} gives
\begin{subequations}
\begin{align}
\label{disk-1st-var1}
\mu(\lambda_Z - \lambda_Z^{-2}) + \frac{1}{2} S_R \lambda_Z^ {-3/2} + \varepsilon_0 \lambda_Z^{-3}{\tilde E}_0^2 + \lambda_Z^{-2} {\tilde E}_0 \tilde P & = 0, \\
\label{disk-1st-var2}
\frac{{\tilde P}}{\varepsilon - \varepsilon_0} - \lambda_Z^{-1} {\tilde E}_0 & = 0. 
\end{align}
\end{subequations}
Equation \eqref{disk-1st-var2} directly gives the relation
\begin{equation} \label{disk-eqi-pola}
\tilde P = (\varepsilon - \varepsilon_0) \lambda_Z^{-1} {\tilde E}_0.
\end{equation}
Substituting Eq.~\eqref{disk-eqi-pola} into Eq.~\eqref{disk-1st-var1}, we have
\begin{equation} \label{disk-eqi-eq}
\mu(\lambda_Z - \lambda_Z^{-2}) + \frac{1}{2} S_R \lambda_Z^ {-3/2} + \varepsilon \lambda_Z^{-3}{\tilde E}_0^2  = 0.
\end{equation}

\begin{figure}[h] %
\centering
\subfigure[]{%
\includegraphics[width=2.8in]{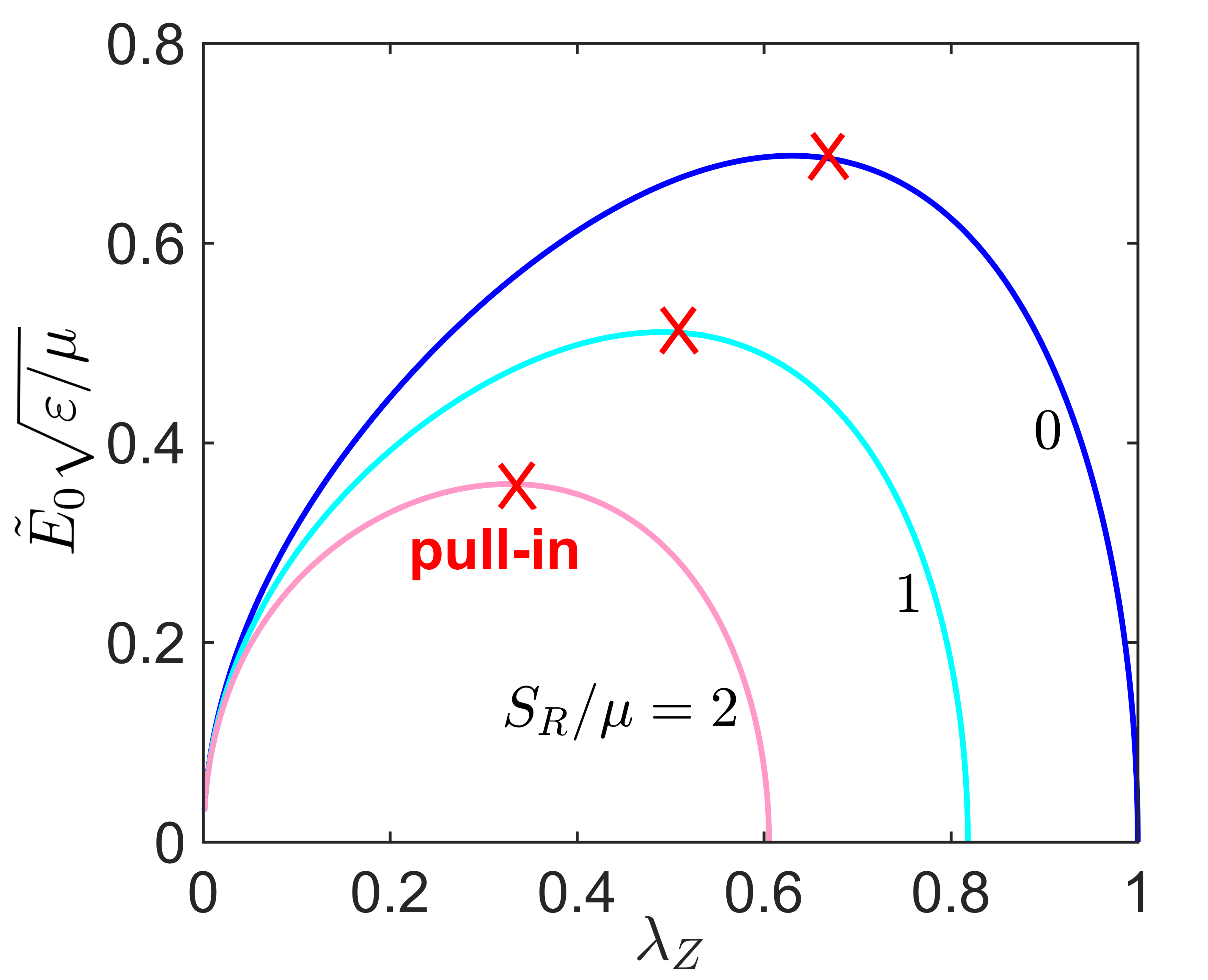}
\label{disk-2a}}
\subfigure[]{%
\includegraphics[width=2.8in]{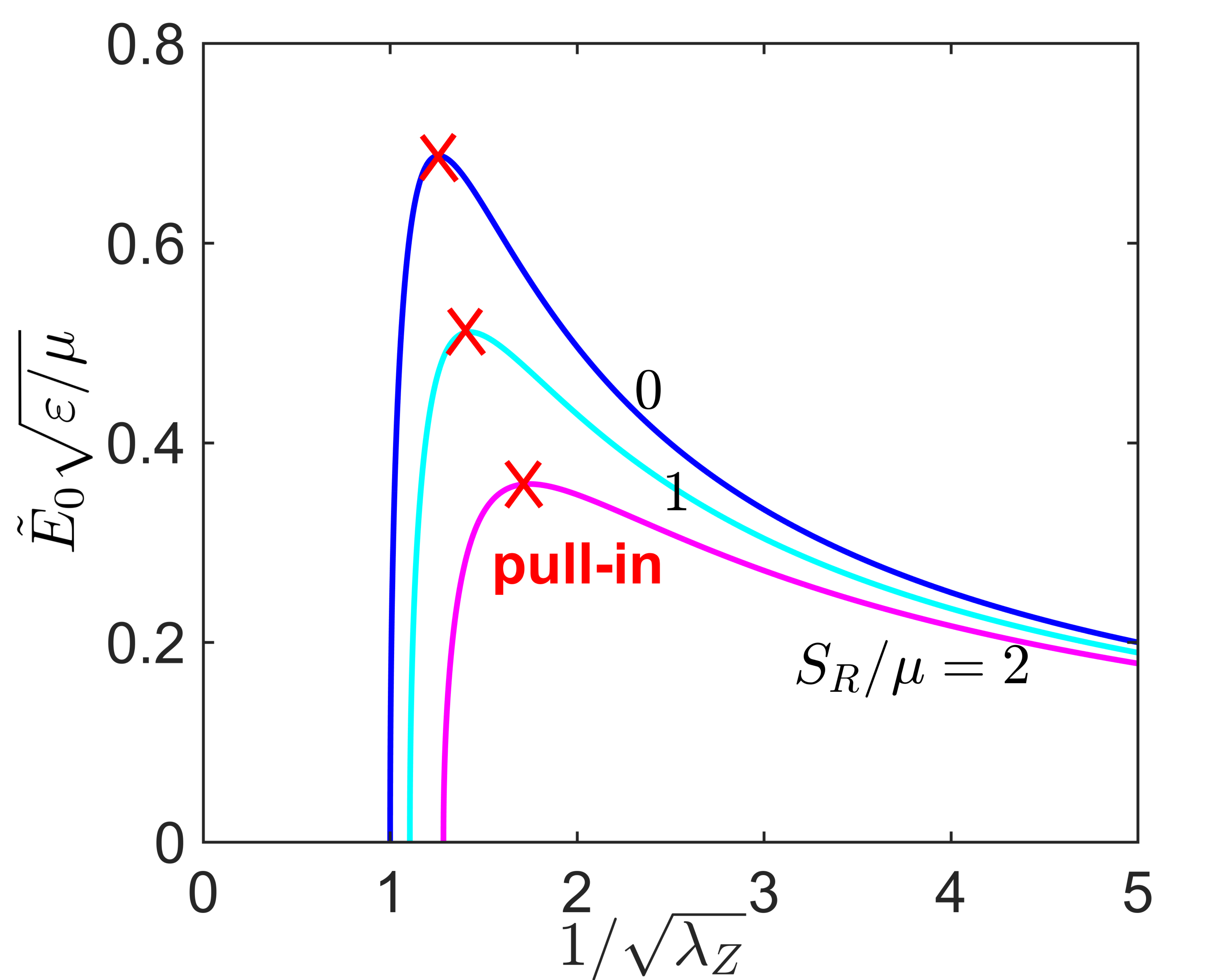}
\label{disk-2b}}
\subfigure[]{%
\includegraphics[width=2.8in]{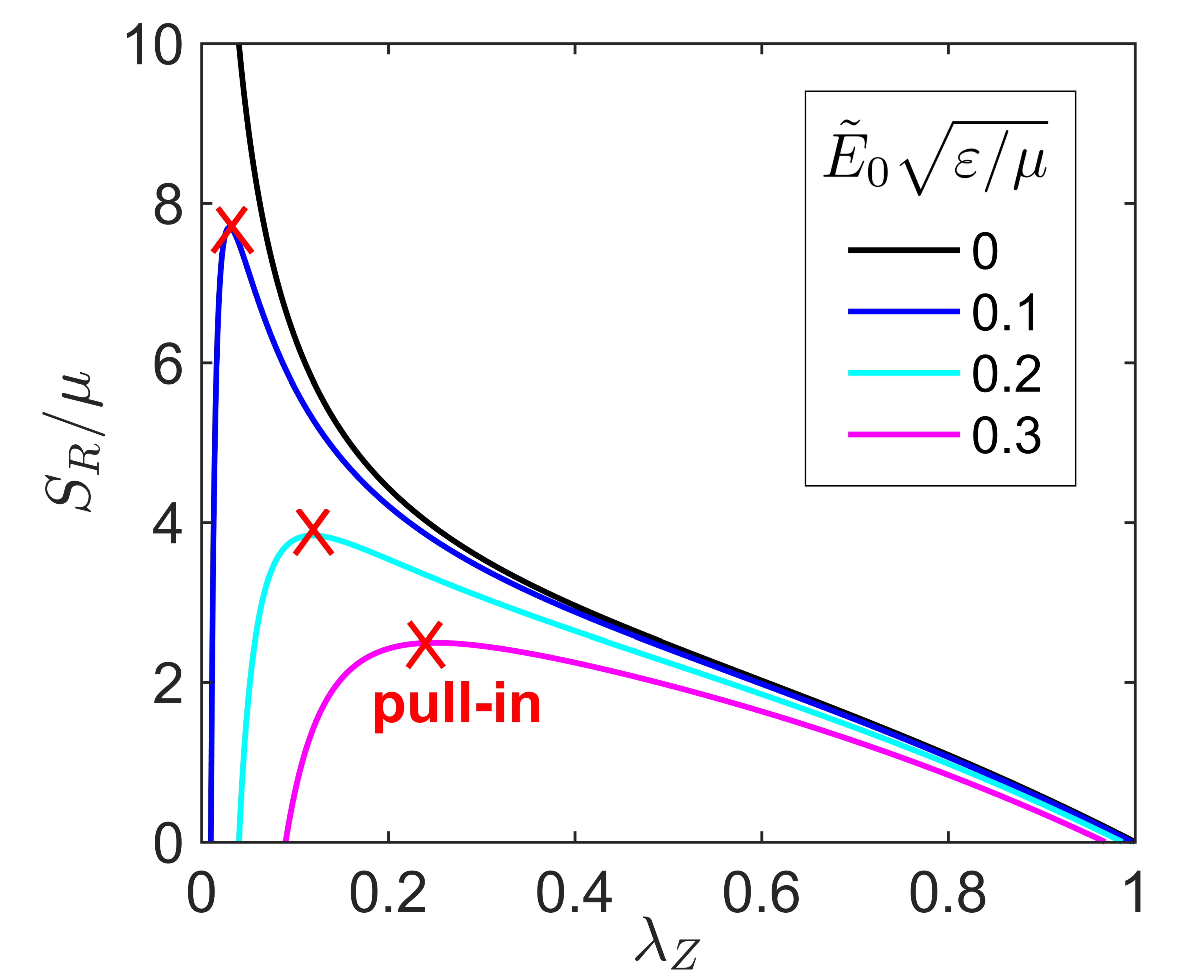}
\label{disk-2c}}
\subfigure[]{%
\includegraphics[width=2.8in]{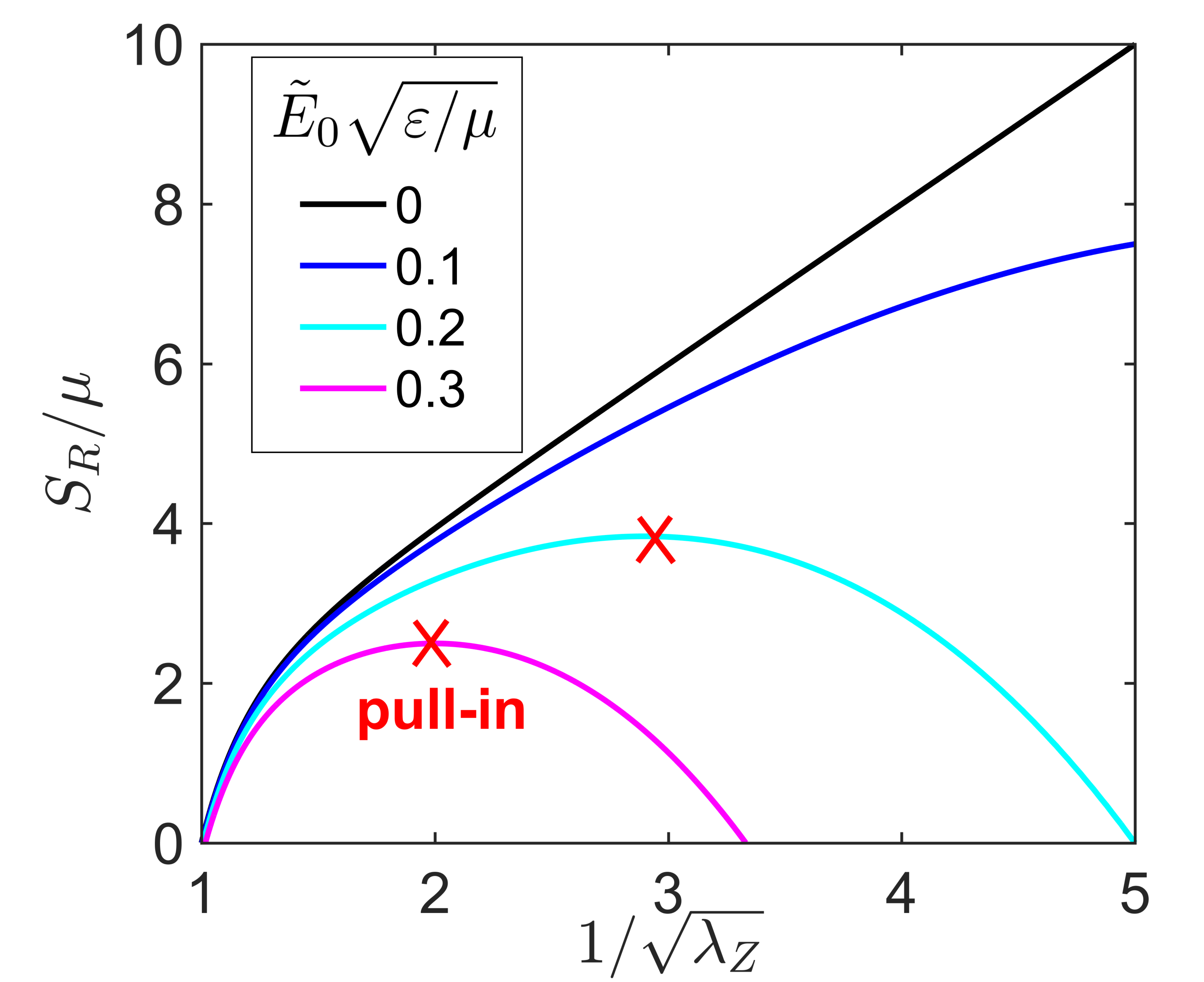}
\label{disk-2d}}
\caption{Deformation of the dielectric disc subject to the electric voltage and the surrounding force. (a) Stretch $\lambda_Z$ in the thickness direction {\it vs.} the nominal electric field under several surrounding forces $S_R/\mu$. (b) Stretch $1/\sqrt{\lambda_Z}$ in the radial direction {\it vs.} the nominal electric field under several surrounding forces $S_R/\mu$. (c) Stretch $\lambda_Z$ in the thickness direction {\it vs.} the surrounding force $S_R/\mu$ under several nominal electric fields. (d) Stretch $1/\sqrt{\lambda_Z}$ in the radial direction {\it vs.} the surrounding force $S_R/\mu$ under several nominal electric fields. The critical point for the onset of pull-in instability is denoted by a cross in each curve.}
\label{disk-2}
\end{figure}

Equation \eqref{disk-eqi-eq} is an algebraic equation of $\lambda_Z$ with two controlled parameters $S_R$ and $\tilde E_0$. For a given pair of $(S_R, \tilde E_0)$, there may exist many cases of the roots of $\lambda_Z$ in Eq.~\eqref{disk-eqi-eq}, for example, multiple real and complex roots. The detailed discussion of the properties of roots of $\lambda_Z$ in Eq.~\eqref{disk-eqi-eq} is not addressed here. The solution for realistic physical situations of the stretch $\lambda_Z$, a real value and $0<\lambda_Z \le 1$, for a given pair of $(S_R, \tilde E_0)$ is shown in Fig.~\ref{disk-2}. \\

Figure \ref{disk-2a} shows the equilibrium solution of the stretch $\lambda_Z$ in Eq.~\eqref{disk-eqi-eq} for a given pair of $(S_R, \tilde E_0)$. Without the dead load $S_R = 0$, the stretch $\lambda_Z$ decreases from $1$ as the nominal electric field $\tilde E_0$ increases. With the increase of the electric field, the disc decreases its thickness to the threshold at which pull-in instability occurs. The onset of pull-in instability is at the peak of each curve. Later we will show that the peak corresponds to a zero determinant of the Hessian matrix at equilibrium. With the increase of the dead load $S_R$, the peak in each curve decreases, which means that it is easier to induce pull-in instability at a larger dead load. After the peak, the disc expands and the thickness decreases rapidly and simultaneously until the onset of electric breakdown. \\

In contrast to the deformation in the thickness direction, Fig.~\ref{disk-2b} shows the deformation in the radial direction with the stretch $1/\sqrt{\lambda_Z}$. The curves in Figs.~\ref{disk-2b} and \ref{film-2a} have the same trends as the increase of the applied dead load and the electric field. \\

Figure \ref{disk-2c} shows the variation of the stretch $\lambda_Z$ in the thickness direction with respect to the dead load $S_R$ for varying electric fields. Interestingly, the curve with a zero electric field has no peak, and the applied dead load increases monotonically with the decrease of the thickness ($\lambda_Z$). \\

Figure \ref{disk-2d} shows the variation of the stretch $1/\sqrt{\lambda_Z}$ of the radial deformation with respect to the dead load $S_R$ under several electric fields. At a zero dead load $S_R =0$, it is obvious that the stretches correspond to different electric fields are almost the same and are close to 1, which indicates that the disc deformation is less insensitive to the electric field compared to the dead load. However, the trend of each curve changes significantly with the increase of the applied dead load, especially the peak of each curve. This implies that for a given electric field, the increase of the applied dead load will eventually make pull-in instability occur. For a larger electric field, the peak of the curve occurs at a smaller dead load. \\

All the curves in Fig.~\ref{disk-2} simply emerge from the equilibrium equation \eqref{disk-eqi-eq}, and a point on the curve corresponds to an equilibrium state. The stability of the equilibrium states has not verified so far. When we described the trend of each curve in Fig.~\ref{disk-2}, we claim that pull-in instability occurs at the peak of each curve and the disc is stable before the point reaches the peak. We will verify this claim in the following stability analysis.

\subsubsection{Stability analysis and actuation strain}%
In this section, we discuss the stability and (pull-in) instability regions of the dielectric disc subjected to both the dead load and the applied voltage. We also discuss the critical stretch and the critical actuation stretch (the maximum actuation stretch) at which pull-in instability occurs. \\

The stability of the equilibrium solutions in Eqs.~\eqref{disk-eqi-pola} and \eqref{disk-eqi-eq}, see Fig.~\ref{disk-2}, can be determined by examining the positive definiteness of the Hessian matrix at equilibrium. By Eq.~\eqref{disk-eqi-pola} and the second variation of the free energy \eqref{disk-free-den1}, we have the Hessian matrix
\begin{equation} \label{disk-eqi-H}
{\bf H} = 
\begin{pmatrix}
\vspace{0.1in}
\mu(1 +2 \lambda_Z^{-3}) - \displaystyle \frac{3}{4} S_R \lambda_Z^ {-5/2} - (\varepsilon_0 + 2\varepsilon) \lambda_Z^{-4}{\tilde E}_0^2 & \lambda_Z^{-2} {\tilde E}_0\\
\lambda_Z^{-2} {\tilde E}_0 & \displaystyle \frac{1}{\varepsilon -\varepsilon_0}
\end{pmatrix}
\end{equation}
with the determinant
\begin{equation} \label{disk-eqi-det}
\det {\bf H} = \frac{1}{\varepsilon - \varepsilon_0} \left ( \mu(1 +2 \lambda_Z^{-3}) - \frac{3}{4} S_R \lambda_Z^ {-5/2} - 3 \varepsilon \lambda_Z^{-4}{\tilde E}_0^2 \right ).
\end{equation}

The stability of the homogeneous deformation requires that the $2 \times 2$ Hessian matrix \eqref{disk-eqi-H} should be positive definite at the equilibrium solution \eqref{disk-eqi-eq}, i.e., all the principal minors of Eq.~\eqref{disk-eqi-H} are positive subject to the equilibrium equation \eqref{disk-eqi-eq}. The $1 \times 1$ principal minors are the diagonal entries ${\bf H}_{11}$ and ${\bf H}_{22}$, and the only $2 \times 2$ principal minor is the determinant \eqref{disk-eqi-det}. Since $\varepsilon$ is greater than $\varepsilon_0$, the entry ${\bf H}_{22}$ is always positive. More importantly, compared to the entry ${\bf H}_{11}$, the determinant \eqref{disk-eqi-det} is more likely to become zero from positive for an equilibrium solution of $\lambda_Z$ in Eq.~\eqref{disk-eqi-eq}. Therefore, in the case of the homogeneous deformation of a dielectric disc, from the Hessian method the equilibrium solution of $\lambda_Z$ from Eq.~\eqref{disk-eqi-eq} becomes unstable when the determinant \eqref{disk-eqi-det} decreases to zero at the equilibrium solution $\lambda_Z$, that is $\det {\bf H} = 0$, namely 
\begin{equation} \label{disk-unstable}
 \mu(1 +2 \lambda_Z^{-3}) - \frac{3}{4} S_R \lambda_Z^ {-5/2} - 3 \varepsilon \lambda_Z^{-4}{\tilde E}_0^2 = 0, \quad ({\rm instability\ condition})
\end{equation} 
where $\lambda_Z$ is the solution of the equilibrium equation \eqref{disk-eqi-eq} for a given pair of $(S_R, \tilde E_0)$. \\

On each equilibrium curve in Fig.~\ref{disk-2}, the occurrence of pull-in instability is marked by a cross at which both the equilibrium equation \eqref{disk-eqi-eq} and the instability condition \eqref{disk-unstable} are satisfied. Through the peak is determined by the solution of both Eqs.~\eqref{disk-eqi-eq} and \eqref{disk-unstable}, it can be understood from the conventional extremization problem of a scalar function with a scalar variable. \\

Again, Eqs.~\eqref{disk-eqi-eq} and \eqref{disk-unstable} together define the threshold (either the critical electric field ${\tilde E}_0^c$ or the critical dead load) of pull-in instability. For example, for a given dead load $S_R$, the solution of the equilibrium equation \eqref{disk-eqi-eq} and the instability condition \eqref{disk-unstable} is $\lambda_Z$ and ${\tilde E}_0^c$. For any electric field with a magnitude smaller than ${\tilde E}_0^c$, there is no pull-in instability.\\

\begin{figure}[h] %
\centering
\subfigure[]{%
\includegraphics[width=2.8in]{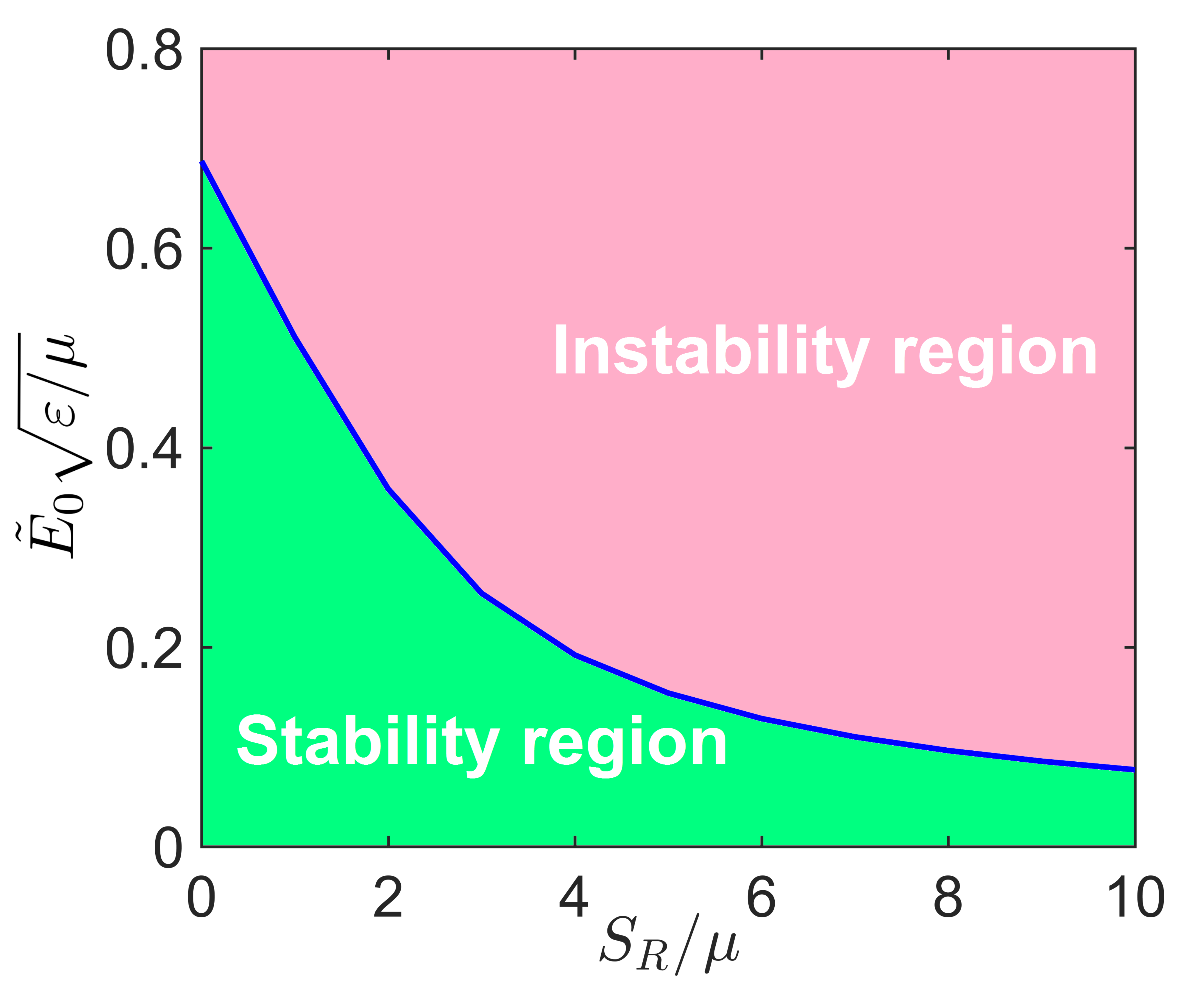}
\label{disk-3a}}
\subfigure[]{%
\includegraphics[width=2.8in]{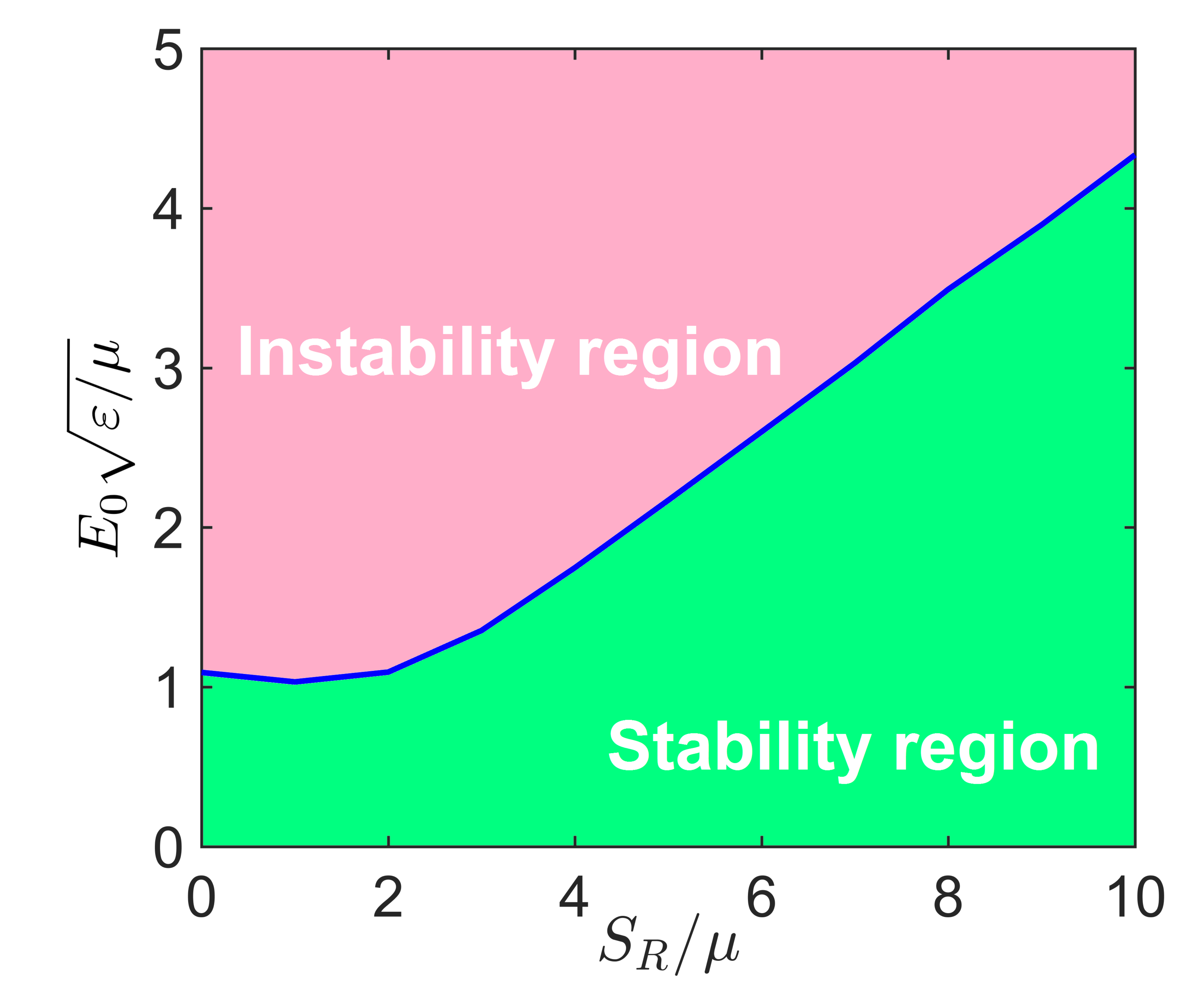}
\label{disk-3b}}
\caption{The stability and instability regions in the dead load - the electric field plane. (a) The dead load $S_R$ {\it vs}. the nominal electric field $\tilde E_0$. (b) The dead load $S_R$ {\it vs}. the true electric field $E_0$. The critical true electric field of pull-in instability is essential for the analysis of electric breakdown of the dielectric disc.}
\label{disk-3}
\end{figure}

Figure \ref{disk-3} plots the stability and instability regions in the dead load - the electric field plane. The stability and instability regions are separated by a blue curve, which is the solution ${\tilde E}_0^c$ of the equilibrium equation \eqref{disk-eqi-eq} and the instability condition \eqref{disk-unstable} for an applied dead load $S_R$. For a given dead load $S_R$, if the applied electric field is less than ${\tilde E}_0^c$, there is no pull-in instability and the dielectric disc is stable. When the electric field increases to ${\tilde E}_0^c$ at which the instability condition \eqref{disk-unstable} is satisfied, pull-in instability occurs. Then the dielectric disc expands its area rapidly, and finally, electric breakdown occurs at a relatively high true electric field. \\

\begin{figure}[h] %
\centering
\subfigure[]{%
\includegraphics[width=2.8in]{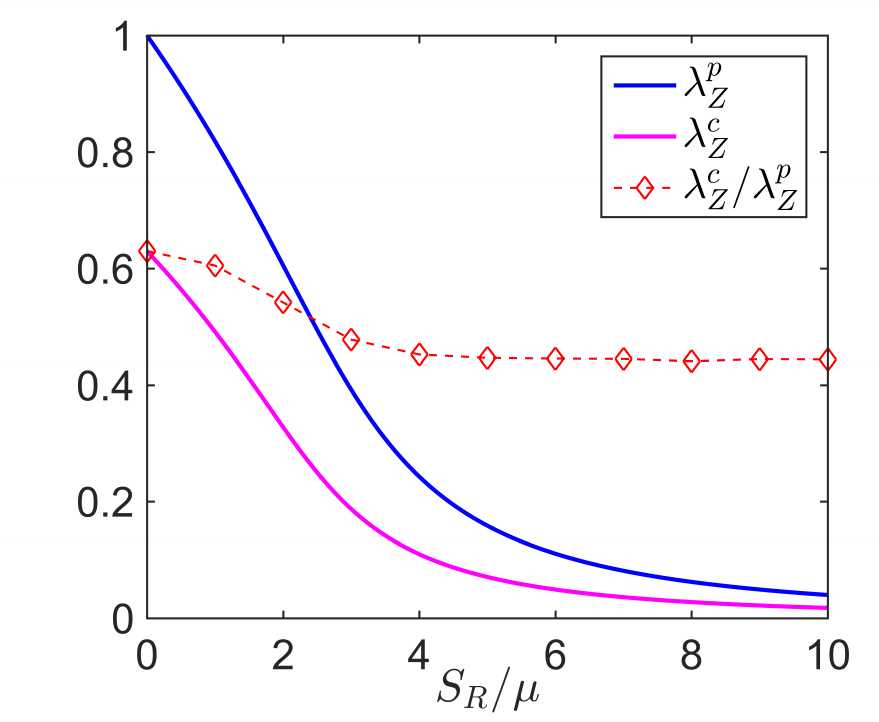}
\label{disk-4a}}
\subfigure[]{%
\includegraphics[width=2.8in]{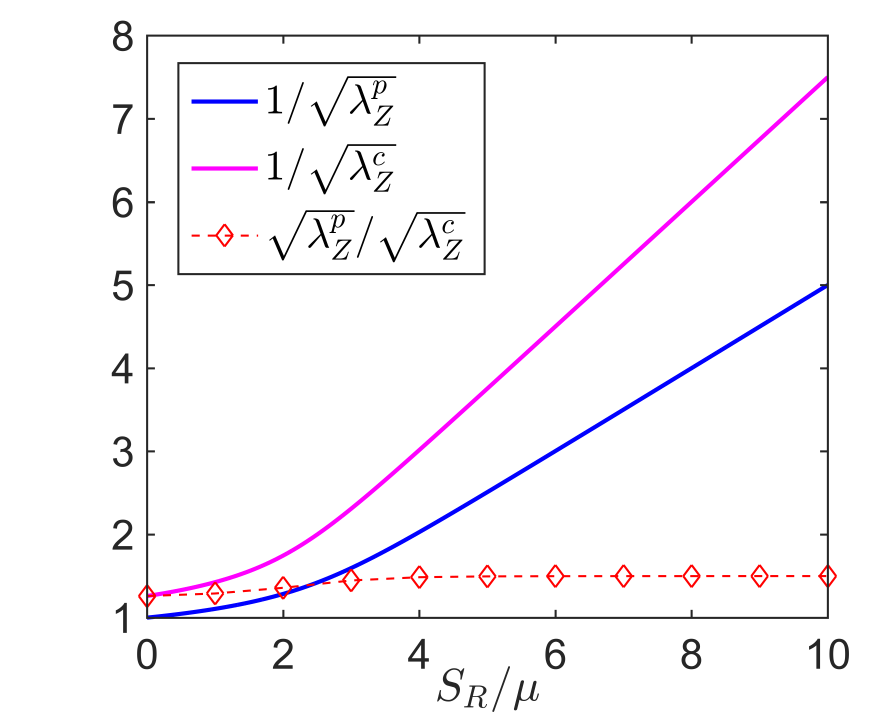}
\label{disk-4b}}
\caption{Effects of the surrounding force $S_R/\mu$ on (a) the prestretch $\lambda_Z^p$, the critical stretch $\lambda_Z^c$ and the actuation stretch $\lambda_Z^c / \lambda_Z^p$ in the thickness direction and (b) the prestretch $1/\sqrt{\lambda_Z^p}$, the critical stretch $1/\sqrt{\lambda_Z^c}$ and the actuation stretch $\sqrt{\lambda_Z^p} / \sqrt{\lambda_Z^c}$ in the radial direction. }
\label{disk-4}
\end{figure}

The actuation stretch problem at equilibrium has been discussed in the dielectric film problem in Fig.~\ref{film-2b}. A natural question to ask here is how large can the actuation stretch be achieved before the occurrence of pull-in instability. In Fig.~\ref{film-2b}, we know that the maximum actuation stretch is obtained at the peak point (the cross) of each curve. In this dielectric disc problem, we plot the variation of the maximum actuation stretch with the increase of the dead load in Fig.~\ref{disk-4}. Also, the prestretch and the total stretch are plotted for direct comparison. \\

In Fig.~\ref{disk-4a}, we plot the maximum actuation stretch $\lambda_Z^c / \lambda_Z^p$ in the thickness direction ($Z-$direction). Since both the dead load and the electric field decrease the thickness of the disc, the prestretch $\lambda_Z^p$, the critical stretch $\lambda_Z^c$, and the maximum actuation stretch $\lambda_Z^c / \lambda_Z^p$ are all less than 1. With the increase of the dead load, both the prestretch $\lambda_Z^p$ and the critical stretch $\lambda_Z^c$ decrease rapidly to zero. However, the maximum actuation stretch $\lambda_Z^c / \lambda_Z^p$ decreases slowly and then reaches a plateau as the dead load increases further. \\

In contrast, we plot the maximum actuation stretch $\sqrt{\lambda_Z^p} / \sqrt{\lambda_Z^c}$ in the radial direction in Fig.~\ref{disk-4b}. Both the prestretch $1/\sqrt{\lambda_Z^p}$ and the critical stretch $1/\sqrt{\lambda_Z^c}$ increase monotonically with the increase of the dead load. However, the maximum actuation stretch is almost constant around the value 1.5. This suggests a useful guideline for the design of a dielectric disc actuator i.e., the maximum actuation stretch is insensitive to the applied dead load.

\subsubsection{Pull-in instability {\it vs}. electric breakdown}%

\begin{figure}[h] %
\centering
\includegraphics[width=4in]{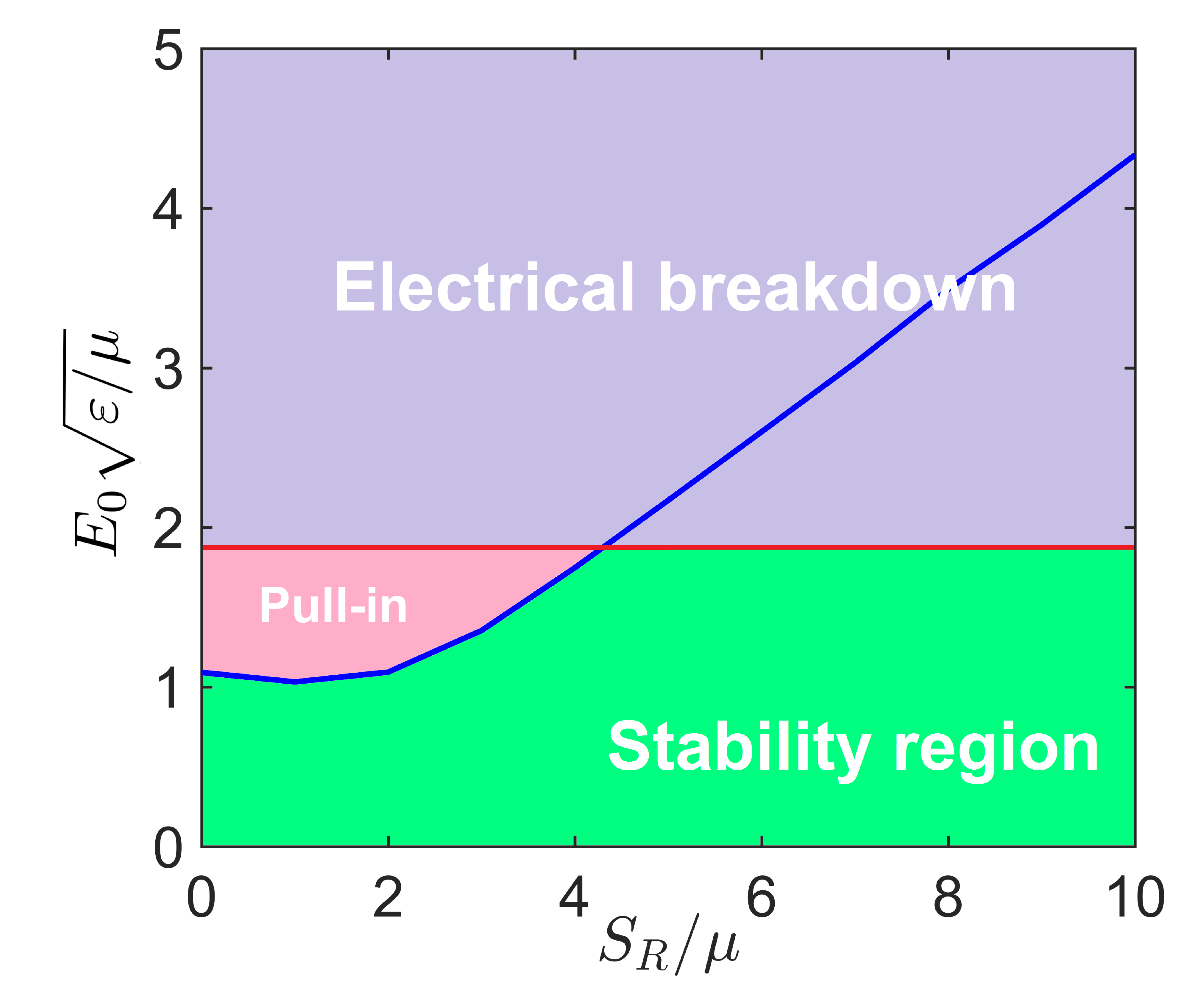}
\caption{The competition between pull-in instability and electric breakdown in the dead load - the true electric field plane. }
\label{disk-5}
\end{figure}

Electric breakdown occurs in the dielectric elastomer when the true electric field reaches a relatively high value \cite{pelrine2000high, koh2009maximal, yang2017avoiding}. Consider a true electric field $E_{\rm EB} = 3 \times 10^8 \ {\rm V \ m}^{-1}$ at which electric breakdown (EB) occurs. Other material parameters used in the numerical calculations
are $\mu = 10^6 \ {\rm N \ m}^{-2}$ and $\varepsilon = 3.54 \times 10^{-11} \ {\rm F \ m}^{-1}$. Then the dimensionless true electric field is given by $E_{\rm EB} \sqrt{\varepsilon/\mu} = 1.785$. Thus, the stability and (pull-in) instability regions in Fig.~\ref{disk-3} are altered.   \\

Since $E_{\rm EB} \sqrt{\varepsilon/\mu} = 1.785$ is a constant,\footnote{The true electric field for the onset of electric breakdown may slightly vary with the deformation of a dielectric elastomer. The constant true electric field $E_{\rm EB}$ is just an approximation used here for the illustration of the competition between pull-in instability and electric breakdown.} the threshold of electric breakdown is just a horizontal line, i.e., the true electric field $E_0 \sqrt{\varepsilon/\mu} = 1.785$ in Fig.~\ref{disk-5}. Above the horizontal line, electric breakdown would occur in the dielectric disc. In addition, below the horizontal line, there is no electric breakdown, and the disc can be either stable or vulnerable to pull-in instability. Thus, there exists a competition between electric breakdown and pull-in instability. This is graphically depicted in Fig.~\ref{disk-5}.

\section{Electromechanical cavitation and the snap-through instability in soft hollow dielectric spheres} \label{section-cavitation}

In this section, we study the nonlinear electromechanical behavior of soft hollow dielectric spheres. For purely mechanical cases, the problem can be reduced to the well-studied phenomenon of cavitation in nonlinear elasticity studied by Ball \cite{ball1982discontinuous}. Electro-cavitation in a sphere of dielectric elastomers is recently studied by Chen and Yang \cite{CHEN2022109995}. Another related problem is the inflation of a soft balloon within the framework of nonlinear elasticity. The inflation process accompanies the snap-through instability that can be harnessed to achieve giant voltage-triggered deformation of dielectric balloons \cite{keplinger2012harnessing, li2013giant}. Nonlinear coupling problem in spherical shells of dielectrics can also be found in other works \cite{dorfmann2006nonlinear, he2011characteristics, dorfmann2014nonlinear-response, xie2016bifurcation, liang2017new, wang2017anomalous, CHEN2022109995}, for example, the snap-through instability of a thick-wall electro-active balloon \cite{rudykh2012snap} and multilayer electro-active spherical balloons \cite{bortot2017analysis}.

\subsection{Formulation }%
Consider a spherical dielectric shell with thickness $H$ and inner radius $R_a$ and outer radius $R_b$. Spherical coordinates $(R, \varTheta, \varPhi)$ with corresponding local base unit vectors $({\bf e}_R, {\bf e}_\varTheta, {\bf e}_\varPhi)$ will be employed. The domain occupied by the undeformed spherical dielectric (see Fig.~\ref{ball-0}) is represented by
\begin{equation} \label{shell}
\Omega_R = \left\{ (R, \varTheta, \varPhi) \in \mathbb{R}^3: R_a \le R \le R_b, 0 \le \varTheta < \pi, 0 \le \varPhi < 2\pi \right\}.
\end{equation}
The boundary of $\Omega_R$ is denoted by $\partial \Omega_R$ consisting of the inner and outer surfaces:
\begin{equation} \label{inner_outer_surfaces}%
\partial \Omega_{R_a} = \left\{ {\bf X} \in \Omega_R: |{\bf X}| = R_a \right\} \quad {\rm and} \quad
\partial \Omega_{R_b} = \left\{ {\bf X} \in \Omega_R: |{\bf X}| = R_b \right\}.
\end{equation}

\begin{figure}[t] 
\centering
\subfigure[]{%
\includegraphics[width=2.5in]{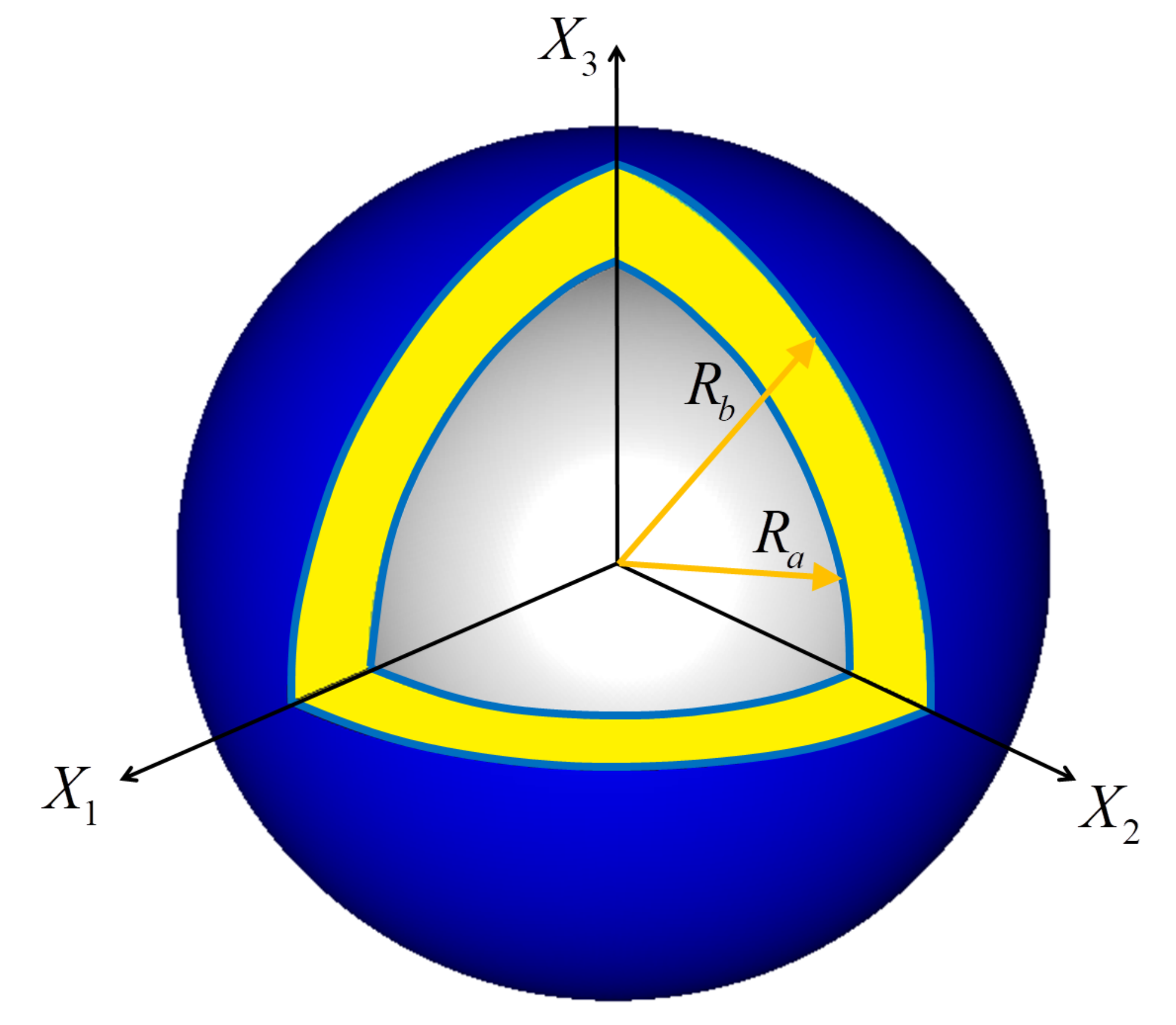}
\label{ball-0}}
\\
\subfigure[]{%
\includegraphics[width=2.5in]{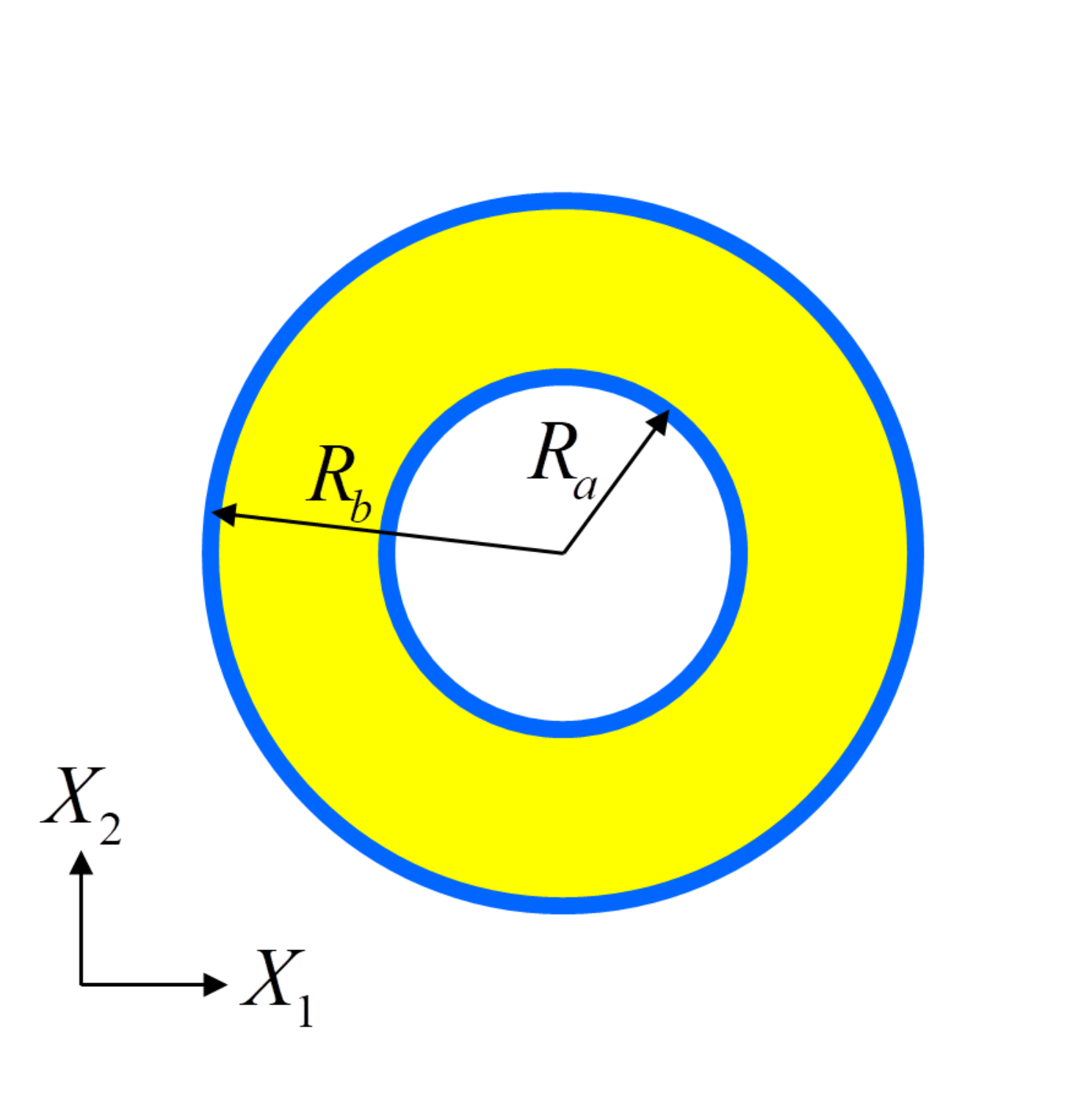}
\label{ball-0a}}
\subfigure[]{%
\includegraphics[width=2.5in]{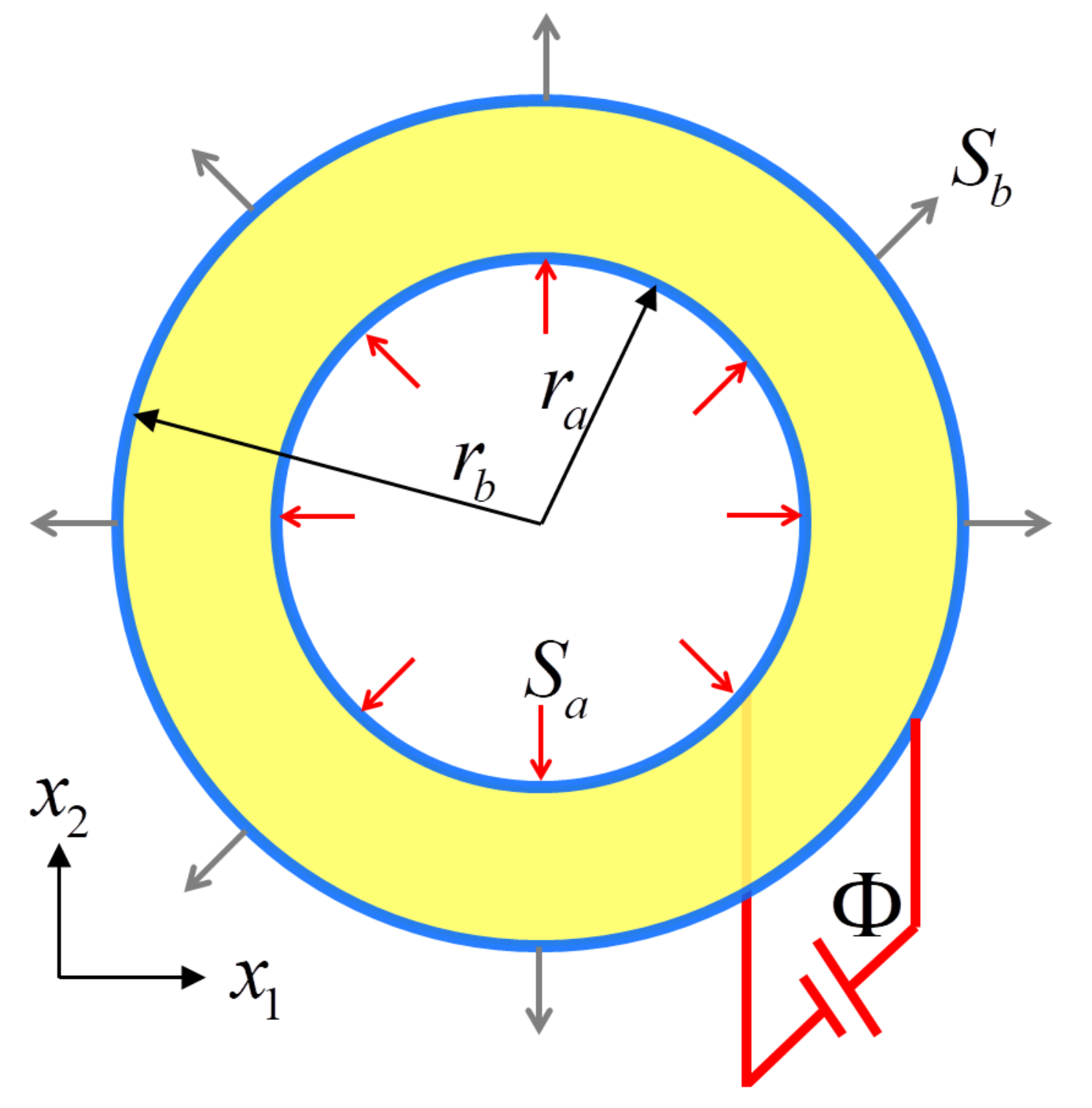}
\label{ball-0b}}
\caption{Schematic of the deformation of a hollow sphere of soft dielectrics subjected to the dead loads $S_a$ and $S_b$ on the inner and outer surfaces, and an applied voltage $\Phi$. The inner and outer surfaces are coated with compliant electrodes. (a) 3D view of an undeformed hollow sphere with inner radius $R_a$ and outer radius $R_b$. (b) 2D view of an undeformed hollow sphere. (c) 2D view of a deformed hollow sphere with inner radius $r_a$ and outer radius $r_b$.}
\label{ball-0}
\end{figure}

\subsubsection{Energy of the system}
The free energy of the system consists of two parts \cite{liu2014energy, yang2017revisiting}:
\begin{equation} \label{free_energy}
{\mathcal F}(\chi, {\tilde {\mathbf P}}) = {\mathcal U}^{\rm mech} (\chi) + {\mathcal E}^{\rm elct} (\chi, {\tilde {\mathbf P}}).
\end{equation}
The term ${\mathcal U}^{\rm mech} (\chi)$ is the mechanical part 
\begin{equation} \label{mech_energy}
{\mathcal U}^{\rm mech} (\chi) = \int_{\Omega_R} W^e (\nabla \chi) - \int_{\partial \Omega_R} {\bf t}_R \cdot \chi
\end{equation}
while the term ${\mathcal E}^{\rm elct} (\chi, {\tilde {\mathbf P}})$ is the electric part, for linear dielectrics, it can be expressed as
\begin{equation} \label{elect_energy}
{\mathcal E}^{\rm elct} (\chi, {\tilde {\mathbf P}}) = \frac{\varepsilon_0}{2}\int_{\Omega_R} J \big|{\bf F}^{-T}\nabla \xi \big|^2 + \int_{\partial \Omega_R} \xi \Big( - \varepsilon_0 J {\bf C}^{-1} \nabla \xi + {\bf F}^{-1} {\tilde {\bf P}} \Big)\cdot {\bf N} + \int_{\Omega_R} \frac{|{\tilde {\mathbf P}}|^2}{2 J (\varepsilon -\varepsilon_0)},
\end{equation}
where ${\bf N}$ is the outer unit normal to $\partial \Omega_R$. The Maxwell equation is
\begin{equation} \label{MW_eq}
{\rm Div} \Big( - \varepsilon_0 J {\bf C}^{-1} \nabla \xi + {\bf F}^{-1} {\tilde {\bf P}} \Big)=0 \quad {\rm in} \ \Omega_R
\end{equation}
and the electric boundary conditions are
\begin{equation} \label{Elec_BC}
\xi = \xi_a \quad {\rm on} \ \partial\Omega_{R_a} \quad {\rm and} \quad \xi = \xi_b \quad {\rm on} \ \partial\Omega_{R_b}.
\end{equation}

\subsubsection{First variation of the energy functional}
Since the deformation gradient ${\mathbf F} = \nabla \chi$ and the polarization ${\tilde {\mathbf P}}$ are independent, the first variation of the free energy \eqref{free_energy} with respect to ${\tilde {\mathbf P}}$ and ${\mathbf F}$, respectively, gives (see the Appendix in \cite{yang2017revisiting} for details)
\begin{equation} 
\label{P_phi_rela}
 \tilde {\bf P} = - (\varepsilon - \varepsilon_0) J \mathbf{F}^{-T} \nabla \xi, \quad  {\rm Div} \ {\bf T} = {\bf 0} \qquad {\rm in} \ \Omega_R, 
\end{equation}
\begin{equation}
\label{Nat_BC}
 {\bf T} {\bf N} = {\bf t}_R \ \quad {\rm on} \ \partial \Omega_R,
\end{equation}
where the total nominal stress ${\bf T}$ is
\begin{equation} \label{T_stress}
{\bf T} = \frac{\partial W^e}{\partial {\bf F}} + {\bf T}^M - \mathcal L_a {\bf F}^{-T}.
\end{equation}
Here $\mathcal L_a$ serves as the Lagrange multiplier, ${\bf T}^M$ denotes the Maxwell stress
\begin{equation} \label{MW_stress}
{\bf T}^M = \frac{1}{\varepsilon J} ({\bf F} {\tilde {\bf D}}) \otimes {\tilde {\bf D}} - \frac{1}{2 \varepsilon J} |{\bf F} {\tilde {\bf D}}|^2 {\bf F}^{-T},
\end{equation}
and %
\begin{equation} \label{nom_displace}
{\tilde {\bf D}} = - \varepsilon J {\bf C}^{-1} \nabla \xi. 
\end{equation}

Note that the total nominal stress ${\bf T}$ and the Maxwell stress ${\bf T}^M$ can also be found in Eqs.~\eqref{suo-08-T-E-3a} and \eqref{def-MW-E-nominal-zhao}, respectively. The polarization ${\tilde {\mathbf P}}$ and the electric displacement ${\tilde {\bf D}}$ are consistent with Eqs.~\eqref{PE-rela-refer} and \eqref{elec-cons-ref} for linear dielectrics. Now we have formed a boundary-value problem $\eqref{MW_eq}-\eqref{Nat_BC}$.

\subsubsection{Spherical deformation}
A deformation ${\bf x} = \chi ({\bf X})$ is spherical if it has the following component form
\begin{equation} \label{S_def}
r = r (R), \quad \vartheta = \varTheta, \quad \varphi = \varPhi,
\end{equation}
where $r$, $\vartheta$ and $\varphi$ are the spherical coordinates. The spherical deformation implies that the hollow sphere also admits a spherical shape after the deformation (see Fig.~\ref{ball-0}). \\

In the spherical coordinates, the deformation gradient of the spherical
deformation \eqref{S_def} has the following component form
\begin{equation} \label{DF_S_def}
{\mathbf F} :=  {\rm diag} \, (\lambda_1, \lambda_2, \lambda_3),
\end{equation}
where 
\begin{equation} \label{DF_S_def-9-10-1}
\lambda_1 =r' , \quad \lambda_2=\lambda_3=\frac{r}{R}.
\end{equation}
In this section, a prime denotes the derivative with respect to $R$. Consequently, 
\begin{equation} \label{Con_DF}
{\mathbf F}^{-1} = {\mathbf F}^{-T} :=  {\rm diag} \, (\lambda_1^{-1}, \lambda_2^{-1}, \lambda_3^{-1}), \quad
{\mathbf C}^{-1} :=  {\rm diag} \, (\lambda_1^{-2}, \lambda_2^{-2}, \lambda_3^{-2}).
\end{equation}

In the spherical deformation of a hollow dielectric sphere with an applied voltage on the inner and outer surfaces, the electric potential $\xi$ is only a function of $R$, i.e., $\xi = \xi (R)$. And the boundary conditions \eqref{Elec_BC} can be expressed as $\xi_a = \xi (R_a)$ and $\xi_b = \xi (R_b)$. Hence, the gradient of the potential in the spherical coordinates in the reference configuration reads
\begin{equation} \label{diff_phi}
\nabla \xi (R) = \xi' {\bf e}_{R}.
\end{equation}

With Eqs.~\eqref{Con_DF} and \eqref{diff_phi}, the nominal electric displacement ${\tilde {\bf D}}$ in Eq.~\eqref{nom_displace}, the Maxwell equation \eqref{MW_eq}, and the Maxwell stress ${\bf T}^M$ in Eq.~\eqref{MW_stress} become
\begin{equation} \label{re_nom_displace}
{\tilde {\bf D}} = - \varepsilon J \lambda_1^{-2} \xi' {\bf e}_{R},
\end{equation}
\begin{equation} \label{sim_MW_eq}
\varepsilon J \Big( (\lambda_1^{-2} \xi')' + \frac{2 \lambda_1^{-2} \xi'}{R} \Big) = 0,
\end{equation}
and 
\begin{equation} \label{re_MW_stress}
{\bf T}^M := 
\frac{\varepsilon J \lambda_1^{-2} (\xi')^2}{2} {\rm diag} \, (\lambda_1^{-1}, - \lambda_2^{-1}, -\lambda_3^{-1}).
\end{equation}

Invoking isotropy, the strain-energy function $W^e$ and the tensor $\displaystyle\frac{\partial W^e}{\partial {\bf F}}$ depend on ${\bf F}=\nabla \chi$ through the principal stretches $\lambda_1$, $\lambda_2$ and $\lambda_3$ (see Eqs.~\eqref{elastic-F-stretch-9-11-1} and \eqref{elastic-PK-s-stretch}). The total nominal stress \eqref{T_stress} is represented by
\begin{equation} \label{SD_T_stress}
{\bf T} := {\rm diag} \, (T_{11}, T_{22}, T_{33}),
\end{equation}
where
\begin{equation} \label{SD_T_stress-9-11-1}
T_{ii} = \frac{\partial W^e}{\partial \lambda_i} - \mathcal L_a \lambda_i^{-1} + T_{ii}^M, \quad i = 1,2,3 \ (\text {no sum}).
\end{equation}

By Eq.~\eqref{SD_T_stress}, the equilibrium equation $\eqref{P_phi_rela}_2$ gives the only non-trivial equation 
\begin{equation} \label{SD_EL_1}
\frac{\partial T_{11}}{\partial R} + \frac{2 T_{11}}{R} - \frac{T_{22} + T_{33}}{R} = 0,
\end{equation}
and the natural boundary conditions \eqref{Nat_BC} read
\begin{equation} \label{SD_EL_1BC}
T_{11} = S_a \quad {\rm at} \ R=R_a \quad {\rm and} \quad 
T_{11} = S_b \quad {\rm at} \ R=R_b.
\end{equation}
Note that $\lambda_2 = \lambda_3$ in Eq.~\eqref{DF_S_def} and then $T_{22} = T_{33}$ in Eq.~\eqref{SD_EL_1}.
\subsubsection{Ideal dielectric elastomer}
By the strain-energy function of incompressible neo-Hookean solids in Eq.~\eqref{ex-neo-H-strain-energy}, Eq.~\eqref{SD_T_stress-9-11-1} becomes
\begin{equation} \label{SD_T_stress--neo}
T_{ii} = \mu \lambda_i - \mathcal L_a \lambda_i^{-1} + T_{ii}^M, \quad i = 1,2,3 \ (\text {no sum}),
\end{equation}
and the unit Jacobian $J = \det {\bf F} =1$ yields
\begin{equation} \label{incom_eq}
\frac{r' r^2}{R^2} = \lambda_1 \lambda_2 \lambda_3 = 1.
\end{equation}

Let us introduce a new variable
\begin{equation} \label{def-lambda}
\lambda = \lambda_2 = \lambda_3 = \frac{r (R)}{R}.
\end{equation}

It follows from Eqs.~\eqref{DF_S_def-9-10-1} and \eqref{incom_eq} that
\begin{equation} \label{eq330}
r' = \lambda_1 = \lambda_2^{-1} \lambda_3^{-1} = \lambda^{-2}.
\end{equation}

Then, the total nominal stress \eqref{SD_T_stress--neo} is reduced to
\begin{equation} \label{SD_T_stress--neo2}
T_{11}  = \Big( \mu + \frac{\varepsilon  ({\xi'})^2}{2} \lambda^8 \Big) \lambda^{-2} - \mathcal L_a \lambda^2, \quad
T_{22} = T_{33} = \Big( \mu - \frac{\varepsilon  (\xi')^2}{2} \lambda^{2} \Big) \lambda - \mathcal L_a \lambda^{-1}.
\end{equation}

The traction boundary conditions \eqref{SD_EL_1BC} become

\begin{equation} \label{SD_EL_1BC_neo}
\Big( \mu + \frac{\varepsilon  ({\xi'})^2}{2} \lambda^8 \Big) \lambda^{-2} - \mathcal L_a \lambda^2 = S_a \quad {\rm at} \quad R=R_a,
\end{equation}
\begin{equation} \label{SD_EL_2BC_neo}
\Big( \mu + \frac{\varepsilon  ({\xi'})^2}{2} \lambda^8 \Big) \lambda^{-2} - \mathcal L_a \lambda^2 = S_b \quad {\rm at} \quad R=R_b.
\end{equation}

By the chain-rule, we have
\begin{equation} \label{CP8-chain-rule-1}
\frac{\partial}{\partial R} =  \frac{\partial \lambda}{\partial R}  \frac{\partial}{\partial \lambda} = \frac{1}{R} (\lambda^{-2} - \lambda) \frac{\partial}{\partial \lambda}.
\end{equation}

By Eqs.~\eqref{CP8-chain-rule-1}, \eqref{sim_MW_eq} and \eqref{SD_T_stress--neo2}, the equilibrium equation \eqref{SD_EL_1} can be reduced to
\begin{equation} \label{re-eq-eq3}
2 \mu (1- \lambda^{-3}) - \lambda^2  \frac{\partial \mathcal L_a}{\partial \lambda} = 0.
\end{equation}

\subsubsection{Solution of the boundary-value problem}
We now have a boundary-value problem consisting of the equilibrium equation \eqref{re-eq-eq3}, the traction boundary conditions \eqref{SD_EL_1BC_neo} and \eqref{SD_EL_2BC_neo}, and the electric boundary conditions \eqref{Elec_BC}. Integration of Eq.~\eqref{re-eq-eq3} with respect to $\lambda$ gives
\begin{equation} \label{sol-BVP-1}
\mathcal L_a = 2 \mu \Big( \frac{1}{4} \lambda^{-4} - \lambda^{-1} \Big) + C_{cons},
\end{equation}
where $C_{cons}$ is an integral constant. By Eq.~\eqref{incom_eq}, the integration of Eq.~\eqref{sim_MW_eq} gives 
\begin{equation}\label{sim_diff_phi}
\xi' = C_0 \frac{r'}{r^2}, \quad
\xi (R) = - \frac{C_0}{r(R)} + C_1,
\end{equation}
where $C_0$ and $C_1$ are the integration constants. With the electric boundary conditions $\xi_a = \xi (R_a)$ and $\xi_b = \xi (R_b)$, we have 
\begin{equation} \label{sol-BVP-6}
C_0 = \frac{r_a r_b}{r_b - r_a} (\xi_b - \xi_a) = \frac{r_a r_b}{r_b - r_a} \Phi, \quad C_1 = \frac{r_b \xi_b - r_a \xi_a}{r_b - r_a}.
\end{equation}

With Eqs.~\eqref{incom_eq} and \eqref{eq330}, $\xi'$ in Eq.~$\eqref{sim_diff_phi}_1$ can be recast as $\xi' = C_0 \lambda^{-4}/ R^{2}$. By Eq.~\eqref{sol-BVP-1}, the boundary conditions \eqref{SD_EL_1BC_neo} and \eqref{SD_EL_2BC_neo} are equal to
\begin{equation} \label{sol-BVP-9a}
\frac{1}{2}\mu \lambda_a^{-4} + 2 \mu \lambda_a^{-1} - C_{cons} + \frac{\varepsilon  C_0^2}{2 R_a^4} \lambda_a^{-4} = S_a \lambda_a^{-2},
\end{equation}
\begin{equation} \label{sol-BVP-9b}
\frac{1}{2}\mu \lambda_b^{-4} + 2 \mu \lambda_b^{-1} - C_{cons} + \frac{\varepsilon  C_0^2}{2 R_b^4} \lambda_b^{-4} = S_b \lambda_b^{-2}.
\end{equation}

Deleting the constant $C_{cons}$ in Eqs.~\eqref{sol-BVP-9a} and \eqref{sol-BVP-9b}, we finally have the relation
\begin{equation} \label{sol-BVP-11}
\frac{1}{2}\mu (\lambda_b^{-4} - \lambda_a^{-4}) + 2 \mu (\lambda_b^{-1} - \lambda_a^{-1}) +  \frac{\varepsilon  C_0^2}{2} \Big( \frac{\lambda_b^{-4}}{R_b^4} - \frac{\lambda_a^{-4}}{R_a^4} \Big) = S_b \lambda_b^{-2} - S_a \lambda_a^{-2},
\end{equation}
where
\begin{equation} \label{sol-BVP-10}
C_0 = \frac{r_a r_b}{r_b - r_a} \Phi = \frac{\lambda_a \lambda_b R_a R_b}{\lambda_b R_b - \lambda_a R_a} \Phi, \quad \lambda_a = \frac{r_a}{R_a} \quad {\rm and} \quad \lambda_b = \frac{r_b}{R_b} = \frac{\sqrt[3] {r_a^3 + R_b^3 - R_a^3}}{R_b}.
\end{equation}
\subsection{Electromechanical cavitation and the snap-through instability}

\subsubsection{Ball's classic cavitation problem}

In Ball's cavitation problem \cite{ball1982discontinuous}, there is no electric field within the hollow sphere, that is $C_0 =0$ in Eq.~\eqref{sol-BVP-11}. Also, the traction force on the inner surface is zero, $S_a =0$. Thus, Eq.~\eqref{sol-BVP-11} reduces to
\begin{equation} \label{ball-2}
\frac{S_b}{2 \mu} = \lambda_b + \frac{1}{4} \lambda_b^{-2} - \Big( \lambda_a^{-1} + \frac{1}{4} \lambda_a^{-4}\Big) \lambda_b^{2}.
\end{equation}
With $\lambda_a$ and $\lambda_b$ in Eq.~\eqref{sol-BVP-10}, Eq.~\eqref{ball-2} is given by
\begin{equation} \label{ball-3}
\frac{S_b}{2 \mu} = \Big( \bar r_a^3 + 1 - \bar R_a^3 \Big)^{1/3} + \frac{1}{4} \Big( \bar r_a^3 + 1 - \bar R_a^3 \Big)^{-2/3} - \Big( \bar r_a^{-1} \bar R_a + \frac{1}{4} \bar r_a^{-4} \bar R_a^4 \Big) \Big( \bar r_a^3 + 1 - \bar R_a^3 \Big)^{2/3},
\end{equation}
where the nondimensional variables are
\begin{equation} \label{ball-4}
\bar r_a = \frac{r_a}{R_b}, \quad \bar R_a = \frac{R_a}{R_b}.
\end{equation}

In the following, some limiting cases based on the geometry and the material properties of the hollow sphere are discussed. In the case of a sufficiently small hollow, i.e., $\bar R_a \to 0$, at fixed $\bar r_a$, the third term on the RHS of Eq.~\eqref{ball-3} vanishes and the other two terms give
\begin{equation} \label{ball-6}
\frac{S_b}{2\mu} \to \frac{5/4 + \bar r_a^3}{(1+  \bar r_a^3)^{2/3}},
\end{equation}
which implies $\frac{S_b}{2\mu} \to \frac{5}{4}$ as $\bar r_a \to 0$. The limiting case can be found in the numerical plots in Fig.~\ref{ball-B-a}. In Ball's paper \cite{ball1982discontinuous}, 

\begin{figure}[h] 
\centering
\subfigure[]{%
\includegraphics[width=2.9in]{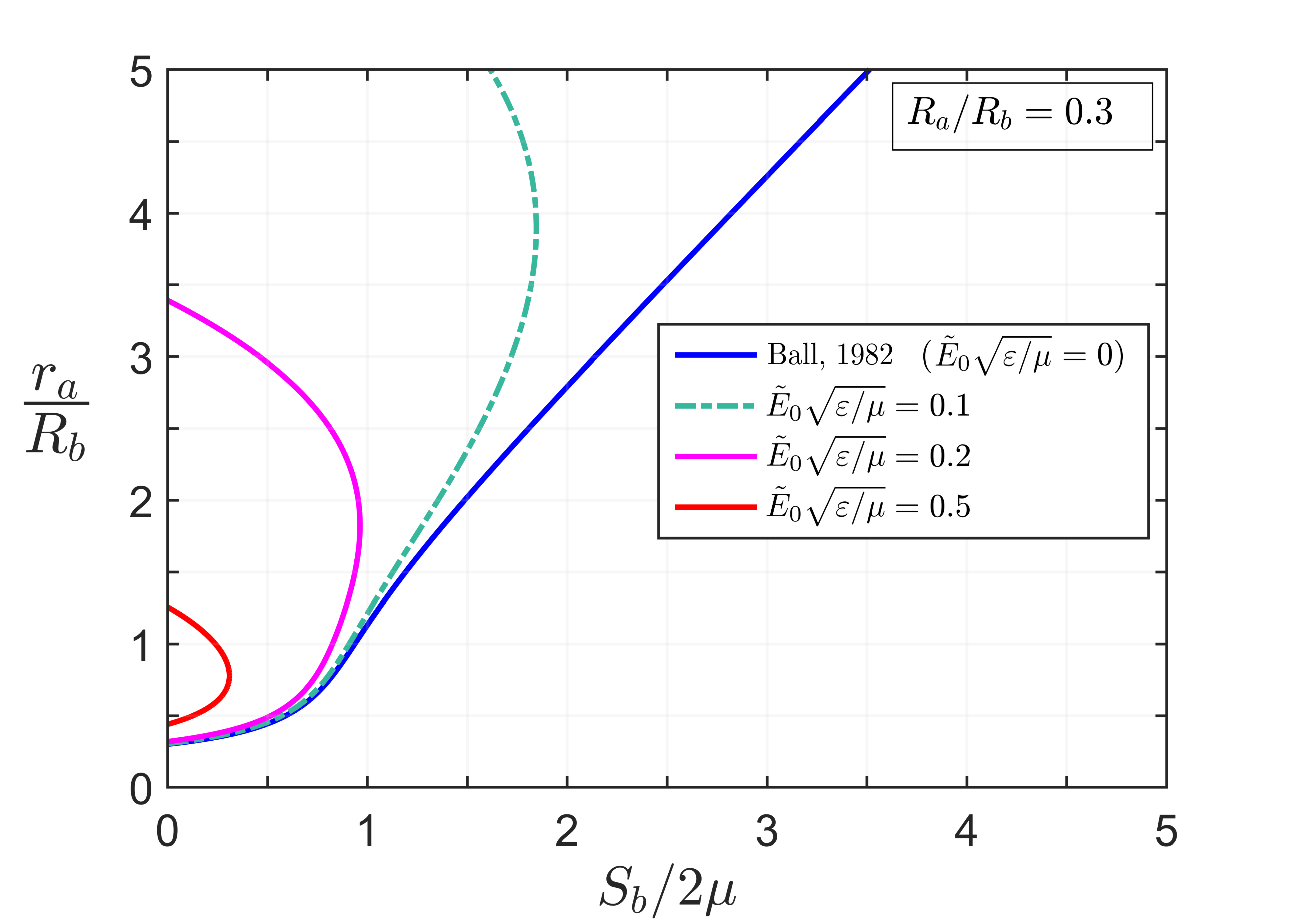}
\label{ball-A-a}}
\subfigure[]{%
\includegraphics[width=2.9in]{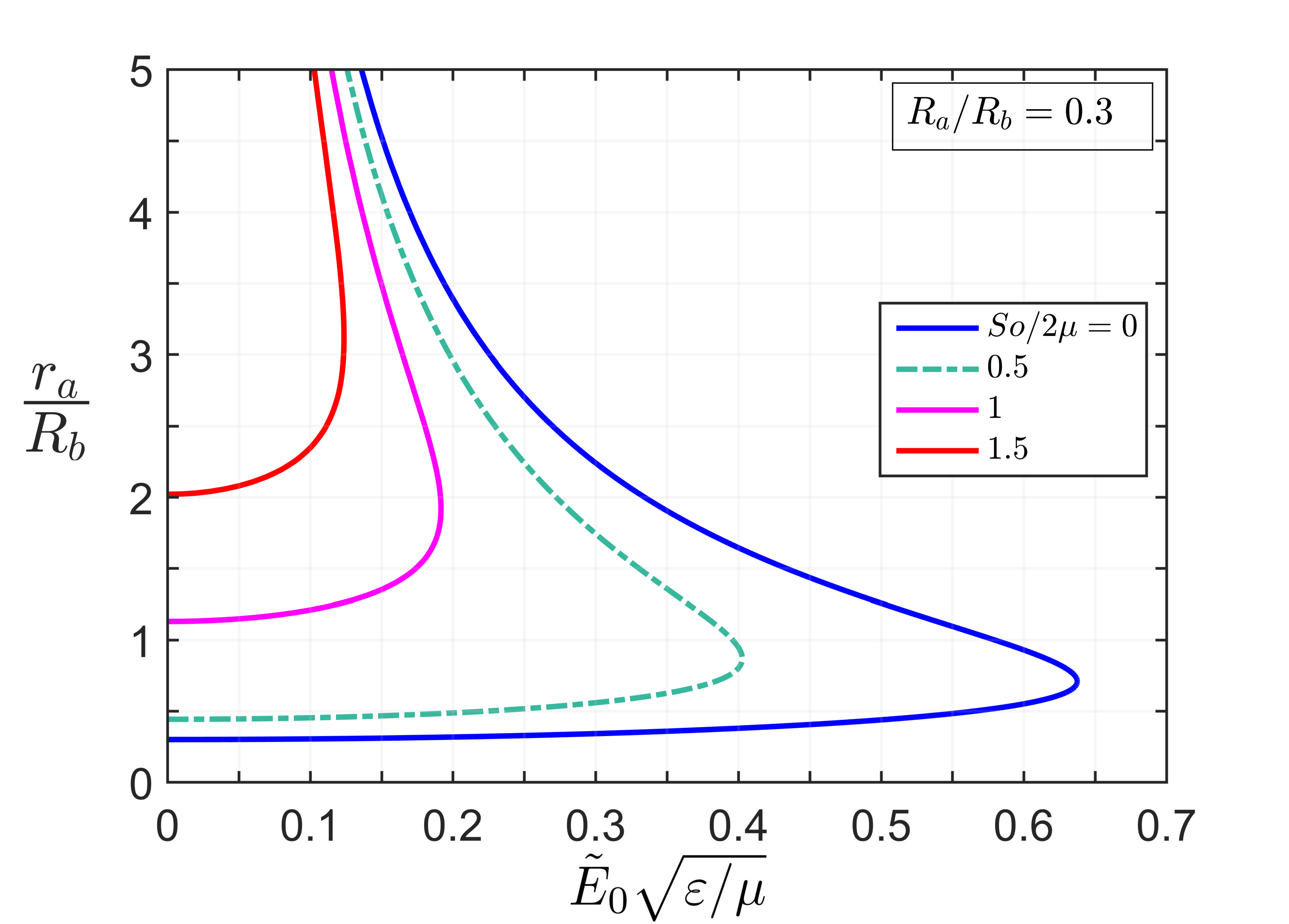}
\label{ball-A-b}}
\caption{Variation of the radius $r_a$ of the cavity with the applied stress $S_b$ on the outer surface and the applied electric field $E_0$ in the radial direction of the hollow sphere. The figures have been drawn for the ratio $R_a/R_b =0.3$. (a) Variation of the cavity radius with the applied dead under several electric fields. (b) Variation of the cavity radius with the applied electric field under several applied dead. A snap-through instability phenomenon can be observed by increasing the applied electric field to the threshold.}
\label{ball-A}
\end{figure}

\begin{figure}[h] 
\centering
\subfigure[]{%
\includegraphics[width=2.9in]{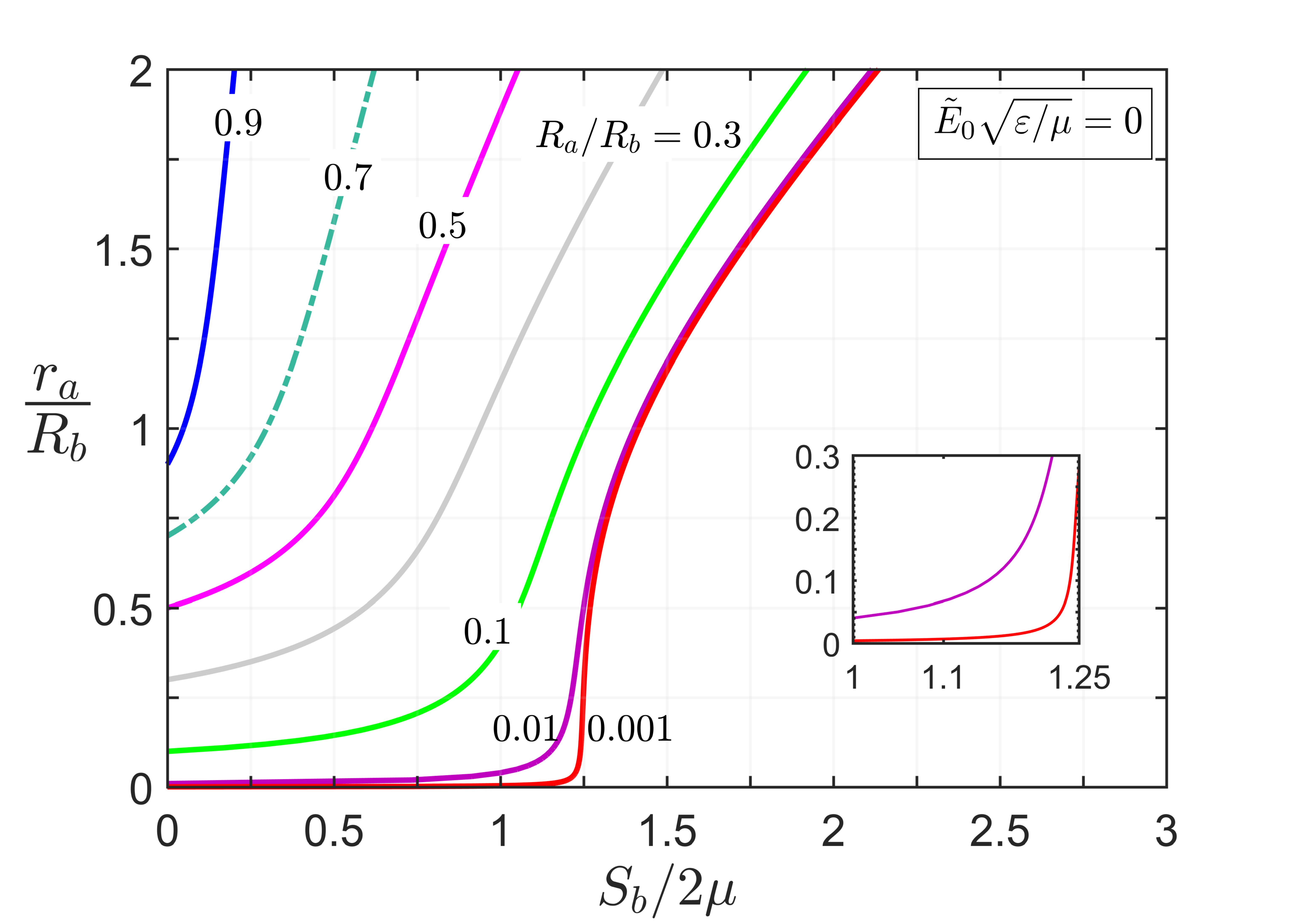}
\label{ball-B-a}}
\subfigure[]{%
\includegraphics[width=2.9in]{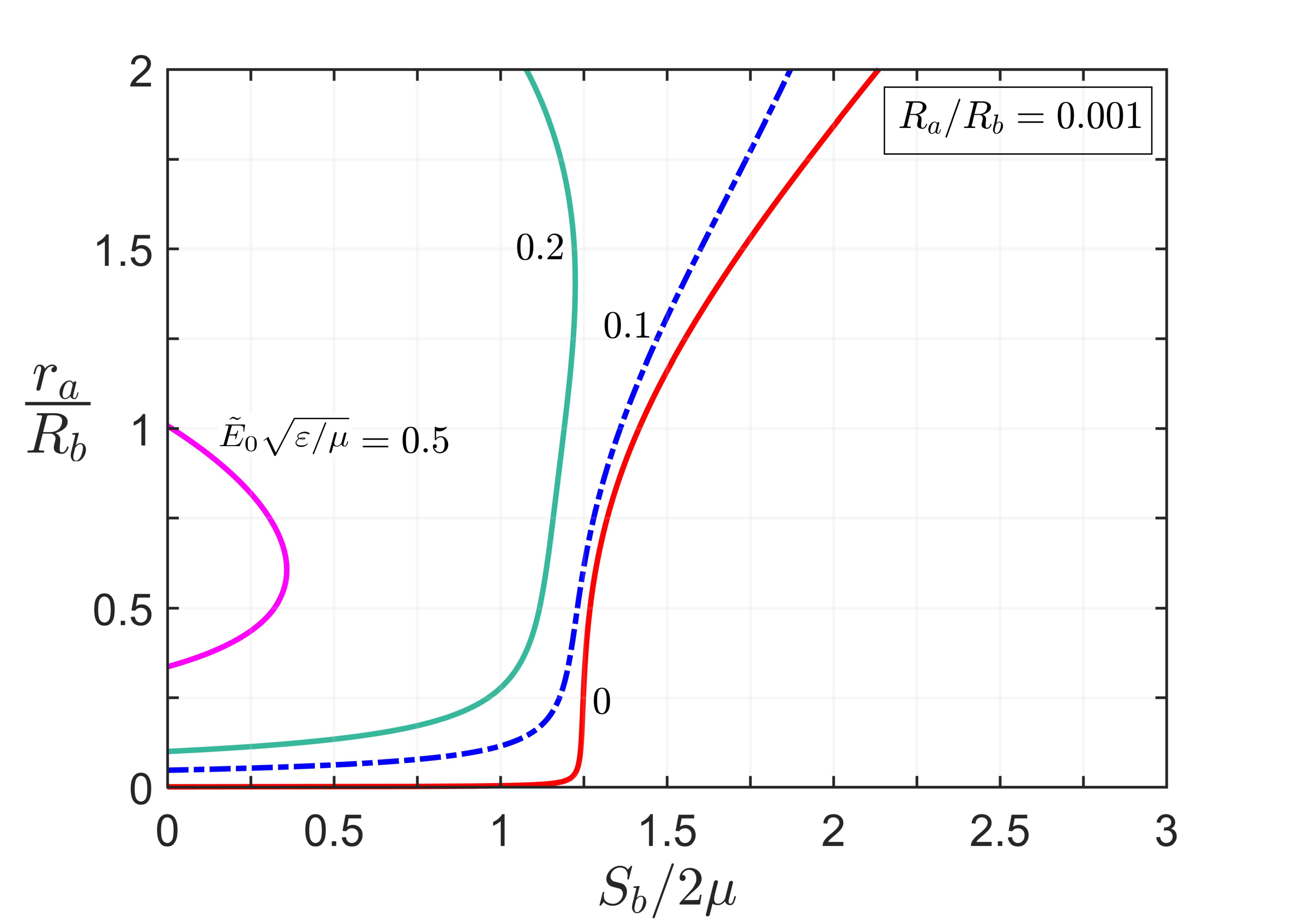}
\label{ball-B-b}}
\caption{Variation of the radius $r_a$ of the cavity with an applied stress $S_b$ on the outer surface of the hollow sphere and an applied electric field $E_0$ in the radial direction. (a) Variation of the cavity radius with the applied dead under several ratios $R_a/R_b$ without electric fields. (b) Variation of the cavity radius with the applied dead under several electric fields at a ratio $R_a/R_b = 0.001$.}
\label{ball-B}
\end{figure}

\subsubsection{Effects of the electric field on the cavitation and the snap-through instability}

Consider only the applied dead load on the outer surface and the applied voltage, i.e., $S_a = 0$ in Eq.~\eqref{sol-BVP-11}. Then, Eq.~\eqref{sol-BVP-11} reduces to
\begin{equation} \label{ball-9}
\frac{S_b}{2 \mu} = \lambda_b + \frac{1}{4} \lambda_b^{-2} - \Big( \lambda_a^{-1} + \frac{1}{4} \lambda_a^{-4}\Big) \lambda_b^{2} +  \frac{\varepsilon  C_0^2}{4 \mu} \Big( \frac{\lambda_b^{-2}}{R_b^4} - \frac{\lambda_a^{-4}  \lambda_b^{2}}{R_a^4} \Big) .
\end{equation}
By $\bar R_a$ in Eq.~\eqref{ball-4} and the nominal electric field $\tilde E_0 = \Phi/(R_b - R_a) = \Phi/H$, Eq.~\eqref{ball-9} becomes
\begin{equation} \label{ball-15}
\frac{S_b}{2 \mu} = \lambda_b + \frac{1}{4} \lambda_b^{-2} - \Big( \lambda_a^{-1} + \frac{1}{4} \lambda_a^{-4}\Big) \lambda_b^{2} + \frac{\lambda_a^{2} {\bar R_a}^2 (1 - {\bar R_a})^2 (1 - \lambda_a^{-4} \bar R_a^{-4} \lambda_b^{4})}{4 (\lambda_b - \lambda_a \bar R_a)^2} \Big( \frac{\tilde E_0}{\sqrt{\mu/\varepsilon}}\Big)^2.
\end{equation}

Thus, for a given dead load $S_b/2\mu$, an applied electric field $\tilde E_0 \sqrt{\varepsilon/\mu}$, and a given geometry of the hollow sphere $\bar R_a = R_a/ R_b$, Eq.~\eqref{ball-15} determines the deformed inner radius $\bar r_a =r_a/R_b$. The snap-through instability phenomenon can be observed by evaluating Eq.~\eqref{ball-15} numerically (see Fig.~\ref{ball-A-b}). A similar form of Eq.~\eqref{ball-15} can also be found in other works \cite{rudykh2012snap, bortot2017analysis}.

\section{Post-buckling analysis: ``Lyapunov-Schmidt-Koiter'' approach} \label{LSK-section}%
For most of the stability problems in nonlinear elasticity, the analytical solution of the stability condition is extremely difficult. Furthermore, even the existence of a solution is not assured. Beginning with the pioneering work by Koiter \cite{koiter1945}, the asymptotic expansion technique that follows the Lyapunov-Schmidt decomposition has become a powerful tool to analyze the bifurcation and the (initial) post-bifurcation of perfect or imperfect systems \cite{van2009wt, budiansky1974theory, triantafyllidis2011stability, hutchinson1970postbuckling, Casciaro2005, koiter1981elastic, simitses1986buckling, triantafyllidis1992stability, peek1993postbuckling, cedolin2010stability}. This procedure of the analysis of stability is known as the ``Lyapunov-Schmidt-Koiter'' approach. \\

In this tutorial, we will go through the basic idea of the ``Lyapunov-Schmidt-Koiter'' approach and then incorporate it into the post-buckling analysis of a dielectric elastomer by simple examples. Through the ``Lyapunov-Schmidt-Koiter'' approach is fruitfully applied to study the initial post-bifurcation stage and investigate the purely mechanical instability, only a few works \cite{pampolini2013continuum} have reported the use of the LSK approach in a post-bifurcated solution of the electromechanical coupling problem. \\

We just consider the case of only one generalized state variable $u$ of a 1D system $\Omega^1 = \{x \in \mathbb R: 0 \le x \le l \}$, i.e., $u \in \mathcal C^2 ([0, l]; \mathbb R^n)$. For example, $u$ can be the displacement of deformable solid.\footnote{In electrostatics of deformable media, there are always two state variables: one corresponds to the deformation and the other corresponds to the electric quantity (the electric displacement \cite{suo2008nonlinear}, the electric field \cite{dorfmann2005nonlinear} or the polarization \cite{liu2014energy}). However, the central idea of the LSK approach for multiple state variables is the same. Accordingly, for simplicity, we illustrate with just one generalized state variable $u$.} Thus, the potential energy $\mathcal P$ of the system is a functional of $u: [0, l] \to \mathbb R^n$. In addition, $\mathcal P$ depends on the load parameter applied to the system. Here we assume the load parameter is a single scalar variable $\lambda \in \mathbb R$ that corresponds to the work done by the external load. The potential energy of the conservative system can be written as
\begin{equation} \label{PBA-1-1}
{\mathcal P} [u; \lambda].
\end{equation}
At a given load parameter $\lambda \in \mathbb R$, an equilibrium state $u: [0, l] \to \mathbb R^n$ requires that 
\begin{equation} \label{PBA-1-2}
\delta {\mathcal P} [u; \lambda] = \frac{\partial {\mathcal P} [u + \tau \delta u; \lambda]}{\partial \tau} \Big|_{\tau=0} \equiv {\mathcal P}' [u; \lambda] \delta u = 0
\end{equation}
for all kinematically admissible variations $\delta u  \in \mathcal C^2 ([0, l]; \mathbb R^n)$. \\


For convenience, we introduce the following notations of the derivatives of functionals \cite{budiansky1974theory}:
\begin{equation} \label{PBA-1-2a}
\left.
\begin{aligned}
{\mathcal P}'[u; \lambda]u_1 & \equiv \left.\frac{\partial \mathcal P [u + \tau_1 u_1; \lambda]}{\partial \tau_1}\right|_{\tau_1=0}, \\
{\mathcal P}''[u; \lambda]u_1 u_1  & \equiv {\mathcal P}''[u; \lambda]u_1^2 \equiv \left.\frac{\partial^2 \mathcal P [u + \tau_1 u_1; \lambda]}{\partial^2 \tau_1}\right|_{\tau_1=0},\\
{\mathcal P}''[u; \lambda]u_1 u_2 & \equiv {\mathcal P}''[u; \lambda]u_2 u_1 \equiv \left.\frac{\partial^2 \mathcal P [u + \tau_1 u_1 + \tau_2 u_2; \lambda]}{\partial \tau_1 \partial \tau_2}\right|_{\tau_1=\tau_2=0}, \\
{\mathcal P}'''[u; \lambda]u_1 u_2 u_3 & \equiv \left.\frac{\partial^3 \mathcal P [u + \tau_1 u_1 + \tau_2 u_2 + \tau_3 u_3; \lambda]}{\partial \tau_1 \partial \tau_2 \partial \tau_3}\right|_{\tau_1=\tau_2=\tau_3=0}, \\
& \ \vdots \\
{\mathcal P}^{(n)}[u; \lambda]u_1 u_2 \cdots u_n & \equiv  \frac{\partial^{(n)} \mathcal P \Big[u + \displaystyle\sum_{i=1}^n \tau_i u_i; \lambda \Big]}{\partial \tau_1 \partial \tau_2 \cdots \partial \tau_n}\Bigg|_{\tau_1=\tau_2=\cdots=\tau_n=0},
\end{aligned}
\right\}
\end{equation}
where $u  \in \mathcal C^2 ([0, l]; \mathbb R^n)$, $u_1, u_2, u_3, \dots, u_n  \in \mathcal C^2 ([0, l]; \mathbb R^n)$, and $\tau_1, \tau_2, \dots, \tau_n, \lambda \in \mathbb R$.

\subsection{Bifurcation analysis}
We assume that there exists a principal (trivial) branch $u_0$ that varies smoothly with the load parameter $\lambda$ as the load increases from zero. The equilibrium of the principal branch $u_0$, from Eq.~\eqref{PBA-1-2}, reads
\begin{equation} \label{PBA-1-3}
\delta {\mathcal P} [u_0 (\lambda); \lambda] \equiv {\mathcal P}' [u_0 (\lambda); \lambda] \delta u = 0.
\end{equation}

As the load parameter $\lambda$ increases from zero to the threshold $\lambda_c$, there exists a bifurcated branch $u(\lambda)$ that bifurcates from the principal branch $u_0 (\lambda)$, such that
\begin{equation} \label{PBA-1-4}
u (\lambda) = u_0 (\lambda) + v (\lambda).
\end{equation}

The bifurcated branch $u (\lambda)$ intersects the trivial branch $u_0 (\lambda)$ at $\lambda_c$, namely
\begin{equation} \label{PBA-1-5}
\lim_{\lambda \to \lambda_c} v (\lambda) = 0.
\end{equation}

The bifurcation buckling mode can be defined by \cite{budiansky1974theory}
\begin{equation} \label{PBA-1-6}
u_1 = \lim_{\lambda \to \lambda_c} \frac{v (\lambda)}{\Vert v \Vert},
\end{equation}
where $\Vert \cdot \Vert$ represents a suitable norm, for example, the inner product form $\Vert u_1 \Vert =\langle u_1, u_1 \rangle ^{1/2}$. The norm of the buckling mode $u_1$ defined in Eq.~\eqref{PBA-1-6} is unity, i.e., $\Vert u_1 \Vert = 1$. In this tutorial, we only consider the case of a single buckling mode at $\lambda=\lambda_c$. A classic example of the single mode at the threshold is the buckling of Euler's column. In contrast, surface instability of a compressed elastic half-space is the case of the multiple buckling modes \cite{biot1963surface, cao2012wrinkles, chen2018surface}, and the LSK approach is used to study the initial post-bifurcation behavior of surface wrinkling of a compressed half-space \cite{cao2012wrinkles} and a neo-Hookean bilayer \cite{hutchinson2013role}, etc.\\

Similar to Eq.~\eqref{PBA-1-3}, the equilibrium equation \eqref{PBA-1-4} of the bifurcated branch is
\begin{equation} \label{PBA-1-7}
{\mathcal P}' [u_0 (\lambda) + v (\lambda); \lambda] \delta u = 0.
\end{equation}

The Taylor series expansion of Eq.~\eqref{PBA-1-7} is 
\begin{equation} \label{PBA-1-8}
{\mathcal P}' [u_0 ; \lambda] \delta u + {\mathcal P}'' [u_0 ; \lambda] v \delta u + \frac{1}{2}{\mathcal P}''' [u_0 ; \lambda] v^2 \delta u + \cdots = 0.
\end{equation}

The first term vanishes from Eq.~\eqref{PBA-1-3}. In the neighborhood of $\lambda_c$, $|\lambda - \lambda_c| \ll 1$, dividing the remaining terms in Eq.~\eqref{PBA-1-8} by $\Vert v \Vert$ and letting $\lambda \to \lambda_c$, we have the bifurcation equation
\begin{equation} \label{PBA-1-9}
{\mathcal P}'' [u_0 (\lambda_c) ; \lambda_c] u_1 \delta u = 0,
\end{equation}
where $u_1$ is the buckling mode defined in Eq.~\eqref{PBA-1-6}. \\

For convenience, we introduce the following notations:
\begin{equation} \label{PBA-1-10}
u_c \equiv u_0 (\lambda_c) \qquad {\rm and} \qquad {\mathcal P}''_c \equiv {\mathcal P}'' [u_0 (\lambda_c) ; \lambda_c],
\end{equation}
then the bifurcation equation \eqref{PBA-1-9} adopts a more compact form
\begin{equation} \label{PBA-1-11}
{\mathcal P}''_c u_1 \delta u = 0.
\end{equation}

\subsection{Initial post-buckling analysis}
Recall the principal branch $u_0 (\lambda)$ and the bifurcated branch $u_0 (\lambda) + v (\lambda)$. Let us decompose $v(\lambda)$ in the following form \cite{budiansky1974theory}:
\begin{equation} \label{PBA-2-1}
v (\lambda) = \zeta u_1 + {\tilde v} (\lambda),
\end{equation}
where $u_1$ is the buckling mode in Eq.~\eqref{PBA-1-6} and the bilinear inner product is $\langle u_1, \tilde v \rangle = 0$. Then the scalar parameter $\zeta$ can be represented by $\zeta = \langle u_1, v \rangle$, which is a measure of the buckling mode $u_1$ contained in the difference $v (\lambda) = u(\lambda) - u_0 (\lambda)$. \\

Similar to Eqs.~\eqref{PBA-1-2a} and \eqref{PBA-1-10}, we introduce the following notations for convenience:
\begin{equation} \label{PBA-2-2}
\left.
\begin{aligned}
& {\mathcal P}_0^{(n)} \equiv {\mathcal P}^{(n)} [u_0 (\lambda); \lambda], \qquad {\mathcal P}_c^{(n)} \equiv {\mathcal P}_0^{(n)} \big|_{\lambda=\lambda_c}, \\
&  \dot{\mathcal P}_0^{(n)} \equiv \frac{\partial}{\partial \lambda} {\mathcal P}_0^{(n)} \equiv \frac{\partial}{\partial \lambda} {\mathcal P}^{(n)} [u_0 (\lambda); \lambda],  \qquad  \dot{\mathcal P}_c^{(n)} \equiv \dot{\mathcal P}_0^{(n)} \big|_{\lambda=\lambda_c},\\
&  \ddot{\mathcal P}_0^{(n)} \equiv \frac{\partial^2}{\partial \lambda^2} {\mathcal P}_0^{(n)} \equiv \frac{\partial^2}{\partial \lambda^2} {\mathcal P}^{(n)} [u_0 (\lambda); \lambda], \qquad \ddot{\mathcal P}_c^{(n)} \equiv \ddot{\mathcal P}_0^{(n)} \big|_{\lambda=\lambda_c}.
\end{aligned}
\right\}
\end{equation}

Consider the Taylor series expansion of Eq.~\eqref{PBA-1-8} at $\lambda = \lambda_c$. By Eq.~\eqref{PBA-1-3}, the expansion reads
\begin{equation} \label{PBA-2-3}
\begin{aligned}
& \left\{ {\mathcal P}''_c + (\lambda -\lambda_c) \dot{\mathcal P}''_c + \frac{1}{2}(\lambda -\lambda_c)^2 \ddot{\mathcal P}''_c + \cdots \right\} v \delta u \\
& \quad + \frac{1}{2} \left\{ {\mathcal P}'''_c + (\lambda -\lambda_c) \dot{\mathcal P}'''_c + \frac{1}{2}(\lambda -\lambda_c)^2 \ddot{\mathcal P}'''_c + \cdots \right\} v^2 \delta u + \cdots = 0.
\end{aligned}
\end{equation}

The asymptotic expansions of the load parameter $\lambda$ and $\tilde v (\lambda)$ in Eq.~\eqref{PBA-2-1} for small $\zeta$ are
\begin{equation} \label{PBA-2-4}
\lambda = \lambda_c + \lambda_1 \zeta + \lambda_2 \zeta^2 + \cdots \quad {\rm and} \quad
\tilde v (\lambda) = \zeta^2 u_2 + \zeta^3 u_3 + \cdots,
\end{equation}
and the generalized state variable $u=u_0 +v=u_0+\zeta u_1 + \tilde v $ in Eq.~\eqref{PBA-1-4} becomes
\begin{equation} \label{PBA-2-5}
u = u_0 + \zeta u_1 + \zeta^2 u_2 + \zeta^3 u_3 + \cdots.
\end{equation}

Substituting the expansions Eq.~\eqref{PBA-2-4} into Eq.~\eqref{PBA-2-3}, we have
\begin{equation} \label{PBA-2-5a}
\begin{aligned}
0 & = \zeta^2 \left\{{\mathcal P}''_c u_2 + \lambda_1 \dot{\mathcal P}''_c  u_1 + \frac{1}{2}{\mathcal P}'''_c u_1^2 \right\} \delta u \\
   & \quad + \zeta^3 \left\{ 
   \begin{aligned}
   & {\mathcal P}''_c u_3 + \lambda_1 \dot {\mathcal P}''_c u_2 + \lambda_2 \dot {\mathcal P}''_c u_1 + \frac{1}{2} \lambda_1^2 \ddot {\mathcal P}''_c u_1 \\
   & + {\mathcal P}'''_c u_1 u_2 + \frac{1}{2}\lambda_1 \dot {\mathcal P}'''_c u_1^2 + \frac{1}{6} {\mathcal P}^{(4)}_c u_1^3 
   \end{aligned}
   \right\} \delta u  + o(\zeta^3).
\end{aligned}
\end{equation}

It follows that the coefficients of $\zeta^2$, $\zeta^3$,$\dots$ in Eq.~\eqref{PBA-2-5a} must vanish, by letting $\delta u = u_1$, together with the equality \eqref{PBA-1-11}, we have the expressions of $\lambda_1$, $\lambda_2$, $\dots$, such that\footnote{See Equations (4.21) and (4.22) in the work \cite{budiansky1974theory}. The sign of $\dot{\mathcal P}''_c u_1^2$ will be discussed later and the case $\dot{\mathcal P}''_c u_1^2 = 0$ is out of the scope of this tutorial. }
\begin{subequations}
\begin{equation} \label{PBA-2-6a}
\lambda_1 = -\frac{1}{2}{\mathcal P}'''_c u_1^3 \Big/ \dot{\mathcal P}''_c u_1^2,
\end{equation}
\begin{equation} \label{PBA-2-6b}
\lambda_2 = - \left( \frac{1}{6}{\mathcal P}^{(4)}_c u_1^4 + {\mathcal P}'''_c u_1^2 u_2 + \lambda_1 [\dot {\mathcal P}''_c u_2 u_1 + \frac{1}{2} \dot{\mathcal P}'''_c u_1^3] + \frac{1}{2}\lambda_1^2 \ddot{\mathcal P}''_c u_1^2\right)\Big/ \dot{\mathcal P}''_c u_1^2.
\end{equation}
\end{subequations}

In particular, for the symmetric bifurcation in many (possibly most) buckling problems \cite{hutchinson1970postbuckling, budiansky1974theory}, the coefficient $\lambda_1=0$ and $\lambda_2$ in Eq.~\eqref{PBA-2-6b} reduces to 
\begin{equation} \label{PBA-2-7}
\lambda_2 = - \left( \frac{1}{6}{\mathcal P}^{(4)}_c u_1^4 + {\mathcal P}'''_c u_1^2 u_2 \right)\Big/ \dot{\mathcal P}''_c u_1^2.
\end{equation}

\subsection{Stability analysis}
The difference between the energies of the state $u+v$ and all the neighborhood states $u+v+\delta u$ at a given load parameter $\lambda$ can be written as
\begin{equation} \label{PBA-3-1}
\Delta {\mathcal P} = \mathcal P [u_0 + v + \delta u; \lambda] -  \mathcal P [u_0 + v; \lambda] = \frac{1}{2} {\mathcal P}'' [u_0 + v; \lambda] (\delta u)^2 + o (\Vert \delta u \Vert^2),
\end{equation}
where the first-order disappears due to the equilibrium and ${\mathcal P}'' [u_0 + v; \lambda] (\delta u)^2$ is defined in Eq.~\eqref{PBA-1-2a}. \\

Note that the minimum of the second variation is positive, the state $u+v$ of the system is stable at a given load parameter $\lambda$. Thus the sign of the second variation is of high interest. If the minimum of the second variation is positive, then the second variation is positive for any perturbation in the neighborhood of state $u+v$ at $\lambda$. However, the extremum of the second variation may not exist due to the unbound domain of the arbitrary perturbation $\delta u$. To ensure the existence of the minimum, we discuss the minimum of the form\footnote{The extremization of this form is identical to the extremization of the second variation ${\mathcal P}'' [u_0 + v; \lambda] (\delta u)^2$ subjected to the constraint (or the so-called normalization condition) $\Vert\delta u\Vert^2 = 1$. The equivalence of these two procedures can refer to Chapter 1.18 in the book \cite{riley2011essential}. In the pioneering work \cite{koiter1945}, Koiter used the form Eq.~\eqref{PBA-3-2} presented in this article.}
\begin{equation} \label{PBA-3-2}
\frac{{\mathcal P}'' [u_0 + v; \lambda] (\delta u)^2}{\Vert\delta u\Vert^2},
\end{equation}
which does not change the sign of the second variation and $\Vert\delta u\Vert = \langle \delta u, \delta u\rangle^{1/2}$. \\

We assume that the minimum of Eq.~\eqref{PBA-3-2} is denoted by $\varpi$ that exists for $\delta u = w$, such that
\begin{equation} \label{PBA-3-3}
\min \frac{{\mathcal P}'' [u_0 + v; \lambda] (\delta u)^2}{\Vert\delta u\Vert^2} = \frac{{\mathcal P}'' [u_0 + v; \lambda] w^2}{\Vert w \Vert^2} = \varpi.
\end{equation}

Therefore, any variation $w + \tau \bar w$ admits the following inequality
\begin{equation} \label{PBA-3-4}
\frac{{\mathcal P}'' [u_0 + v; \lambda] (w + \tau \bar w)^2}{\Vert w + \tau \bar w\Vert^2} \ge \frac{{\mathcal P}'' [u_0 + v; \lambda] w^2}{\Vert w \Vert^2} = \varpi,
\end{equation}
alternatively,
\begin{equation} \label{PBA-3-5}
{\mathcal P}'' [u_0 + v; \lambda] w^2 + 2 \tau {\mathcal P}'' [u_0 + v; \lambda] w \bar w + \tau^2 {\mathcal P}'' [u_0 + v; \lambda] {\bar w}^2 \ge \varpi \left\{ \Vert w \Vert^2 + 2\tau \langle w, \bar w \rangle + \tau^2 \Vert \bar w \Vert^2 \right\}.
\end{equation}

With Eq.~\eqref{PBA-3-3}, the inequality \eqref{PBA-3-5} reduces to
\begin{equation} \label{PBA-3-6}
2 \tau \left\{ {\mathcal P}'' [u_0 + v; \lambda] w \bar w - \varpi \langle w, \bar w \rangle \right\} + \tau^2 \left\{ {\mathcal P}'' [u_0 + v; \lambda] {\bar w}^2 - \varpi \Vert \bar w \Vert^2\right\} \ge 0.
\end{equation}
Since $\tau \in \mathbb R$ is arbitrary, the inequality \eqref{PBA-3-6} holds if and only if 
\begin{equation} \label{PBA-3-7}
{\mathcal P}'' [u_0 + v; \lambda] w \bar w - \varpi \langle w, \bar w \rangle =0.
\end{equation}
For $\lambda=\lambda_c$ and $v=0$ and with Eq.~\eqref{PBA-1-9} (or Eq.~\eqref{PBA-1-11}), it is clear that $w=u_1$ implies $\varpi =0$ in Eq.~\eqref{PBA-3-7}. \\

Similar to the asymptotic expansions \eqref{PBA-2-4}, in the neighborhood of $\lambda=\lambda_c$, we expand
\begin{equation} \label{PBA-3-8}
w = u_1 + \zeta w_1 + \zeta^2 w_2 + \cdots \quad {\rm and} \quad
\varpi = \zeta \varpi_1 + \zeta^2 \varpi_2 + \cdots
\end{equation}
with $\langle u_1, w_n \rangle=0, n = 1,2,3,\dots$. With the expansions in Eqs.~\eqref{PBA-2-4} and \eqref{PBA-3-8}, we can expand Eq.~\eqref{PBA-3-7} at $\lambda=\lambda_c$ and $u_0 (\lambda_c)$, namely
\begin{equation} \label{PBA-3-9}
\begin{aligned}
0 & = \zeta \left\{ \big({\mathcal P}''_c w_1 + \lambda_1 \dot{\mathcal P}''_c  u_1 + {\mathcal P}'''_c u_1^2 \big) \bar w -  \varpi_1 \langle u_1, \bar w \rangle \right\} \\
   & \quad + \zeta^2 \left\{ \left( 
   \begin{aligned}
   & {\mathcal P}''_c w_2 + \lambda_1 \dot {\mathcal P}''_c w_1 + \lambda_2 \dot {\mathcal P}''_c u_1 + \frac{1}{2} \lambda_1^2 \ddot {\mathcal P}''_c u_1 \\
   & + {\mathcal P}'''_c u_1 w_1 +\lambda_1 \dot {\mathcal P}'''_c u_1^2 + {\mathcal P}'''_c u_2 u_1 + \frac{1}{2} {\mathcal P}^{(4)}_c u_1^3 
   \end{aligned}
   \right) \bar w  - \varpi_1 \langle w_1, \bar w \rangle - \varpi_2 \langle u_1, \bar w\rangle \right\} + o(\zeta^2).
\end{aligned}
\end{equation}
Vanishing of the coefficients of $\zeta$, $\zeta^2$, $\dots$ gives
\begin{subequations}
\begin{equation} \label{PBA-3-10a}
\big({\mathcal P}''_c w_1 + \lambda_1 \dot{\mathcal P}''_c  u_1 + {\mathcal P}'''_c u_1^2 \big) \bar w -  \varpi_1 \langle u_1, \bar w \rangle = 0,
\end{equation}
\begin{equation} \label{PBA-3-10b}
\left( 
   \begin{aligned}
   & {\mathcal P}''_c w_2 + \lambda_1 \dot {\mathcal P}''_c w_1 + \lambda_2 \dot {\mathcal P}''_c u_1 + \frac{1}{2} \lambda_1^2 \ddot {\mathcal P}''_c u_1 \\
   & + {\mathcal P}'''_c u_1 w_1 +\lambda_1 \dot {\mathcal P}'''_c u_1^2 + {\mathcal P}'''_c u_2 u_1 + \frac{1}{2} {\mathcal P}^{(4)}_c u_1^3 
   \end{aligned}
   \right) \bar w  - \varpi_1 \langle w_1, \bar w \rangle - \varpi_2 \langle u_1, \bar w\rangle = 0.
\end{equation}
\end{subequations}

If we choose $\bar w = u_1$ in Eq.~\eqref{PBA-3-10a}, then we have ${\mathcal P}''_c w_1 u_1 = {\mathcal P}''_c u_1 w_1= 0$ from Eq.~\eqref{PBA-1-11}, together with $\Vert u_1 \Vert = \langle u_1, u_1 \rangle^{1/2} =1$, we obtain
\begin{equation} \label{PBA-3-11}
\varpi_1 = \lambda_1 \dot{\mathcal P}''_c  u_1^2 + {\mathcal P}'''_c u_1^3,
\end{equation}
which, with Eq.~\eqref{PBA-2-6a}, can be further reduced to
\begin{equation} \label{PBA-3-12}
\varpi_1 = -\lambda_1 \dot{\mathcal P}''_c  u_1^2.
\end{equation}

Similarly, if we choose $\bar w = u_1$ in Eq.~\eqref{PBA-3-10b}, and with ${\mathcal P}''_c w_2 u_1 = {\mathcal P}''_c u_1 w_2 = 0$ from Eq.~\eqref{PBA-1-11} and $\langle w_1, u_1\rangle =0$ in Eq.~\eqref{PBA-3-8}, we can get
\begin{equation} \label{PBA-3-13}
\varpi_2 = 
\left( 
 \lambda_1 \dot {\mathcal P}''_c w_1 + \lambda_2 \dot {\mathcal P}''_c u_1 + \frac{1}{2} \lambda_1^2 \ddot {\mathcal P}''_c u_1 + {\mathcal P}'''_c u_1 w_1 +\lambda_1 \dot {\mathcal P}'''_c u_1^2 + {\mathcal P}'''_c u_2 u_1 + \frac{1}{2} {\mathcal P}^{(4)}_c u_1^3 
   \right) u_1.
\end{equation}

For the case of the asymmetric bifurcation $\lambda_1 \neq 0$, with the expansions \eqref{PBA-2-4} and \eqref{PBA-3-8} as well as the value $\varpi_1$ in Eq.~\eqref{PBA-3-12}, the minimum $\varpi$ in Eq.~\eqref{PBA-3-3} related to the second variation can be written as
\begin{equation} \label{PBA-3-14}
\varpi = - (\lambda - \lambda_c) \dot{\mathcal P}''_c  u_1^2 + o (|\lambda-\lambda_c|)  \qquad \text{for asymmetric bifurcation} \ \lambda_1 \neq 0.
\end{equation}
If $\varpi >0$, it corresponds to the stability of the bifurcated branch; on the contrary, the $\varpi <0$ indicates the instability of the bifurcated branch. \\

Again, for the case of the symmetric bifurcation $\lambda_1=0$, it implies $\varpi_1 =0$ in Eq.~\eqref{PBA-3-12} and $\varpi_2$ in Eq.~\eqref{PBA-3-13} becomes $\varpi_2 = 
\left( 
\lambda_2 \dot {\mathcal P}''_c u_1 + {\mathcal P}'''_c u_1 w_1 + {\mathcal P}'''_c u_2 u_1 + \frac{1}{2} {\mathcal P}^{(4)}_c u_1^3 
   \right) u_1$. Also, the zero coefficient of $\zeta^2$ in Eq.~\eqref{PBA-2-5a} gives $\left\{{\mathcal P}''_c u_2 + \frac{1}{2}{\mathcal P}'''_c u_1^2 \right\} \delta u =0$ while Eq.~\eqref{PBA-3-10a} becomes $\big({\mathcal P}''_c w_1 + {\mathcal P}'''_c u_1^2 \big) \bar w = 0$ for $\lambda_1=0$ and $\varpi_1 =0$.  By comparing these two reduced equations and letting $\delta u = \bar w$, we then have $2 u_2 = w_1$. Now $\varpi_2$ further becomes $\varpi_2 = 
\left( 
\lambda_2 \dot {\mathcal P}''_c u_1 + 3{\mathcal P}'''_c u_2 u_1 + \frac{1}{2} {\mathcal P}^{(4)}_c u_1^3 
   \right) u_1$. Recalling the form of $\lambda_2$ in Eq.~\eqref{PBA-2-7} for the symmetric bifurcation $\lambda_1 =0$, we finally have
\begin{equation} \label{PBA-3-15}
\varpi_2 = -2\lambda_2 \dot {\mathcal P}''_c u_1^2.
\end{equation}
With Eqs.~\eqref{PBA-2-4}, \eqref{PBA-3-8} and \eqref{PBA-3-15}, the minimum $\varpi$ in Eq.~\eqref{PBA-3-3} becomes
\begin{equation} \label{PBA-3-16}
\varpi = - 2(\lambda - \lambda_c) \dot{\mathcal P}''_c  u_1^2 + o (|\lambda-\lambda_c|) \qquad \text{for symmetric bifurcation} \ \lambda_1=0.
\end{equation}

Finally, we would like to discuss the sign of $\dot{\mathcal P}''_c  u_1^2$ in Eqs.~\eqref{PBA-2-6a}, \eqref{PBA-2-7}, \eqref{PBA-3-12}, \eqref{PBA-3-14}, and \eqref{PBA-3-16}. Since ${\mathcal P}''_0 u_1^2$ is positive for $\lambda < \lambda_c$ based on the stability of the principal branch $u_0$ below the threshold $\lambda_c$, and ${\mathcal P}''_0  u_1^2 = 0$ at $\lambda=\lambda_c$ from the equality \eqref{PBA-1-11}, it follows that $\dot{\mathcal P}''_c  u_1^2 \le 0$. The discussion of the special case $\dot{\mathcal P}''_c  u_1^2 = 0$ is out of the scope of this tutorial and therefore the case $\dot{\mathcal P}''_c  u_1^2 = 0$ is not included here. More discussion of the sign of $\dot{\mathcal P}''_c  u_1^2$ can be found in the work \cite{budiansky1974theory}. \\

Based on the above discussion, the negative sign of $\dot{\mathcal P}''_c  u_1^2$ makes the stability conditions \eqref{PBA-3-14} and \eqref{PBA-3-16} in alternative forms that only depend on the sign of $\lambda_1$ and $\lambda_2$. Dividing Eqs.~\eqref{PBA-3-14} and \eqref{PBA-3-16} by the positive value, $-\dot{\mathcal P}''_c  u_1^2 > 0$, and with the expansion of $\lambda$ in Eq.~\eqref{PBA-2-4}, we have 
\begin{equation} \label{PBA-3-17}
\varpi \Rightarrow \frac{\varpi}{-\dot{\mathcal P}''_c  u_1^2} = (\lambda - \lambda_c) + o (|\lambda-\lambda_c|) = \lambda_1 \zeta + o(\zeta)  \qquad \text{for asymmetric bifurcation} \ \lambda_1 \neq 0
\end{equation}
and
\begin{equation} \label{PBA-3-18}
\varpi \Rightarrow \frac{\varpi}{-\dot{\mathcal P}''_c  u_1^2} = 2(\lambda - \lambda_c) + o (|\lambda-\lambda_c|) = 2 \lambda_2 \zeta^2 + o(\zeta^2) \qquad \text{for symmetric bifurcation} \ \lambda_1=0.
\end{equation}

\begin{figure}[h] 
\centering
\subfigure[]{%
\includegraphics[width=0.3\textwidth]{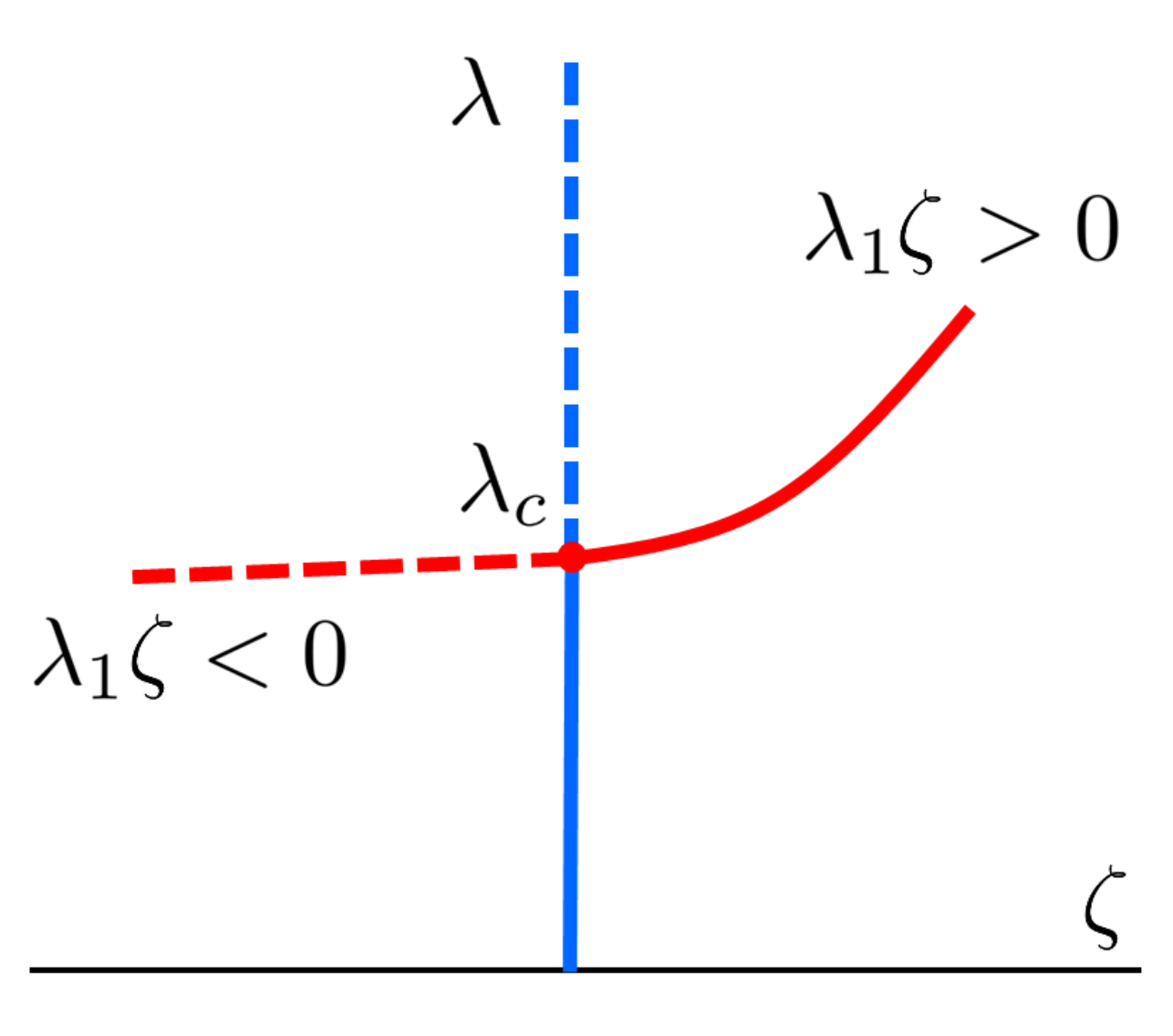}
\label{PBA-1a}}
\subfigure[]{%
\includegraphics[width=0.3\textwidth]{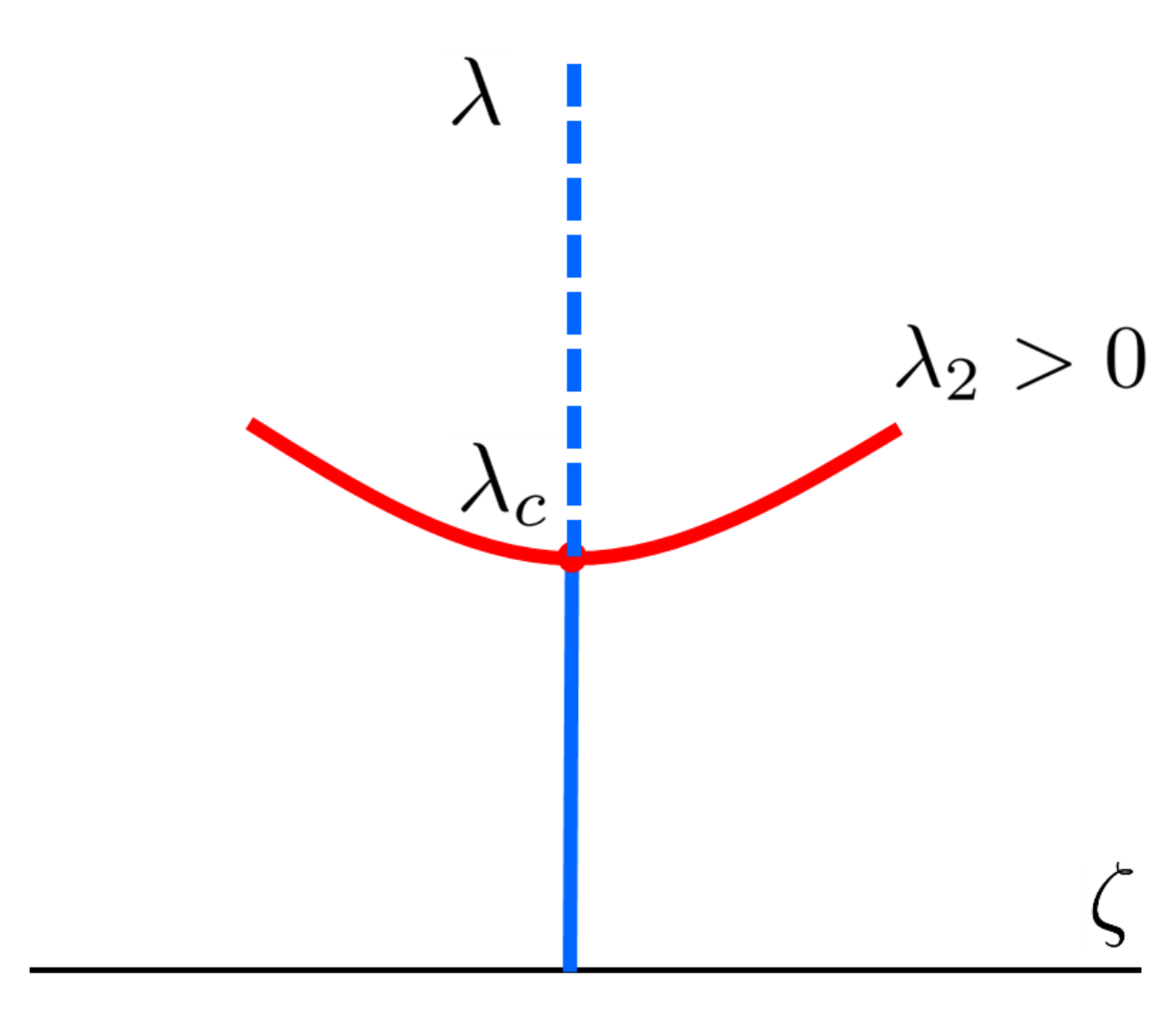}
\label{PBA-1b}}
\subfigure[]{%
\includegraphics[width=0.3\textwidth]{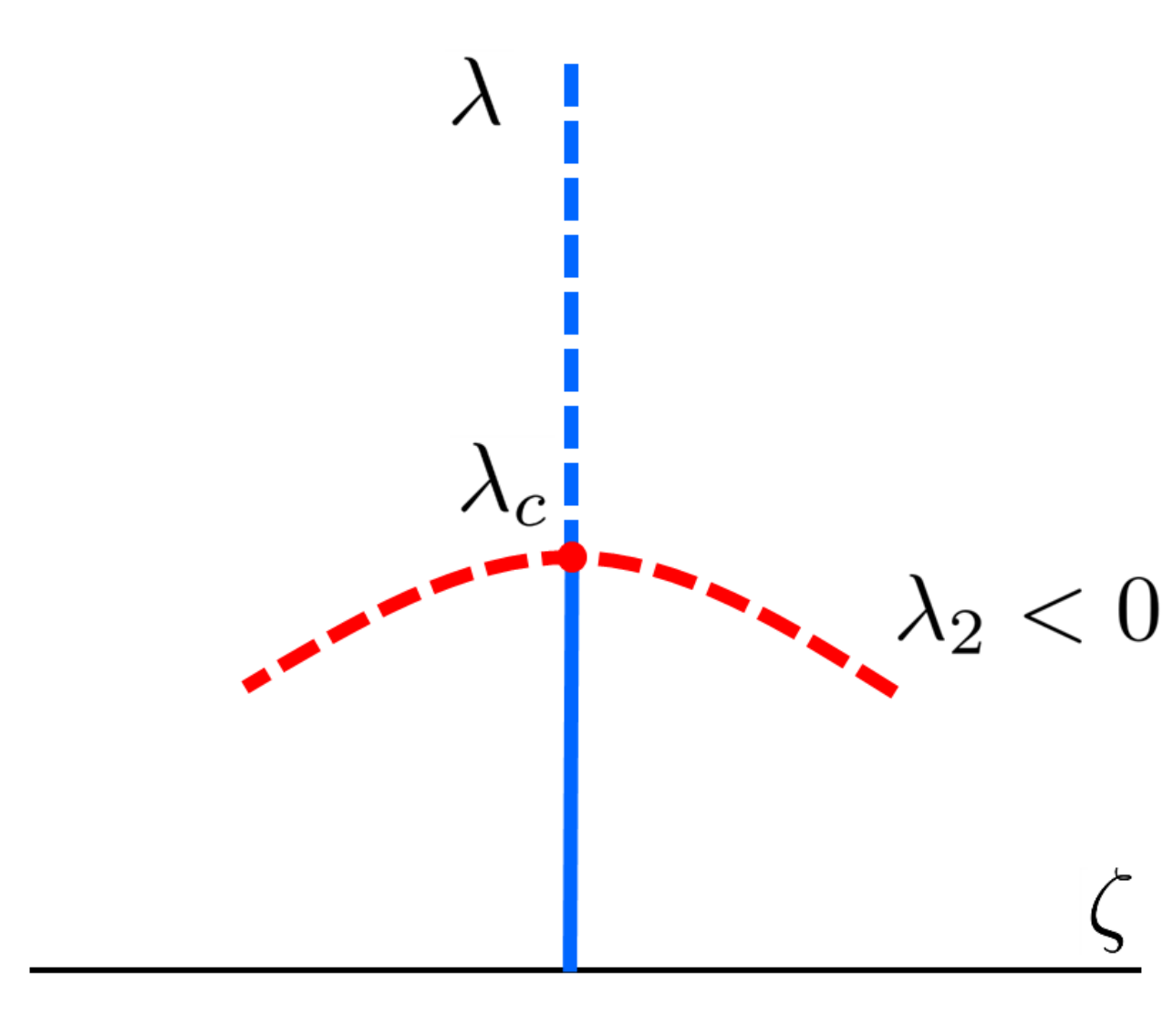}
\label{PBA-1c}}
\caption{Equilibrium paths and their stabilities. Stable paths are denoted by solid curves while unstable paths are drawn by dashed curves. The relation between the sign of the second variation and the values $\lambda_1$ and $\lambda_2$ are given in Eqs.~\eqref{PBA-3-17} and \eqref{PBA-3-18}, respectively. (a) Asymmetric bifurcation with $\lambda_1 \neq 0$. (b) Symmetric bifurcation with $\lambda_2 > 0$ and $\lambda_1 =0$. (c) Symmetric bifurcation with $\lambda_2 < 0$ and $\lambda_1 =0$.}
\label{PBA-1}
\end{figure}

Again, the value of $\lambda_1$ in Eq.~\eqref{PBA-3-17} is given by Eq.~\eqref{PBA-2-6a} while the value of $\lambda_2$ in Eq.~\eqref{PBA-3-18} is given by Eq.~\eqref{PBA-2-7}. Since most of the bifurcations are symmetric, then the sign of the value $\lambda_2$ in Eq.~\eqref{PBA-3-18} is of interest. For a positive $\lambda_2$, the bifurcated branch is stable. On the contrary, a negative $\lambda_2$ corresponds to an unstable bifurcated branch. The stable and unstable branches corresponding to the criteria \eqref{PBA-3-17} and \eqref{PBA-3-18} are shown in Fig.~\ref{PBA-1} in the case of a single buckling mode.

\subsection{Example: Buckling of Euler's column subjected to an electric field}
\begin{figure}[h] %
    \centering
    \includegraphics[width=4in]{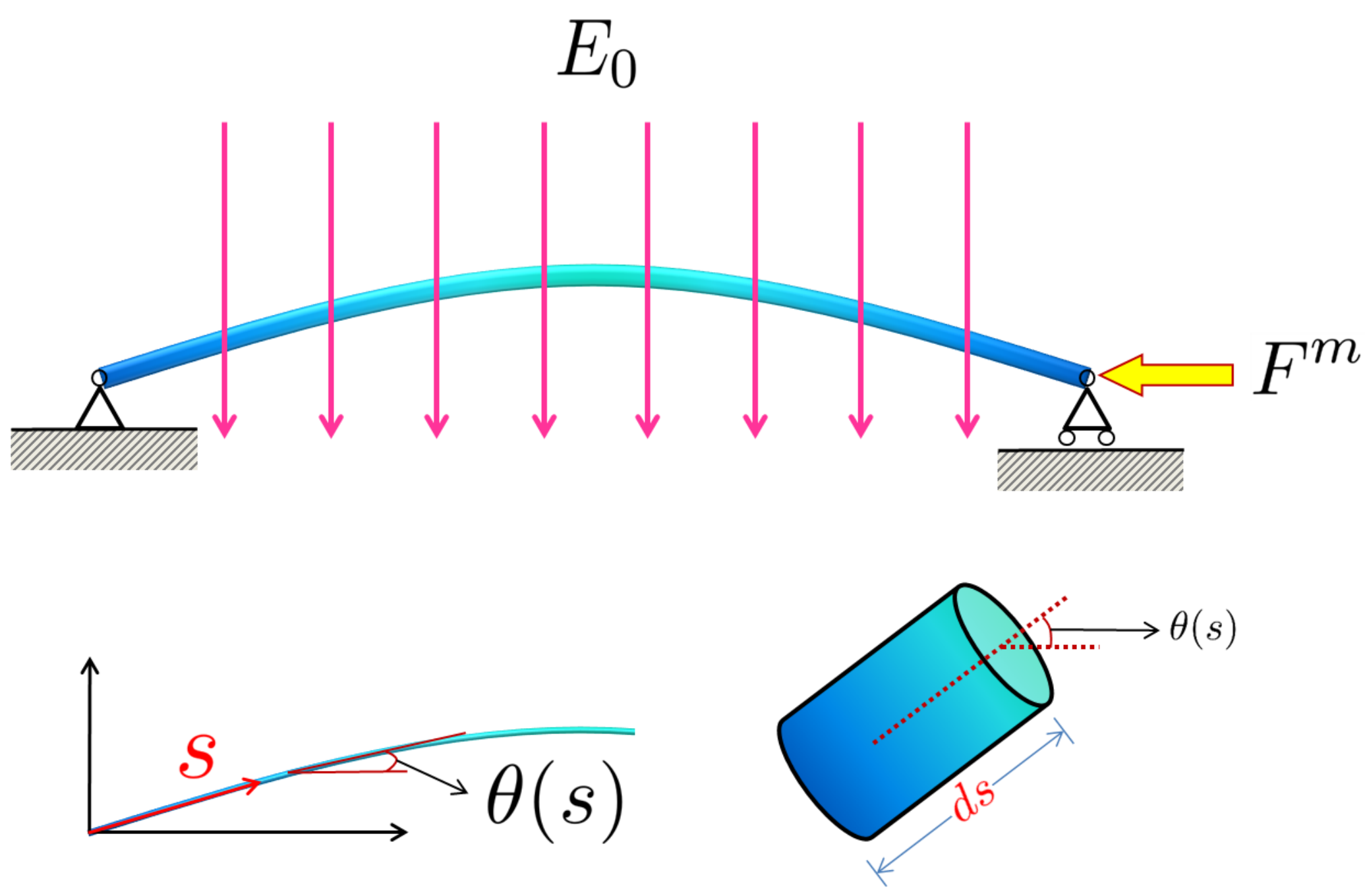}
    \caption{Schematic of the buckling of an incompressible dielectric column with pinned-pinned ends subjected to an external electric field $E_0$ (in the vertical direction) and an applied force $F^m$ (in the horizontal direction). The column length is $L$ while the cross-sectional radius is $r$, $r \ll L$. The arc length coordinate is denoted by $s$. The scalar function $\theta (s)$ is the rotation (angle) of the column cross-section from its initial horizontal direction.}
    \label{LSK-ex-fig1}
\end{figure}

Let us consider the buckling and the post-buckling of an incompressible dielectric column subject to both mechanical and the electric loads (see Fig.~\ref{LSK-ex-fig1}). The analysis of the purely mechanical buckling of Euler's column can be found in many works, just name a few \cite{timoshenko1961theory, cedolin2010stability, budiansky1974theory, triantafyllidis2011stability, van2009wt, antman2013nonlinear}. \\

The potential energy of the electromechanical system shown in Fig.~\ref{LSK-ex-fig1} is given by\footnote{For the purely mechanical part, the form of the potential energy can refer to many works, e.g., \cite{budiansky1974theory, triantafyllidis2011stability, van2009wt}. The second term on the RHS of Eq.~\eqref{LSK-ex-1} comes from the electric energy, which has the form $- \int_0^L \pi r^2 \left( \int_{\boldsymbol 0}^{\boldsymbol E_0} {\boldsymbol D} \cdot d{\boldsymbol E} \right) ds$ for a dielectric column subjected to an external electric field shown in Fig.~\ref{LSK-ex-fig1}. The electric field ${\boldsymbol E}$ within the column can be decomposed into the one in the arc length and the one perpendicular to the arc length, such that ${\boldsymbol E} = (E_s, E_s^{\perp}) = (E \sin \theta, \frac{\varepsilon_0}{\varepsilon} E \cos \theta)$. Such kinds of components come from the jump conditions of the electric field on the surface between the vacuum and the dielectric column. For the detailed jump conditions, one can refer to the review \cite{dorfmann2017nonlinear}. Also, the electric displacement ${\boldsymbol D}$ in the column has the component form ${\boldsymbol D} = (D_s, D_s^{\perp}) = (\varepsilon E_s,  \varepsilon E_s^{\perp}) = (\varepsilon E \sin \theta, \varepsilon_0 E \cos \theta)$. It follows that $- \int_0^L \pi r^2 \left( \int_{\boldsymbol 0}^{\boldsymbol E_0} {\boldsymbol D} \cdot d{\boldsymbol E} \right) ds = - \int_0^L \pi r^2 \left[ \int_0^{E_0} \left(\varepsilon E \sin^2\theta + \frac{\varepsilon_0^2}{\varepsilon}E \cos^2\theta \right) dE \right] ds = - \int_0^L \frac{\pi r^2}{2} \varepsilon E_0^2 \left( \sin^2 \theta + \frac{\varepsilon_0^2}{\varepsilon^2} \cos^2 \theta \right) ds$.}
\begin{equation} \label{LSK-ex-1}
\mathcal P [\theta(s); E_0, F^m] = \int_0^L \frac{1}{2} B \left (\frac{\partial \theta}{\partial s} \right)^2 ds - \int_0^L \frac{\pi r^2}{2} \varepsilon E_0^2 \left( \sin^2 \theta + \frac{\varepsilon_0^2}{\varepsilon^2} \cos^2 \theta \right) ds - F^m \int_0^L (1-\cos \theta) ds,
\end{equation}
where $\theta$ is the rotation of the cross-section, $s$ is the arc length coordinate, $B$ is the bending stiffness, $\varepsilon_0$ is the vacuum permittivity, $\varepsilon$ is the material permittivity, $E_0$ is the external electric field, and $F^m$ is the horizontal load. The state variable in Eq.~\eqref{LSK-ex-1} is regarded as the rotation $\theta$, which can describe the response of the column subjected to both the electric and the mechanical loads. \\

Consider the variation $\theta (s) \to \theta (s) + \tau \bar\theta (s)$. Here $\tau$ is a sufficiently small parameter and $\bar\theta (s)$ is an arbitrary function that satisfies all the kinematically admissible deformations. Then the first variation of the functional \eqref{LSK-ex-1} is given by
\begin{equation} \label{LSK-ex-2}
\begin{aligned}
\delta \mathcal P & = \left. \frac{d}{d \tau} \mathcal P [\theta(s) + \tau \bar\theta (s); E_0, F^m] \right|_{\tau=0}\\
& = \int_0^L B \left (\frac{\partial \theta}{\partial s} \right) \left (\frac{\partial \bar\theta}{\partial s} \right) ds - \int_0^L {\pi r^2} \varepsilon E_0^2 \left( 1 - \frac{\varepsilon_0^2}{\varepsilon^2}\right) \bar\theta \sin \theta \cos \theta ds - F^m \int_0^L \bar\theta \sin\theta ds \\
& = \left. B \left (\frac{\partial \theta}{\partial s} \right) \bar\theta \right|_0^L -\int_0^L \left[ B \left (\frac{\partial^2 \theta}{\partial s^2} \right) + F^e \sin \theta \cos \theta + F^m \sin\theta \right] \bar\theta ds,
\end{aligned}
\end{equation}
where $F^e$ is related to the electric load and defined by
\begin{equation} \label{LSK-ex-L-1a}
F^e = {\pi r^2} \varepsilon E_0^2 \left( 1 - \frac{\varepsilon_0^2}{\varepsilon^2}\right).
\end{equation}
The first variation condition, $\delta \mathcal P = 0$ in Eq.~\eqref{LSK-ex-2}, gives the Euler-Lagrange equation 
\begin{equation} \label{LSK-ex-3}
B \left (\frac{\partial^2 \theta}{\partial s^2} \right) + F^e \sin \theta \cos \theta + F^m \sin\theta = 0
\end{equation}
and the natural boundary condition
\begin{equation} \label{LSK-ex-4}
\frac{\partial \theta}{\partial s} = 0 \quad {\rm at} \ s=0 \, \& \, L.
\end{equation}
A trivial solution (unbuckling) to the boundary-value problem is  
\begin{equation} \label{LSK-ex-5}
\theta(s) = \theta_0 (s) = 0 \quad {\rm for} \ 0 \le s \le L.
\end{equation}

\subsubsection{Linear bifurcation analysis}
Let $\theta^{\star}(s)$ be the increment of the rotation $\theta(s)$. The linearized versions of Eqs.~\eqref{LSK-ex-3} and \eqref{LSK-ex-4} at the trivial solution \eqref{LSK-ex-5} are
\begin{equation} \label{LSK-ex-L-1}
B \left (\frac{\partial^2 \theta^{\star}}{\partial s^2} \right) + (F^e + F^m) \theta^{\star} = 0
\end{equation}
and
\begin{equation} \label{LSK-ex-L-2}
\frac{\partial \theta^{\star}}{\partial s} = 0 \quad {\rm at} \ s=0 \, \& \, L.
\end{equation}

Nonzero solution $\theta^{\star}(s)$ of Eqs.~\eqref{LSK-ex-L-1} and \eqref{LSK-ex-L-2} gives the condition\footnote{Note that the linear bifurcation analysis only gives the necessary condition for the existence of non-trivial solutions.}
\begin{equation} \label{LSK-ex-L-3}
F^e+F^m = B \frac{n^2 \pi^2}{L^2}
\end{equation}
and
\begin{equation} \label{LSK-ex-L-4}
\theta_n (s) = \cos\frac{n\pi s}{L} \quad {\rm for} \ 0 \le s \le L,
\end{equation}
where $n \in \mathbf Z^*$ is a non-zero integer and the normalization is $\langle \theta_n, \theta_n \rangle \equiv \frac{2}{L} \int_0^L \theta_n^2 (s) ds = 1$. When $(F^e + F^m)$ increases from zero to $B \pi^2/L^2$, the first buckling mode $\theta_1 (s)= \cos\frac{\pi s}{L}$ occurs.

\subsubsection{Energy stability criterion}
In contrast to the linear bifurcation analysis, we study the stability of the electromechanical system by the energy method. Thus we have to examine the second variation condition. Similar to the first variation \eqref{LSK-ex-2}, the second variation of the potential energy \eqref{LSK-ex-1} is 
\begin{equation} \label{LSK-ex-E-1}
\begin{aligned}
\delta^2 \mathcal P & = \left. \frac{d^2}{d \tau^2} \mathcal P [\theta(s) + \tau \bar\theta (s); E_0, F^m] \right|_{\tau=0}\\
& = \int_0^L B \left (\frac{\partial \bar\theta}{\partial s} \right)^2 ds - F^e \int_0^L \bar\theta^2 \cos (2\theta) ds - F^m \int_0^L \bar\theta^2 \cos\theta ds.
\end{aligned}
\end{equation}
At the trivial solution $\theta_0 (s) = 0$ in Eq.~\eqref{LSK-ex-5}, the second variation \eqref{LSK-ex-E-1} reads
\begin{equation} \label{LSK-ex-E-2}
\delta^2 \mathcal P|_{\theta_0 =0} = \int_0^L \left[ B \left (\frac{\partial \bar\theta}{\partial s} \right)^2 - (F^e + F^m) \bar\theta^2 \right] ds.
\end{equation}

The integral \eqref{LSK-ex-E-2} is a continuous functional of the smooth variation $\bar\theta (s)$. If the minimum value of the functional \eqref{LSK-ex-E-2} is non-negative, then the equilibrium solution $\theta_0 (s) = 0$ in Eq.~\eqref{LSK-ex-5} is stable. To ensure the existence of the minimum value of the second variation $\delta^2 \mathcal P|_{\theta_0 =0}$ in Eq.~\eqref{LSK-ex-E-2}, we introduce a compact subset of the domain of the functional, which does not change the positive definiteness of the functional. Consider the following normalization condition
\begin{equation} \label{LSK-ex-E-3}
\Vert \bar \theta \Vert^2 = \langle \bar\theta, \bar\theta \rangle \equiv \frac{2}{L} \int_0^L \bar\theta^2 ds = 1.
\end{equation}
Based on the boundary condition \eqref{LSK-ex-4}, we consider the Fourier cosine series of the kinematically admissible variation
\begin{equation} \label{LSK-ex-E-4}
\bar\theta (s) = \sum_{n=0}^{\infty} a{_n} \cos \frac{n\pi s}{L}.
\end{equation}
The normalization condition \eqref{LSK-ex-E-3}, with the orthogonality condition, now reads
\begin{equation} \label{LSK-ex-E-5}
\sum_{n=0}^{\infty} a_n^2 =1.
\end{equation}
With Eqs.~\eqref{LSK-ex-E-3}$-$\eqref{LSK-ex-E-5}, the integral \eqref{LSK-ex-E-2} is bounded and becomes
\begin{equation} \label{LSK-ex-E-6}
\delta^2 \mathcal P|_{\theta_0 =0} = \frac{L}{2} \sum_{n=0}^{\infty} \left\{ a_n^2 \left[ B \left(\frac{n \pi}{L}\right)^2 - (F^e + F^m) \right] \right\} \ge \varpi,
\end{equation}
where $\varpi$ is the minimum of the integral, namely
\begin{equation} \label{LSK-ex-E-7}
\varpi = {\rm min} \ \delta^2 \mathcal P|_{\theta_0 =0} = \frac{L}{2}  \left[ B \left(\frac{\pi}{L}\right)^2 - (F^e + F^m) \right] \quad {\rm at} \ \bar\theta = \cos\frac{\pi s}{L}.
\end{equation}

Apparently, the minimum value $\varpi$ in Eq.~\eqref{LSK-ex-E-7} is positive for $0<(F^e + F^m) < B \left(\frac{\pi}{L}\right)^2$, which indicates that the trivial solution $\theta_0 (s) =0$ is stable. On the contrary, the trivial solution $\theta_0 (s) =0$ becomes unstable for $(F^e + F^m) > B \left(\frac{\pi}{L}\right)^2$ since the minimum $\varpi$ is negative.

\subsubsection{Post-buckling analysis: LSK approach}
By the notations in Eq.~\eqref{PBA-1-2a}, the first variation of the potential energy \eqref{LSK-ex-1} is
\begin{equation} \label{LSK-ex-Po-1}
{\mathcal P}'[\theta; E_0, F^m]\theta_1 = \int_0^L B \left (\frac{\partial \theta}{\partial s} \right) \left (\frac{\partial \theta_1}{\partial s} \right) ds - F^e \int_0^L \theta_1 \sin \theta \cos \theta ds - F^m \int_0^L \theta_1 \sin\theta ds,
\end{equation}
and the higher-order variations are
\begin{align} 
\label{LSK-ex-Po-2}
& {\mathcal P}''[\theta; E_0, F^m]\theta_1 \theta_2  = \int_0^L B \left (\frac{\partial \theta_2}{\partial s} \right) \left (\frac{\partial \theta_1}{\partial s} \right) ds - F^e \int_0^L \theta_1 \theta_2 \cos (2\theta) ds - F^m \int_0^L \theta_1 \theta_2 \cos\theta ds, \\
\label{LSK-ex-Po-3}
& {\mathcal P}'''[\theta; E_0, F^m]\theta_1 \theta_2 \theta_3  = 2 F^e \int_0^L \theta_1 \theta_2 \theta_3 \sin (2\theta) ds + F^m \int_0^L \theta_1 \theta_2 \theta_3 \sin\theta ds, \\
\label{LSK-ex-Po-4}
& {\mathcal P}^{(4)}[\theta; E_0, F^m]\theta_1 \theta_2 \theta_3 \theta_4  = 4 F^e \int_0^L \theta_1 \theta_2 \theta_3 \theta_4 \cos (2\theta) ds + F^m \int_0^L \theta_1 \theta_2 \theta_3 \theta_4 \cos\theta ds.
\end{align}

Note that the generalized displacement $u$ in Eq.~\eqref{PBA-1-2a} is the rotation $\theta (s)$ while the load parameter $\lambda$ is the horizontal load $F^m$. The electric field $E_0$ is regarded as a constant in this example for simplicity. Thus we only have one control parameter $F^m$ here. \\

Consider the trivial solution $\theta (s) = \theta_0 (s) =0$, i.e., the prebuckling solution of the column. With the second variation \eqref{LSK-ex-Po-2}, the bifurcation equation \eqref{PBA-1-9} becomes
\begin{equation} \label{LSK-ex-Po-5}
0 = {\mathcal P}''[0; E_0, F_c^m]\theta_1 \delta \theta = \int_0^L B \left (\frac{\partial \delta \theta}{\partial s} \right) \left (\frac{\partial \theta_1}{\partial s} \right) ds - F^e \int_0^L \theta_1 \delta \theta ds - F_c^	m \int_0^L \theta_1 \delta \theta ds,
\end{equation}
which gives similar results as that of the linear bifurcation analysis, namely
\begin{equation} \label{LSK-ex-Po-6}
B \left (\frac{\partial^2 \theta_1}{\partial s^2} \right) + (F^e + F_c^m) \theta_1 = 0
\end{equation}
and the natural boundary condition
\begin{equation} \label{LSK-ex-Po-7}
\frac{\partial \theta_1}{\partial s} = 0 \quad {\rm at} \ s=0 \, \& \, L.
\end{equation}
The lowest eigenvalue corresponds to the critical load, namely\footnote{For simplicity, we just consider the case in which the electric load $F^e$ is $0< F^e < \frac{B \pi^2}{L^2}$. And there is no buckling if the applied mechanical load $F^m$ is below the critical load $F_c^m$.}
\begin{equation} \label{LSK-ex-Po-8}
{F^e + F_c^m} = \frac{B \pi^2}{L^2},
\end{equation}
and the normalized eigenfunction $\theta_1 (s)$, with the normalization $\langle \theta_1, \theta_1\rangle \equiv \frac{2}{L} \int_0^L \theta_1^2 ds =1$ in Eq.~\eqref{LSK-ex-E-3}, is  
\begin{equation} \label{LSK-ex-Po-9}
\theta_1 = \cos\frac{\pi s}{L}.
\end{equation}
With the third and fourth variations \eqref{LSK-ex-Po-3} and \eqref{LSK-ex-Po-4} as well as Eq.~\eqref{LSK-ex-Po-9}, we obtain
\begin{subequations} \label{LSK-ex-Po-10}
\begin{equation} \label{LSK-ex-Po-10a}
{\mathcal P}'''_c \theta_1^3 = {\mathcal P}'''[0; E_0, F_c^m]\theta_1^3 = 0, 
\end{equation}
\begin{equation} \label{LSK-ex-Po-10b}
{\mathcal P}^{(4)}_c \theta_1^4 = {\mathcal P}^{(4)}[0; E_0, F_c^m]\theta_1^4 = (4F^e + F_c^m) \int_0^L \theta_1^4 ds = \frac{3}{8} (4F^e +F_c^m) L.
\end{equation}
By Eq.~\eqref{LSK-ex-Po-3} and $\theta(s) =\theta_0 (s) =0$ for the principal branch (unbuckle), we have 
\begin{equation} \label{LSK-ex-Po-10c}
{\mathcal P}'''_c \theta_1^2 \theta_2 = {\mathcal P}'''[0; E_0, F_c^m]\theta_1^2 \theta_2  = 0.
\end{equation}
Also, with Eqs.~\eqref{LSK-ex-Po-2} and \eqref{LSK-ex-Po-9}, we have the partial derivative
\begin{equation} \label{LSK-ex-Po-10d}
\dot {\mathcal P}''_c \theta_1^2 = \left.\frac{\partial }{\partial F^m} {\mathcal P}''[\theta] \theta_1^2 \right|_{(\theta_0, F_c^m)} = - \int_0^L \theta_1^2 ds = - \frac{L}{2}.
\end{equation}
\end{subequations}
Now from Eq.~\eqref{LSK-ex-Po-10}, we can get the values of $\lambda_1$ in Eq.~\eqref{PBA-2-6a} and $\lambda_2$ in Eq.~\eqref{PBA-2-7}, such that
\begin{subequations} \label{LSK-ex-Po-11}
\begin{equation} \label{LSK-ex-Po-11a}
\lambda_1 = -\frac{1}{2}{\mathcal P}'''_c \theta_1^3 \Big/ \dot{\mathcal P}''_c \theta_1^2 = 0, 
\end{equation}
\begin{equation} \label{LSK-ex-Po-11b}
\quad \lambda_2 = - \left( \frac{1}{6}{\mathcal P}^{(4)}_c \theta_1^4 + {\mathcal P}'''_c \theta_1^2 \theta_2 \right)\Big/ \dot{\mathcal P}''_c \theta_1^2 = \frac{1}{8} (4 F^e + F_c^m).
\end{equation}
\end{subequations}

Since $\lambda_1 =0$ and $\lambda_2 > 0$, from Eq.~\eqref{PBA-3-18} and the schematic in Fig.~\ref{PBA-1b}, the bifurcation branch is symmetric and stable after the horizontal load $F^m$ slightly exceeds the threshold $F_c^m$. More discussions are also listed here. With Eq.~\eqref{LSK-ex-Po-11}, the expansion of the load parameter $F^m$ is
\[F^m = F_c^m + \lambda_1 \zeta + \lambda_2 \zeta^2 + o(\zeta^2) = F_c^m + \frac{1}{8}(4F^e +F_c^m) \zeta^2 + o(\zeta^2).\]
Since the small scalar $\zeta$ is the amplitude of the buckling mode $\theta_1$ in Eq.~\eqref{LSK-ex-Po-9}, we have
\begin{equation} \label{LSK-ex-Po-12}
\frac{F^m}{F_c^m} = 1 + \frac{1}{8} \left(1+ 4 \frac{F^e}{F_c^m}\right) \zeta^2
\end{equation}
and the load-shortening distance is
\begin{equation} \label{LSK-ex-Po-13}
\triangle L = \int_0^L \Big\{ 1 - \cos \big[\zeta \theta_1 (s) \big] \Big\} ds \cong \int_0^L \frac{\zeta^2 \theta_1^2 (s)}{2} ds = \frac{\zeta^2 L}{4}.
\end{equation}

By Eq.~\eqref{LSK-ex-Po-13}, Eq.~\eqref{LSK-ex-Po-12} can be rewritten as
\begin{equation} \label{LSK-ex-Po-14}
\frac{F^m}{F_c^m} = 1 + \frac{1}{2} \frac{\triangle L}{L} \left(1+ 4 \frac{F^e}{F_c^m}\right).
\end{equation}

From $F^e + F_c^m = \frac{B \pi^2}{L^2}$ in Eq.~\eqref{LSK-ex-Po-8}, we introduce the ratio $\beta = F^e \Big/ \frac{B \pi^2}{L^2}$, and then $F_c^m =(1-\beta) \frac{B \pi^2}{L^2}$. It follows that
\begin{equation} \label{LSK-ex-Po-15}
\frac{F^m}{F_c^m} = 1 + \frac{1}{2} \frac{\triangle L}{L} \left( 1+ 4 \frac{\beta}{1-\beta} \right).
\end{equation}

For $\beta=0$, it is the classic result of the post-buckling analysis of Euler's column subjected to purely mechanical loading \cite{van2009wt, triantafyllidis2011stability}.

\section{Concluding remarks and future study} \label{section-conclusion}

In what is intended to be a simple tutorial article, we can hope to be neither comprehensive nor as rigorous as we would like. As highlighted earlier in the introduction, the reader should take the current tutorial merely as a starting point before delving into the rather rich literature that already exists on the topic of instability, bifurcation, and electromechanical coupling. Further study of these topics could be taken in several directions which we outline below along with some selected references. \\

While the first section of our tutorial presented a broad overview of the field of stability and bifurcation analysis, our primary focus, and all the relevant examples, are in the context of \emph{elastic} materials that do not exhibit dissipation. In soft materials that exhibit some sort of electromechanical coupling, dissipation can emanate from diffusion of ions, electronic conduction, formation of defects, or viscous effects in the mechanical behavior of the polymer. Invoking viscoelasticity, current leakage and dielectric relaxation, stability analysis of dielectric elastomers was conducted by Chiang et. al. \cite{chiang2012model, chiang2012performance}. A detailed examination of the effect of viscoelasticity on electromechanical instabilities of dielectric elastomers was conducted by Wang et al. \cite{WANG2016213} where a finite element implementation was also developed.\\

A comprehensive theoretical framework for dissipation due to diffusing ions in dielectrics was created by Xiao and Bhattacharya \cite{xiao2008continuum} while Darbaniyan et. al. \cite{darbaniyan2019designing} provide a model with thermal effects. Although other works exist this topic, these two aforementioned references (and citations therein) may be used to acquire a broad overview of modeling efforts in diffusion and thermal related dissipation. However, neither of these two works focus on instabilities and remains a (relatively) understudied field within soft matter. On a related note, materials that have embedded charges or dipoles i.e. electrets, also provide a facile route to create soft electromechanical materials \cite{rahmati2019nonlinear, deng2014electrets}. Stability issues in such materials have received little attention.\\

Another important aspect pertaining to instabilities in soft matter is related to extreme deformations where a singularity may exist, for example, electro-creasing instability involves a singularity. Similar to \cite{wang2012bursting}, electro-creasing could have  applications including capacitors and energy harvesting; however, an analytical solution of creasing instability is still an open problem. A related example is the problem related to the crumpling of thin sheets \cite{cerda1999conical, cerda1998conical, kodali2017crumpling, wang2019flexoelectricity} that may be harnessed for energy harvesting in wearable electronics. \\

A significant line of research is ongoing on developing soft composites that exhibit unusual macroscopic electromechanical behavior. Specifically, germane to the current article, several researchers have investigated how instabilities at the microscale may influence the overall response of the heterogeneous materials. For example, Ponte Casta\~neda and Siboni \cite{PONTECASTANEDA2012293} proposed a homogenization framework
for electro-elastic composite materials at finite strains incorporating nonlinear dielectric behavior. Using this homogenization framework, electromechanical instabilities in fiber-constrained, dielectric-elastomer composites subjected to electromechanical loadings were further studied \cite{SIBONI2014211, siboni2014electromechanical, SIBONI2019}. In addition, electromechanical instabilities in soft heterogeneous dielectric elastomers were investigated in multi-layered dielectric composites at finite strains, eg., Bertoldi and Gei \cite{katia2011instabilities}, Rudykh and deBotton \cite{stephan2011stability}, Rudykh et al. \cite{rudykh2014multiscale}, Spinelli and Lopez-Pamies \cite{SPINELLI201515}.
\\

We remark that the solution for most nonlinear problems in electroelasticity, and certainly that involve stability and bifurcation analysis, necessarily require computational approaches. In our tutorial, in order to highlight the basic concepts, we have restricted ourselves to examples that are amenable to analytical solutions. In order to explore the computational literature, we refer the reader to Vu et al. \cite{vu2007numerical} who presented a finite element approach to simulate the nonlinear coupling behavior of electroactive polymers under electric stimulation. To investigate the influence of the free space on the electric field and on the deformation field, Vu and Steinmann \cite{VU201282} employed a coupled BEM-FEM approach to exploit the advantage of the finite element method in solving nonlinear problems in nonlinear electroelasticity. For dissipative electro-magneto-mechanics, Miehe et al. \cite{Miehe2011Variational}, based on incremental variational principles, presented a general framework for the macroscopic, continuum-based formulation and numerical implementation of dissipative functional materials with electro-magneto-mechanical couplings. For dissipative electro-mechanics, Z$\ddot{\text a}$h and Miehe \cite{ZAH2013487} developed an approach for computational homogenization based on rigorous exploitation of rate-type and incremental variational principles. In finite magneto-electro-elasticity, Miehe et al. \cite{Miehe2015Homogenization, MIEHE2016294} outlined a variational-based framework for homogenization and multiscale stability analysis, which allows tracking of post-critical solution paths such as those related to pull-in instabilities. Polukhov et al. \cite{POLUKHOV2018165} investigated the computational stability analysis of electroactive polymer composites by implementing Bloch-Floquet wave analysis in the context of a finite element discretization, with a particular focus on the investigation of macroscopic loss of strong ellipticity and microscopic bifurcation-type instabilities. \\

The confluence of stability and electromechanical effects plays a significant role in biological cells also---arguably an important subset of soft matter. Among other consequences of this, such an interaction between cells and electric fields involves changes in the shapes of the cells as well as formation of pores (i.e., electroporation). Aside from the classical works by Helfrich \cite{winterhalter1988deformation, kummrow1991deformation}, the following are some recent representative works on this topic which also contain citations to the vast literature on this subject \cite{yamamoto2010stability, portet2012destabilizing, vlahovska2019electrohydrodynamics}.\\

Finally, we remark that despite numerous works in the purely mechanical case, very few have addressed post-buckling of soft dielectrics.

\section*{Acknowledgments}
Support from the University of Houston and the M. D. Anderson Professorship is gratefully acknowledged. S.Y. would like to acknowledge the National Natural Science Foundation of China (Grants No. 12202249).


\bibliographystyle{ieeetr}

\bibliography{yangAMR}

\end{document}